\documentclass{aa}  
\usepackage{graphicx}
\usepackage{txfonts}
\usepackage{topcapt}
\begin{document}

   \title{DEATHSTAR: Nearby AGB stars with the Atacama Compact Array}

   \subtitle{I. CO envelope sizes and asymmetries: A new hope for accurate mass-loss-rate estimates}

   \author{S.~Ramstedt\inst{1} \and W.~H.~T. Vlemmings\inst{2} \and L.~Doan\inst{1} \and T.~Danilovich\inst{3} \and M.~Lindqvist\inst{2} \and M.~Saberi\inst{2} \and  H.~Olofsson\inst{2} \and E.~De Beck\inst{2} \and M.~A.~T.~Groenewegen\inst{4} \and S.~H\"ofner\inst{1} \and J.~H.~Kastner\inst{5} \and F.~Kerschbaum\inst{6} \and T.~Khouri\inst{2} \and M.~Maercker\inst{2} \and R.~Montez\inst{7} \and G.~Quintana-Lacaci\inst{8} \and R.~Sahai\inst{9} \and D.~Tafoya\inst{2} \and A.~Zijlstra\inst{10}}

   \institute{Theoretical Astrophysics, Division for Astronomy and Space Physics, Department of Physics and Astronomy, Uppsala University, Box 516, SE-751 20 Uppsala, Sweden\\
              \email{sofia.ramstedt@physics.uu.se}
         \and
         Department of Space, Earth and Environment, Chalmers University of Technology, Onsala Space Observatory, 439 92 Onsala, Sweden \and
Department of Physics and Astronomy, Institute of Astronomy, KU Leuven, Celestijnenlaan 200D, 3001 Leuven, Belgium
\and  Koninklijke Sterrenwacht van Belgi\"e, Ringlaan 3, B-1180 Brussels, Belgium \and Rochester Institute of Technology, Rochester, NY, USA \and Department of Astrophysics, University of Vienna, T\"urkenschanzstr. 17, 1180 Vienna, Austria \and Smithsonian Astrophysical Observatory, 60 Garden Street, Cambridge, MA 02138, USA \and Instituto de F\'isica Fundamental (IFF-CSIC), Serrano 123, Madrid, CP 28006 , Spain \and Jet Propulsion Laboratory, MS 183-900, California Institute of Technology, Pasadena, CA 91109, USA \and Jodrell Bank Centre for Astrophysics, Alan Turing Building, University of Manchester, Manchester M13 9PL, UK
}   \date{}

\abstract{This is the first publication from the DEATHSTAR project. The overall goal of the project is to reduce the uncertainties of the observational estimates of mass-loss rates from evolved stars on the Asymptotic Giant Branch (AGB).}{The aim in this first publication is to constrain the sizes of the $^{12}$CO emitting region from the circumstellar envelopes around 42 mostly southern AGB stars, of which 21 are M-type and 21 are C-type, using the Atacama Compact Array (ACA) at the Atacama Large Millimeter/submillimeter Array (ALMA). The symmetry of the outflows is also investigated.}{Line emission from $^{12}$CO $J$=2$\rightarrow$1 and 3$\rightarrow$2 from all of the sources were mapped using the ACA. In this initial analysis, the emission distribution was fit to a Gaussian distribution in the $uv$-plane. A detailed radiative transfer analysis will be presented in a future publication. The major and minor axis of the best-fit Gaussian at the line center velocity of the $^{12}$CO $J$=2$\rightarrow$1 emission gives a first indication of the size of the emitting region. Furthermore, the fitting results, such as the Gaussian major and minor axis, center position, and the goodness of fit across both lines, constrain the symmetry of the emission distribution. For a subsample of sources, the measured emission distribution is compared to predictions from previous best-fit radiative transfer modeling results. }{We find that the CO envelope sizes are, in general, larger for C-type than for M-type AGB stars, which is as expected if the CO/H$_{2}$ ratio is larger in C-type stars. Furthermore, the measurements show a relation between the measured (Gaussian) $^{12}$CO $J$=2$\rightarrow$1 size and circumstellar density that, while in broad agreement with photodissociation calculations, reveals large scatter and some systematic differences between the different stellar types. For lower mass-loss-rate irregular and semi-regular variables of both M- and C-type AGB stars, the $^{12}$CO $J$=2$\rightarrow$1 size appears to be independent of the ratio of the mass-loss rate to outflow velocity, which is a measure of circumstellar density. For the higher mass-loss-rate Mira stars, the $^{12}$CO $J$=2$\rightarrow$1 size clearly increases with circumstellar density, with larger sizes for the higher CO-abundance C-type stars. The M-type stars appear to be  consistently smaller than predicted from photodissociation theory. The majority of the sources have CO envelope sizes that are consistent with a spherically symmetric, smooth outflow, at least on larger scales. For about a third of the sources, indications of strong asymmetries are detected. This is consistent with what was found in previous interferometric investigations of northern sources. Smaller scale asymmetries are found in a larger fraction of sources.}{These results for CO envelope radii and shapes can be used to constrain detailed radiative transfer modeling of the same stars so as to determine mass-loss rates that are independent of photodissociation models. For a large fraction of the sources, observations at higher spatial resolution will be necessary to deduce the nature and origin of the complex circumstellar dynamics revealed by our ACA observations.} 
 
\titlerunning{DEATHSTAR: Nearby AGB stars with ALMA ACA}
\authorrunning{Ramstedt et al.}
   \maketitle

\begin{figure*}[h]
\begin{center}
\includegraphics[height=5.15cm]{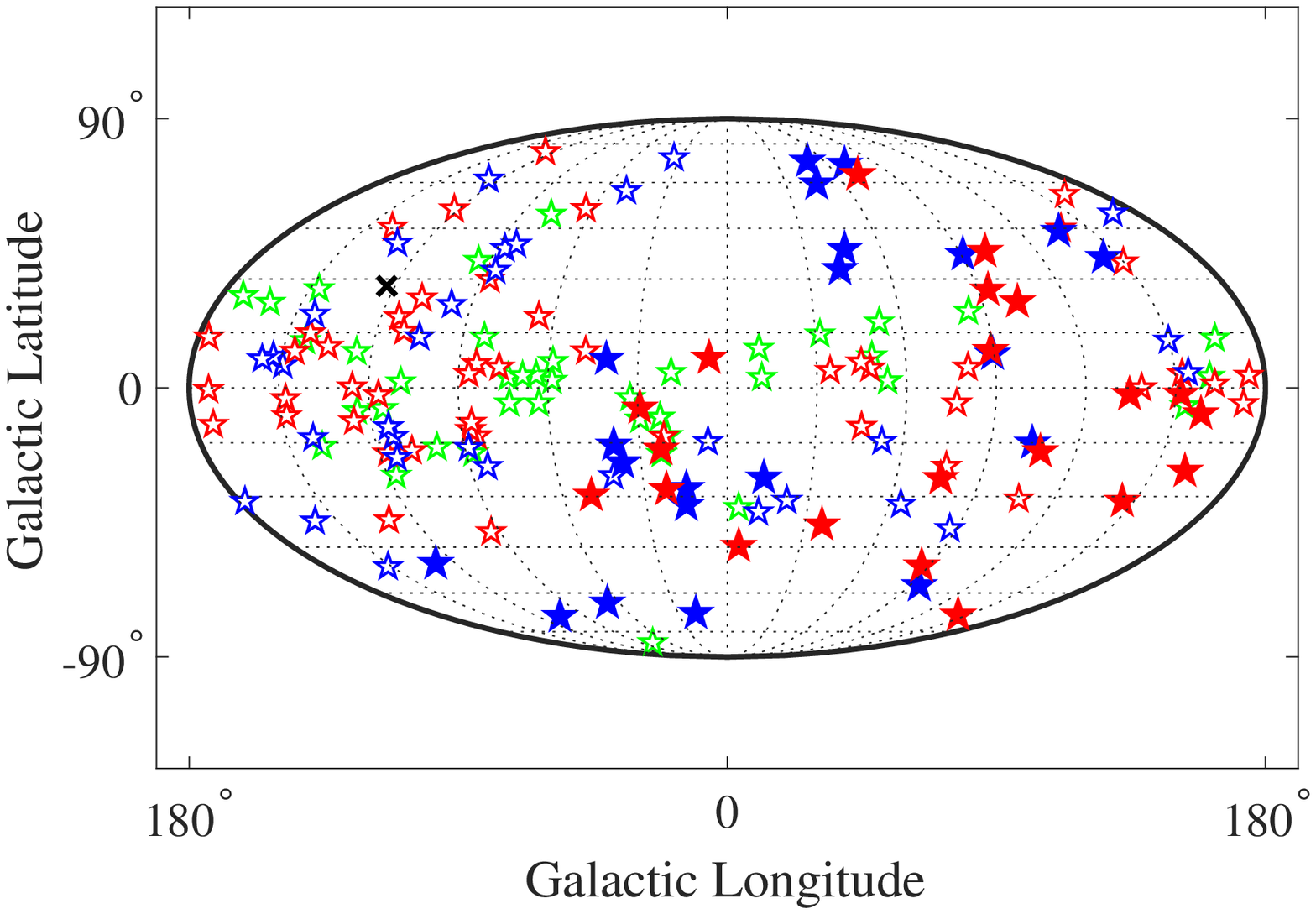}
\includegraphics[height=5.25cm]{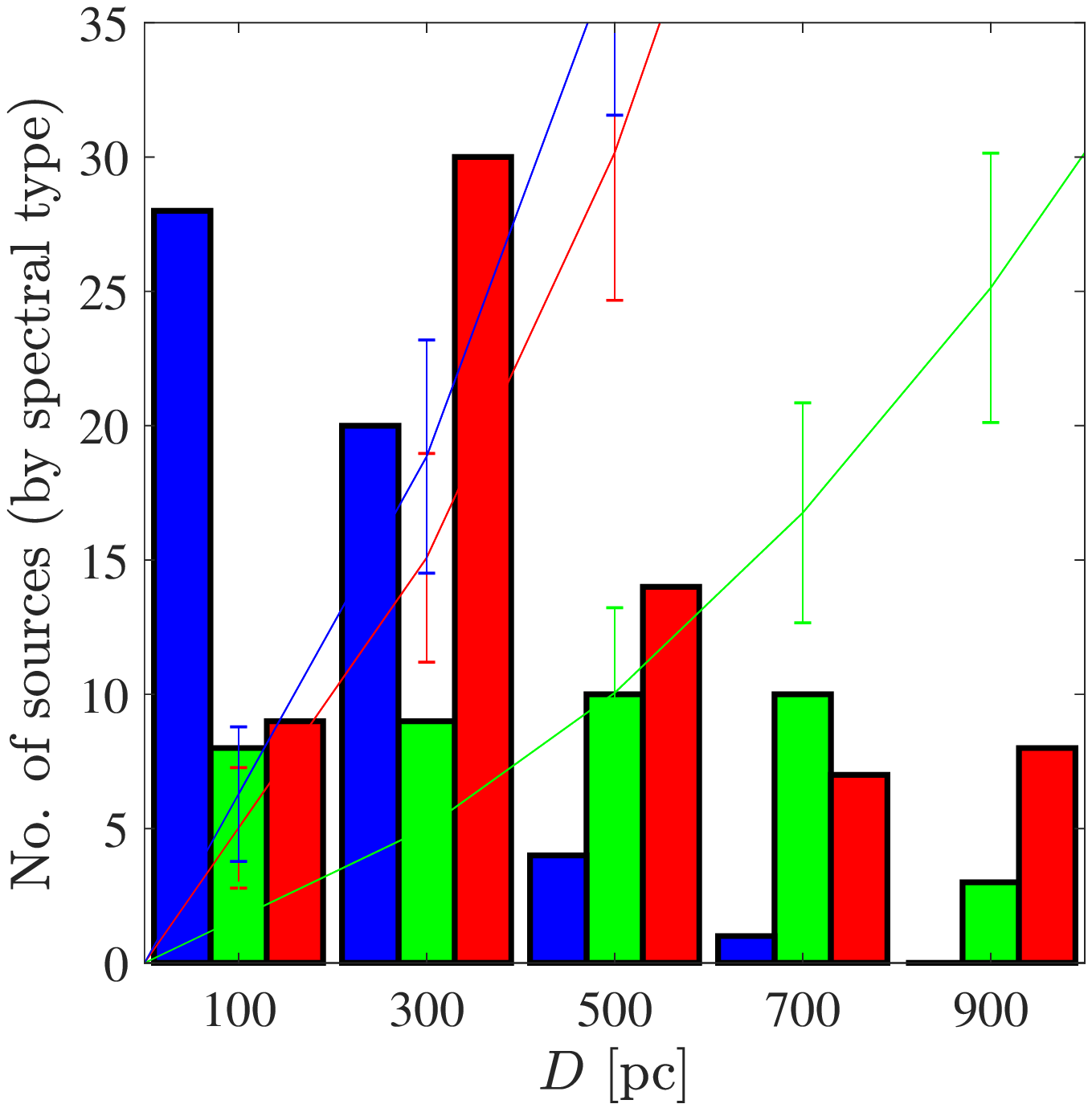}
\includegraphics[height=5.25cm]{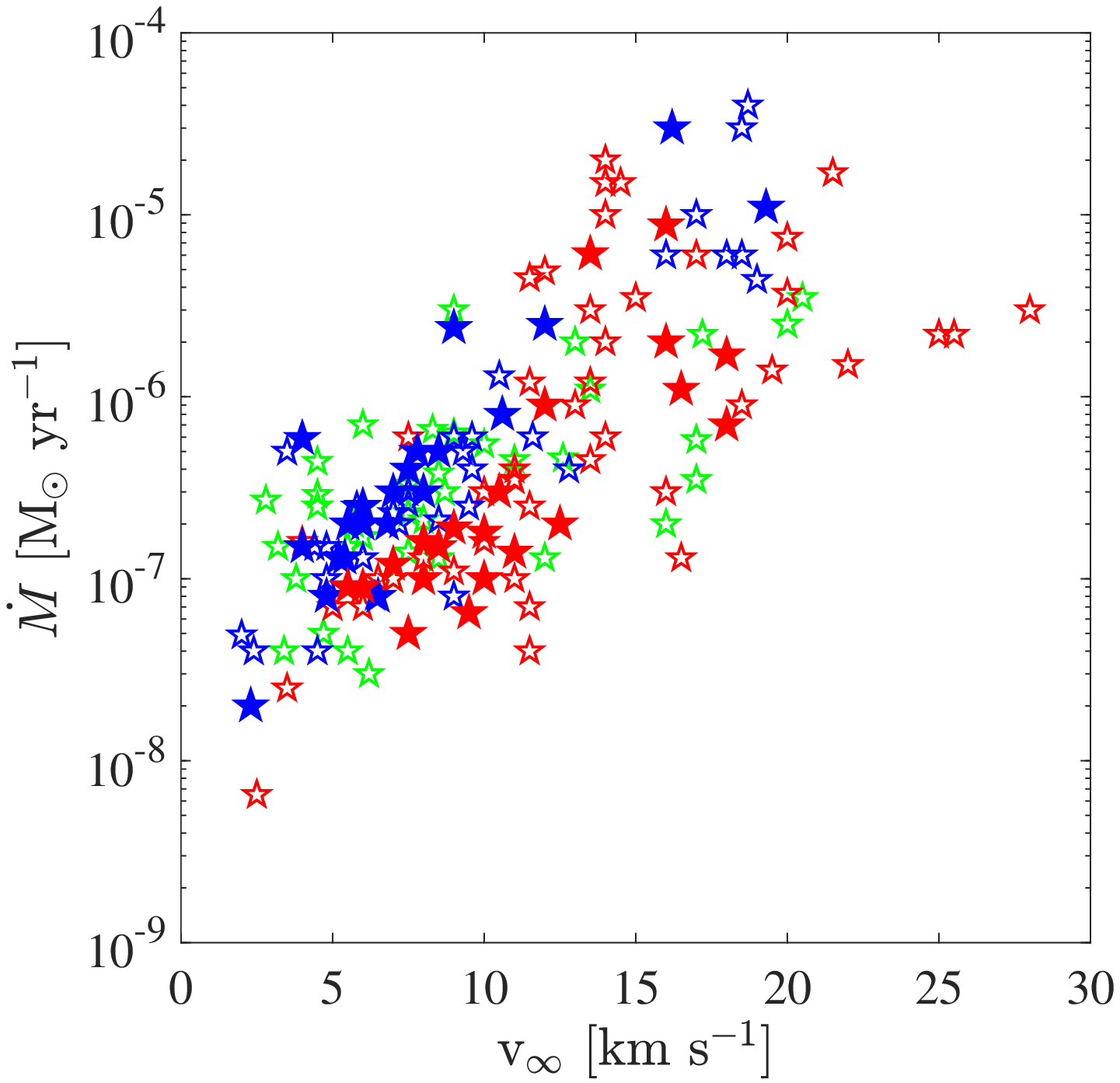}

\smallskip \smallskip
\includegraphics[height=5.25cm]{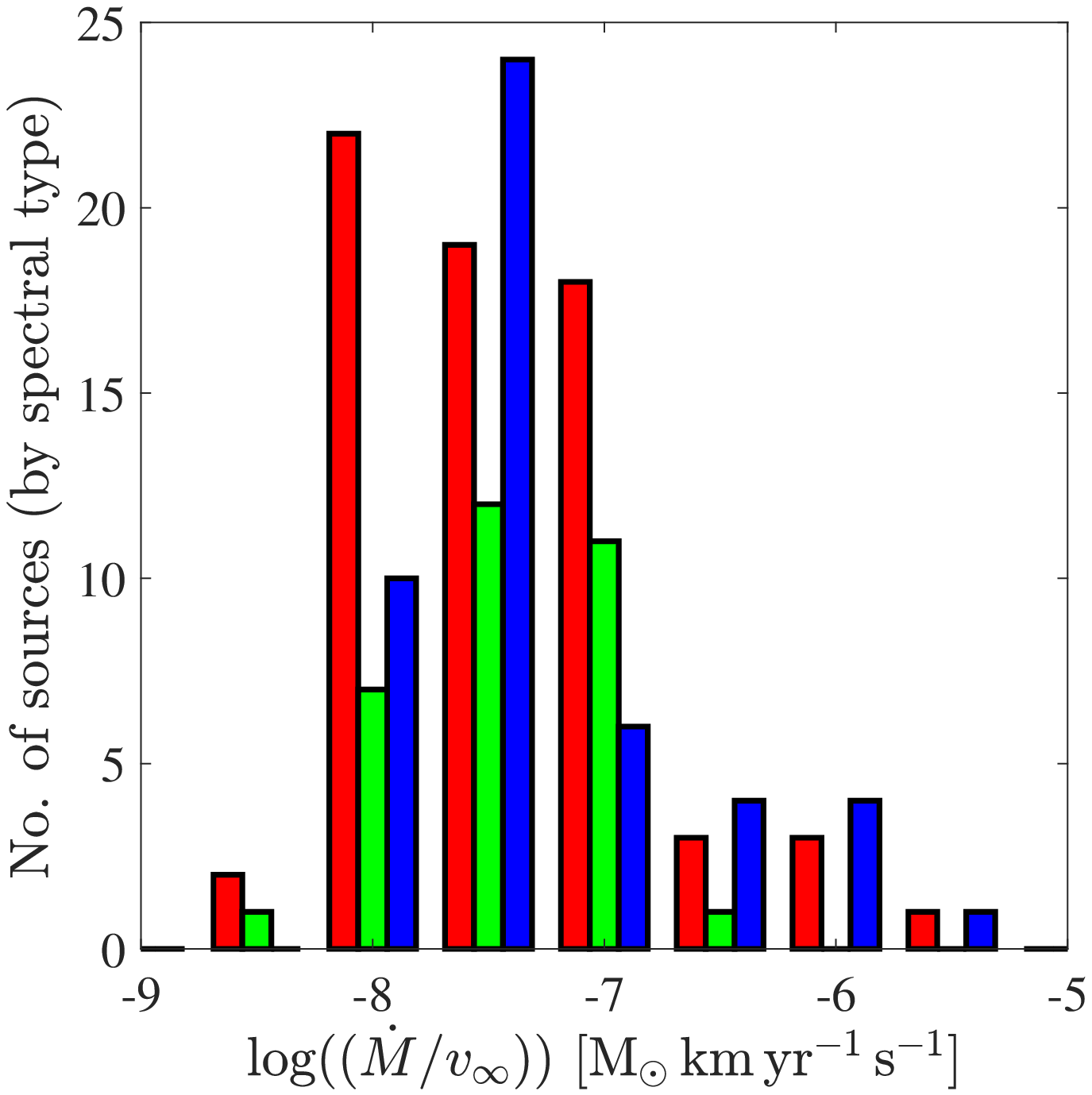} \hspace{0.7cm}
\includegraphics[height=5.25cm]{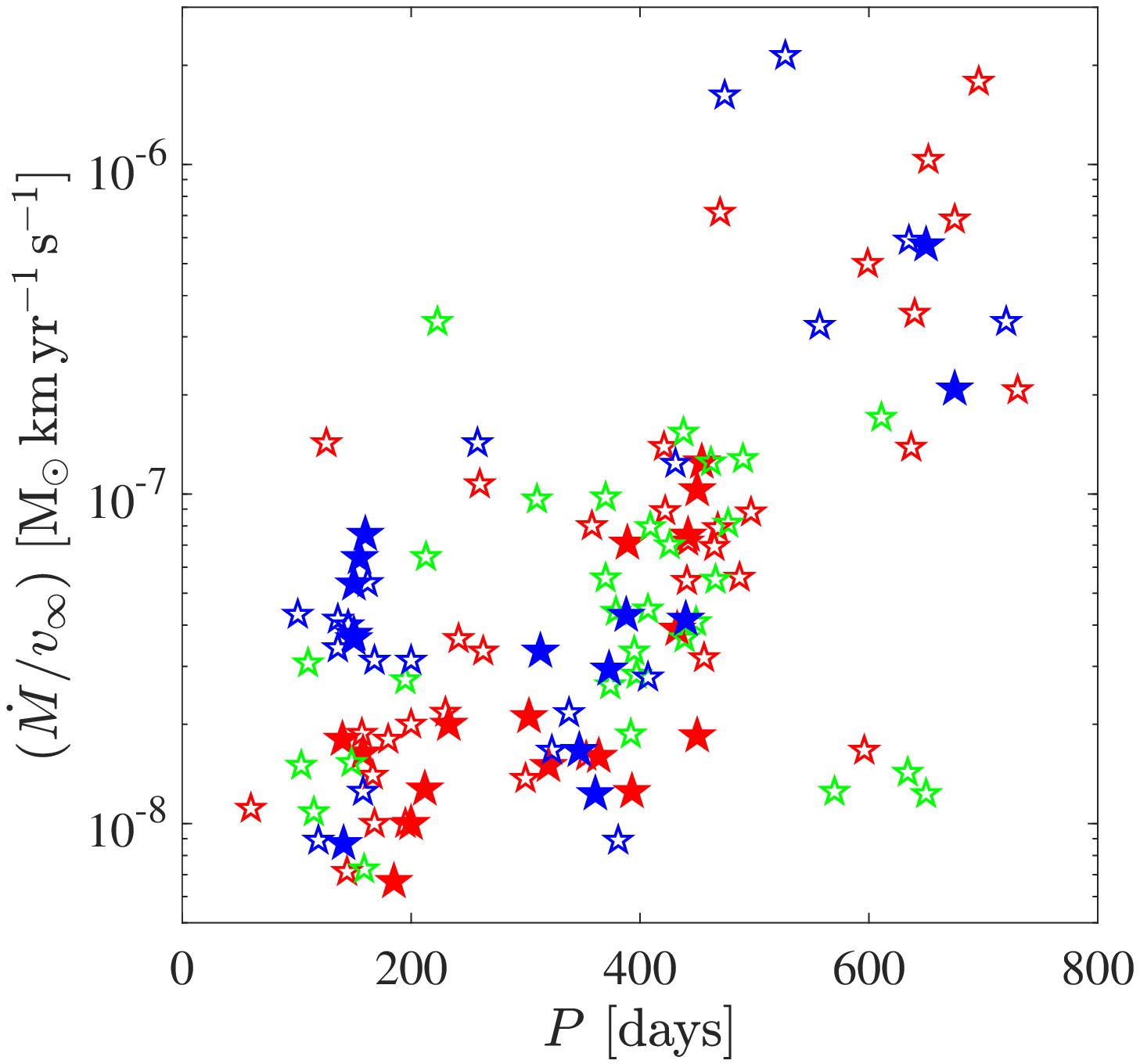} \hspace{0.7cm}
\includegraphics[height=5.25cm]{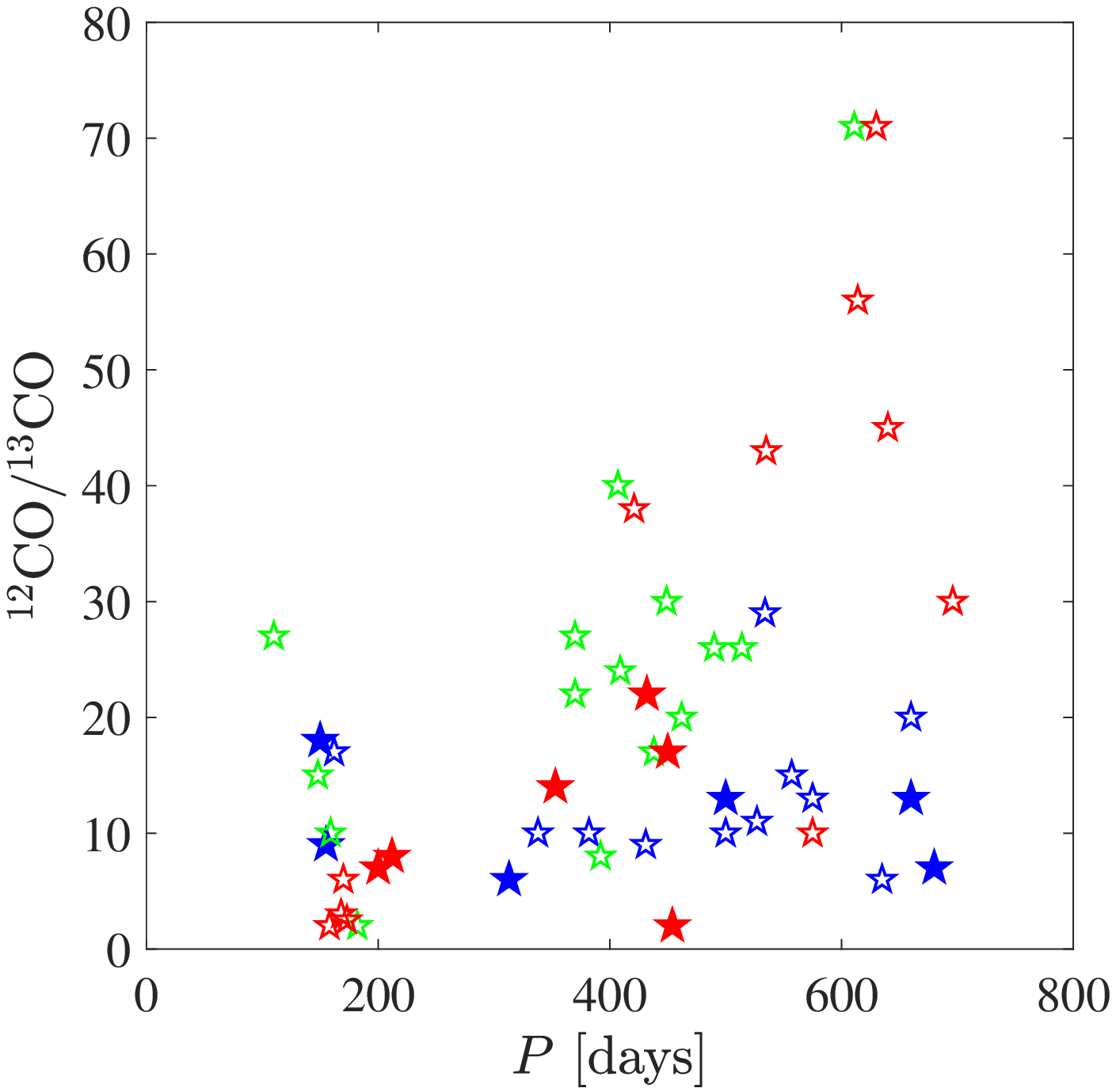}
\caption{Sample statistics of the full $\sim$180 star DEATHSTAR sample. M-type stars are blue, C-type stars are red, and S-type stars are green. Filled symbols mark the stars included in this paper. {\bf Upper left:} Galactic distribution of the full sample. {\bf Upper middle:} Distance, $D$, distribution. These are preliminary distances from previous publications (see text and Table~\ref{sample}). The solid lines show the expected distributions for each spectral type with Poisson errors \citep{jura90,juraklei92,juraetal93}. {\bf Upper right:} Wind properties (mass-loss rate, $\dot{M}$, and terminal expansion velocity, $v_{\infty}$) from previous publications. See text for references. {\bf Lower left:} The $\dot{M}$/$v_{\infty}$ distribution for the full sample. We note that $\dot{M}$/$v_{\infty}$ is a proxy for the wind density. {\bf Lower middle:} $\dot{M}$/$v_{\infty}$ as a function of the pulsational period. {\bf Lower left:} $^{12}$CO/$^{13}$CO ratio for the sample sources \citep{ramsolof14} is plotted against the pulsational period. Both parameters are expected to increase as the stars evolve. See text for a further discussion.}
\label{sample_f}
\end{center}
\end{figure*}

\begin{table}[]
   \caption{Forty-two sources observed with the ACA by spectral type and variability type listed in ascending mass-loss-rate order.}  
     \resizebox{\columnwidth}{!}{\begin{tabular}{l|ccccc} 
      \hline \hline
      Source  & Variability & $\dot{M}$\tablefootmark{a} & $v_{\infty}$\tablefootmark{a} & $D$ & $D_{\rm{Gaia}}$ \\
        & type & [M$_{\odot}$\,yr$^{-1}$] & [km s$^{-1}$] & [pc] & [pc] \\
      \hline \hline 
      \noalign{\vskip 1mm}
         \multicolumn{5}{l}{ {\it M-type semi-regular and irregular stars:}} \\
         \noalign{\vskip 1mm}
         \object{L$_{2}$~Pup}   & SRb & $2\phantom{.}\times10^{-8}$ & \phantom{0}2.3 & \phantom{11}85 & 140 \\
      \object{W~Hya}                    & SRa & $8\phantom{.}\times10^{-8}$ & \phantom{0}6.5 & \phantom{11}65 & 160 \\
      \object{T~Mic}                    & SRb & $8\phantom{.}\times10^{-8}$ & \phantom{0}4.8 & \phantom{1}130 & 190 \\
      \object{Y~Scl}                    & SRb & $1.3\times10^{-7}$\phantom{.} & \phantom{0}5.2 & \phantom{1}330 & 310 \\
      \object{V1943~Sgr}                & Lb   & $1.3\times10^{-7}$\phantom{.} & \phantom{0}5.4 & \phantom{1}150 & 665 \\
      \object{BK~Vir}                   & SRb & $1.5\times10^{-7}$\phantom{.} & \phantom{0}4.0 & \phantom{1}190 & 235 \\
      \object{V~Tel}                    & SRb & $2\phantom{.}\times10^{-7}$ & \phantom{0}6.8 & \phantom{1}290 & 400 \\
      \object{SU~Vel}                   & SRb & $2\phantom{.}\times10^{-7}$ & \phantom{0}5.5 & \phantom{1}250 & 290 \\
      \object{UY~Cet}                   & SRb & $2.5\times10^{-7}$\phantom{.} & \phantom{0}6.0 & \phantom{1}300 & 320 \\
      \object{SV~Aqr}                   & Lb & $3\phantom{.}\times10^{-7}$ & \phantom{0}8.0 & \phantom{1}470 & 390 \\
      \object{SW~Vir}                   & SRb & $4\phantom{.}\times10^{-7}$ & \phantom{0}7.5 & \phantom{1}120 & 300 \\
      \object{CW~Cnc}                   & Lb & $5\phantom{.}\times10^{-7}$ & \phantom{0}8.5 & \phantom{1}280 & 210 \\
      \object{RT~Vir}                   & SRb & $5\phantom{.}\times10^{-7}$ & \phantom{0}7.8 & \phantom{1}170 & 490 \\
      \object{R~Crt}                    & SRb & $8\phantom{.}\times10^{-7}$ & 10.6 & \phantom{1}170 & 240 \\
      \noalign{\vskip 1mm}
         \multicolumn{3}{l}{ {\it M-type Mira stars:}} \\
         \noalign{\vskip 1mm}
         \object{R~Leo}                 & M & $2\phantom{.}\times10^{-7}$ & \phantom{0}6.0 & \phantom{1}130 & 70 \\
         \object{R~Hya}                 & M & $3\phantom{.}\times10^{-7}$ & \phantom{0}7.0 & \phantom{1}150 & 225\\
         \object{R~Hor}                 & M & $5.9\times10^{-7}$\phantom{.} & \phantom{0}4.0 &  \phantom{1}310 & 310 \\
         \object{RR~Aql}                        & M & $2.4\times10^{-6}$\phantom{.} & \phantom{0}9.0 & \phantom{1}530 & 320 \\
         \object{IRC-10529}             & M & $2.5\times10^{-6}$\phantom{.} & 12.0 & \phantom{1}270 & - \\
         \object{WX~Psc}                        & M & $1.1\times10^{-5}$\phantom{.} & 19.3 & \phantom{1}600 & - \\
         \object{IRC+10365}             & M & $3\phantom{.}\times10^{-5}$ & 16.2 & \phantom{1}750 & 360 \\
         \noalign{\vskip 1mm}
         \multicolumn{3}{l}{ {\it C-type semi-regular and irregular stars:}} \\
         \noalign{\vskip 1mm}
         \object{TW~Oph}                        & SRb & $5\times10^{-8}$ & \phantom{0}7.5 & \phantom{1}280\tablefootmark{b} & 590 \\      
         \object{NP~Pup}                        & Lb & $6.5\times10^{-8}$ & \phantom{0}9.5 & \phantom{1}420\tablefootmark{b} & 470 \\
         \object{TW~Hor}                        & SRb & $9\times10^{-8}$ & \phantom{0}5.5 & \phantom{1}400\tablefootmark{b} & 420 \\
         \object{T~Ind}                 & SRb & $9\times10^{-8}$ & \phantom{0}6.0 & \phantom{1}570\tablefootmark{b} & 610 \\
         \object{RT~Cap}                        & SRb & $1\times10^{-7}$ & \phantom{0}8.0 & \phantom{1}450\tablefootmark{c} & 430 \\
         \object{AQ~Sgr}                        & SRb & $2.5\times10^{-7}$ & 10.0 & \phantom{1}420\tablefootmark{d} & 560 \\
         \object{U~Hya}                 & SRb & $1.2\times10^{-7}$ & \phantom{0}7.0 & \phantom{1}160\tablefootmark{b} & 170 \\
         \object{W~Ori}                 & SRb & $1.4\times10^{-7}$ & 11.0 & \phantom{1}220\tablefootmark{b} & 1010 \\
         \object{V~Aql}                 & SRb & $1.5\times10^{-7}$ & \phantom{0}8.5 & \phantom{1}370\tablefootmark{b} & 400 \\
         \object{Y~Pav}                 & SRb & $1.6\times10^{-7}$ & \phantom{0}8.0 & \phantom{1}360\tablefootmark{c} & - \\
         \object{X~Vel}                 & SR & $1.8\times10^{-7}$ & 10.0 & \phantom{1}310\tablefootmark{c} & 640 \\
         \object{Y~Hya}                 & SRb & $1.9\times10^{-7}$ & \phantom{0}9.0 & \phantom{1}350\tablefootmark{b} & 475 \\
         \object{SS~Vir}                        & SRa & $2\times10^{-7}$ & 12.5 & \phantom{1}540\tablefootmark{c} & 670 \\
         \object{W~CMa}                 & Lb & $3\times10^{-7}$ & 10.5 & \phantom{1}450\tablefootmark{d} & 555 \\
         \noalign{\vskip 1mm}
         \multicolumn{3}{l}{ {\it C-type Mira stars:}} \\
         \noalign{\vskip 1mm}
         \object{R~Lep}                 & M & $7\times10^{-7}$ & 18.0 & \phantom{1}250\tablefootmark{b} & 420 \\
         \object{CZ~Hya}                        & M & $9\times10^{-7}$ & 12.0 & \phantom{1}960\tablefootmark{c} & 3300 \\
         \object{R~For}                 & M & $1.1\times10^{-6}$ & 16.5 & \phantom{1}610\tablefootmark{c} & 630 \\
         \object{R~Vol}                 & M & $1.7\times10^{-6}$ & 18.0 & \phantom{1}730\tablefootmark{c} & 840 \\
         \object{RV~Aqr}                        & M & $2\times10^{-6}$ & 16.0 & \phantom{1}670\tablefootmark{c} & 860 \\
         \object{V688~Mon}              & M & $6.1\times10^{-6}$ & 13.5 & 1770\tablefootmark{e} & 500 \\
         \object{V1259~Ori}             & M & $8.8\times10^{-6}$ & 16.0 & 1600\tablefootmark{e} & -\\
         \hline \hline
   \end{tabular}}
   \tablefoot{
The columns give wind properties (mass-loss rate, $\dot{M}$, and final expansion velocity, $v_{\infty}$), and preliminary distance, $D$, from previous publications (see text for details). The final column gives the distances from Gaia data release 2 \citep{bailjone18} for comparison purposes (see also Fig.~\ref{gaia}).\\
   \tablefoottext{a}{From the previous analysis. See text for references.}
   \tablefoottext{b}{Hipparcos parallax.} 
   \tablefoottext{c}{Period-Luminosity relation \citep{groewhit96}.}
   \tablefoottext{d}{Assuming 4000\,L$_{\odot}$.}
   \tablefoottext{e}{Adopted from \citet{menzetal06}.}}
   \label{sample}
\end{table}


\section{Introduction}
Stars with zero-age-main-sequence masses in the range of $\sim$0.8--8\,M$_{\odot}$ evolve into asymptotic giant branch (AGB) stars during the late stages of their evolution. The heavy mass loss during the AGB makes the stars major contributors of newly synthesized elements and dust to their surroundings. Understanding the mass-loss process is crucial for comprehending the evolution of stars in this mass range, but also for extragalactic population studies. Mass-loss rates, $\dot{M}$, on the AGB are found to range from $\sim$10$^{-8}$--10$^{-4}$\,M$_{\odot}$\,yr$^{-1}$ \citep[e.g.,][and references therein]{hofnolof18}. It is challenging to find reliable observational methods to measure mass-loss rates covering this wide range \citep{ramsetal08}, but it is crucial since the measurements will provide key constraints for theoretical models \citep[e.g.,][]{eriketal14,marietal16,bladetal19}. Wind formation is studied using dynamical wind models \citep[e.g.,][]{hofn08, eriketal14,bladetal15,hofnetal16} with the ultimate goal of developing a predictive theory of AGB mass loss. This will permit reliable estimates of dust production and, for example, the intrinsic reddening of distant galaxies \citep[e.g.,][]{conr13}. These models are constrained using observed wind properties, that is, $\dot{M}$ and wind velocities, $v_{\infty}$. However, with the derived $\dot{M}$ uncertainties reaching as high as a factor of three \citep[within the adopted spherically symmetric model,][]{ramsetal08}, the dynamical wind models cannot be sufficiently well constrained (e.g., Fig.~7 in the recent paper by \citealt{bladetal19} shows how the wind parameters vary with model input parameters).  

Observations of $^{12}$CO radio line emission (originating in the circumstellar envelope, CSE, which is created by the wind), in combination with detailed radiative transfer, is considered to be the most reliable method for determining AGB wind properties \citep[e.g.,][and references therein]{hofnolof18}. The poorly constrained size of the $^{12}$CO envelope is a remaining, significant source of uncertainty for the mass-loss rates estimated using this method. The generally adopted strategy is to use an envelope-sized estimate based on the photodissociation model by \citet{mamoetal88}. Major uncertainties are that this model assumed a standard interstellar radiation field \citep{drai78}, and that numerical methods and shielding functions have been updated since then \citep[e.g.,][]{groe17,sabeetal19}. In radiative transfer models that are intended to determine wind parameters from $^{12}$CO line observations, the $^{12}$CO envelope size is estimated using a functional fit to the Mamon et al. photodissociation model results. The envelope size is a function of the wind parameters, including the $^{12}$CO abundance, which increase with density, that is, {$\dot{M}/v_{\rm{e}}$. A more exact way to determine the $^{12}$CO envelope size (e.g., independent of our knowledge of shielding functions) is to constrain it directly using interferometry. Direct observations will further improve the accuracy, since the radiation environment changes from source to source . 

A pioneering interferometric survey of AGB (and post-AGB) CO CSEs was performed by \citet{nerietal98}. Forty-six sources were mapped in $^{12}$CO $J$=1\,$\rightarrow$\,0 (from now on CO(1-0), etc.) emission with three antennas on the Plateau de Bure interferometer combined with the IRAM 30\,m telescope. The sample was selected on evolutionary status, declination, IR color, and CO line strength. They found a good agreement between measured CO(1-0) sizes and photodissociation radii based on the model by \citet{mamoetal88}; however, as expected, they observed significant scatter. They also concluded that about 30\% of the envelopes show significant asymmetry. This investigation was later followed up with considerably improved capabilities at the Plateau de Bure by the COSAS program \citep{castetal10}. The goal was to investigate the morphologies of the envelopes, in particular during the transition to post-AGB, and this sample contained several sources at later evolutionary stages than the AGB. \citet{castetal10} presented detailed maps of the CO(1-0) and (2-1) emission from 16 sources and thoroughly discussed each source. The synthesized beams were typically of the order 3-5\arcsec~and 1-2\arcsec~for CO(1-0) and (2-1), respectively. They found that the measured envelope sizes were, on average, slightly larger than expected from the photodissociation model by \citet{mamoetal88}. They further concluded that the AGB envelopes generally show round shapes and approximately isotropic expansion, while most later sources, that is post-AGB, exhibit axial symmetry and fast bipolar flows.

We have started a new project called DEATHSTAR\footnote{www.astro.uu.se/deathstar} (DEtermining Accurate mass-loss rates for THermally pulsing AGB STARs) in which the overall aim is to better constrain the measurements of AGB mass-loss rates. Observations of a large sample of "typical" AGB envelopes (sources with known strong deviations from spherical symmetry have been omitted), covering the full range of AGB stellar and wind parameters, will be analyzed in detail in future work. The already-available data base of CO lower-$J$ transitions will be modeled, together with available and new interferometric observations using updated radiative transfer models. In this first paper, new interferometric data of CO(2-1) and (3-2) emission obtained with the Atacama Compact Array (ACA) at the Atacama Large Millimeter/submillimeter Array (ALMA) are presented. The sample selection is explained in Sect.~\ref{sample_s}. The observations, data reduction, and data analysis are outlined in Sect.~\ref{obs}. The results with an analysis of the CO line emission distributions (size and morphology) and an overview of the detections of emission from other molecular species are presented in Sect.~\ref{results}. Finally, the results are discussed and summarized in Sects.~\ref{dis} and \ref{sum}.


\section{The sample}
\label{sample_s}
The full sample for which the circumstellar CO line emission will be modeled consists of the $\sim$180 C-, M-, and S-type AGB stars analyzed in \citet{schoolof01}, \citet{gonzetal03}, and \citet{ramsetal06} together with additional sources presented in \citet{danietal15}. In this initial paper, the new interferometric data for the southern M- and C-type stars are presented. Some of the available sample statistics for the full $\sim$180 star DEATHSTAR sample are shown in Fig.~\ref{sample_f}. The distance distribution (Fig.~\ref{sample_f}, middle) is compared with the estimated distribution in the solar neighborhood \citep{jura90,juraklei92,juraetal93}. The estimated distribution is derived from 2MASS and ground-based observations \citep{jura90b} and assumes a smooth distribution of 40 C-type stars kpc$^{-2}$, a scale height of 200\,pc, and that there are a third as many S-type as C-type stars. For the full $\sim$180 star DEATHSTAR sample, the C-type stars from \citet{schoolof01} are all brighter than $K$\,=\,2~mag. The M-type sample consists of the non-Mira stars from the General Catalog of Variable Stars \citep[GCVS;]{samuetal17} with quality flag 3 in the IRAS 12, 25, and 60\,$\mu$m bands and 60\,$\mu$m flux $\gtrsim$\,3\,Jy with the addition of the Mira stars in \citet{gonzetal03}. The S-type sample also consists of stars that have good quality flux measurements in the IRAS 12, 25, and 60\,$\mu$m bands, that are found in the General Catalog of Galactic S stars, and that are detected in Tc and are hence intrinsic. The completeness of the S-type sample is discussed in \citet{ramsetal09} and it is thought to be complete out to 600\,pc. Furthermore, stars of all three spectral types are only included in the sample if they are detected in CO radio line emission, which could be reproduced under the assumption of spherical symmetry. Sources that show strongly asymmetric line profiles when observed with single-dish telescopes, or with known CO-detached shells, are hence not included (e.g.,~R~Scl, U~Ant, EP~Aqr, and $\pi^{1}$~Gru). In this paper, we also exclude stars that have previously been observed with ALMA; however, they will be included in the future analysis. The lower panel of Fig.~\ref{sample_f} shows that the stellar and wind parameters of the full $\sim$180 star sample cover the ranges expected for AGB stars. As expected, fewer stars are found at the high end. The sample is biased to mass-losing stars since only stars that are previously detected in CO radio line emission are included. It is also likely that the full range of AGB masses is not covered simply because higher mass stars are rare. It is our assessment that the full $\sim$180 star DEATHSTAR sample is representative of Galactic mass-losing AGB stars and covers the relevant ranges of wind and stellar parameters to provide the necessary constraints for theoretical models. 

The 42 stars (21 M-type, 21 C-type), which were observed with ALMA ACA in Cycle 4 and for which the data are presented in this paper, are listed in Table~\ref{sample} with the variability type (as listed in the GCVS), mass-loss rate, final wind velocity, distance according to the previous analysis in \citet{schoolof01}, \citet{gonzetal03}, and \citet{danietal15}, and Gaia data release 2 (DR2) \citep{gaia1,gaia2} distance from \citet{bailjone18}. The mass-loss rates and wind velocities were estimated by reproducing several low-$J$ CO lines for each source using the non-LTE, nonlocal, spherically symmetric radiative transfer code that was first presented in \citet{schoolof01}. The CO/H$_{2}$ abundance ratio, which is necessary to derive the total gas mass-loss rate, is assumed to be $2\times10^{-4}$ for the M-type stars and $1\times10^{-3}$ for the C-type stars. The results from the photodissociation model by \citet{mamoetal88} was used in the radiative transfer modeling. In order to give consistent mass-loss rates and distances, the distances adopted here (and listed in Table~\ref{sample}) are the same as those used in the papers listed above. The distances will be updated as part of the future radiative transfer analysis planned for the full sample. For the semi-regular M-type stars, a stellar bolometric luminosity of 4000\,L$_{\odot}$ was adopted. For M-type Mira variables and some C-type stars, the period-luminosity relations from \citet{whitetal94} and \citet{groewhit96} were used to estimate the luminosity, respectively. From the luminosity, the distances were determined by either using two blackbodies or by using the dust radiative transfer code DUSTY \citep{ivezetal99} to model the spectral energy distribution (SED). For the remaining C-type stars, the distance was estimated directly from the original Hipparcos\nocite{hipp} parallax (The HIPPARCOS and TYCHO catalogs 1997) or adopted from \citet{menzetal06}. The method used for each C-type star is noted in Table~\ref{sample}. Figure~\ref{gaia} shows the comparison between the adopted distances from a previous analysis and the new Gaia DR2 distances for the full sample \citep{bailjone18}. The spread is very large and the uncertainties affecting the Gaia DR2 distances for these types of stars are further discussed in Appendix~\ref{distances}. The Gaia DR2 distances are typically larger than the adopted distances. The mass-loss rates given in Table~\ref{sample} scale with distance as $D^{n}$ where 1.4\,$\lesssim n \lesssim$\,1.9 \citep{ramsetal08} and would, in general, be larger than those in Table~\ref{sample} if Gaia DR2 distances are used in the analysis.

\begin{figure}
\begin{center}
\includegraphics[width=\columnwidth]{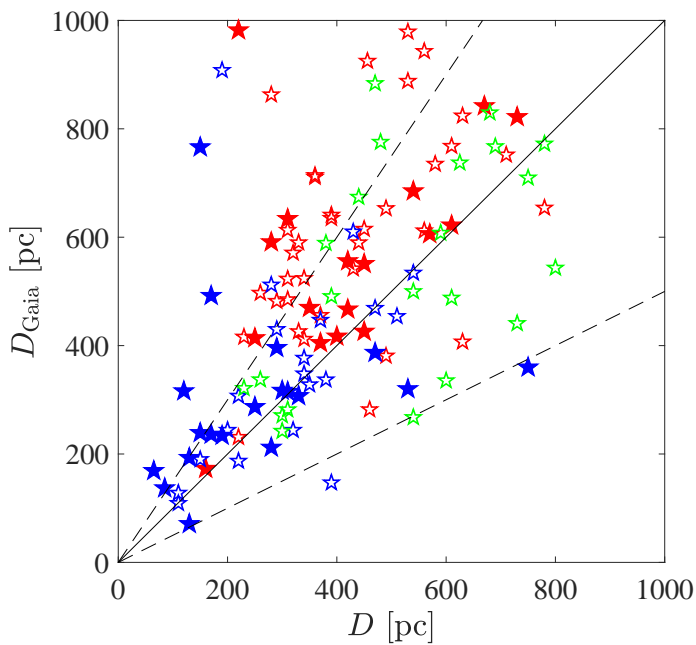}
\caption{Comparison between distances from Gaia DR2 parallaxes and the adopted distances from the previous analysis (see text for details) for the full $\sim$180 star DEATHSTAR sample. Red symbols are carbon stars, blue symbols are M-type stars, and green symbols are S-type stars. Filled symbols are the sources included in this paper. The solid, black line marks a one-to-one correlation, while the dashed lines show the range of $\pm$50\%. }
\label{gaia}
\end{center}
\end{figure}


\section{Observations, data reduction, and analysis}
\label{obs}
\subsection{Observations with the ACA}
The 42 sample sources listed in Table~\ref{sample} were observed with the ACA in stand-alone mode in Cycle~4 in Bands 6 and 7 (project codes: 2016.2.00025.S and 2017.1.00595.S). The correlator was set up with four spectral windows in each band. In Band 6 the windows were centered on 216.4, 218.3, 230.7, and 232.1\,GHz. In Band 7 they were centered on 330.75, 332.25, 343.52, and 345.6\,GHz. Line emission from $^{12}$CO $J=2\rightarrow1$, $3\rightarrow2$, and $^{13}$CO $J=3\rightarrow2$ were covered in this setup, as well as emission from a large number of other chemically interesting molecules, including SiO, SiS, and CS, for example. The spectral resolution of the imaged data was set to 0.75\,km\,s$^{-1}$ in the $^{12}$CO and $^{13}$CO windows and to 1.0 and 1.5\,km\,s$^{-1}$ in the other spectral windows in Band 6 and 7, respectively.

All data were calibrated using the standard pipeline scripts and imaged using the Common Astronomy Software Applications package \citep[CASA;][]{mcmuetal07}. Self-calibration was performed using a small number of channels, typically two, across the peak $^{12}$CO line emission and applied to all sources in both bands. This improved the signal-to-noise ratio by 10-15\% on average. All data were cleaned initially using 10000 iterations. For sources brighter than $\sim$20\,Jy, we found that further cleaning iterations was necessary to recover the signal and finally 20000 iterations were applied to all bright ($>$20\,Jy) sources. The processed data have improved the quality compared to the data products provided in the ALMA archive. Each spectral window in both bands were imaged separately and can be accessed through CDS (see also www.astro.uu.se/deathstar).

The full width at half-maximum (FWHM) beam widths and root mean square (rms) noise levels at a velocity resolution of 0.75 \,km\,s$^{-1}$, which were measured in the emission-free channels in the $^{12}$CO line window of the final images in both bands, are given in Table~\ref{fwhm_rms} for all sources. The science goal beam widths and rms values at 1.5\,km\,s$^{-1}$ were 5.5\arcsec~and 40\,mJy\,beam$^{-1 }$ as well as 3-4\arcsec~and 50\,mJy\,beam$^{-1}$ in Bands 6 and 7, respectively. For a significant fraction of the data sets, the beams are larger than what was aimed for as well as elongated; this is an effect of the project being partly observed as a filler program with some of the sources at low elevation. The maximum recoverable scale will cover a range depending on the exact antenna configuration and frequency, but on average it was 25\arcsec$\pm\,4\arcsec$ and 18\arcsec$\pm\,2\arcsec$ for Band 6 and 7, respectively. The data quality is sufficient for the DEATHSTAR project goals; however, as is discussed below, there are compelling reasons to follow up on a majority of the sources with higher-spatial-resolution observations.


\subsection{Fitting the emission distribution in the $uv$-plane}
\label{uvfit}
The first step is to fit the visibility data in the CO(2-1) and (3-2) measurement sets with Gaussian distributions for all sources using the CASA task UVMULTIFIT \citep{mavietal14}. The least-square fit gives the center position coordinates, the major axis and the axis ratio of the best-fit Gaussian, and the position angle of the major axis for each channel over which the fit has been performed. This provides initial estimates for the sizes of the emitting regions and indications of deviations from symmetry (as described below). The next step will be to perform detailed radiative transfer modeling that is constrained by the available multitransition single-dish, including the CO(1-0) line, and interferometry data for each source to determine the size of the CO envelope and to fit the emission distributions. This will be presented in a future publication.


\subsection{Emission distributions from 1D radiative transfer models}
\label{1d}
Emission distributions for the CO(2-1) and (3-2) lines were calculated from our previous best-fit radiative transfer model results for a subsample of sources. This is a first step toward full radiative transfer modeling in order to reproduce the data available for each source (planned for a future publication). The subsample consists of six stars: three M-type and three C-type stars of which there is one low-, one intermediate-, and one high-mass-loss-rate source for each type. The models were directly adapted from the previous best-fit models \citep[][see also Table~\ref{sample}]{danietal15} and no attempt has been made to improve the fit to the ALMA ACA data. Instead, the purpose is to start evaluating the validity of the photodissociation model used to estimate the size of the circumstellar CO envelopes in the models. For these sources, the original radiative transfer used the photodissociation radius from Mamon et al. (1988). The output from the best-fit radiative transfer models of the six sources was used to create image cubes with an imaging ray-tracing program that is part of the radiative transfer package developed by Sch\"{o}ier (2000)\nocite{schoPhD}. The image cubes were created with 32$\times$32 pixels with sizes of 0\farcs8 and 0\farcs7 for each channel across the CO(2-1) and (3-2) line, respectively. Once image cubes were created, they were run through the ALMA simulator in CASA to simulate observations with the same configuration and under the same conditions as during the ACA observations of each respective source.


\section{Results}
\label{results}
\subsection{Line profiles}
The CO $J$\,=\,2$\rightarrow$1 and 3$\rightarrow$2 line profiles were generated from the image cubes across the lines and are presented in Figs~\ref{linesM_SR}--\ref{linesC_M}. A circular aperture of 18\arcsec~was used for both transitions because this covered the emitting region without reaching lower-fidelity regions that are close to the image edges. Table~\ref{fwhm_rms} gives the peak flux, the center velocity, and the total velocity width of each line. The peak value was measured at the maximum point across the line. The center velocity was measured at the center between the two points where the flux reaches 5\% of the peak value at the extreme velocities. The total width is the velocity width between the two 5\%-peak-flux-value points. The line profiles show an interesting structure that is likely indicative of circumstellar dynamics and which was revealed due to the very high signal-to-noise ratio attained. For example, some lines are distinctly asymmetric (see e.g., R~Hya and SS~Vir). Furthermore, some lines seem to show extended wings (see e.g., L$^2$~Pup and TW~Hor). Indications of circumstellar inhomogeneities and nonisotropic kinematics are further discussed below (Sect.~\ref{structure}).


\subsection{Emission distribution}
\subsubsection{Gaussian distribution fits}
\label{uvfit_res}
The results of fitting the data using Gaussian visibility distributions are given in Table~\ref{visfit}. The second column gives two times the predicted photodissociation radius, $R_{\rm{p}}$, over the distance, $D$ (from Table~\ref{sample}), as a measure of the expected size of the CO line emitting region. The photodissociation radius, which was measured as the radius where the CO abundance has dropped to half of its initial value, was calculated using Eqns. 10 and 11 from \citet{schoolof01}, which were derived \citep{stanetal95} by fitting the results of the CO photodissociation model by \citet{mamoetal88}. The parameter values given in Table~\ref{sample} were used for the calculation. Table~\ref{visfit} also lists the beam-deconvolved FWHM major axis and the axis ratio of the best-fit Gaussian at the center velocity (from Table~\ref{fwhm_rms}). The error is the average error of the fits in the two channels adjacent to the center velocity. When the error is large when the center channels are strongly affected by resolved-out flux, for example (see discussion in Sect.~\ref{structure} and Figs~\ref{uvM_SR}--\ref{uvC_M}), the results are given without errors and marked by a colon (:).

There are good reasons \citep[from previous radiative transfer models, e.g.,][]{ramsolof14} to expect that the CO(2-1) line will be excited almost throughout the entire CSE. However, without detailed models for individual sources, we cannot estimate how well the fitted Gaussian semi-major axis will correspond to the photodissociation radius, or by what factor it may deviate (see Sect.~\ref{compmod}). In Fig.~\ref{gaussfit}, we illustrate the reliability of using the measured Gaussian size as a proxy for the size of the emitting region and for the photodissociation radius, $R_p$, based on the recent models by \citet{sabeetal19}. These models include a more complete treatment of CO self-shielding and yield radii that are mostly similar to \citet{mamoetal88}, but they can be up to $\sim40\%$ smaller for some combinations of mass-loss rates and initial CO abundances. We used the LIME 1.9.5 radiative transfer code \citep{brinch10} to produce radiative transfer models of the photodissociation models from \citet{sabeetal19}, using the CSE parameters that were adopted in their chemical models. Subsequently, we fit a Gaussian profile to the CO(2-1) emission at the systemic velocity using a 1~km~s$^{-1}$ channel width using the same method as in Sect.~\ref{uvfit}. Figure~\ref{gaussfit} shows the CO(2-1) line intensity distributions for three M-star models with different mass-loss rates. Each model intensity distribution has been scaled to the fitted Gaussian. For intermediate- to high-mass-loss-rate sources in particular, the intensity distribution is significantly non-Gaussian, and the determination of a Gaussian FWHM is strongly dependent on the exact shape of the intensity distribution in the inner region. Since low-angular-resolution observations lack the uv-coverage to extract this information and as irregular structures are especially common in these areas, the Gaussian radii determined for the CSEs with the highest density, $\dot{M}/v_{\infty}\gtrsim 10^{-7}$~M$_\odot$~km~s$^{-1}$~yr$^{-1}$, are less reliable. The figure also shows the calculated photodissociation radius from \citet{sabeetal19} for each model (vertical lines) scaled with the Gaussian FWHM radius. For the densest CSEs, the CO(2-1) line does not extend to the photodissociation radius and the Gaussian radii are therefore not a measure for $R_p$.

Based on the formula and the distance estimates from \citet{stanetal95} in Table~\ref{sample}, the CO(2-1) major axis is a factor of two smaller than $2\times R_{\rm{p}}/D$ for the semi- and irregular variables, and a factor of three smaller for the Mira variables, regardless of chemical type. This is consistent with the expectations from the modeling (both chemical and radiative transfer). Furthermore, for a small majority of the sources (26/42), the CO(2-1) axis ratio deviates $\leq$10\% from one, meaning that slightly more than half of the sources are close to being circular. For the more aspherical sources, the photodissociation calculations that assume a standard spherically expanding CSE introduce further uncertainties.

The major axes of the Gaussian fits to the CO(2-1) emission at the center velocity channels (third column of Table~\ref{visfit} multiplied with the distance from Table~\ref{sample}) are plotted against $\dot{M}/v_{\infty}$ in Fig.~\ref{bredd}. The uncertainty of the major axis is very small in arcseconds (see Figs~\ref{uvM_SR}--\ref{uvC_M}). Instead, the error bars in the y-direction are dominated by the distance uncertainty, which cannot be evaluated easily. The figure also shows the photodissociation diameter for the M-type and C-type stars, respectively. The photodissociation radii were calculated across the range of $\dot{M}/v_{\infty}$ for $v_{\infty}=7.5$~km~s$^{-1}$. We used the LIME 1.9.5 radiative transfer code again
\citep{brinch10} to produce radiative transfer models of the \citet{sabeetal19} photodissociation model grid, and we used the same Gaussian fitting procedure as mentioned above to measure the predicted CO(2-1) size. A spline fit to the expected CO(2-1) size is shown by the solid lines in Fig.~\ref{bredd}, which thus indicates the dependence of the measured size of the CO(2-1) emitting region on the photodissociation radius as well as on the circumstellar density. The different results for higher- and lower-mass-loss-rate sources are apparent in the figure: The measured size of the CO(2-1) emitting region for lower-mass-loss-rate M-type stars appears to show a rather weak dependence on the circumstellar density, which is also seen in the dependence of photodissociation diameter and the modeled Gaussian diameter on the density. For higher-mass-loss rate M-type Mira sources, the dependence on the density is much steeper, with a slope similar to what is expected from the photodissociation models. In almost all cases, however, the measured diameter is smaller than the expected CO(2-1) diameter based on the chemical and radiative transfer models. The lower-mass-loss-rate C-type stars show no significant dependence on circumstellar density, or at least a large scatter around the expected relation. Essentially all of the observed low-mass-loss-rate C-type stars have CO(2-1)-emitting major axes of 1000--2000\,AU. The higher-mass-loss-rate C-type Mira stars, on the other hand, show a dependence on density that appears to be somewhat steeper than for the M-type Miras. For all C-type Miras, the measured diameters are larger than expected.

Although the measured size of the CO(2-1) region is smaller than the calculated photodissociation diameter for essentially all sources, it is apparent from Fig.~\ref{bredd} that the correlation between the two is not straightforward. There are a variety of possible explanations for the deviations between the observations and the photodissociation models. The most obvious one is the uncertainty of the distance, which introduces significant scatter in the size determinations. Furthermore, as shown in \citet{sabeetal19}, changes in initial CO abundance, CSE temperature profiles, and the interstellar radiation field can have significant effects. Finally, different density profiles, dust properties, or dust-to-gas ratios from what is assumed in the photodissociation models can all change the estimated radius. 

The full results of the fitting of Gaussian emission distributions to the ALMA ACA data across all of the CO line channels are displayed in Figs \ref{uvM_SR}--\ref{uvC_M}. The major and minor axis of each channel, as well as the RA and Dec offsets of the center of the best-fit Gaussian relative to the SIMBAD J2000 coordinates of each source, are plotted against lsr-velocity for each line and source as denoted in each plot. The interpretation of these plots is further discussed in Sect.~\ref{structure}.

\begin{figure}[h]
\includegraphics[width=\columnwidth]{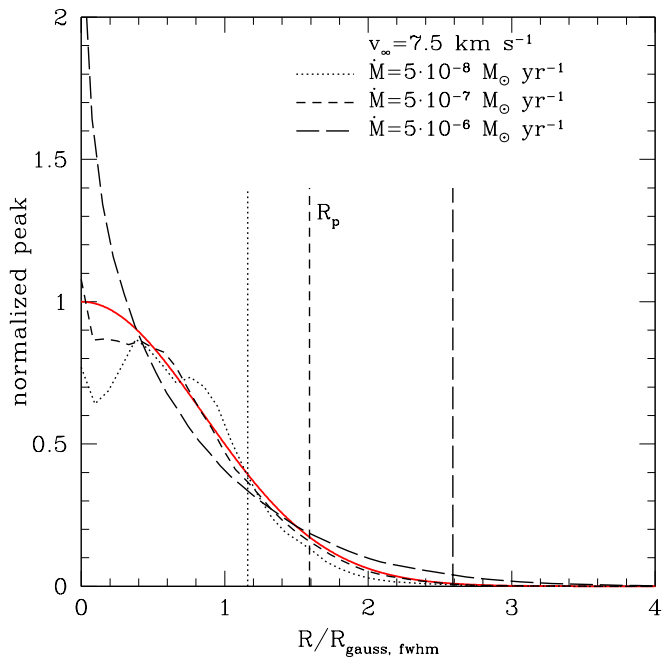}
\caption{CO(2-1) intensity distribution at the systemic velocity as a function of radius for three M-star models from \citet{sabeetal19}. The models were normalized to the peak of the fitted Gaussian and scaled to the FWHM radius of that Gaussian for each model separately. The normalized Gaussian is indicated by the solid red line in the figure. For each model, the vertical lines indicate the photodissociation radius $R_p$, which was also scaled to the Gaussian FWHM radius for each model. See text for a further explanation.}
\label{gaussfit}
\end{figure}

\begin{figure}[h]
\includegraphics[width=\columnwidth]{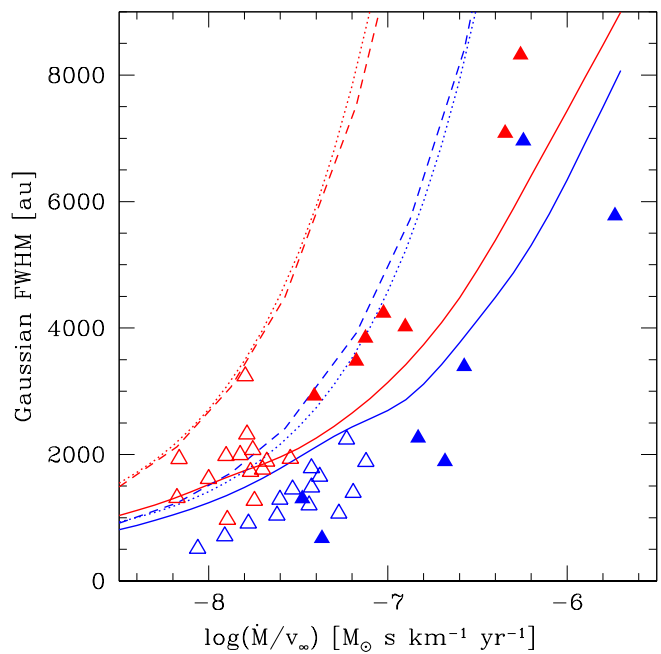}
\caption{Full-width half-maximum (FWHM) major axis of the best-fit Gaussian at the center velocity channel of the CO(2-1) emission (marked by triangles and in astronomical units by multiplying by the distance from Table~\ref{sample}) as a function of circumstellar density (as measured by $\dot{M}$/$v_{\infty}$ from Table~\ref{sample}). M-type stars are blue. C-type stars are red. Mira-type variables are marked with solid symbols and other variables appear with open symbols. The dashed lines show the photodissociation diameter for M-type (blue) and C-type (red) stars from the model grid by \citet{sabeetal19} (calculated with $v_{\infty}$\,=\,7.5\,km\,s$^{-1}$). The dotted lines are the parametrized fits to the results from \citet{mamoetal88}. The solid lines show a spline-fit to the expected Gaussian FWHM determined from radiative transfer modeling of the models from \citet{sabeetal19}. See text for a further explanation. }
\label{bredd}
\end{figure}

\begin{table}[h]
   \caption{Results from the Gaussian visibility distribution fitting. }   
     \resizebox{\columnwidth}{!}{\begin{tabular}{l|c|cc|cc|c} 
      \hline \hline
            & \underline{2$R_{\rm{p}}$} & \multicolumn{2}{c}{CO(2-1)} & \multicolumn{2}{c}{CO(3-2)} & Asym. \\
 \cline{3-4} 
 \cline{5-6} 
      Source  & $D$ & Major & Ratio & Major & Ratio & Feat. \\
         & [\arcsec] & [\arcsec] &  & [\arcsec] & &\\
      \hline \hline 
      \noalign{\vskip 1mm}
         \multicolumn{7}{l}{ {\it M-type semi-regular and irregular stars:}} \\
         \noalign{\vskip 1mm}
         \object{L$_{2}$~Pup}   & 10.6 & \phantom{1}6.04$\pm$0.08 & 0.85$\pm$0.02 & 5.9: & 1.0: & 2, 5, 8  \\
      \object{W~Hya}                    & 22.6 & 10.92$\pm$0.12 & 0.75$\pm$0.06 & 5.59$\pm$0.06 & 0.89$\pm$0.01 &  \\
      \object{T~Mic}                    & 12.0 & \phantom{1}7.04$\pm$0.12 & 0.99$\pm$0.02 & 3.48$\pm$0.08 & 0.97$\pm$0.05 & 6, 7  \\
      \object{Y~Scl}                    & \phantom{1}6.0 & \phantom{1}3.88$\pm$0.20 & 0.89$\pm$0.05 & 3.02$\pm$0.16 & 0.75$\pm$0.05 & 8 \\
      \object{V1943~Sgr}                & 13.0 & \phantom{1}6.95$\pm$0.12 & 0.94$\pm$0.03 & 7.9: & 0.9: & 2 \\
      \object{BK~Vir}                   & 12.2 & \phantom{1}5.84$\pm$0.15 & 0.89$\pm$0.04 & 3.46$\pm$0.04 & 0.90$\pm$0.02 & 5, 6 \\
      \object{V~Tel}                    & \phantom{1}8.0 & \phantom{1}5.00$\pm$0.08 & 0.95$\pm$0.02 & 3.22$\pm$0.07 & 0.94$\pm$0.03 & 1 \\
      \object{SU~Vel}                   & \phantom{1}9.8 & \phantom{1}4.85$\pm$0.07 & 0.76$\pm$0.01 & 3.37$\pm$0.05 & 0.75$\pm$0.02 & \\
      \object{UY~Cet}                   & \phantom{1}9.0 & \phantom{1}5.48$\pm$0.13 & 0.82$\pm$0.03 & 3.04$\pm$0.08 & 0.86$\pm$0.03 & 4 \\
      \object{SV~Aqr}                   & \phantom{1}5.8 & \phantom{1}3.83$\pm$0.17 & 0.94$\pm$0.06 & 2.36$\pm$0.18 & 0.98$\pm$0.14 & 6 \\
      \object{SW~Vir}                   & 27.0 & \phantom{1}8.92$\pm$0.13 & 1.00$\pm$0.02 & 6.7: & 1.0: & 2, 6 \\
      \object{CW~Cnc}                   & 12.6 & \phantom{1}8.00$\pm$0.15 & 0.88$\pm$0.02 & 3.92$\pm$0.13 & 0.92$\pm$0.04 & 7 \\
      \object{RT~Vir}                   & 21.2 & \phantom{1}8.18$\pm$0.14 & 0.89$\pm$0.02 & 5.25$\pm$0.06 & 0.82$\pm$0.01 & 7 \\
      \object{R~Crt}                    & 25.0 & \phantom{1}7.72$\pm$0.08 & 1.0: & 5.5: & 0.9: & 2 \\
      \noalign{\vskip 1mm}
         \multicolumn{7}{l}{ {\it M-type Mira stars:}} \\
         \noalign{\vskip 1mm}
         \object{R~Leo}                 & 18.2 & 10.00$\pm$0.11 & 0.88$\pm$0.02 & 4.70$\pm$0.08 & 1.0: & 2, 5, 8 \\
         \object{R~Hya}                 & 18.8 & \phantom{1}4.5: & 1.0:  & 4.5: & 0.8: & 6, 7, 8 \\
         \object{R~Hor}                 & 16.8 & \phantom{1}7.28$\pm$0.04 & 0.84$\pm$0.01 & 4.1: & 0.8: & 2, 4 \\
         \object{RR~Aql}                        & \phantom{1}4.5 & \phantom{1}6.39$\pm$0.07 & 0.95$\pm$0.01 & 7.2: & 0.9: & 2, 5 \\
         \object{IRC-10529}             & 28.6 & \phantom{1}7.00$\pm$0.06 & 0.93$\pm$0.01 & 8.2: & 0.6: & 2, 3 \\
         \object{WX~Psc}                        & 25.4 & 11.6: & 0.9: & 8.5: & 0.7: & 2, 3 \\
         \object{IRC+10365}             & 40.8 & \phantom{1}7.73$\pm$0.08 & 0.93$\pm$0.01 & 4.18$\pm$0.05 & 0.92$\pm$0.02 & 3 \\
         \noalign{\vskip 1mm}
         \multicolumn{7}{l}{ {\it C-type semi-regular and irregular stars:}} \\
         \noalign{\vskip 1mm}
         \object{TW~Oph}                        & \phantom{1}8.0 & \phantom{1}4.68$\pm$0.26 & 1.00$\pm$0.08 & 2.57$\pm$0.09 & 0.96$\pm$0.04 & \\      
         \object{NP~Pup}                        & \phantom{1}5.8 & \phantom{1}4.63$\pm$0.27 & 0.93$\pm$0.08 & 3.16$\pm$0.32 & 0.87$\pm$0.12 & 3 \\
         \object{TW~Hor}                        & \phantom{1}8.4 & \phantom{1}5.77$\pm$0.06 & 0.66$\pm$0.01 & 2.65$\pm$0.05 & 0.94$\pm$0.02 & 5, 6 \\
         \object{T~Ind}                 & \phantom{1}5.6 & \phantom{1}3.50$\pm$0.14 & 0.83$\pm$0.06 & 2.14$\pm$0.28 & 0.91$\pm$0.13 & 1, 3 \\
         \object{RT~Cap}                        & \phantom{1}7.0 & \phantom{1}4.42$\pm$0.23 & 0.89$\pm$0.05 & 2.25$\pm$0.20 & 1.00$\pm$0.15 &  1 \\
         \object{AQ~Sgr}                        & \phantom{1}9.0 & \phantom{1}4.93$\pm$0.11 & 0.90$\pm$0.02 & 1.4: & 1.0: & \\
         \object{U~Hya}                 & 17.8 & 10.8: & 0.8: & 6.2: & 0.7: & 2 \\
         \object{W~Ori}                 & \phantom{1}9.2 & \phantom{1}4.38$\pm$0.07 & 0.98$\pm$0.03 & 3.28$\pm$0.07 & 0.92$\pm$0.02 & 7, 8 \\
         \object{V~Aql}                 & 10.6 & \phantom{1}5.65$\pm$0.09 & 0.97$\pm$0.02 & 3.22$\pm$0.05 & 0.94$\pm$0.02 & \\
         \object{Y~Pav}                 & 11.2 & \phantom{1}4.89$\pm$0.10 & 0.99$\pm$0.03 & 3.04$\pm$0.06 & 0.94$\pm$0.03 & \\
         \object{X~Vel}                 & 13.0 & \phantom{1}4.07$\pm$0.10 & 0.99$\pm$0.03 & 2.63$\pm$0.09 & 0.92$\pm$0.04 & 3 \\
         \object{Y~Hya}                 & 12.2 & \phantom{1}5.39$\pm$0.15 & 1.0: & 2.63$\pm$0.10 & 0.89$\pm$0.06 & \\
         \object{SS~Vir}                        & \phantom{1}7.4 & \phantom{1}5.88$\pm$0.86 & 0.57$\pm$0.13 & 3.34$\pm$0.25 & 0.70$\pm$0.08 & 5, 6, 8 \\
         \object{W~CMa}                 & 11.6 & \phantom{1}4.26$\pm$0.11 & 1.00$\pm$0.02 & 3.20$\pm$0.08 & 0.98$\pm$0.03 & 6 \\
         \noalign{\vskip 1mm}
         \multicolumn{7}{l}{ {\it C-type Mira stars:}} \\
         \noalign{\vskip 1mm}
         \object{R~Lep}                 & 17.2 & 11.7: & 0.9: & 3.58$\pm$0.07 & 0.99$\pm$0.03 & 2 \\
         \object{CZ~Hya}                        & \phantom{1}9.8 & \phantom{1}3.98$\pm$0.11 & 0.99$\pm$0.04 & 2.47$\pm$0.15 & 0.88$\pm$0.09 & \\
         \object{R~For}                 & 15.8 & \phantom{1}5.69$\pm$0.11 & 0.99$\pm$0.02 & 3.60$\pm$0.07 & 0.93$\pm$0.03 & 3 \\
         \object{R~Vol}                 & 16.6 & \phantom{1}5.83$\pm$0.08 & 0.97$\pm$0.02 & 4.0: & 1.0: & 2, 3 \\
         \object{RV~Aqr}                        & 24.6 & \phantom{1}6.01$\pm$0.05 & 0.99$\pm$0.01 & 4.2: & 0.9: & 2, 3 \\
         \object{V688~Mon}              & 16.3 & \phantom{1}6: & 1: & 2.98$\pm$0.08 & 0.97$\pm$0.03 & 3 \\
         \object{V1259~Ori}             & 20.9 & \phantom{1}5.0: & 1.0: & 3.48$\pm$0.06 & 0.92$\pm$0.02 & 3, 4\\
         \hline \hline
   \end{tabular}}
   \tablefoot{The second column gives the photodissociation radius two times over the distance as a measure of the expected size of the CO emitting region. Columns 3-6 give the major axis and the axis ratio with errors for the best-fit Gaussian at the center velocity of the respective line. The final column gives the asymmetrical features for each source, as is explained in Sect.~\ref{structure}. See text for a further explanation. }
   \label{visfit}
\end{table}


\subsubsection{Comparison with results from spherically symmetric radiative transfer models}
\label{compmod}
Here we present the comparison between the emission distributions from ALMA and those calculated from our previous best-fit radiative transfer model results (Sect.~\ref{1d}). The same analysis was performed on the model results as on the observational data. Figure~\ref{comp} shows the results of the Gaussian emission distribution fitting (Sects~\ref{uvfit} and \ref{uvfit_res}) from the models compared with the observational results for the CO(2-1) emission for each of the six selected sources. For error bars on the observational results, see Figs~\ref{uvM_SR}--\ref{uvC_M}. The results are affected by the difficulties of fitting a Gaussian distribution to the weaker and smaller emission distribution close to the edge of the line. It is obvious that the radiative transfer analysis of the full data set will be necessary before any firm conclusions can be drawn. For the M-type stars, it seems as if the size of the emission might be smaller than expected for the low mass-loss-rate source (R~Leo) and the opposite for the high mass-loss-rate source (IRC-10529). The upcoming analysis will evaluate whether this is a general trend or specific to the selected sources. This trend would indicate an even steeper dependence on the circumstellar density than predicted from the photodissociation model by \citet{mamoetal88} and is contrary to the fit to the M-type Mira stars in Fig.~\ref{bredd}. The results for the C-type stars are less clear, partly because the fitting results to the U Hya data seem strongly affected by resolved out flux (see Sect.~\ref{structure}). It is also possible that the emission from C-type stars is subject to stronger optical depth effects due to the higher CO abundance. This will also be evaluated in the upcoming radiative transfer analysis. 

\begin{figure*}[th]
\includegraphics[height=6cm]{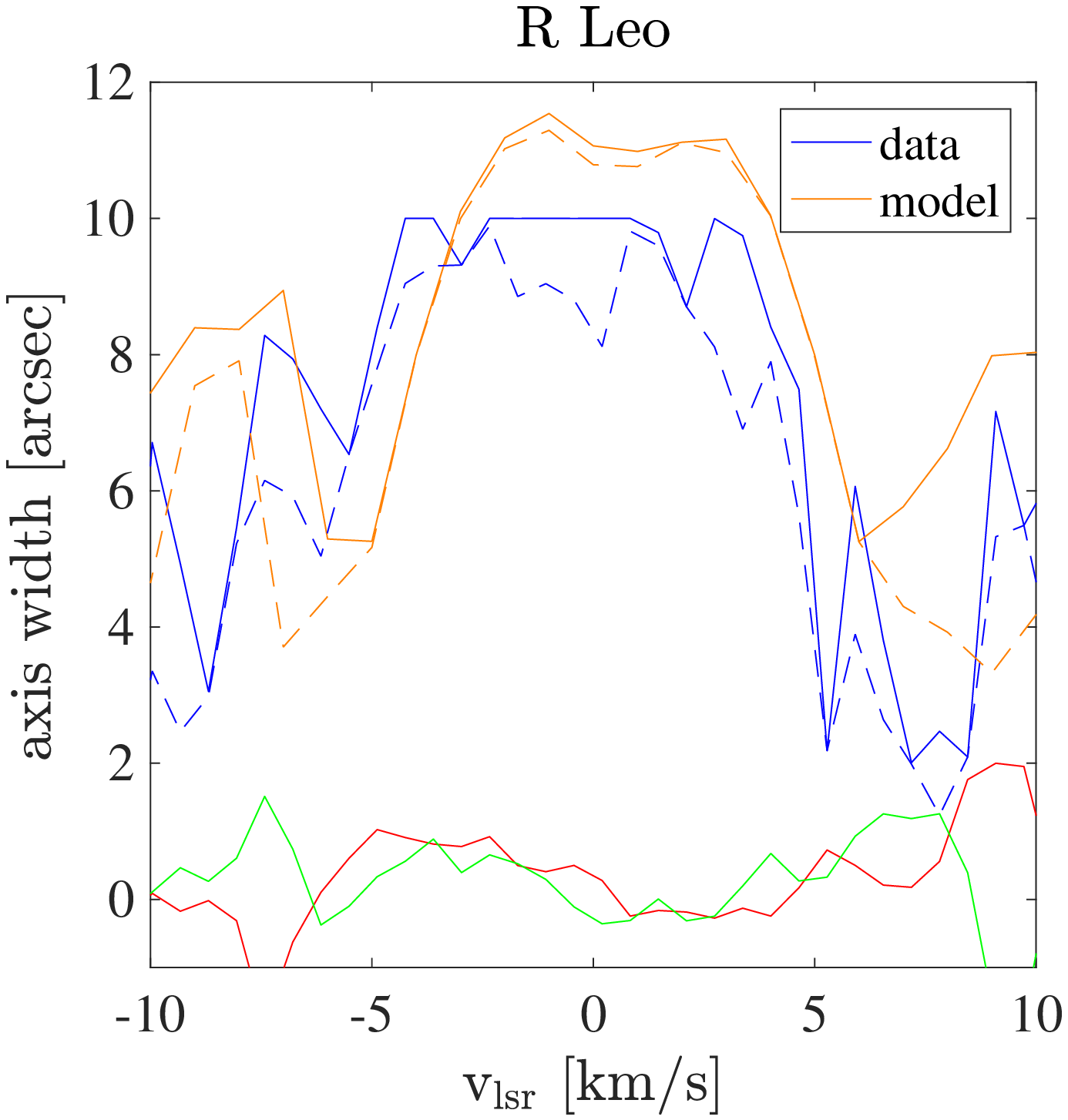}
\includegraphics[height=6cm]{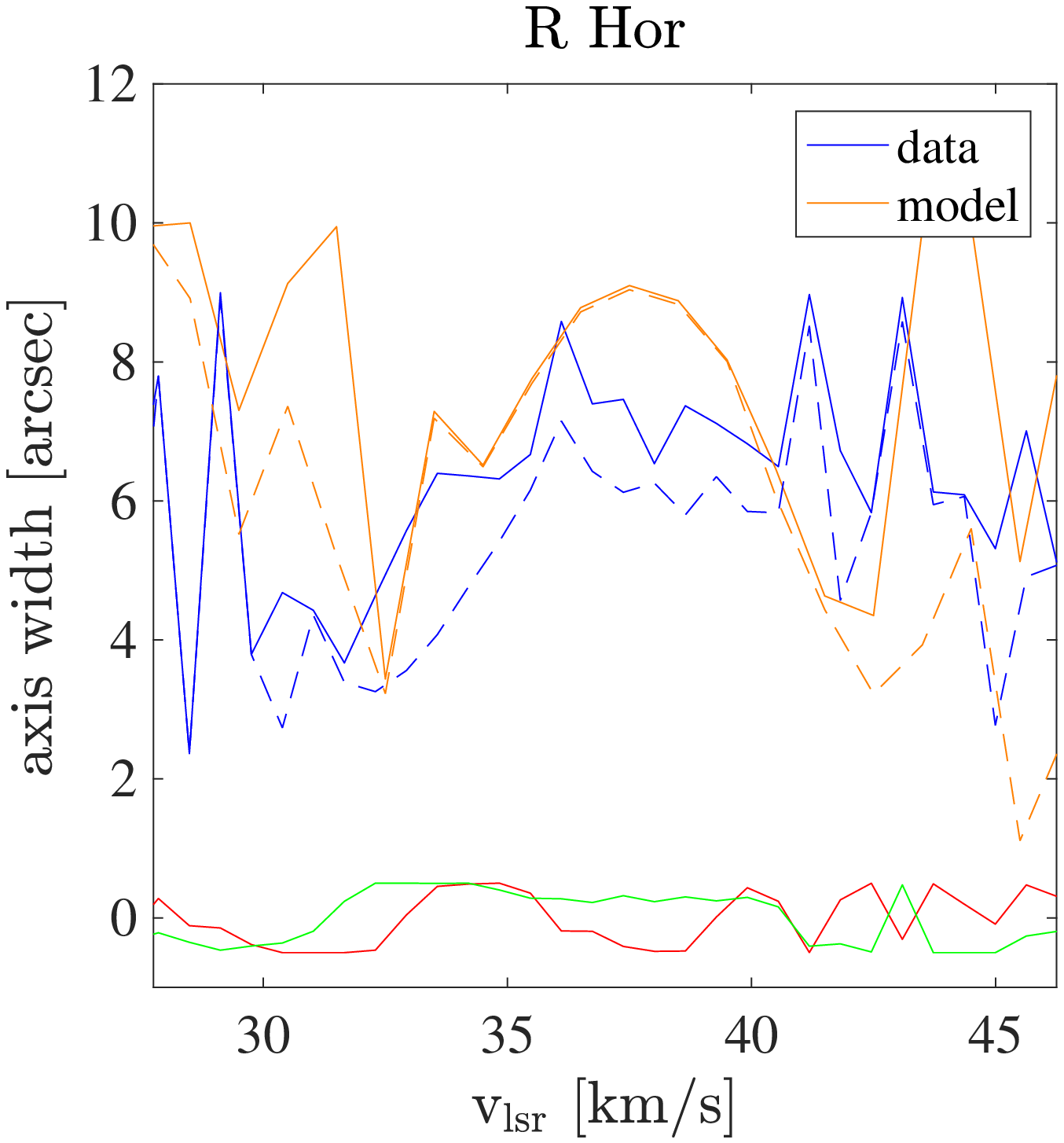}
\includegraphics[height=6cm]{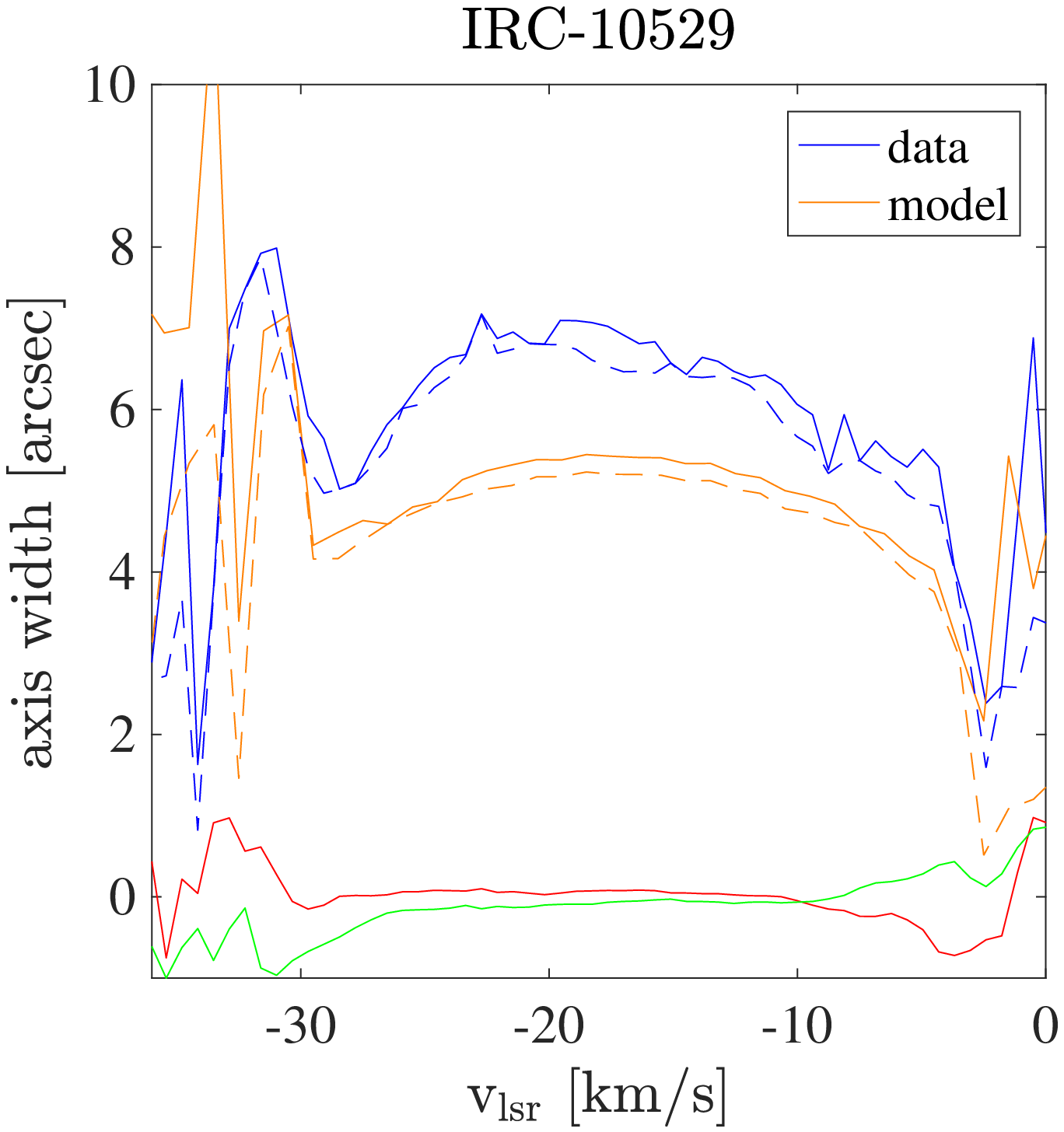}

\smallskip

\includegraphics[height=6.05cm]{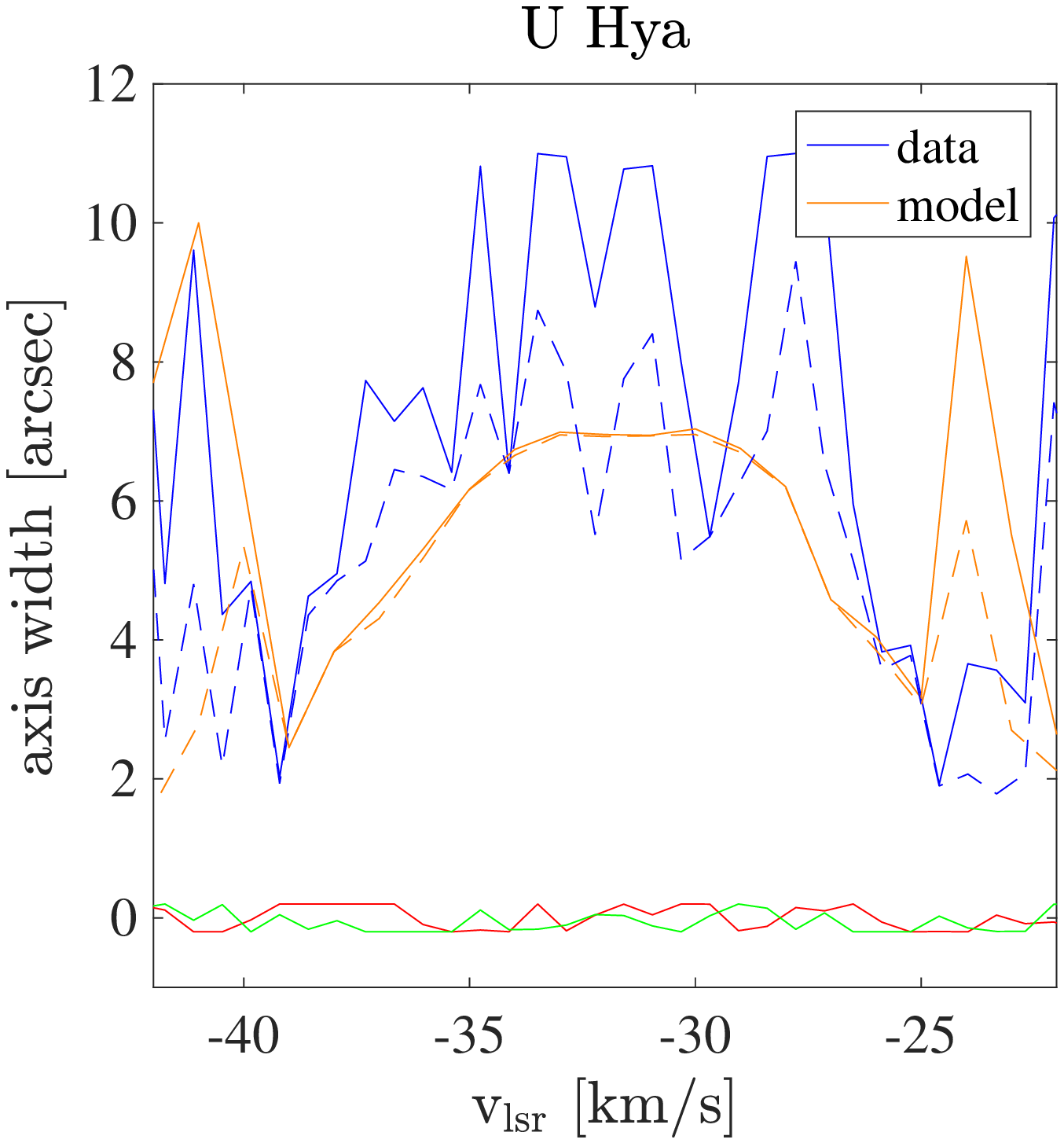}
\includegraphics[height=6.05cm]{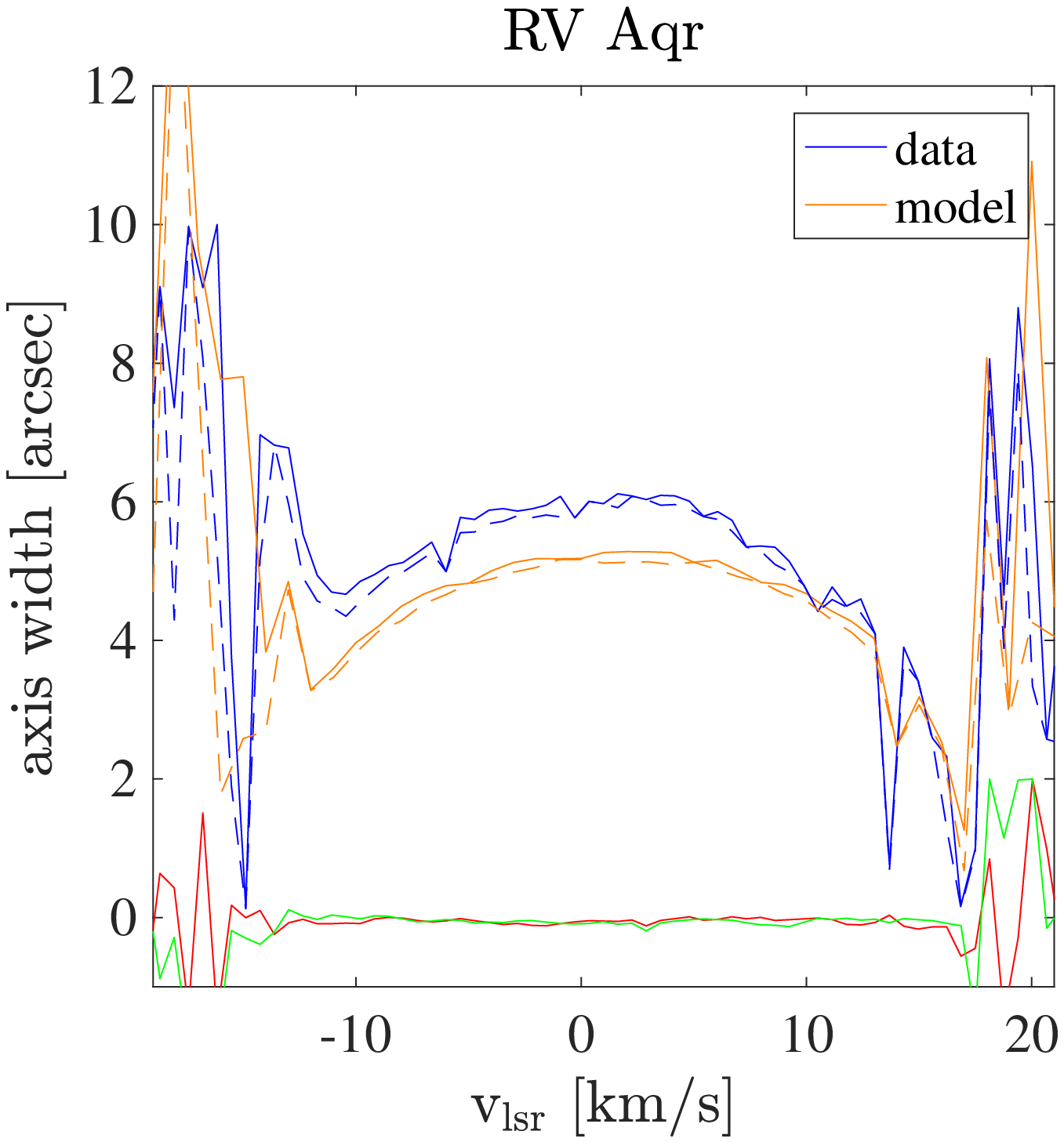}
\includegraphics[height=6.05cm]{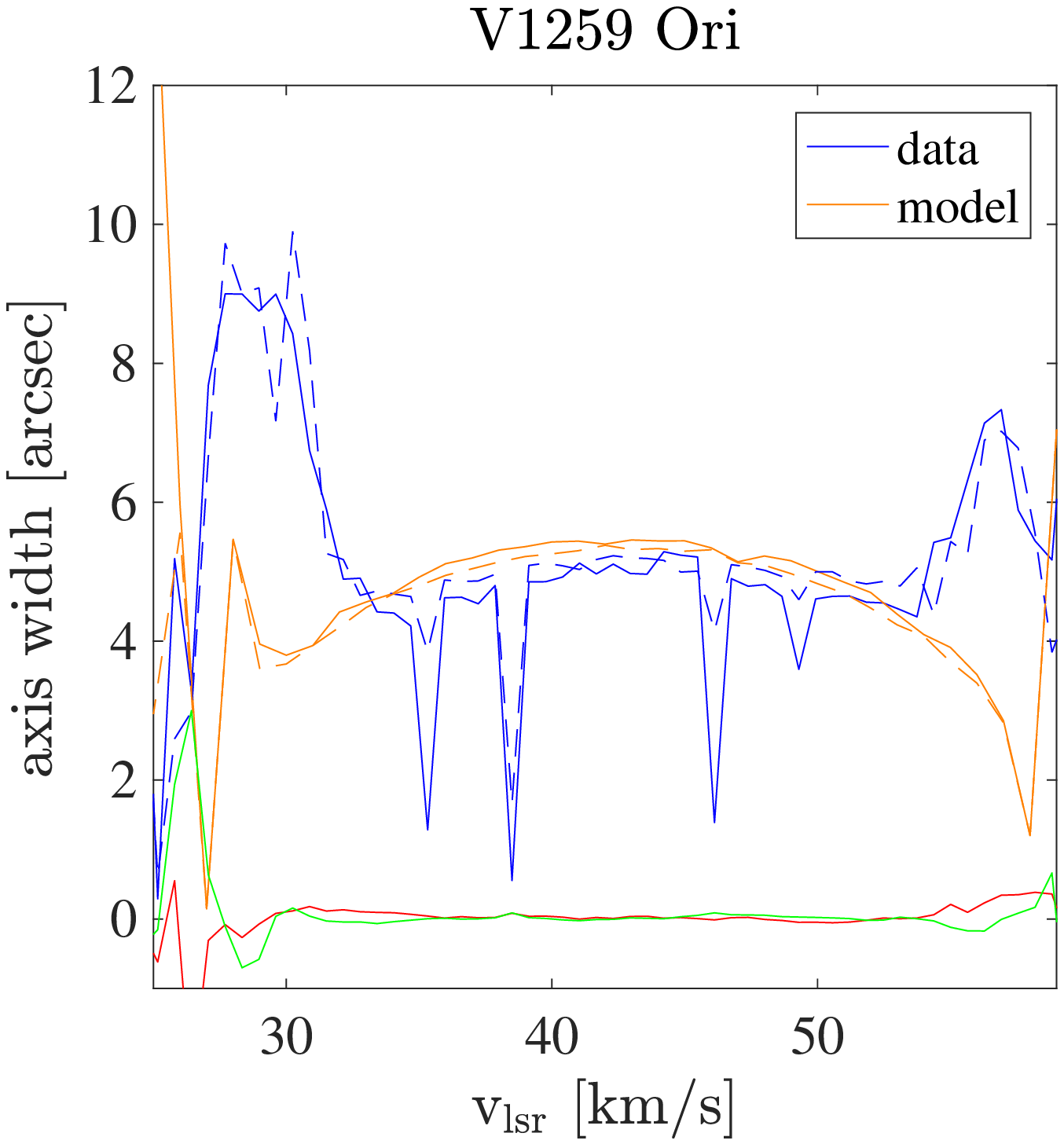}
\caption{Comparison between the results from fitting a Gaussian emission distribution to the CO(2-1) data (blue) and the model results (orange) based on previous best-fit models (see text). The solid lines show the major axis and the dashed lines show the minor axis. The red and green lines show the RA and Dec. offsets for the data, respectively. The top row shows three M-type stars, and the bottom row shows three C-type stars, with an increasing mass-loss rate (Table~\ref{sample}). The results are affected by the difficulties of fitting a Gaussian distribution to the weaker and smaller emission distribution close to the edge of the line. For error bars on the results from the measurements, see Figs~\ref{uvM_SR}--\ref{uvC_M}.}
\label{comp}
\end{figure*}


\subsection{Asymmetrical features}
\label{structure}
Several asymmetrical features can be identified in the observational results presented thus far, that is, the line profiles (Figs~\ref{linesM_SR}--\ref{linesC_M}) and the results from the Gaussian fitting (Figs~\ref{uvM_SR}--\ref{uvC_M}). These features can be divided into two classes: Features that are consistent with a spherically symmetric CSE and those that are not. 


\subsubsection{Consistent with a smooth outflow}
The features that can be explained with a spherically symmetric CSE are as follows.
\begin{itemize}
\item[1.]{An offset between the center position (of the fitted Gaussian) and the source coordinates is explained the most easily by uncertainties in the adopted coordinates (taken from SIMBAD pre-Gaia DR2), when seen consistently in both lines. Examples include V~Tel and RT~Cap, for instance.}
\item[2.]{The interferometer acts as a spatial filter and since only the ACA was used (and not the total-power array), large-scale, smooth emission is resolved out. Resolved-out flux results in a lower peak flux level than when measured with a (similar-sized) single-dish telescope. Furthermore, if a significant fraction of the flux is resolved out, the visibility distribution cannot be well-fitted by a simple Gaussian distribution. This is the most apparent across the center channels where the emission distribution is expected to be the largest. Therefore, it is also a more common problem for the CO(3-2) line where the beam is smaller. Examples of this can be seen in RR~Aql (CO(3-2)), WX~Psc (both lines), and R~Vol (CO(3-2)).}
\item[3.]{Self-absorption on the blue-shifted side of the line can be seen as asymmetry in the line profile where it is sometimes apparent that emission is "missing" on the blue side of the line. Examples include IRC-10529, NP~Pup, and T~Ind. It is even more apparent when looking at how the sizes of the sources change across the channels. Sources with self-absorption show a prominent peak in the size distribution at blue-shifted velocities since these channels probe emission coming from cooler circumstellar layers (and hence a larger region). }
\end{itemize}
These features are relatively easy to explain and related to how the observations were performed for the first and second
points. 


\subsubsection{Indications of circumstellar anisotropic structure}
The next class of features is less straight-forward to interpret and the features are mostly thought to be clearly inconsistent with a spherically symmetric outflow. Further modeling and/or observations will be necessary to explain exactly how they arise. The majority of sources that exhibit these features need to be followed up with higher-spatial-resolution observations. Features that cannot be explained by a smoothly expanding, spherically symmetric CSE are the following.
\begin{itemize}
\item[4.]{Triangular line profiles are not a newly observed feature; however, a satisfactory explanation as to how they arise is lacking. This feature seems to be more common among high-mass-loss-rate sources. Clear examples are R~Hor and V1259~Ori.}
\item[5.]{High-velocity wings are not a new feature either, but they can be detected from more sources in this data set due to the superior sensitivity reached with the ACA observations compared to previous single-dish observations. The line profiles of TW~Hor,  for example, are very similar to those of the known asymmetrical CSE of $\pi^{1}$~Gru \citep[e.g., ][]{doanetal20}. The CO emission from $\pi^{1}$~Gru can be reproduced by a CSE consisting of an expanding equatorial torus (explaining the two central peaks) and a bi-polar faster outflow leading to wide wings \citep{doanetal17}. Other, more tentative examples of line profiles with high-velocity wings include those of L$_2$~Pup, R~Leo, and BK~Vir.}
\item[6.]{The fitting of the visibility distribution gives the center position of the best-fit Gaussian in each channel and the offset relative to the center position is plotted in Figs~\ref{uvM_SR}--\ref{uvC_M}. Several sources show center position gradients. A position gradient with velocity is either indicative of expansion in a higher-density layer (e.g., an equatorial torus such as in the case of $\pi^{1}$~Gru) or of rotation \citep[e.g.,][]{ramsetal18,vlemetal18} depending on the orientation of the sources. Examples of sources with center position gradients are BK~Vir (in RA), TW~Hor (both RA and Dec), SS Vir (both RA and Dec), and W~CMa (in RA).}
\item[7.]{A spherically symmetric CSE with a constant expansion velocity grows monotonically in size from the edge of the line to the center velocity. The effects of self-absorption and resolved-out flux on the size distribution are discussed above. For some sources, the source size changes in an unpredictable way that cannot be easily explained without further modeling and/or observations. Examples include T~Mic, CW~Cnc, RT~Vir, and W~Ori.}
\item[8.]{It is directly apparent from the line profile that the emission cannot come from a standard CSE for a handful of the sample sources. Instead, these line profiles are anomalous. Some are previously known to be anomalous, for example, L$_{2}$~Pup \citep[e.g.,][]{kervetal16} and R~Leo \citep{fonfetal19}, while others have not been studied with a focus on the detailed structure of the CSE, for example, R~Hya and SS~Vir.  }
\end{itemize}


\subsection{Detections of emission from molecules other than $^{12}$CO}
A large number of molecules other than $^{12}$CO have transitions with frequencies within the observed bands. For all sources, the peak fluxes of the detected lines as measured across a 10\arcsec~aperture are listed in Tables~\ref{other_MSR}-\ref{other_CM}. The spectral resolution across each spectral window is the same as the one used for imaging and is given in Sect.~\ref{obs}. The lines listed in Tables~\ref{other_MSR}-\ref{other_CM} are those for which we are confident of detection at the given spectral resolution and aperture. It is possible that further detections could be confirmed with more optimization of the data analysis. 

Lines from SiO, SiS, $^{13}$CO, CS, and H$^{13}$CN are detected in a majority of the sources. Unsurprisingly, oxides such as SO, SO$_{2}$, and H$_{2}$O are only detected in the M-type stars as are H$_{2}$S and less common isotopologues of SiS and NaCl. In the C-type stars, $^{13}$CN and $^{13}$CS are detected, sometimes together with lines from molecules with several carbon atoms, such as HC$_{3}$N, C$_{4}$H, and SiC$_{2}$. A further analysis of these lines is beyond the scope of this paper, but an overview of the firm detections are given here and follow-up studies are encouraged.


\section{Discussion and summary}
\label{dis}
The measurements presented in this paper show that, for this sample, the CO envelope sizes are, in general, larger for C-type than for M-type stars. As explained both above and below, the envelope size depends on many different parameters, but this trend is expected if the CO/H$_{2}$ ratio is larger in C-type stars. Based on chemical equilibrium models, the ratio between the C- and M-type CO-abundance is generally assumed to be equal to five when deriving total gas mass-loss rates from observations of CO emission lines. As part of the analysis planned for the full DEATHSTAR sample, the difference in abundance depending on chemical type (also S-type stars) can now be observationally constrained for a large sample of AGB stars.

Furthermore, the measurements show a relation between the measured (Gaussian) CO(2-1) size and circumstellar density that, while in broad agreement with photodissociation calculations, reveal large scatter and some systematic differences between the different stellar types. A significant amount of scatter arises due to significant distance uncertainties. Part of the differences can also be explained by excitation and optical depth effects that decrease the reliability of the Gaussian size determination for $\dot{M}/v_{\infty}\gtrsim 10^{-7}$~M$_\odot$~km~s$^{-1}$~yr$^{-1}$.  For lower-mass-loss-rate irregular and semi-regular variables of both M- and C-type stars, the CO(2-1) size seems essentially independent of $\dot{M}$/$v_{\infty}$. For the higher-mass-loss-rate Mira stars, the CO(2-1) size clearly increases with circumstellar density, with larger sizes for the higher CO-abundance C-type stars.  The M-type stars appear to be consistently smaller than what was predicted from photodissociation theory. The differences between the estimates and measurements could (as shown in \citet{sabeetal19}) be due to, for example, a systematically overestimated CO abundance, differences in the CSE temperature profile, or differences in the UV-environment. Additionally, a difference in the adopted properties of the dust, which is responsible for much of the shielding in low density CSEs, and/or the adopted dust-to-gas ratio in the models would lead to different sizes. Finally, a difference in the CSE density profile due to changes in velocity and/or the mass loss rate could significantly alter the photodissociation radius. With the size measurements that are now available for the large sample of sources in the DEATHSTAR project, it will be possible to investigate many of these factors in more detail. 

In recent years there has been a strong focus on investigating the isotropy of the mass-loss process in low- to intermediate mass evolved stars, which was partly expedited by the superior imaging capabilities of ALMA. Several AGB sources with previously known complex circumstellar dynamics and/or binary companions have been mapped in detail \citep[e.g.,][]{ramsetal14,decietal15,maeretal16,doanetal17,kimetal17,homaetal18,fonfetal19}. With the DEATHSTAR project, the aim is to get an overview of how widespread and significant the detected asymmetrical features are, from the perspective of measuring the amount of circumstellar wind material. A goal is to eventually be able to evaluate the uncertainties that these features will result in when estimating mass-loss rates from more distant, unresolved sources. 

The results presented here show that the majority of the sources have CO CSEs that are consistent with a spherically symmetric, smooth outflow, at least on larger scales. This is based on the line profile shapes and that the emission distribution across the line channels can be well-fitted with a Gaussian visibility distribution with axis ratios that are close to one for a majority of the sources. For about a third of the sources, indications of strong asymmetries are detected. This is consistent with what was found in previous interferometric investigations of northern sources \citep{nerietal98,castetal10}. In the DEATHSTAR sample, some of the known asymmetric sources have been removed, but the higher sensitivity reached in the ACA observations allowed us to detect relatively weaker features, such as extended line wings, than what was possible in previous investigations. This offers an explanation as to why the same fraction of strong asymmetries is still found. In a large fraction of the sources, some deviation from spherical symmetry is detected, often on smaller scales. Whether these smaller deviations can have a significant effect on the estimated mass-loss rate or not cannot be evaluated without further analysis.

The peak fluxes of lines from molecules other than $^{12}$CO detected within the observed bands are listed in Tables~\ref{other_MSR}-\ref{other_CM}. In general, lines from molecules that are known to be abundant in AGB CSEs (e.g., SiO, SiS, CS) are detected in almost all sources. Some less abundant oxides (e.g., H$_{2}$O and SO) are only detected in the M-type stars, while emission from molecules with several carbon atoms (e.g., SiC$_{2}$) are only detected toward the C-type stars. No further analysis has been attempted. 

\section{Outlook}
\label{sum}
This paper presents the first observational results from the DEATHSTAR project. The CO(2-1) and (3-2) ALMA ACA observations of the southern M- and C-type stars observed in ALMA Cycle 4 are presented and analyzed. In an upcoming publication, the results on the M-, S-, and C-type sources observed in Cycle 5 will be presented. The next step is to perform detailed radiative transfer modeling that is constrained by the presented ACA data together with the previously attained single-dish observations of the lower-$J$ CO transition lines (up to $J$=6$\rightarrow$5) in order to determine mass-loss rates that are independent of photodissociation model results. This will also be presented in future publications. For a large fraction of the sources, observations at higher spatial resolution will be necessary to deduce the nature and origin of the complex circumstellar dynamics revealed by the observations and data analysis presented in this paper.


\begin{acknowledgements}
This paper makes use of the following ALMA data: ADS/JAO.ALMA\#2016.2.00025.S and \#2017.1.00595.S. ALMA is a partnership of ESO (representing its member states), NSF (USA) and NINS (Japan), together with NRC (Canada), MOST and ASIAA (Taiwan), and KASI (Republic of Korea), in cooperation with the Republic of Chile. The Joint ALMA Observatory is operated by ESO, AUI/NRAO and NAOJ. 

This work has made use of data from the European Space Agency (ESA) mission
{\it Gaia} (\url{https://www.cosmos.esa.int/gaia}), processed by the {\it Gaia}
Data Processing and Analysis Consortium (DPAC,
\url{https://www.cosmos.esa.int/web/gaia/dpac/consortium}). Funding for the DPAC
has been provided by national institutions, in particular the institutions
participating in the {\it Gaia} Multilateral Agreement.
\end{acknowledgements}


\begin{appendix}

\section{Parallaxes for AGB stars from Gaia DR2}
\label{distances}

\begin{figure}[h]
\includegraphics[width=\columnwidth]{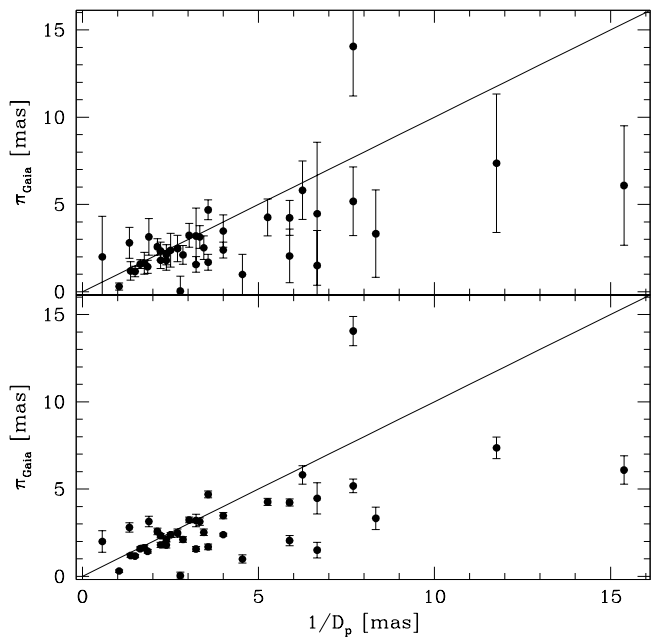}
\caption{Gaia DR2 parallaxes for our sample compared to the distances used in the current mass-loss models. In the bottom panel, the Gaia parallaxes are presented with their formal errors. In the top panel, the {\it astrometric\_excess\_noise} value is added in quadrature to the formal errors following the empirical results from \cite{langevelde18}.}
\label{gaiaplot}
\end{figure}

Since the paper presents mass-loss values based on earlier analysis, we have presented the distances that were used in the mass-loss determinations. Recently, Gaia DR2 astrometric solutions were released for all but one of the stars in our sample. While the formal errors on the Gaia parallaxes are generally small, the reliability of the Gaia parallaxes for AGB stars is under debate \citep[e.g.,][]{langevelde18}. There are several reasons why the Gaia parallaxes to AGB stars might be wrong or have significantly underestimated the assigned errors. Firstly, AGB stars are often larger than the parallax signature itself (with sizes of order one to a few astronomical units) and they have convective surface motions that can cause the photo-center to shift significantly \citep{chiavassa18}. Secondly, AGB pulsations can affect the astrometric measurements. Finally, AGB stars are so bright that Gaia approaches saturation. The Gaia DR2 catalog contains a number of parameters that can provide a measure for the reliability of the astrometry. One of these is the {\it astrometric\_excess\_noise}, which represents modeling errors for sources that do not behave according to the adopted astrometric model of \citet{lindegren12}. In \cite{langevelde18}, an empirical analysis of very long baseline interferometry (VLBI) parallaxes compared to the Gaia results revealed that adding the {\it astrometric\_excess\_noise} to the formal parallax errors in quadrature was needed to reconcile the two methods. In Fig.~\ref{gaiaplot} we show the difference between the formal and excess noise errors. When we adopted this procedure, only four of the AGB stars in our sample have a parallax solution that satisfies $\pi/ \sigma_{\pi}>5$. Recently, another measure of reliability was introduced, namely the RUWE\footnote{Lindegren et al.~2018, Gaia memo; GAIA-C3-TN-LU-LL-124-01} (renormalized unit weighted error). This represents the square-root of the reduced $\chi^2$ value, which was corrected for the strong dependence of the $\chi^2$ value on magnitude and color. For many applications, a RUWE value $<1.4$ is empirically found to represent a good fit. However, as is noted in Lindegren et al.~(2018), the RUWE normalization does not work optimally for the brightest sources $G<12$, which includes all of our AGB stars. In these cases, Lindegren et al.~(2018) stress that a RUWE threshold should be set based on empirical evidence and not a theoretical distribution. For this, we can use the northern AGB star BX~Cam as an example. Recent VLBI observations have revealed a parallax of $\pi_{\rm VLBI}=1.73\pm0.03$~mas \citep{Matsuno2020}. This value is significantly different from the Gaia value of $\pi_{\rm Gaia}=4.13\pm0.25$ even when taking the {\it astrometric\_excess\_noise} of 0.67~mas into account. However, the RUWE of BX~Cam is 1.04, which would have qualified as a good solution in most cases. In comparing the stellar luminosity, \cite{Matsuno2020} conclude that the larger VLBI distance is more reliable since the VLBI distance would imply a luminosity of $\sim4950$~L$_\odot$, while the Gaia distance would result in a luminosity of only $\sim870$~L$_\odot$. This example illustrates that even for AGB stars with an apparent low RUWE value, we should exercise caution when adopting the current Gaia DR2 parallaxes.   


\section{Imaging results}

\begin{table*}[h]
   \caption{Imaging results.} 
   \resizebox{\textwidth}{!}{\begin{tabular}{l|ccc|ccc|ccc|ccc} 
      \hline \hline
            & \multicolumn{3}{c}{Band 6} & \multicolumn{3}{c}{Band 7} & \multicolumn{3}{c}{CO(2-1)} & \multicolumn{3}{c}{CO(3-2)} \\
 \cline{2-4} 
 \cline{5-7} 
 \cline{8-10}
 \cline{11-13}
      \noalign{\vskip 1mm}
      Source  & $\theta$ & P.A. & rms & $\theta$ & P.A. & rms & $F_{\rm{peak}}$ & $v_{\rm{c}}$ & $\Delta v$ & $F_{\rm{peak}}$ & $v_{\rm{c}}$ & $\Delta v$ \\
        & [''] & [$^{\circ}$] & [$\frac{\rm{mJy}}{\rm{beam}}$] & [''] & [$^{\circ}$] & [$\frac{\rm{mJy}}{\rm{beam}}$] & [Jy] & [km\,s$^{-1}$] & [km\,s$^{-1}$] & [Jy] & [km\,s$^{-1}$] & [km\,s$^{-1}$] \\
             \noalign{\vskip 1mm}
      \hline \hline 
      \noalign{\vskip 1mm}
         \multicolumn{13}{l}{ {\it M-type semi-regular and irregular stars:}} \\
         \noalign{\vskip 1mm}
         \object{L$_{2}$~Pup}   & 6.5\,$\times$\,5.4 & \phantom{-}67.0 & \phantom{1}52 & 3.9\,$\times$\,3.5 & -70.1 & 103 & 19.8 & \phantom{-}33.3 & \phantom{1}7.5 & 47.2 & \phantom{-}32.7 & \phantom{1}8.0 \\
      \object{W~Hya}                    & 6.6\,$\times$\,4.4 & -89.1 & \phantom{1}51 & 4.5\,$\times$\,2.9 & -89.9 & 340 & 36.0 & \phantom{-}41.0 & 17.0 & 68.0 & \phantom{-}41.0 & 17.5\\
      \object{T~Mic}                    & 6.3\,$\times$\,4.0 & -78.2 & \phantom{1}76 & 4.9\,$\times$\,2.7 & -79.1 & 168 & 17.5 & \phantom{-}25.3 & 12.5 & 20.0 & \phantom{-}25.0 & 13.0\\
      \object{Y~Scl}                    & 6.5\,$\times$\,4.3 & \phantom{-}81.8 & \phantom{1}71 & 5.1\,$\times$\,2.7 & \phantom{-}89.9 & 138 & \phantom{1}4.9 & \phantom{-}28.8 & 11.5 & \phantom{1}7.8 & \phantom{-}29.0 & 12.0\\
      \object{V1943~Sgr}                & 6.5\,$\times$\,4.2 & -73.1 & \phantom{1}76 & 5.2\,$\times$\,2.7 & -73.5 & 170 & 14.0 & -14.8 & 12.7 & 19.0 & -15.0 & 13.0 \\
      \object{BK~Vir}                   & 7.6\,$\times$\,4.8 & -89.0 & 112 & 4.9\,$\times$\,3.1 & -67.2 & 144 & 12.9 & \phantom{-}16.8 & 11.5 & 25.8 & \phantom{-}17.0 & 11.0 \\
      \object{V~Tel}                    & 6.5\,$\times$\,5.4 & \phantom{-}74.3 & \phantom{1}52 & 4.4\,$\times$\,3.8 & \phantom{-}84.9 & 155 & 11.9 & -32.5 & 16.0 & 21.3 & -32.8 & 15.5 \\
      \object{SU~Vel}                   & 6.3\,$\times$\,4.5 & -78.1 & \phantom{1}48 & 4.3\,$\times$\,3.2 & \phantom{-}85.5 & 136 & 13.5 & \phantom{-1}7.5 & 13.0 & 25.7 & \phantom{-1}7.5 & 13.0 \\
      \object{UY~Cet}                   & 6.2\,$\times$\,4.7 & \phantom{-}88.1 & \phantom{1}72 & 5.2\,$\times$\,3.0 & -69.5 & 142 & \phantom{1}9.5 & \phantom{-1}4.7 & 15.3 & 18.1 & \phantom{-1}4.8 & 15.5 \\
      \object{SV~Aqr}                   & 6.4\,$\times$\,4.5 & -82.9 & \phantom{1}40 & 4.3\,$\times$\,2.8 & -89.9 & 158 & \phantom{1}4.0 & \phantom{-1}5.6 & 18.8 & \phantom{1}5.8 & \phantom{-1}6.8 & 18.5 \\
      \object{SW~Vir}                   & 8.2\,$\times$\,4.6 & -73.2 & \phantom{1}71 & 4.6\,$\times$\,3.3 & \phantom{-}74.7 & 141 & 36.5 & -11.5 & 17.0 & 66.4 & -11.5 & 17.0 \\
     \object{CW~Cnc}                    & 7.3\,$\times$\,4.6 & -75.8 & \phantom{1}53 & 4.2\,$\times$\,3.4 & \phantom{-}83.4 & 161 & \phantom{1}9.5 & \phantom{-}15.0 & 24.0 & 12.8 & \phantom{-}15.0 & 23.3\\
      \object{RT~Vir}                   & 8.8\,$\times$\,4.4 & -70.8 & \phantom{1}72 & 4.7\,$\times$\,3.3 & \phantom{-}68.2 & 145 & 16.9 & \phantom{-}17.0 & 18.0 & 31.0 & \phantom{-}17.3 & 17.5 \\
      \object{R~Crt}                    & 7.1\,$\times$\,4.4 & \phantom{-}81.1 & \phantom{1}52 & 4.4\,$\times$\,2.8 & -87.1 & 153 & 26.3 & \phantom{-}11.3 & 23.5 & 44.3 & \phantom{-}11.3 & 23.5 \\
      \noalign{\vskip 1mm}
         \multicolumn{13}{l}{ {\it M-type Mira stars:}} \\
         \noalign{\vskip 1mm}
         \object{R~Leo}                 & 7.0\,$\times$\,4.7 & -85.8 & \phantom{1}54 & 4.2\,$\times$\,3.2 & \phantom{-}69.5 & 176 & 25.3 & \phantom{1}-0.5 & 15.0 & 57.8 & \phantom{1}-0.5 & 16.0 \\
         \object{R~Hya}                 & 6.6\,$\times$\,4.4 & -89.7 & \phantom{1}51 & 4.4\,$\times$\,2.8 & -87.6 & 123 & 26.0 & \phantom{1}-9.8 & 21.0 & 67.9 & \phantom{1}-9.8 & 21.5 \\
         \object{R~Hor}                 & 6.1\,$\times$\,5.4 & \phantom{-}79.6 & \phantom{1}45 & 3.9\,$\times$\,2.8 & -82.9 & 133 & 54.7 & \phantom{-}37.5 & 13.0 & 76.3 & \phantom{-}37.5 & 13.0 \\
         \object{RR~Aql}                        & 6.2\,$\times$\,5.4 & -73.5 & \phantom{1}50 & 4.9\,$\times$\,3.4 & -68.3 & \phantom{1}97 & 19.0 & \phantom{-}27.8 & 16.5 & 27.1 & \phantom{-}27.8 & 16.5 \\
         \object{IRC-10529}             & 6.4\,$\times$\,5.2 & -71.9 & \phantom{1}52 & 4.6\,$\times$\,3.4 & -67.1 & 101 & 33.3 & -17.8 & 30.5 & 43.4 & -17.5 & 30.0 \\
         \object{WX~Psc}                        & 7.4\,$\times$\,5.4 & \phantom{-}31.1 & \phantom{1}89 & 4.4\,$\times$\,3.2 & \phantom{-}70.5 & 129 & 28.6 & \phantom{-1}9.0 & 39.0 & 38.9 & \phantom{-1}9.5 & 39.0 \\
         \object{IRC+10365}             & 6.9\,$\times$\,5.6 & \phantom{-}89.0 & \phantom{1}62 & 4.1\,$\times$\,3.4 & -74.4 & 149 & 24.4 & -31.3 & 34.5 & 32.4 & -31.3 & 35.5 \\
         \noalign{\vskip 1mm}
         \multicolumn{13}{l}{ {\it C-type semi-regular and irregular stars:}} \\
         \noalign{\vskip 1mm}
         \object{TW~Oph}                        & 10.3\,$\times$\,4.5 & -84.8 & \phantom{1}72 & 4.5\,$\times$\,2.9 & -89.3 & \phantom{1}94 & \phantom{1}6.3 & \phantom{-}28.4 & 17.3 & \phantom{1}9.5 & \phantom{-}28.5 & 17.0 \\     
         \object{NP~Pup}                        & 6.3\,$\times$\,5.2 & \phantom{-}84.9 & \phantom{1}52 & 4.1\,$\times$\,3.5 & -80.6 & 105 & \phantom{1}3.0 & \phantom{-}13.3 & 21.5 & \phantom{1}3.4 & \phantom{-}12.8 & 21.5 \\
         \object{TW~Hor}                        & 6.1\,$\times$\,5.7 & \phantom{-}81.0 & \phantom{1}41 & 4.0\,$\times$\,3.1 & -77.2 & 112 & 13.9 & \phantom{-1}1.3 & 14.5 & 24.1 & \phantom{-1}1.1 & 13.8 \\
         \object{T~Ind}                 & 6.0\,$\times$\,5.2 & -84.9 & \phantom{1}47 & 5.0\,$\times$\,2.7 & -72.7 & 168 & \phantom{1}5.3 & \phantom{-}15.8 & 12.5 & \phantom{1}8.6 & \phantom{-}15.8 & 11.5 \\
         \object{RT~Cap}                        & 6.8\,$\times$\,3.9 & -74.3 & \phantom{1}75 & 5.1\,$\times$\,2.6 & -74.0 & 181 & \phantom{1}5.7 & -18.0 & 16.0 & \phantom{1}6.6 & -18.0 & 15.0 \\
         \object{AQ~Sgr}                        & 6.4\,$\times$\,4.5 & -82.6 & \phantom{1}47 & 6.0\,$\times$\,2.7 & -79.7 & 179 & \phantom{1}8.2 & \phantom{-}21.5 & 22.0 & \phantom{1}9.7 & \phantom{-}21.5 & 23.0 \\
         \object{U~Hya}                 & 6.9\,$\times$\,4.4 & \phantom{-}86.7 & \phantom{1}51 & 4.3\,$\times$\,2.9 & -86.6 & 144 & 43.3 & -31.3 & 14.5 & 63.1 & -31.3 & 14.5 \\
         \object{W~Ori}                 & 6.3\,$\times$\,5.4 & \phantom{-}81.3 & \phantom{1}49 & 4.0\,$\times$\,3.1 & \phantom{-}85.5 & 146 & 11.3 & \phantom{1}-1.5 & 23.0 & 18.0 & \phantom{1}-1.5 & 23.0 \\
         \object{V~Aql}                 & 7.7\,$\times$\,5.0 & -68.7 & \phantom{1}62 & 4.3\,$\times$\,3.6 & -79.6 & 123 & 12.8 & \phantom{-}53.5 & 19.0 & 22.9 & \phantom{-}53.8 & 19.5 \\
         \object{Y~Pav}                 & 9.8\,$\times$\,5.0 & \phantom{-}28.3 & \phantom{1}58 & 4.3\,$\times$\,3.8 & \phantom{-}44.1 & 135 & 12.7 & \phantom{1}-3.8 & 18.5 & 19.3 & \phantom{1}-3.5 & 18.0 \\
         \object{X~Vel}                 & 6.5\,$\times$\,4.4 & -77.5 & \phantom{1}49 & 4.3\,$\times$\,3.2 & \phantom{-}87.3 & 136 & \phantom{1}8.3 & -19.3 & 22.5 & 14.7 & -19.3 & 22.5 \\
         \object{Y~Hya}                 & 7.5\,$\times$\,4.3 & -74.8 & \phantom{1}54 & 4.6\,$\times$\,2.6 & \phantom{-}89.3 & 146 & 10.7 & \phantom{1}-8.5 & 19.0 & 15.0 & \phantom{1}-8.5 & 18.0 \\
         \object{SS~Vir}                        & 7.6\,$\times$\,4.5 & -85.0 & 112 & 4.9\,$\times$\,3.0 & -68.6 & 137 & \phantom{1}3.9 & \phantom{-1}8.5 & 28.0 & \phantom{1}6.8 & \phantom{-1}8.7 & 30.3 \\
         \object{W~CMa}                 & 7.4\,$\times$\,4.2 & -79.8 & \phantom{1}44 & 5.2\,$\times$\,2.6 & \phantom{-}69.9 & 116 & \phantom{1}8.8 & \phantom{1}-0.3 & 21.5 & 11.4 & \phantom{1}-0.3 & 21.5 \\
         \noalign{\vskip 1mm}
         \multicolumn{13}{l}{ {\it C-type Mira stars:}} \\
         \noalign{\vskip 1mm}
         \object{R~Lep}                 & 7.4\,$\times$\,3.8 & -75.2 & 106 & 4.2\,$\times$\,2.7 & \phantom{-}89.4 & 125 & 22.6 & \phantom{-}11.3 & 37.5 & 21.8 & \phantom{-}11.5 & 38.0 \\
         \object{CZ~Hya}                        & 6.9\,$\times$\,4.3 & -73.6 & \phantom{1}52 & 4.6\,$\times$\,2.7 & \phantom{-}84.9 & 154 & \phantom{1}6.5 & \phantom{-}13.5 & 26.0 & \phantom{1}9.4 & \phantom{-}13.3 & 25.5 \\
         \object{R~For}                 & 7.2\,$\times$\,4.0 & -76.8 & \phantom{1}68 & 4.5\,$\times$\,3.1 & \phantom{-}78.9 & 168 & 11.6 & \phantom{1}-2.4 & 34.8 & 20.0 & \phantom{1}-2.3 & 33.5 \\
         \object{R~Vol}                 & 6.3\,$\times$\,6.0 & -48.0 & \phantom{1}72 & 5.0\,$\times$\,3.6 & \phantom{1}-5.0 & 152 & 14.6 & -10.8 & 36.5 & 24.4 & -10.8 & 36.5 \\
         \object{RV~Aqr}                        & 6.5\,$\times$\,5.4 & -67.9 & \phantom{1}50 & 4.5\,$\times$\,3.0 & -73.3 & 153 & 26.1 & \phantom{-1}1.0 & 30.0 & 37.8 & \phantom{-1}1.3 & 30.5 \\
         \object{V688~Mon}              & 6.4\,$\times$\,4.6 & -87.1 & \phantom{1}76 & 4.6\,$\times$\,2.9 & -79.0 & 224 & 25.2 & \phantom{-1}3.0 & 28.0 & 33.7 & \phantom{-1}3.0 & 27.0 \\
         \object{V1259~Ori}             & 6.2\,$\times$\,4.6 & -79.5 & \phantom{1}77 & 5.1\,$\times$\,2.8 & -71.1 & 229 & 28.9 & \phantom{-}43.0 & 31.0 & 48.0 & \phantom{-}43.3 & 28.5 \\
         \hline \hline
   \end{tabular}}
   \tablefoot{Columns 2-7 give the imaging results: Full width at half-maximum beam-widths (major\,$\times$\,minor axis), $\theta$, and position angle, P.A., at the center frequency of $^{12}$CO spectral windows, as well as the rms noise level in both band 6 and 7, respectively, at a velocity resolution of 0.75 \,km\,s$^{-1}$. Columns 8-13 give the CO line parameters: Peak flux, center velocity, and total velocity width of the line profiles of both CO transitions presented in Figs~\ref{linesM_SR}--\ref{linesC_M}.}
   \label{fwhm_rms}
\end{table*}


\section{Detections of emission from molecules other than $^{12}$CO}

\begin{sidewaystable*}[htp]
\caption{Peak flux of detected molecular emission measured within a circular 10\arcsec~aperture centered on the M-type semi-regular and irregular stars. The spectral resolution for each spectral window is given in Sect.~\ref{obs}. The peak flux error is on the order of 20\%.}
\begin{center}
\resizebox{\textwidth}{!}{\begin{tabular}{|l|c|c|c|c|c|c|c|c|c|c|c|c|c|c|c|}
\hline \hline
Line & Frequency & \multicolumn{14}{c}{Peak flux [Jy]} \\
& [GHz] & L$_{2}$~Pup & W~Hya & T~Mic & Y~Scl & V1943~Sgr & BK~Vir & V~Tel & SU~Vel & UY~Cet & SV~Aqr & SW~Vir & CW~Cnc & RT~Vir & R~Crt \\
\hline \hline
\noalign{\vskip 1mm}
\multicolumn{16}{l}{ {\it Band 6, spectral window 1}} \\
\noalign{\vskip 1mm}
SiO ($v$=1, $J$=5-4)  & 215.596 & 1.5 & 96 & 1.6 & - & 1.2 & 0.9 & 0.2 & 0.1 & - & - & 1.4 & 0.1 & 0.5 & 4.1 \\ 
$^{34}$SO (6$ _{5}$-5$_{4}$) & 215.840 & 0.1 & 0.2 & - & - & - & - & - & - & - & - & - & - & 0.1 & 0.2 \\ 
SO$_{2}$ (22$_{2,20}$-22$_{1,21}$) & 216.643 & 0.7 & 0.6 & 0.1 & - & - & - & - & - & - & - & - & - & 0.3 & 0.2 \\ 
H$_{2}$S (2$_{2,0}$-2$_{1,1}$) & 216.710 & - & - & - & - & - & - & - & - & - & - & - & - & - & - \\ 
SiS ($v$=1, $J$=12-11) & 216.758 & 0.1 & 0.2 & - & - & - & - & - & - & - & - & - & - & - & - \\ 
SiO (5-4) & 217.105 & 16.5 & 45 & 7.7 & 2 & 9.8 & 6.1 & 3.2 & 2.6 & 4 & 2.2 & 18.5 & 3.6 & 12 & 19 \\ 
\noalign{\vskip 1mm}
\multicolumn{16}{l}{ {\it Band 6, spectral window 2}} \\
\noalign{\vskip 1mm}
SiS (12-11) & 217.818 & 0.2 & - & - & - & - & - & - & - & - & - & - & - & - & - \\ 
\noalign{\vskip 1mm}
\multicolumn{16}{l}{ {\it Band 6, spectral window 4}} \\
\noalign{\vskip 1mm}
$^{29}$SiS (13-12) & 231.627 & - & - & - & - & - & - & - & - & - & - & - & - & - & - \\ 
SO$_{2}$ ($v_{2}$=1, 14$_{3,11}$-14$_{2,12}$) & 231.981 & 0.1 & 0.1 & - & - & - & - & - & - & - & - & - & - & - & - \\
Si$^{33}$S (13-12) & 232.629 & - & - & - & - & - & - & - & - & - & - & - & - & - & - \\
H$_{2}$O ($v_{2}$=1, 5$_{5,0}$-6$_{4,3}$) & 232.687 & 0.2 & 0.2 & - & - & - & - & - & - & - & - & - & - & - & - \\
\noalign{\vskip 1mm}
\multicolumn{16}{l}{ {\it Band 7, spectral window 1}} \\
\noalign{\vskip 1mm}
$^{13}$CO (3-2) & 330.588 & 6.4 & 1.8 & 0.8 & 1 & 0.9 & 1.8 & 1.1 & 1.6 & 1.6 & 0.5 & 5.5 & 0.7 & 3.0 & 4.0 \\
Na$^{37}$Cl (27-26) & 330.805 & - & - & - & - & - & - & - & - & - & - & - & - & - & - \\
\noalign{\vskip 1mm}
\multicolumn{16}{l}{ {\it Band 7, spectral window 2}} \\
\noalign{\vskip 1mm}
SO$_{2}$ (11$_{6,6}$-12$_{5,7}$) & 331.580 & 0.2 & - & - & - & - & - & - & - & - & - & - & - & - & - \\
SO$_{2}$ (21$_{2,20}$-21$_{1,21}$) & 332.091 & 1.4 & 0.7 & - & - & - & - & - & - & - & - & - & - &  0.5 & 0.5 \\
SO$_{2}$ (4$_{3,1}$-3$_{2,2}$) & 332.505 & 1.4 & 0.6 & - & - & - & - & - & - & - & - & - & - &  0.8 &  0.8 \\
$^{30}$SiS (19-18) & 332.550 & - & - & - & - & - & - & - & - & - & - & - & - & - & - \\
\noalign{\vskip 1mm}
\multicolumn{16}{l}{ {\it Band 7, spectral window 3}} \\
\noalign{\vskip 1mm}
SO$ _{2}$ (34$_{3,31}$-34$_{2,32}$) & 342.762 & 0.2 & 0.5 & - & - & - & - & - & - & - & - & - & - &  0.3 & 0.2 \\
CS (7-6) & 342.883 & - & 0.5 & - & - & - & - & - & - & - & - & - & - & 0.2 & - \\
$^{29}$SiO (8-7) & 342.981 & 1.3 & 30 & 6.3 & 1 & 4.4 & 4.7 & 2.5 & 2.6 & 2.8 & 0.2 & 10 & 2.0 & 7.0 & 8.0 \\
SiS ($v$=1, $J$=19-18) & 343.101 & 0.04 & - & - & - & - & - & - & - & - & - & - & - & - & - \\
SO ($^{3} \Sigma$ $v$=1, 9$_{8}$-8$_{7}$) & 343.829 & - & 0.5 & - & - & - & - & - & - & - & - & - & - & - & - \\
SO (8$_{8}$-7$_{7}$) & 344.311 & 1.4 & 6.5 & 1.9 & 0.2 & 1.2 & 1.4 & 0.4 & 0.4 & 1.2 & 0.1 & 1.8 & 0.5 & 3.7 & 3.0 \\
\noalign{\vskip 1mm}
\multicolumn{16}{l}{ {\it Band 7, spectral window 4}} \\
\noalign{\vskip 1mm}
H$^{13}$CN (4-3) & 345.340 & 3.4 & 8.7 & - & - & 0.4 & - & - & - & 0.3 & - & 0.1 & - & 2.0 & 2.2 \\
\hline \hline
\end{tabular}}
\end{center}
\label{other_MSR}
\end{sidewaystable*}%


\begin{table*}[htp]
\caption{Peak flux of detected molecular emission measured within a circular 10\arcsec~aperture centered on the M-type Mira stars. The spectral resolution for each spectral window is given in Sect.~\ref{obs}. The peak flux error is on the order of 20\%.}
\begin{tabular}{|l|c|c|c|c|c|c|c|c|}
\hline \hline
Line & Frequency & \multicolumn{7}{c}{Peak flux [Jy]} \\
 & [GHz] & R Leo & R  Hya & R Hor & RR Aql & IRC-10529 & WX Psc & IRC+10365 \\ 
 \hline \hline
\noalign{\vskip 1mm}
\multicolumn{9}{l}{ {\it Band 6, spectral window 1}} \\
\noalign{\vskip 1mm}
SiO ($v$=1, $J$=5-4)  & 215.596 & 36 & 6.0 & 0.5 & 0.3 & 2.5 & 17 & 4.0 \\ 
$^{34}$SO (6$ _{5}$-5$_{4}$) & 215.840 & 0.1 & - & - & 0.1 & 0.1 & - & 0.1 \\ 
SO$_{2}$ (22$_{2,20}$-22$_{1,21}$) & 216.643 & 5.7 & - & - & 0.2 & - & - & - \\ 
H$_{2}$S (2$_{2,0}$-2$_{1,1}$) & 216.710 & - & - & - & - & 0.3 & 0.2 & - \\ 
SiS ($v$=1, $J$=12-11) & 216.758 & - & - & - & - & - & 0.1 & - \\ 
SiO (5-4) & 217.105 & 36 & 22.4 & 6.7 & 4.2 & 5.7 & 9.6 & 12 \\ 
\noalign{\vskip 1mm}
\multicolumn{9}{l}{ {\it Band 6, spectral window 2}} \\
\noalign{\vskip 1mm}
SiS (12-11) & 217.818 & - & - & - & - & 1.8 & 2.9 & 0.9 \\ 
\noalign{\vskip 1mm}
\multicolumn{9}{l}{ {\it Band 6, spectral window 4}} \\
\noalign{\vskip 1mm}
$^{29}$SiS (13-12) & 231.627 & - & - & - & - & 0.2 & 0.4 & 0.1 \\ 
SO$_{2}$ ($v_{2}$=1, 14$_{3,11}$-14$_{2,12}$) & 231.981 & - & - & - & - & - & - & - \\ 
Si$^{33}$S (13-12) & 232.629 & - & - & - & - & 0.1 & - & - \\ 
H$_{2}$O ($v_{2}$=1, 5$_{5,0}$-6$_{4,3}$) & 232.687 & - & - & - & - & - & 0.3 & 0.1 \\ 
\noalign{\vskip 1mm}
\multicolumn{9}{l}{ {\it Band 7, spectral window 1}} \\
\noalign{\vskip 1mm}
$^{13}$CO (3-2) & 330.588 & 3.2 & 2.0 & 4.0 & 0.5 & 7.0 & 7.0 & 6.0 \\ 
Na$^{37}$Cl (27-26) & 330.805 & - & - & - & - & - & 0.1 & - \\ 
\noalign{\vskip 1mm}
\multicolumn{9}{l}{ {\it Band 7, spectral window 2}} \\
\noalign{\vskip 1mm}
SO$_{2}$ (11$_{6,6}$-12$_{5,7}$) & 331.580 & - & - & - & - & - & - & - \\ 
SO$_{2}$ (21$_{2,20}$-21$_{1,21}$) & 332.091 & 0.4 & - & - & 0.3 & - & - & - \\ 
SO$_{2}$ (4$_{3,1}$-3$_{2,2}$) & 332.505 & 0.4 & - & - & 1.9 & 1.2 & 1.2 & 0.8 \\ 
$^{30}$SiS (19-18) & 332.550 & - & - & - & - & 0.3 & 0.5 & 0.2 \\ 
\noalign{\vskip 1mm}
\multicolumn{9}{l}{ {\it Band 7, spectral window 3}} \\
\noalign{\vskip 1mm}
SO$ _{2}$ (34$_{3,31}$-34$_{2,32}$) & 342.762 & 0.3 & - & - & - & - & - & - \\ 
CS (7-6) & 342.883 & 0.7 & 0.2 & - & - & 0.7 & 1.4 & 1.0 \\ 
$^{29}$SiO (8-7) & 342.981 & 21.5 & 12.5 & 7.0 & 2.0 & 1.3 & 2.7 & 3.1 \\ 
SiS ($v$=1, $J$=19-18) & 343.101 & - & - & - & - & 0.3 & 0.4 & - \\ 
SO ($^{3} \Sigma$ $v$=1, 9$_{8}$-8$_{7}$) & 343.829 & - & - & - & - & - & - & - \\ 
SO (8$_{8}$-7$_{7}$) & 344.311 & 4 & 1.4 & 1.4 & 1.4 & 0.4 & 0.4 & 0.8 \\ 
\noalign{\vskip 1mm}
\multicolumn{9}{l}{ {\it Band 7, spectral window 4}} \\
\noalign{\vskip 1mm}
H$^{13}$CN (4-3) & 345.340 & 7.7 & 0.9 & - & 1.4 & 0.6 & 1.7 & 1.6 \\ 
\hline \hline
\end{tabular}
\label{other_MM}
\end{table*}%


\begin{sidewaystable*}[htp]
\caption{Peak flux of detected molecular emission measured within a circular 10\arcsec~aperture centered on the C-type semi-regular and irregular stars. The spectral resolution for each spectral window is given in Sect.~\ref{obs}. The peak flux error is on the order of 20\%.}
\begin{center}
\resizebox{\textwidth}{!}{\begin{tabular}{|l|c|c|c|c|c|c|c|c|c|c|c|c|c|c|c|}
\hline \hline
Line & Frequency & \multicolumn{14}{c}{Peak flux [Jy]} \\
 & [GHz] & TW~Oph & NP~Pup & TW~Hor & T~Ind & RT~Cap & AQ~Sgr & U~Hya & W~Ori & V~Aql & Y~Pav & X~Vel & Y~Hya & SS~Vir & W~CMa \\
\hline \hline
\noalign{\vskip 1mm}
\multicolumn{16}{l}{ {\it Band 6, spectral window 1}} \\
\noalign{\vskip 1mm}
SiO (5-4) & 217.105 & 0.3 & - & - & - & 0.1 & 0.1 & 0.1 & 3.6 & 0.4 & 0.1 & 0.3 & 0.4 & - & - \\ 
$^{13}$CN ($N$=2-1)\tablefootmark{a} & 217.287 &  &  &  &  &  &  &  0.1 &  &  &  &  &  &  &   \\
$^{13}$CN ($N$=2-1)\tablefootmark{b} & 217.315 & - & - & - & - & - & - & 0.1 & - & 0.1 & - & - & - & - & - \\
\noalign{\vskip 1mm}
\multicolumn{9}{l}{ {\it Band 6, spectral window 2}} \\
\noalign{\vskip 1mm}
$^{13}$CN ($N$=2-1)\tablefootmark{c}  & 217.467 & - & - & - & - & - & - &  0.1 & - & - & - & - & - & - & -\\
SiS (12-11) & 217.818 & - & - & - & - & - & - & - & - & - & - & - & - & - & - \\
HC$_{3}$N (24-23) & 218.325 & 0.1 & - & - & - & - & - & - & 0.2 & 0.1 & - & 0.1 & 0.2 & - & - \\
C$_{4}$H ($N$=23-22)\tablefootmark{d} & 218.837 & - & - & - & - & - & - & - & - & - & - & - & - & - & - \\
C$_{4}$H ($N$=23-22)\tablefootmark{e} & 218.875 & - & - & - & - & - & - & - & - & - & - & - & - & - & - \\
\noalign{\vskip 1mm}
\multicolumn{9}{l}{ {\it Band 6, spectral window 4}} \\
\noalign{\vskip 1mm}
$^{13}$CS (5-4) & 231.221 & - & - & - & - & - & - & - & 0.1 & 0.1 & - & 0.1 & 0.1 & - & - \\
$^{29}$SiS (13-12) & 231.627 & - & - & - & - & - & - & - & - & - & - & - & - & - & - \\
SiC$_{2}$ (10$_{2,9}$-9$_{2,8}$) & 232.534 & 0.2 & - & - & - & - & - & - & 0.3 & 0.4 & - & 0.2 & 0.6 & - & - \\
\noalign{\vskip 1mm}
\multicolumn{9}{l}{ {\it Band 7, spectral window 1}} \\
\noalign{\vskip 1mm}
$^{13}$CO (3-2) & 330.588 & - & - & 0.8 & - & - & 0.2 & 4.5 & 0.4 & 0.4 & 0.7 & - & - & - & 0.3 \\
SiC$_{2}$ (14$_{6,9}$-13$_{6,8}$) & 330.870 & 0.4 & - & - & - & - & - & - & 0.4 & 0.5 & - & 0.3 & 0.7 & - & - \\
\noalign{\vskip 1mm}
\multicolumn{9}{l}{ {\it Band 7, spectral window 3}} \\
\noalign{\vskip 1mm}
SiC$_{2}$ (15$_{2,14}$-14$_{2,13}$) & 342.805 & 0.3 & - & - & - & - & - & - & 0.1 & 0.5 & - & 0.4 & 0.3 & - & - \\
CS (7-6) & 342.883 & 3.4 & - & - & - & 1.6 & 0.3 & 1.1 & 5.0 & 6.2 & - & 3.6 & 4.5 & - & - \\
$^{29}$SiO (8-7) & 342.981 & - & - & - & - & - & - & - & - & 0.1 & - & - & - & - & - \\
\noalign{\vskip 1mm}
\multicolumn{9}{l}{ {\it Band 7, spectral window 4}} \\
\noalign{\vskip 1mm}
H$^{13}$CN (4-3) 345.340 & 0.8 & - & - & - & 0.8 & 0.3 & 4.0 & 1.9 & 1.5 & 0.5 & 0.3 & 1.2 & 0.1 & - \\
\hline \hline
\end{tabular}}
\end{center}
\tablefoot{
\tablefoottext{a} {J=5/2-3/2, F1=2-1, F=2-2};
\tablefoottext{b} {J=5/2-3/2, F1=2-2, F=2-3};
\tablefoottext{c} {J=5/2-3/2, F1=3-2, F=4-3 };
\tablefoottext{d} {J=47/2-45/2, F=23-22};
\tablefoottext{e} {J=45/2-43/2, F=23-22}
   }
\label{other_CSR}
\end{sidewaystable*}%


\begin{table*}[htp]
\caption{Peak flux of detected molecular emission measured within a circular 10\arcsec~aperture centered on the C-type Mira stars. The spectral resolution for each spectral window is given in Sect.~\ref{obs}. The peak flux error is on the order of 20\%.}
\resizebox{\textwidth}{!}{\begin{tabular}{|l|c|c|c|c|c|c|c|c|}
\hline \hline
 Line & Frequency & \multicolumn{7}{c}{Peak flux [Jy]} \\
  & [GHz] & R Lep & CZ Hya & R For & R Vol & RV Aqr & V688 Mon & V1259 Ori \\ 
\hline \hline
\noalign{\vskip 1mm}
\multicolumn{9}{l}{ {\it Band 6, spectral window 1}} \\
\noalign{\vskip 1mm}
SiO (5-4) & 217.105 & 1.9 & 0.3 & 2 & 1.9 & 3.2 & 1.0 & 0.8 \\ 
$^{13}$CN ($N$=2-1)\tablefootmark{a} & 217.287 & 0.1 & - & 0.1 & - & 0.1 & - & - \\ 
$^{13}$CN ($N$=2-1)\tablefootmark{b} & 217.315 & 0.1 & - & 0.2 & 0.1 & 0.2 & - & - \\
\noalign{\vskip 1mm}
\multicolumn{9}{l}{ {\it Band 6, spectral window 2}} \\
\noalign{\vskip 1mm}
$^{13}$CN ($N$=2-1)\tablefootmark{c}  & 217.467 & 0.1 & - & 0.1 & 0.1 & 0.1 & - & - \\ 
SiS (12-11) & 217.818 & - & - & 0.2 & - & 0.7 & 0.7 & 1.5 \\
HC$_{3}$N (24-23) & 218.325 & 0.2 & - & 0.4 & 0.1 & 0.1 & 0.4 & 0.1 \\
C$_{4}$H ($N$=23-22)\tablefootmark{d} & 218.837 & - & - & - & - & - & 0.2 & 0.4 \\
C$_{4}$H ($N$=23-22)\tablefootmark{e} & 218.875 & - & - & - & - & - & 0.2 & 0.6 \\
\noalign{\vskip 1mm}
\multicolumn{9}{l}{ {\it Band 6, spectral window 4}} \\
\noalign{\vskip 1mm}
$^{13}$CS (5-4) & 231.221 & 0.2 & - & 0.2 & 0.2 & 0.3 & 0.1 & 0.2 \\ 
$^{29}$SiS (13-12) & 231.627 & - & - & - & - & 0.1 & 0.1 & 0.1 \\ 
SiC$_{2}$ (10$_{2,9}$-9$_{2,8}$) & 232.534 & 0.1 & - & 0.3 & 0.2 & 0.3 & 0.4 & 0.4 \\
\noalign{\vskip 1mm}
\multicolumn{9}{l}{ {\it Band 7, spectral window 1}} \\
\noalign{\vskip 1mm}
$^{13}$CO (3-2) & 330.588 & 0.9 & 0.6 & 0.8 & 1.4 & 2.5 & 1.6 & - \\ 
SiC$_{2}$ (14$_{6,9}$-13$_{6,8}$) & 330.870 & - & - & 0.3 & 0.2 & 0.3 & 0.5 & - \\
\noalign{\vskip 1mm}
\multicolumn{9}{l}{ {\it Band 7, spectral window 3}} \\
\noalign{\vskip 1mm}
SiC$_{2}$ (15$_{2,14}$-14$_{2,13}$) & 342.805 & - & - & 0.1 & - & - & - & - \\
CS (7-6) & 342.883 & 5.6 & 0.5 & 6.8 & 6.5 & 11 & 5.6 & - \\
$^{29}$SiO (8-7) & 342.981 & 0.1 & - & 0.5 & 0.3 & 0.5 & - & - \\
\noalign{\vskip 1mm}
\multicolumn{9}{l}{ {\it Band 7, spectral window 4}} \\
\noalign{\vskip 1mm}
H$^{13}$CN (4-3) & 345.340 & 3.1 & 0.4 & 2.3 & 2.5 & 5 & 4 & - \\ 
\hline \hline
\end{tabular}}
\tablefoot{
\tablefoottext{a} {J=5/2-3/2, F1=2-2, F=2-3};
\tablefoottext{b} {J=5/2-3/2, F1=2-1, F=2-2};
\tablefoottext{c} {J=5/2-3/2, F1=3-2, F=4-3 };
\tablefoottext{d} {J=47/2-45/2, F=23-22};
\tablefoottext{e} {J=45/2-43/2, F=23-22}}
\label{other_CM}
\end{table*}%


\section{Line profiles}

\begin{figure*}[t]
\includegraphics[height=4.5cm]{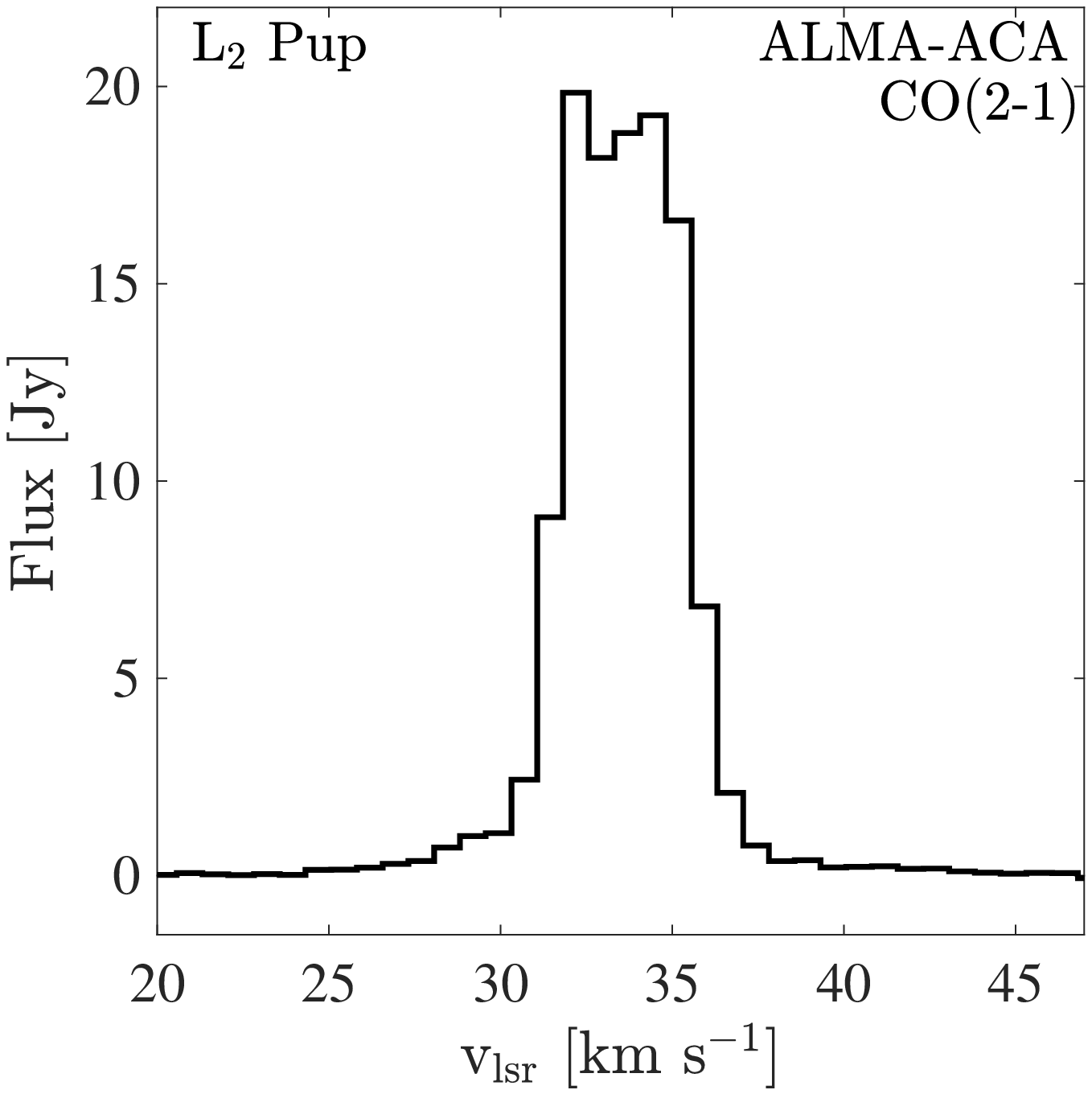}
\includegraphics[height=4.5cm]{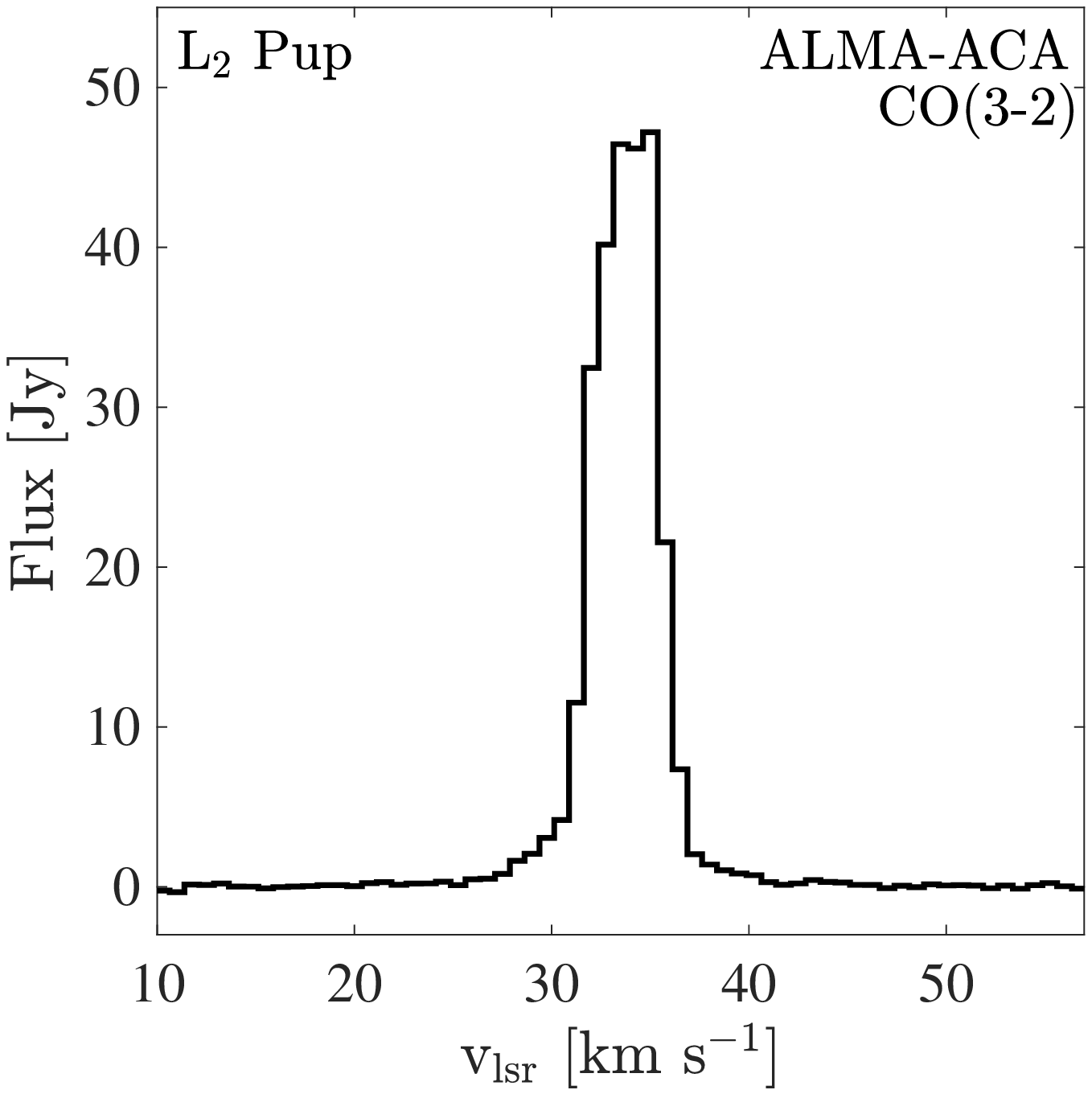}
\includegraphics[height=4.5cm]{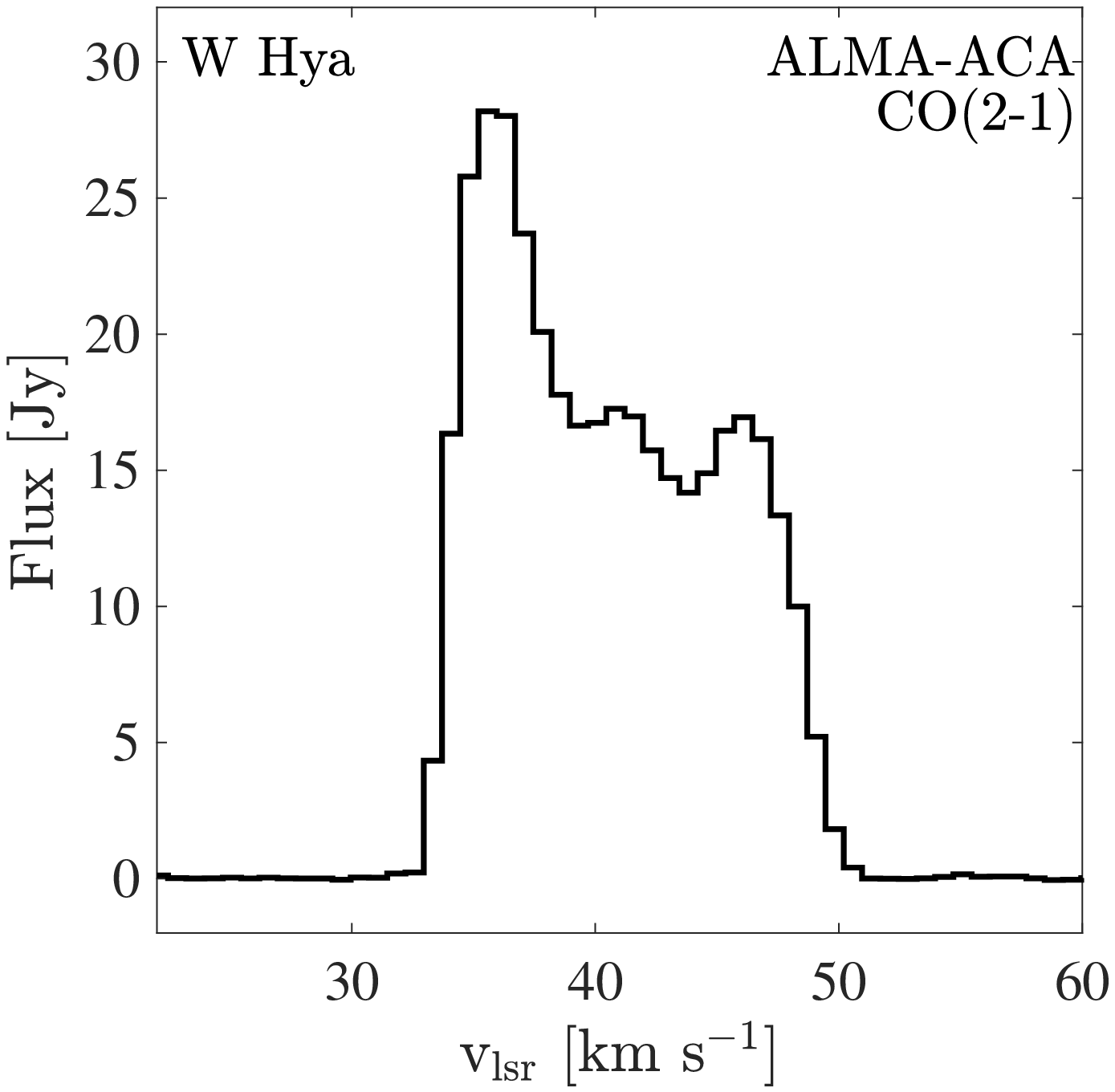}
\includegraphics[height=4.5cm]{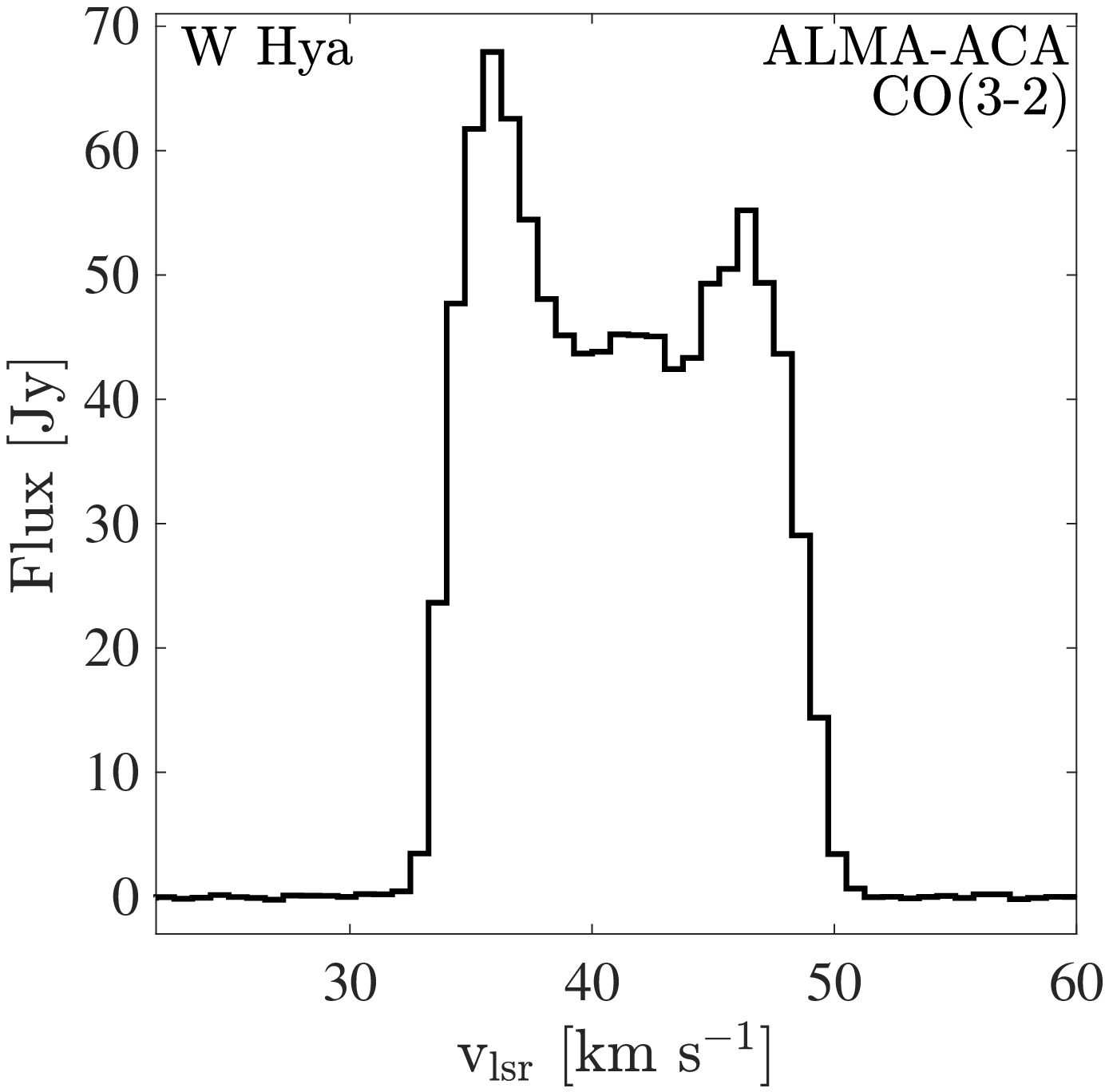}

\includegraphics[height=4.5cm]{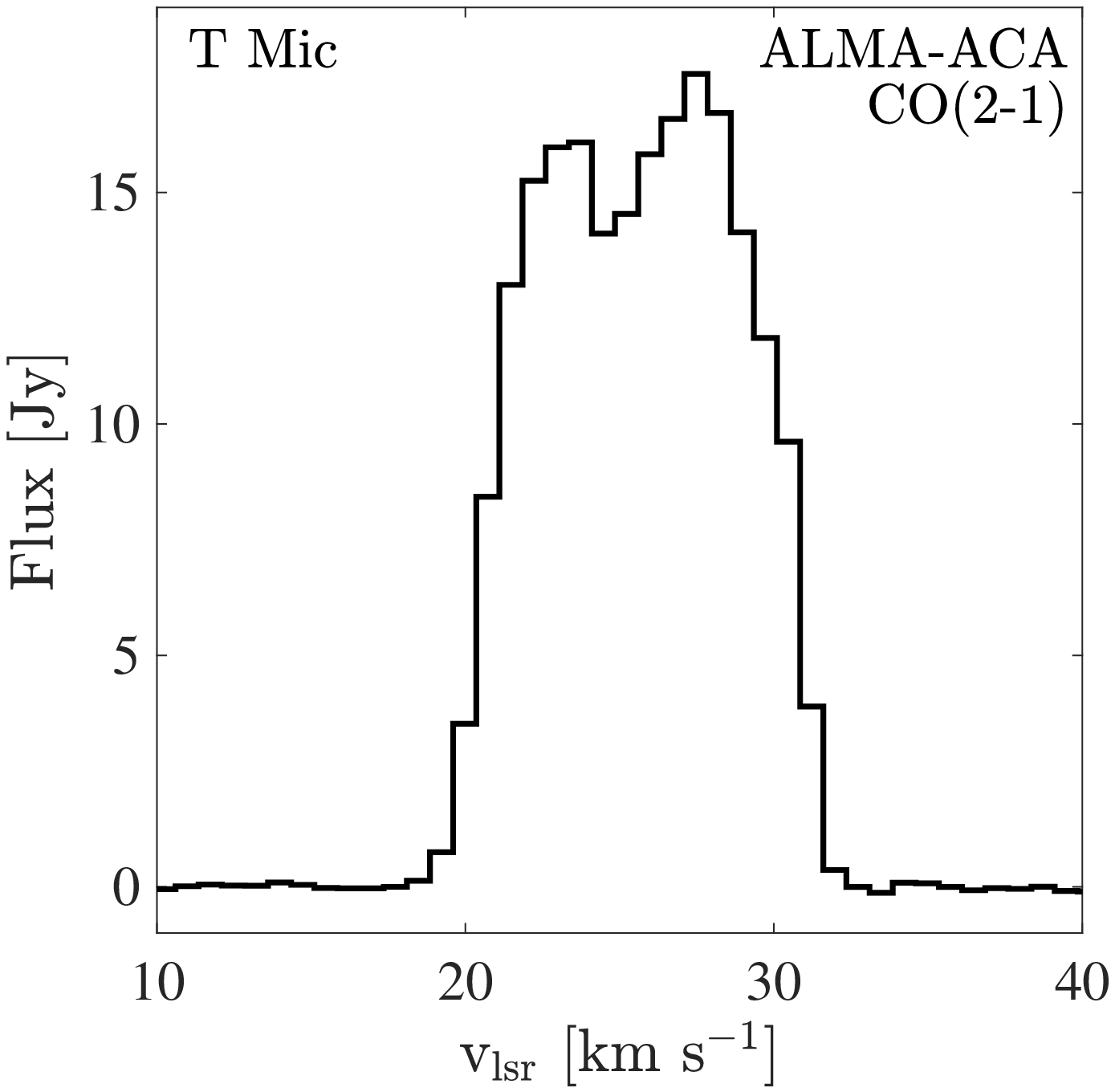}
\includegraphics[height=4.5cm]{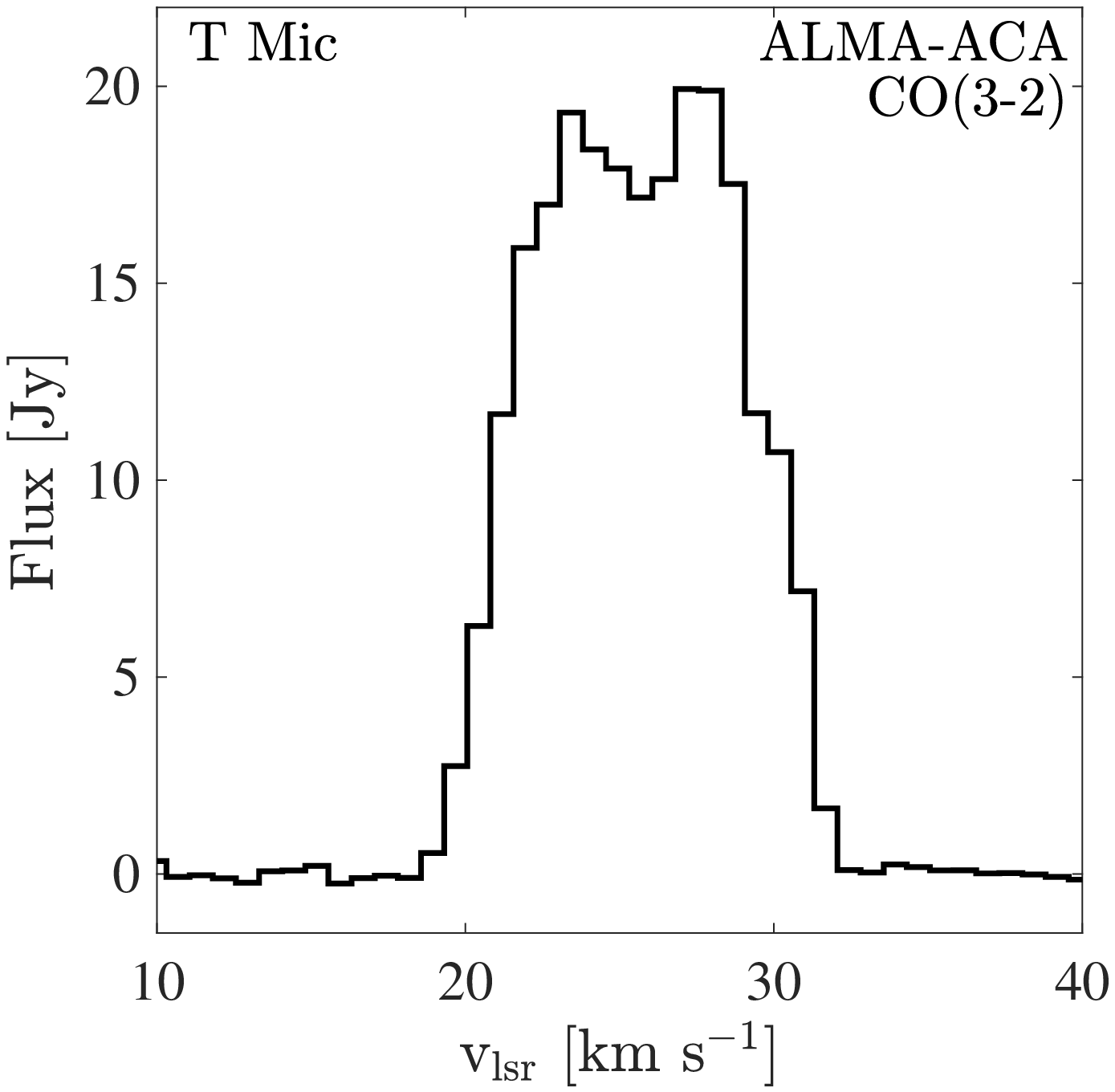}
\includegraphics[height=4.5cm]{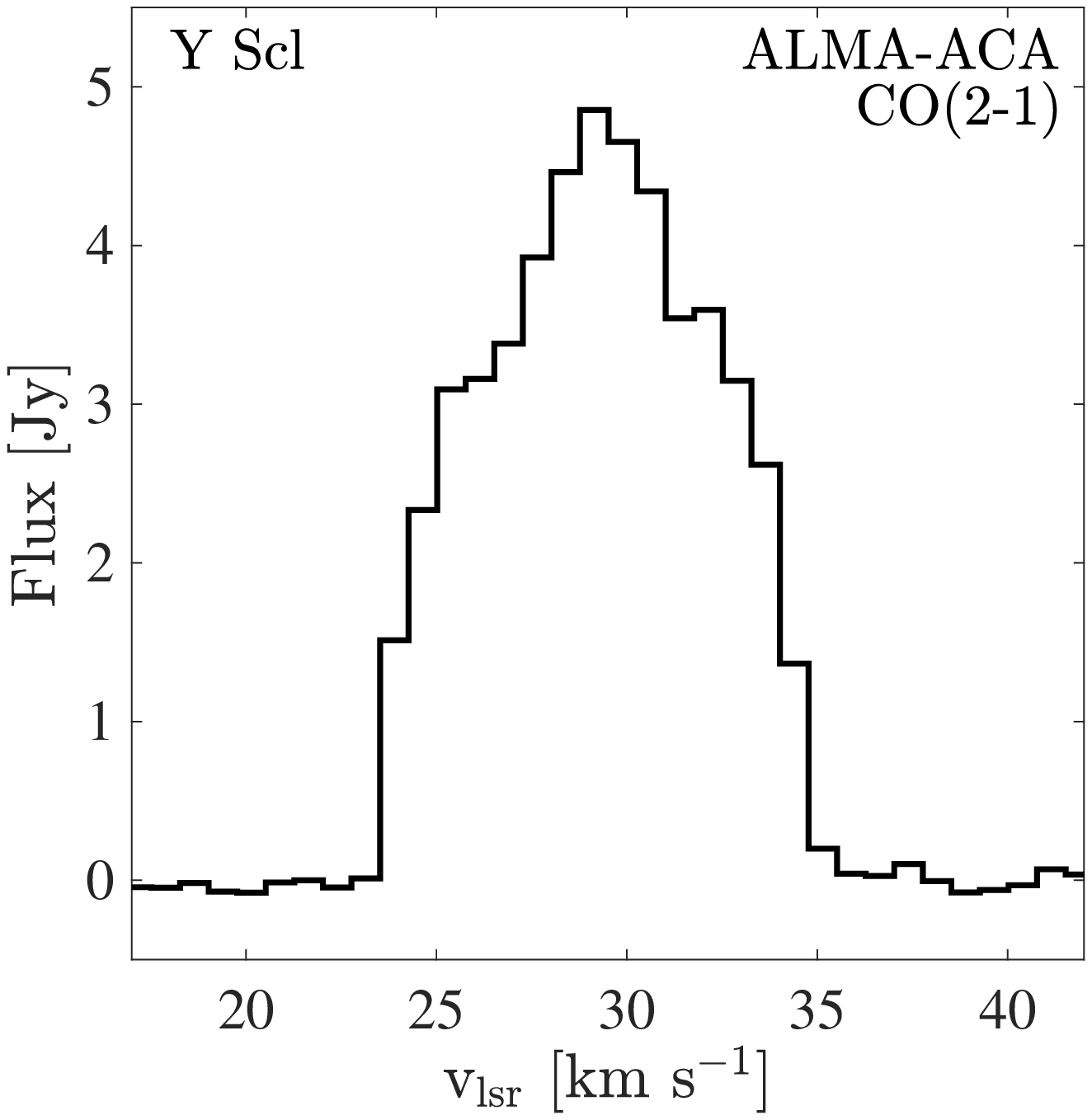}
\includegraphics[height=4.5cm]{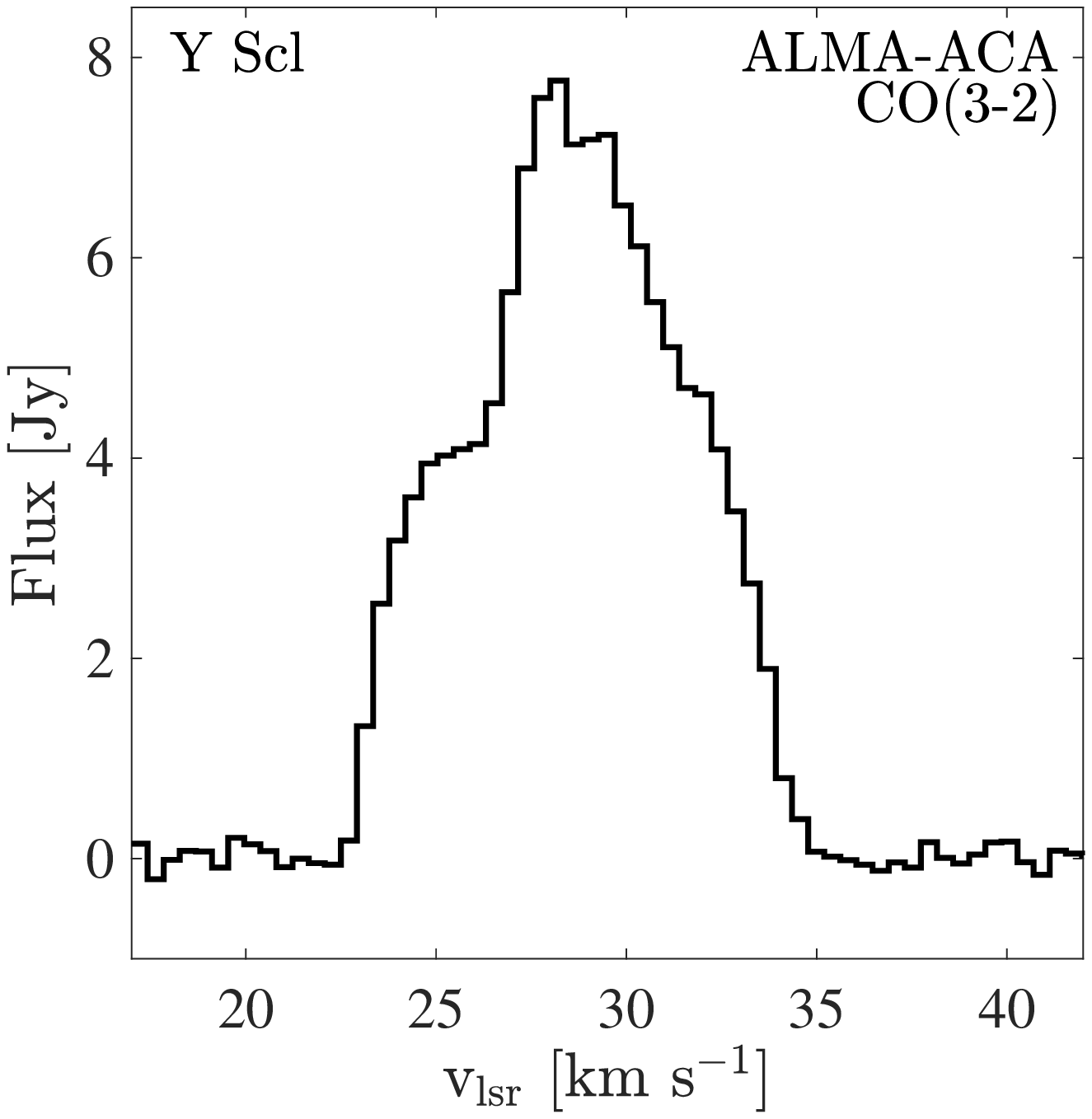}

\includegraphics[height=4.5cm]{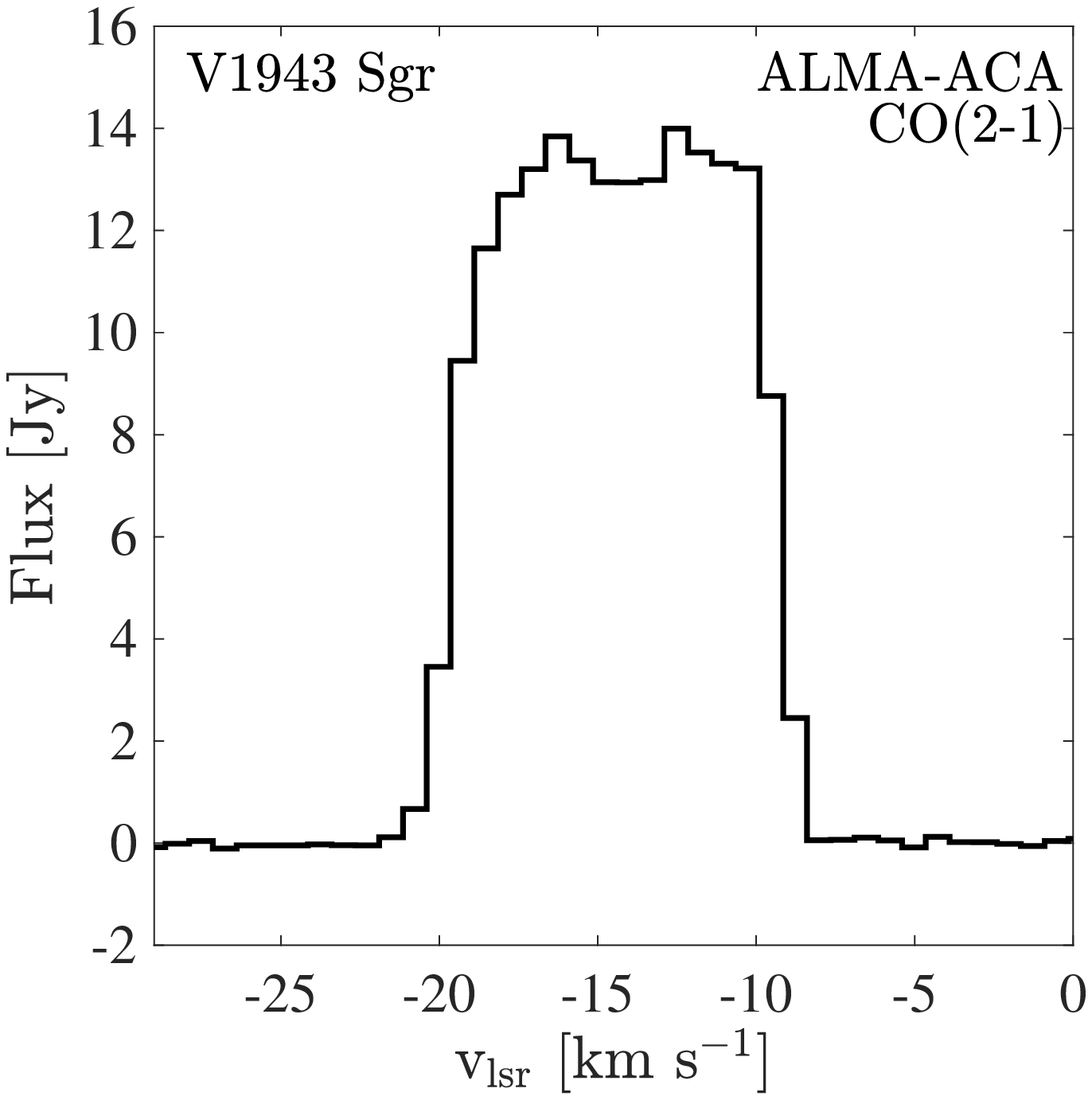}
\includegraphics[height=4.5cm]{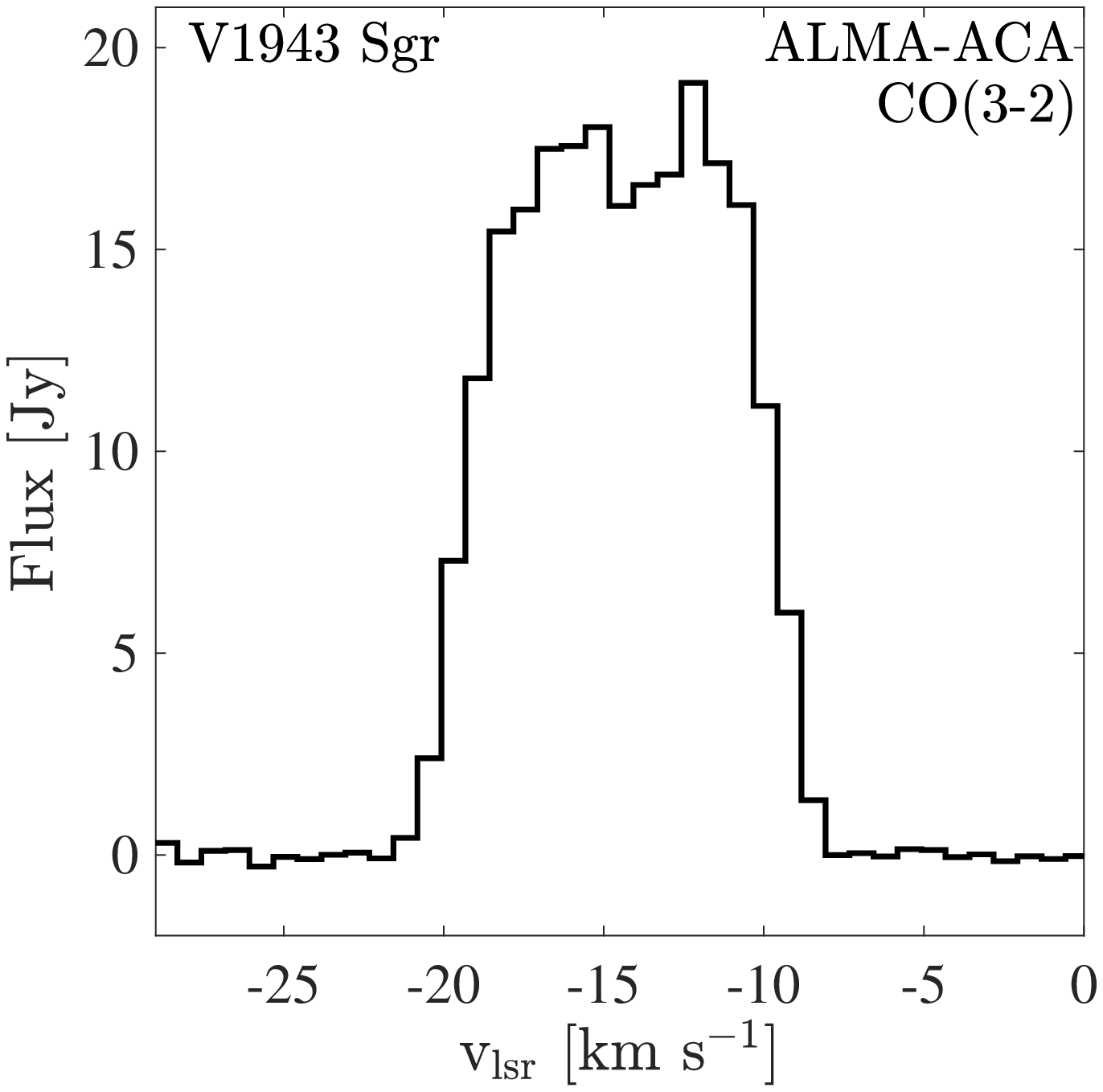}
\includegraphics[height=4.5cm]{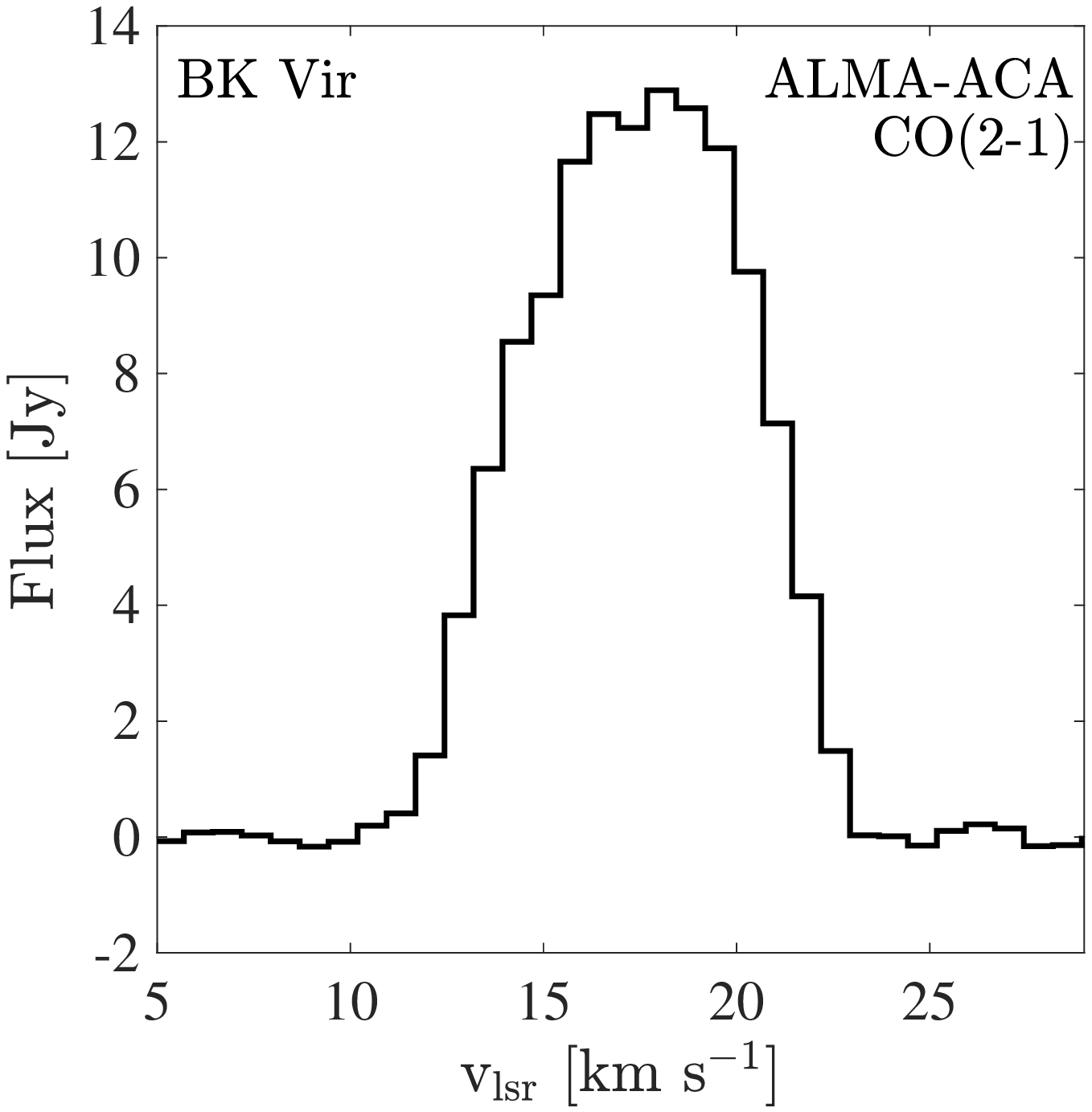}
\includegraphics[height=4.5cm]{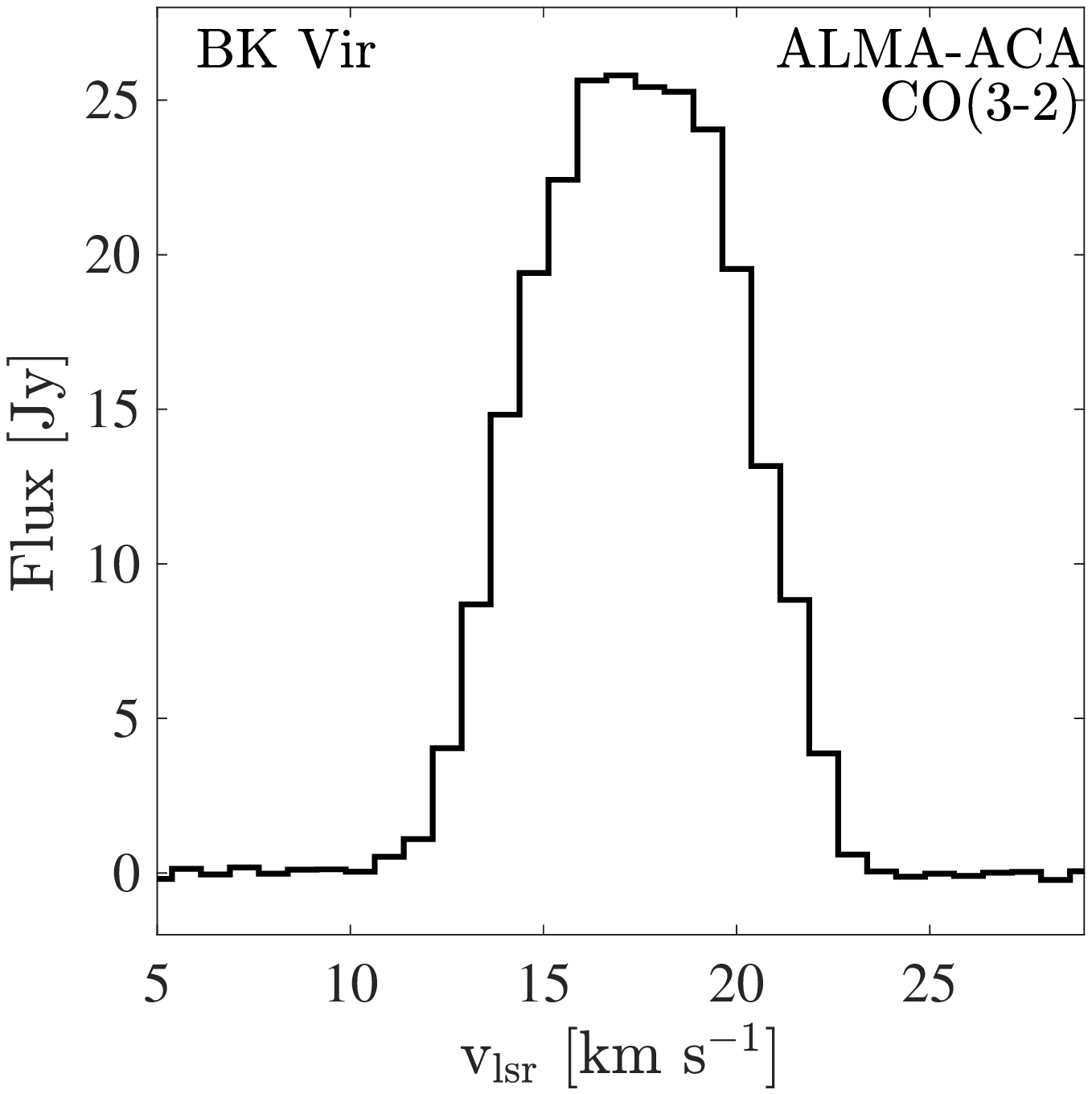}

\includegraphics[height=4.5cm]{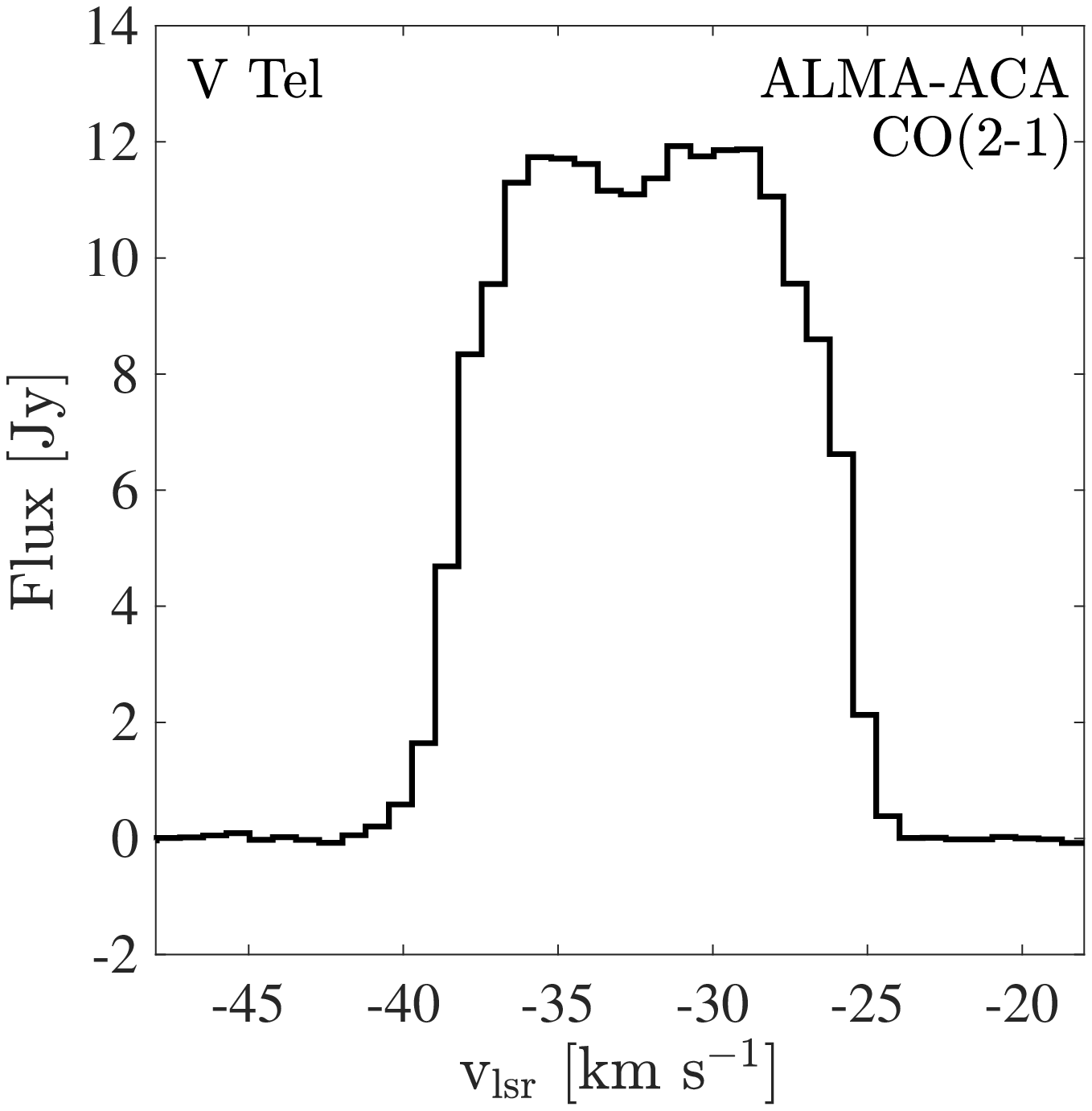}
\includegraphics[height=4.5cm]{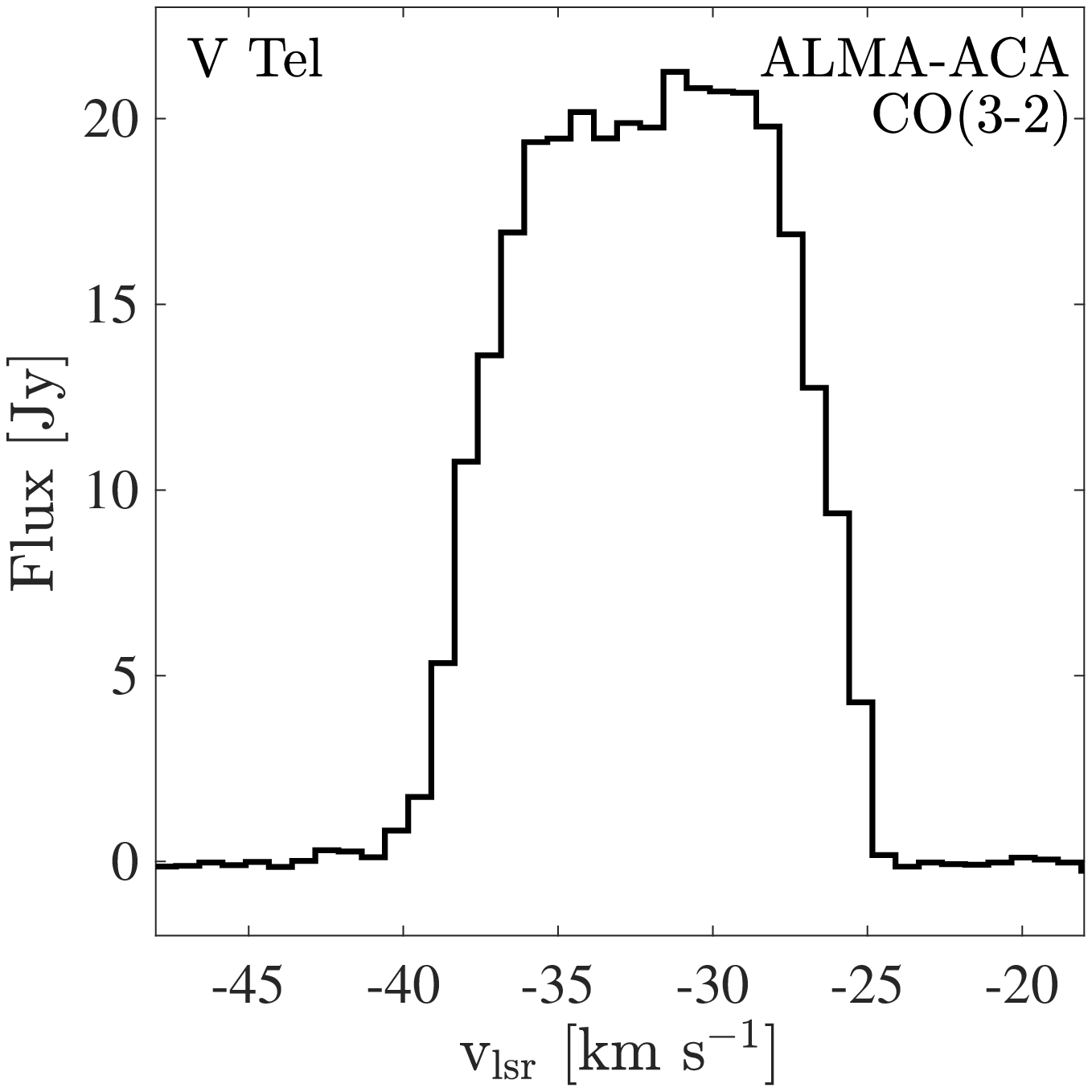}
\includegraphics[height=4.5cm]{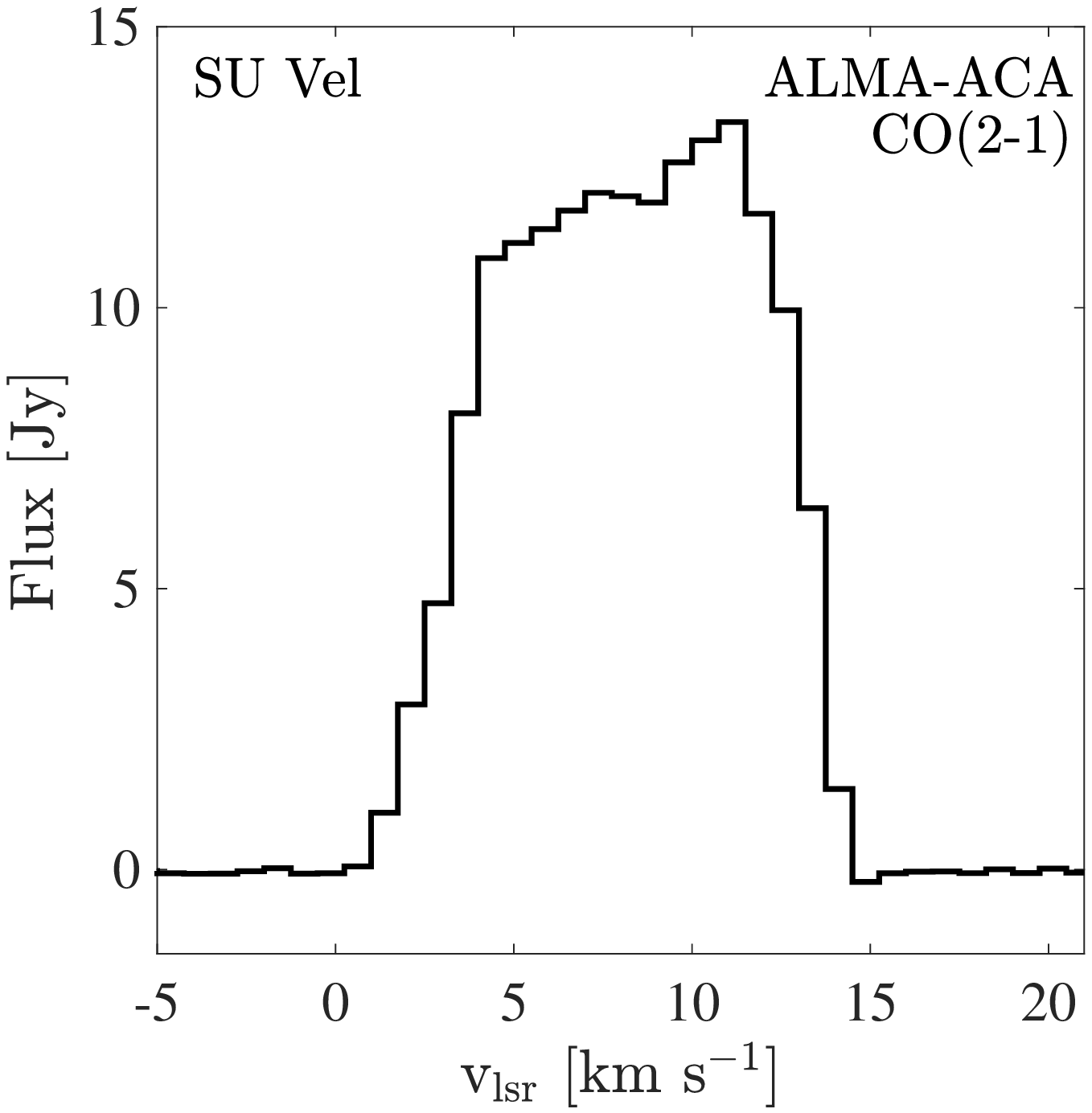}
\includegraphics[height=4.5cm]{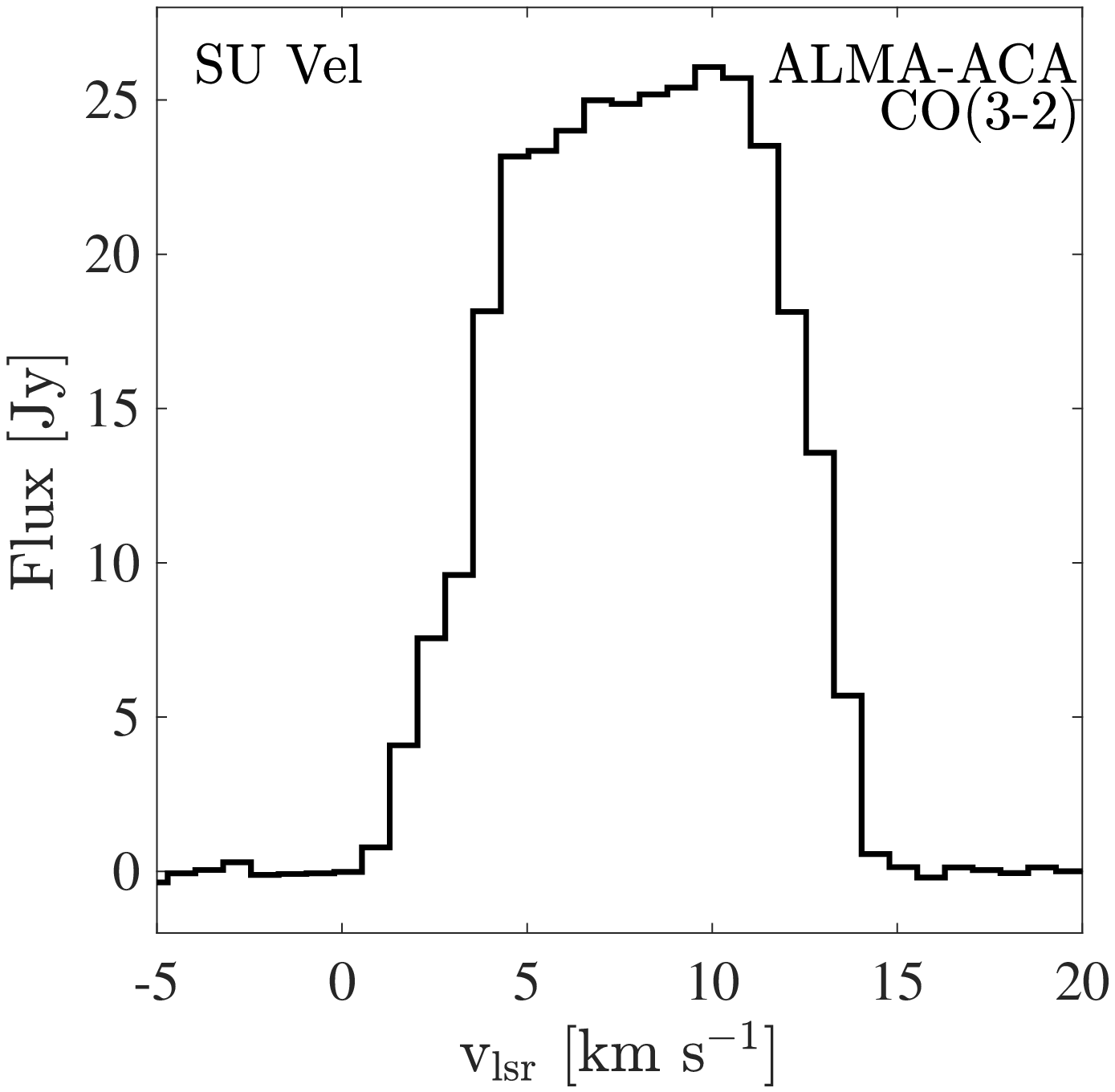}

\includegraphics[height=4.5cm]{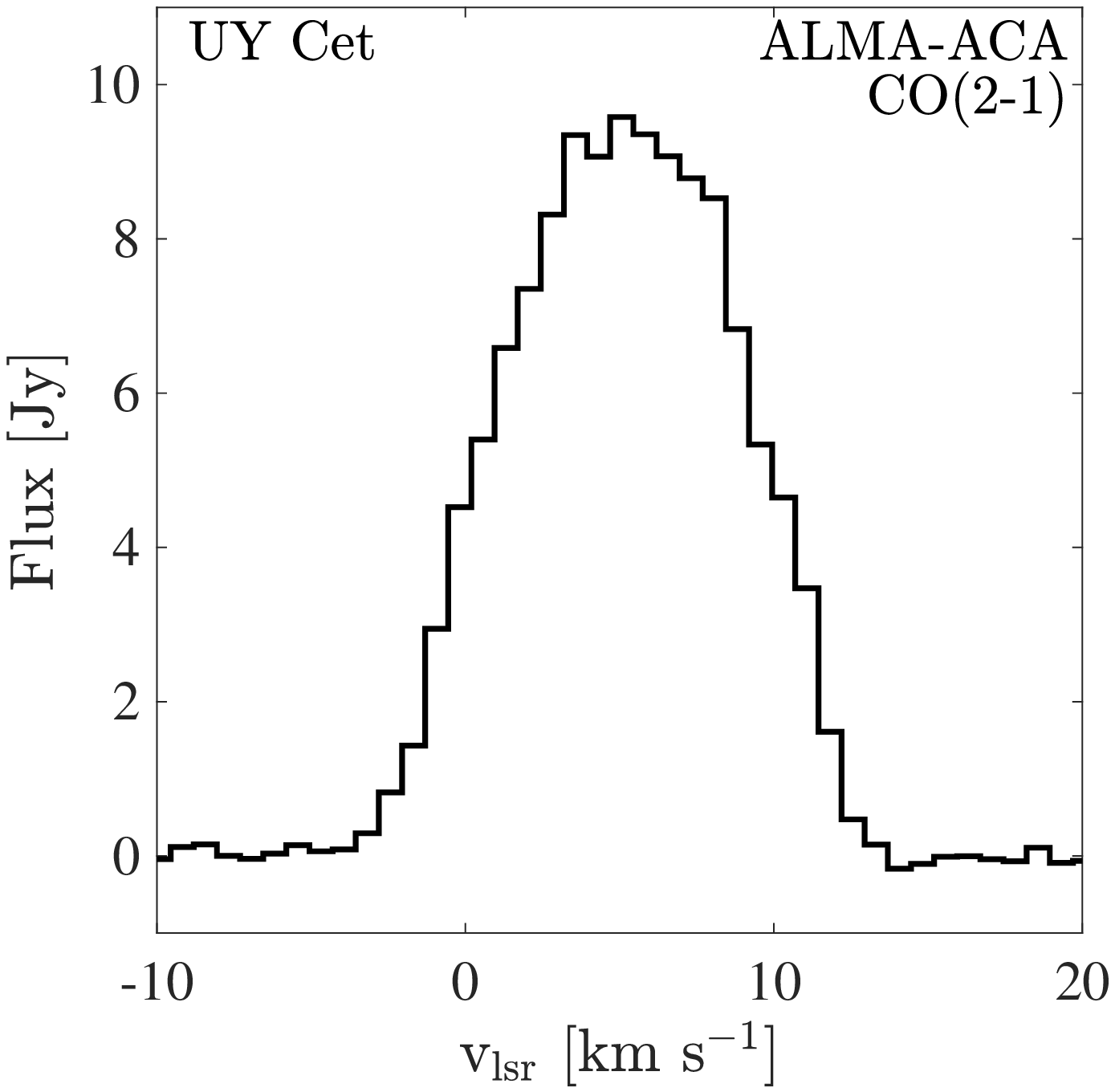}
\includegraphics[height=4.5cm]{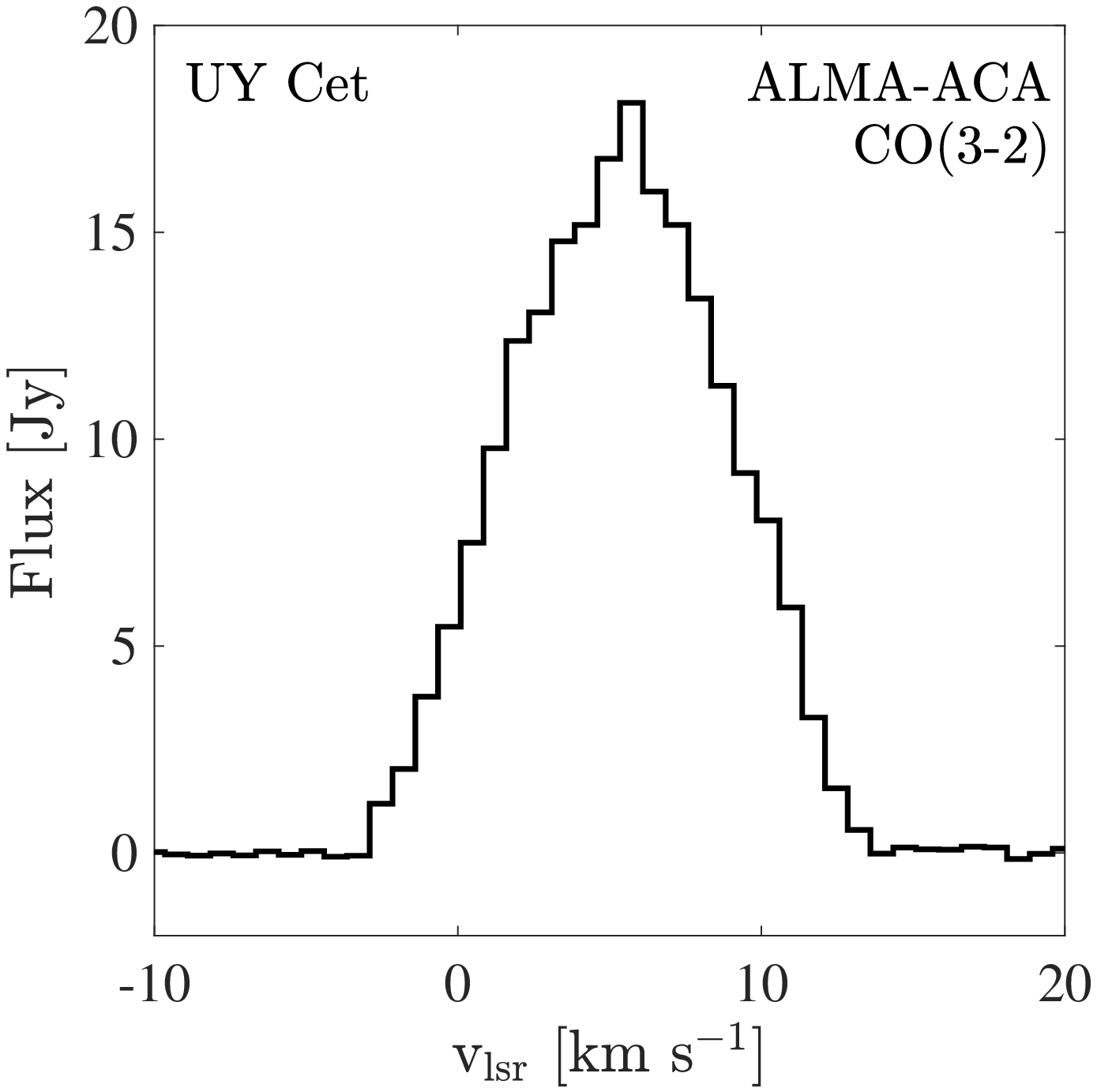}
\includegraphics[height=4.5cm]{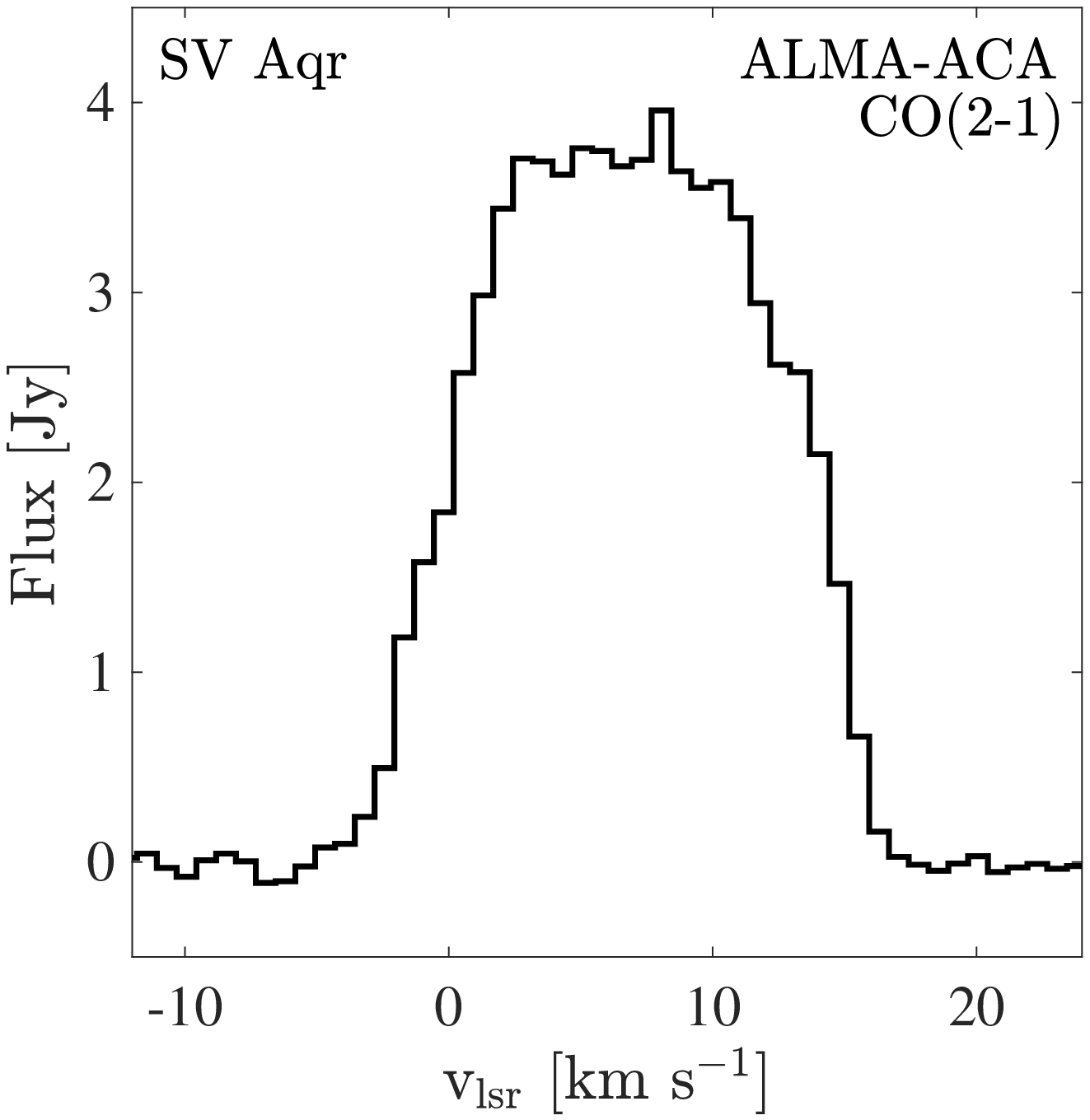}
\includegraphics[height=4.5cm]{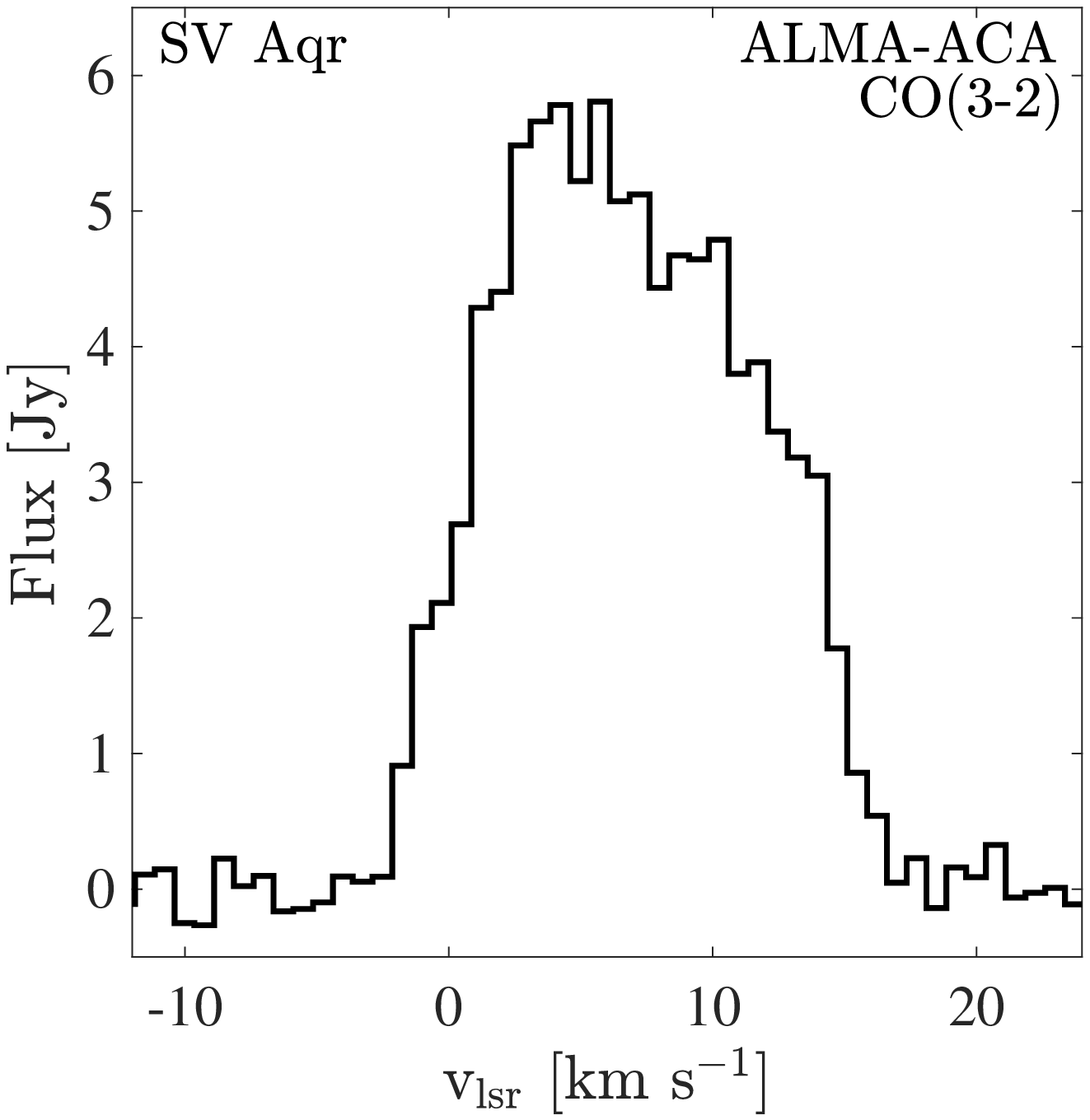}
\caption{ CO $J$\,=\,2$\rightarrow$1 and 3$\rightarrow$2 line profiles measured toward the M-type AGB stars of the sample discussed in this paper. The source name is given in the upper left corner and the transition is in the upper right corner of each plot.}
\label{linesM_SR}
\end{figure*}


\begin{figure*}[t]
\includegraphics[height=4.5cm]{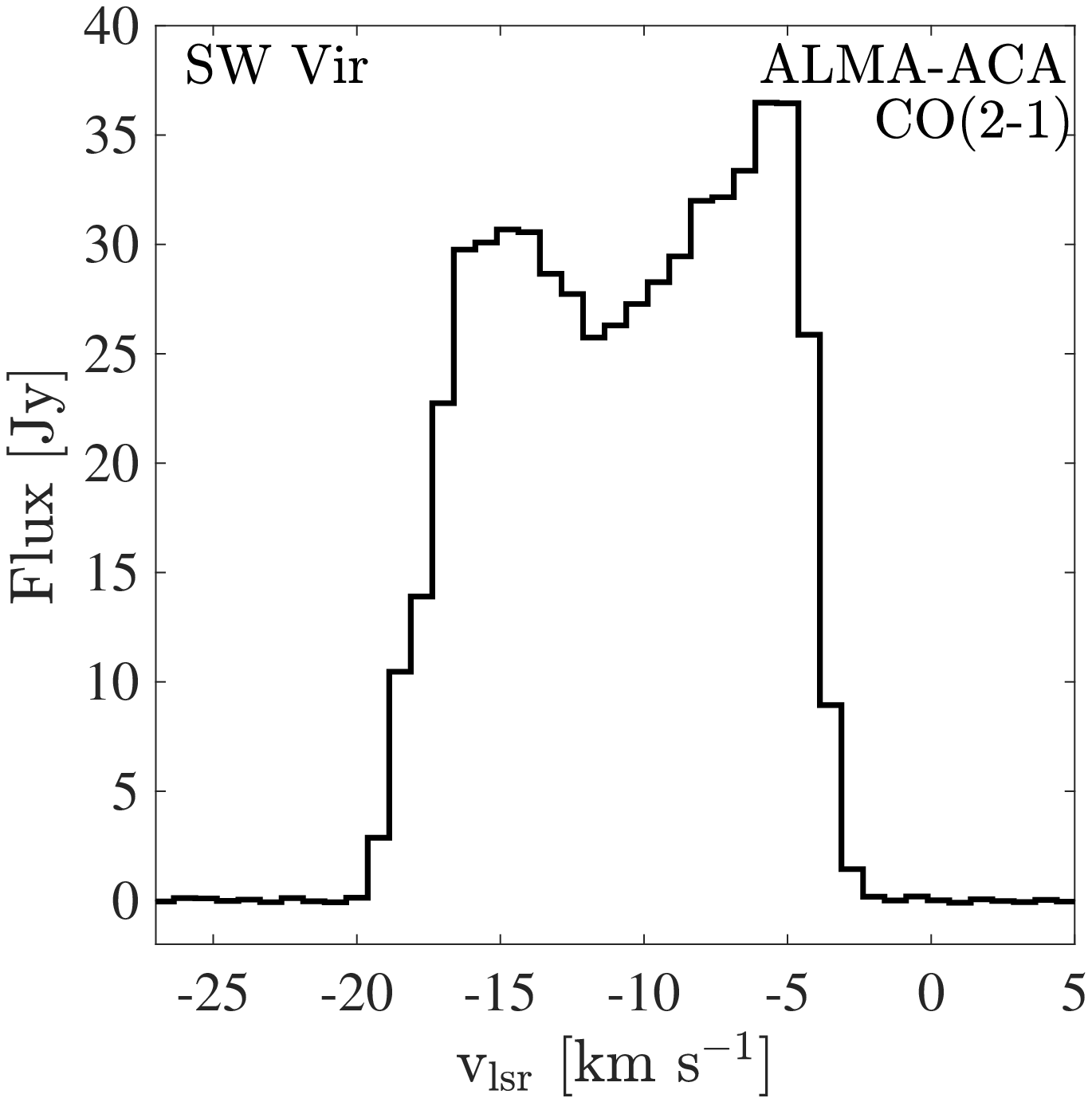}
\includegraphics[height=4.5cm]{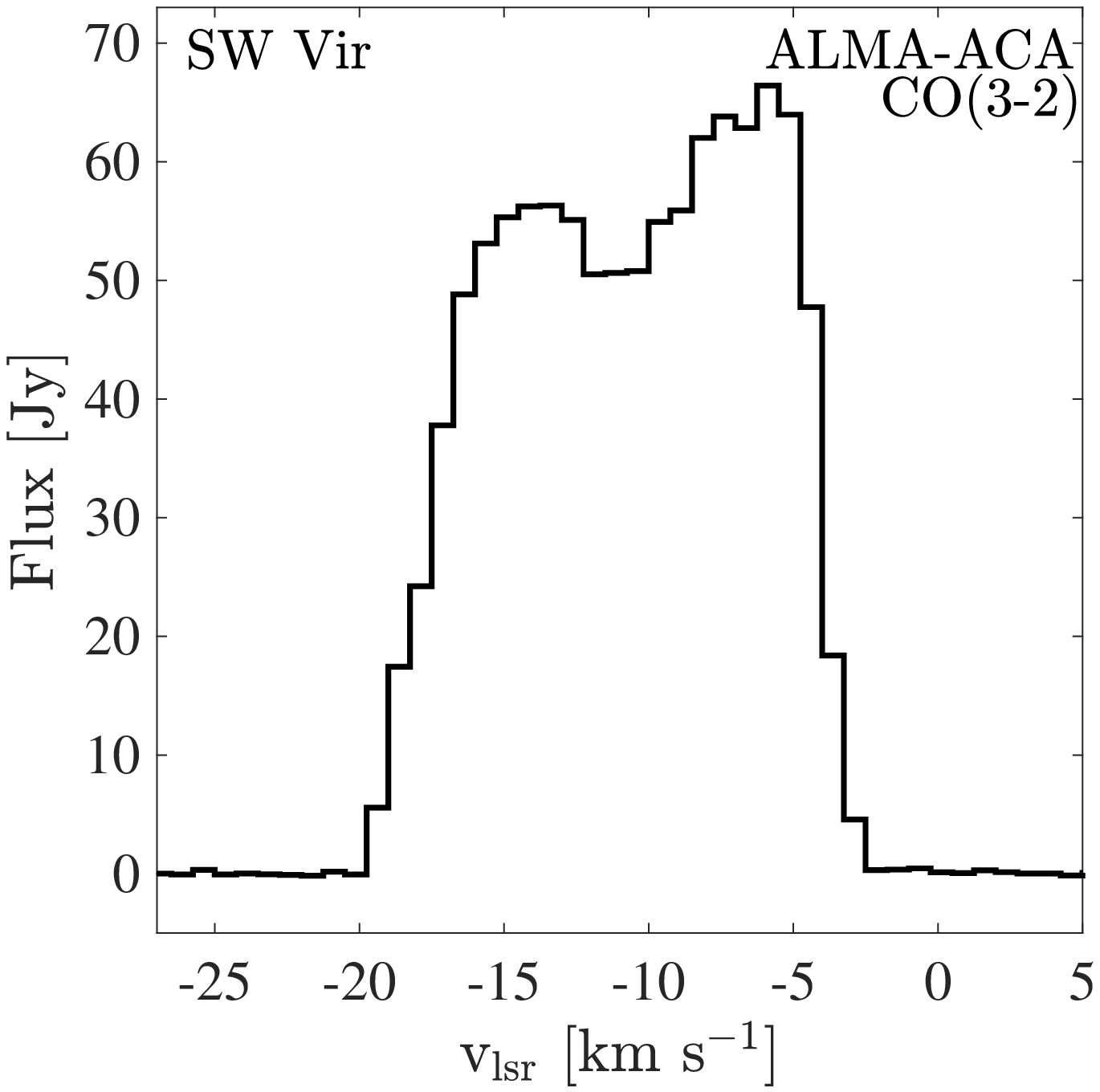}
\includegraphics[height=4.5cm]{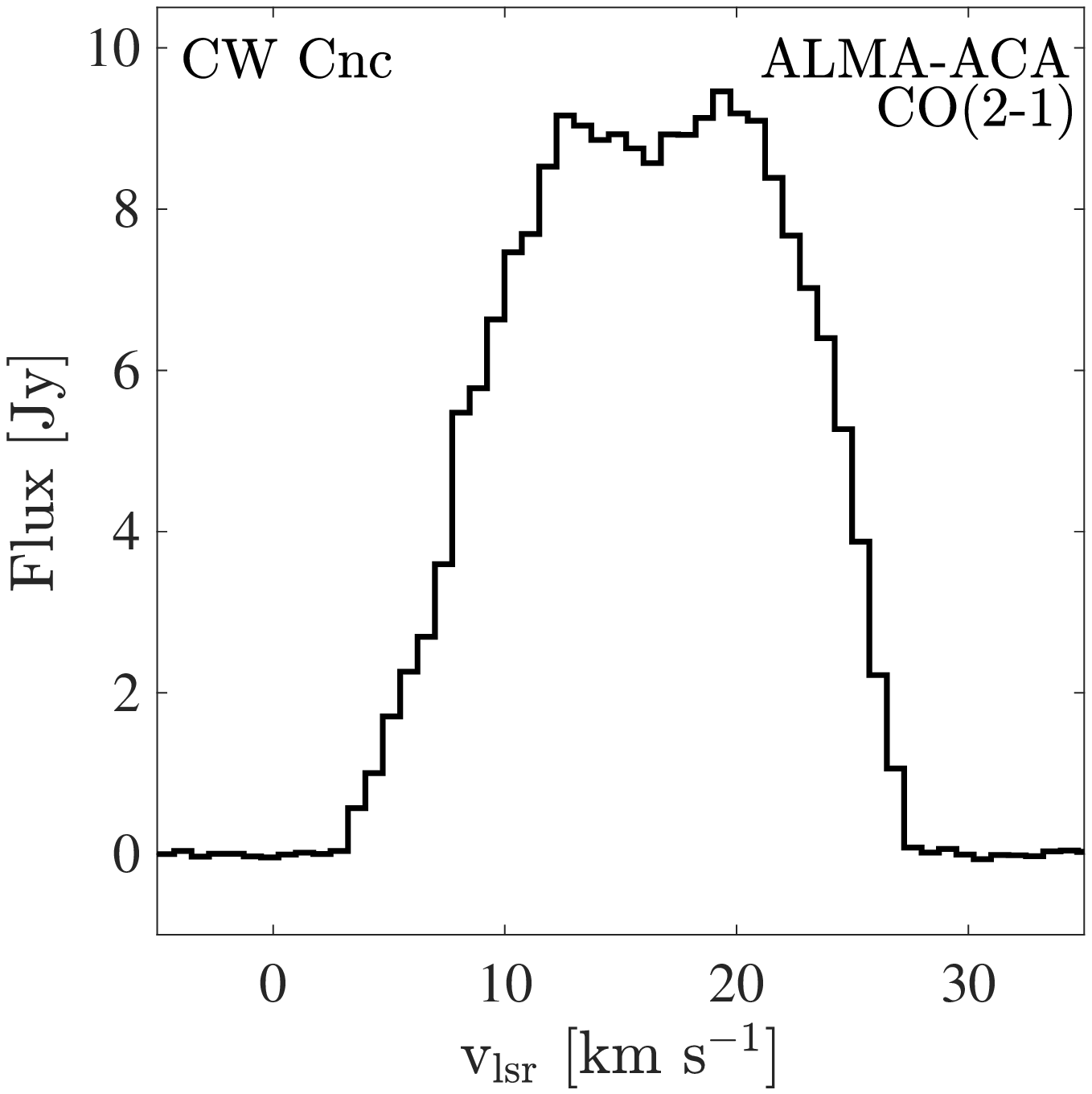}
\includegraphics[height=4.5cm]{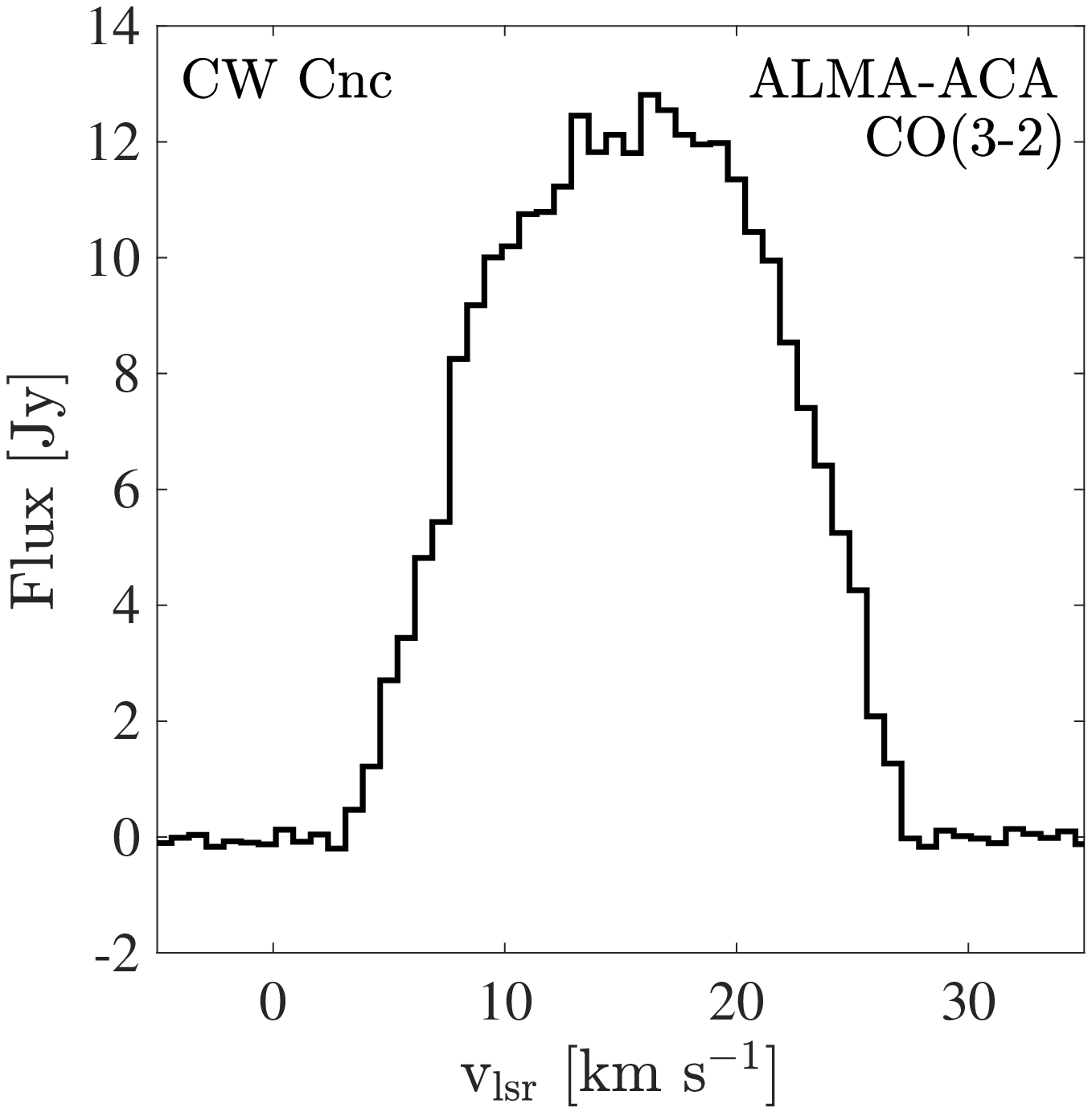}

\includegraphics[height=4.5cm]{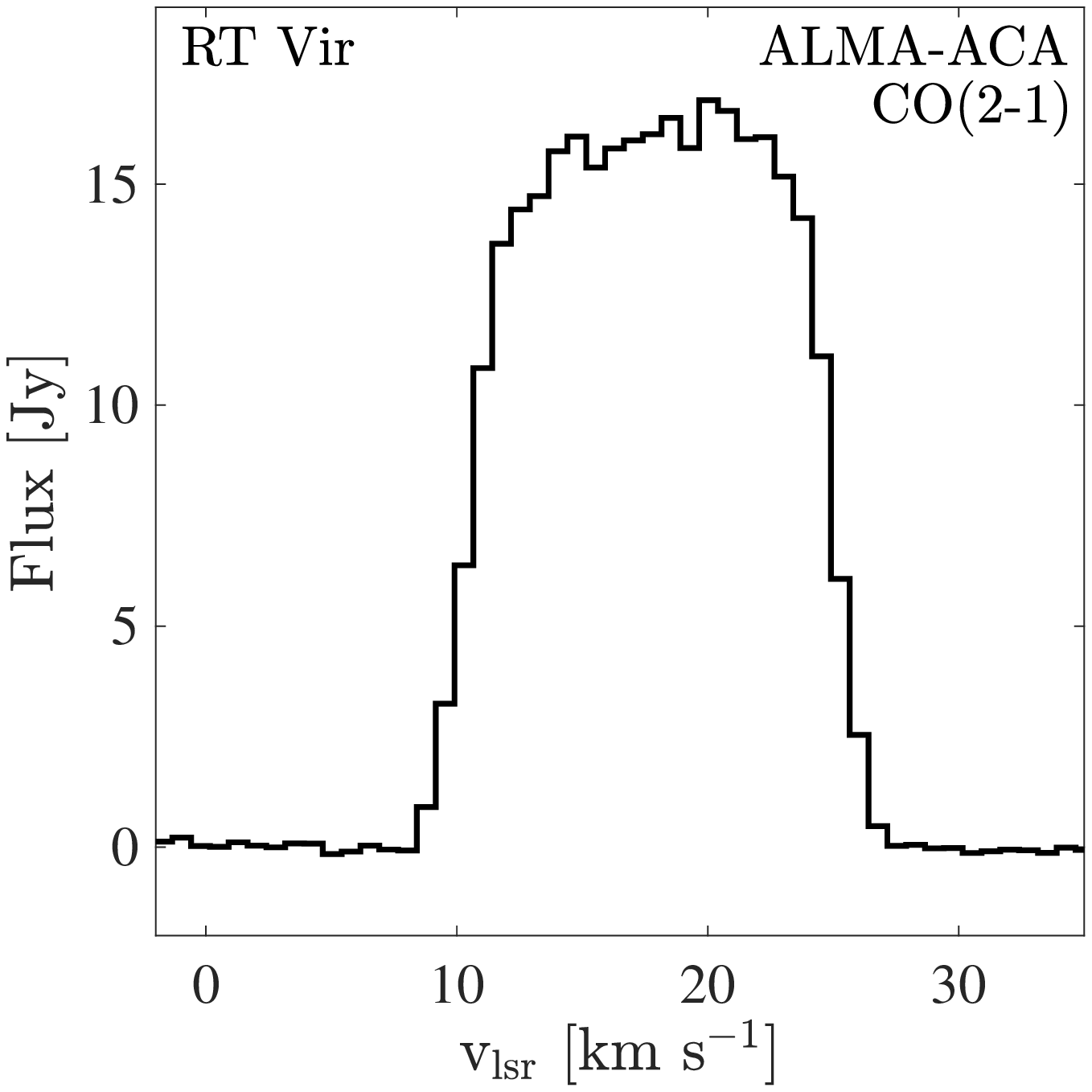}
\includegraphics[height=4.5cm]{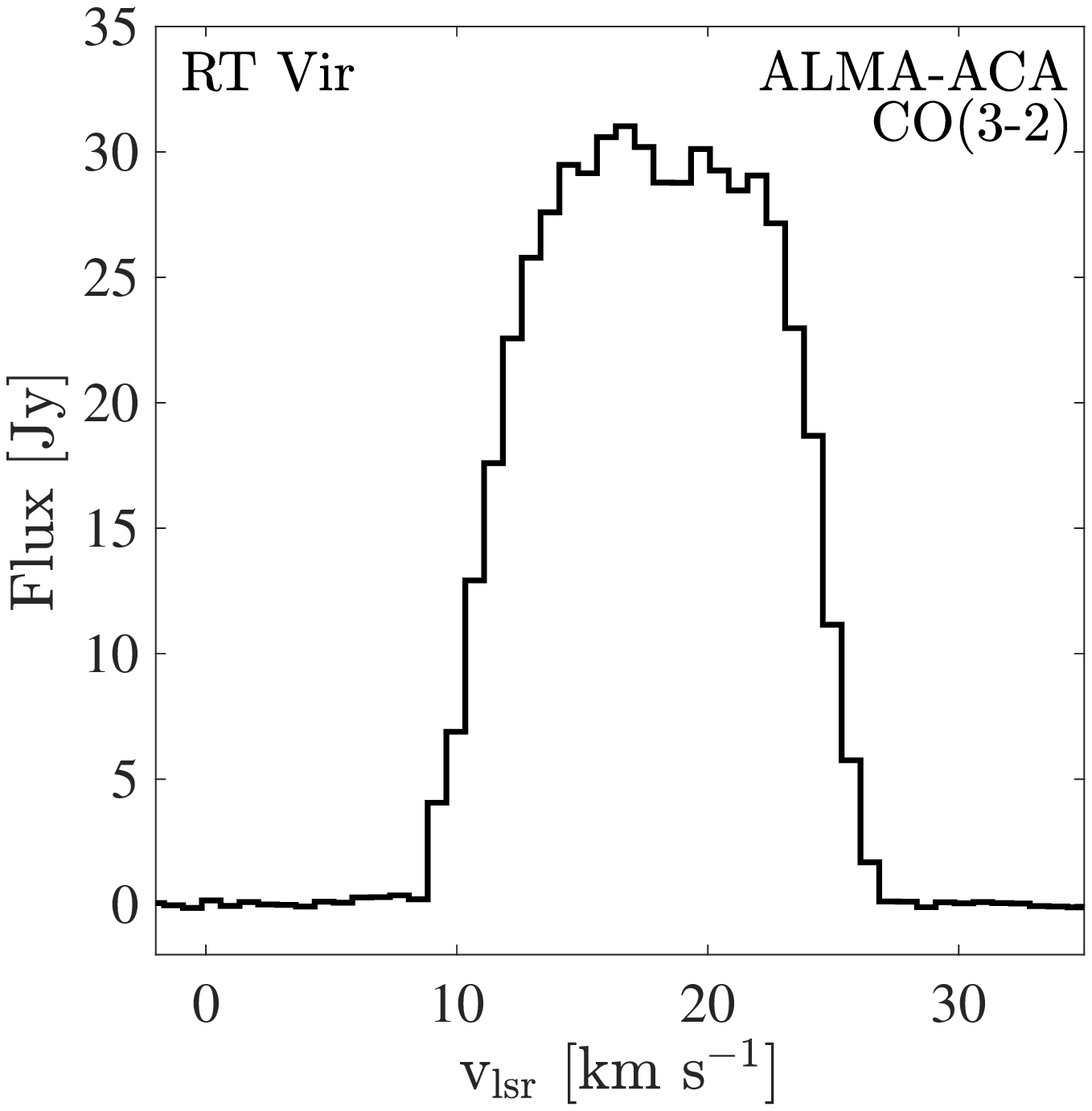}
\includegraphics[height=4.5cm]{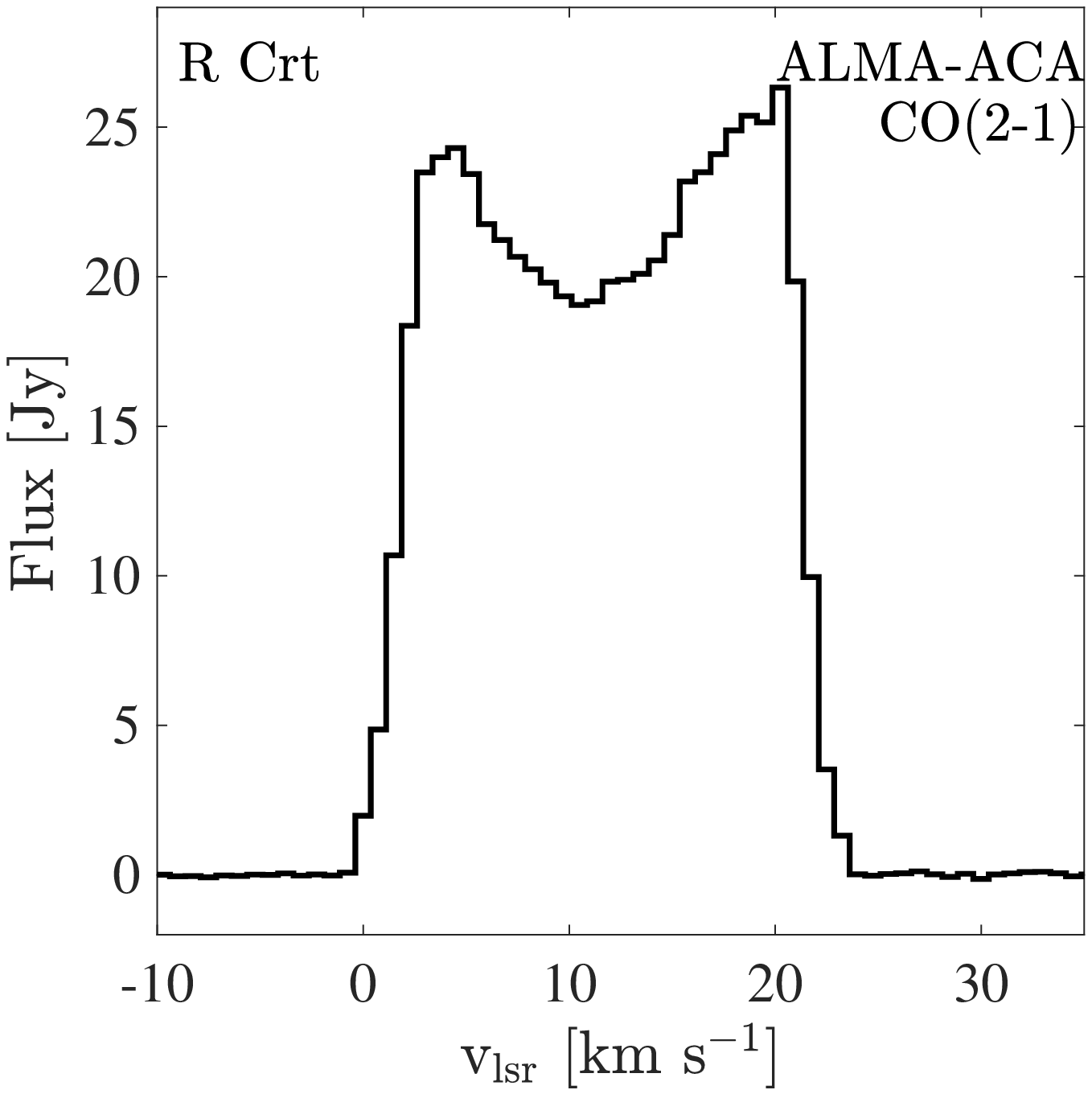}
\includegraphics[height=4.5cm]{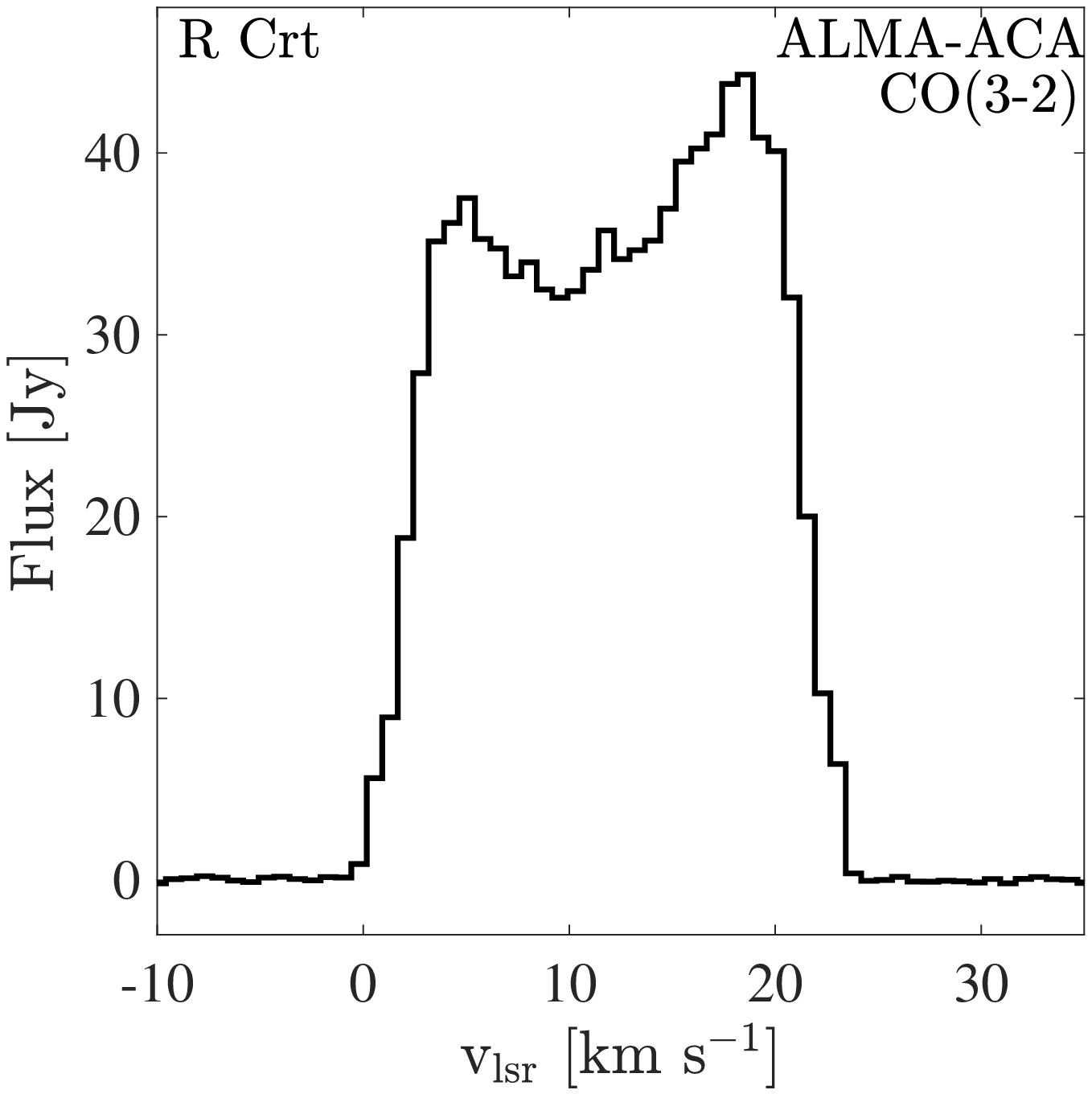}

\includegraphics[height=4.45cm]{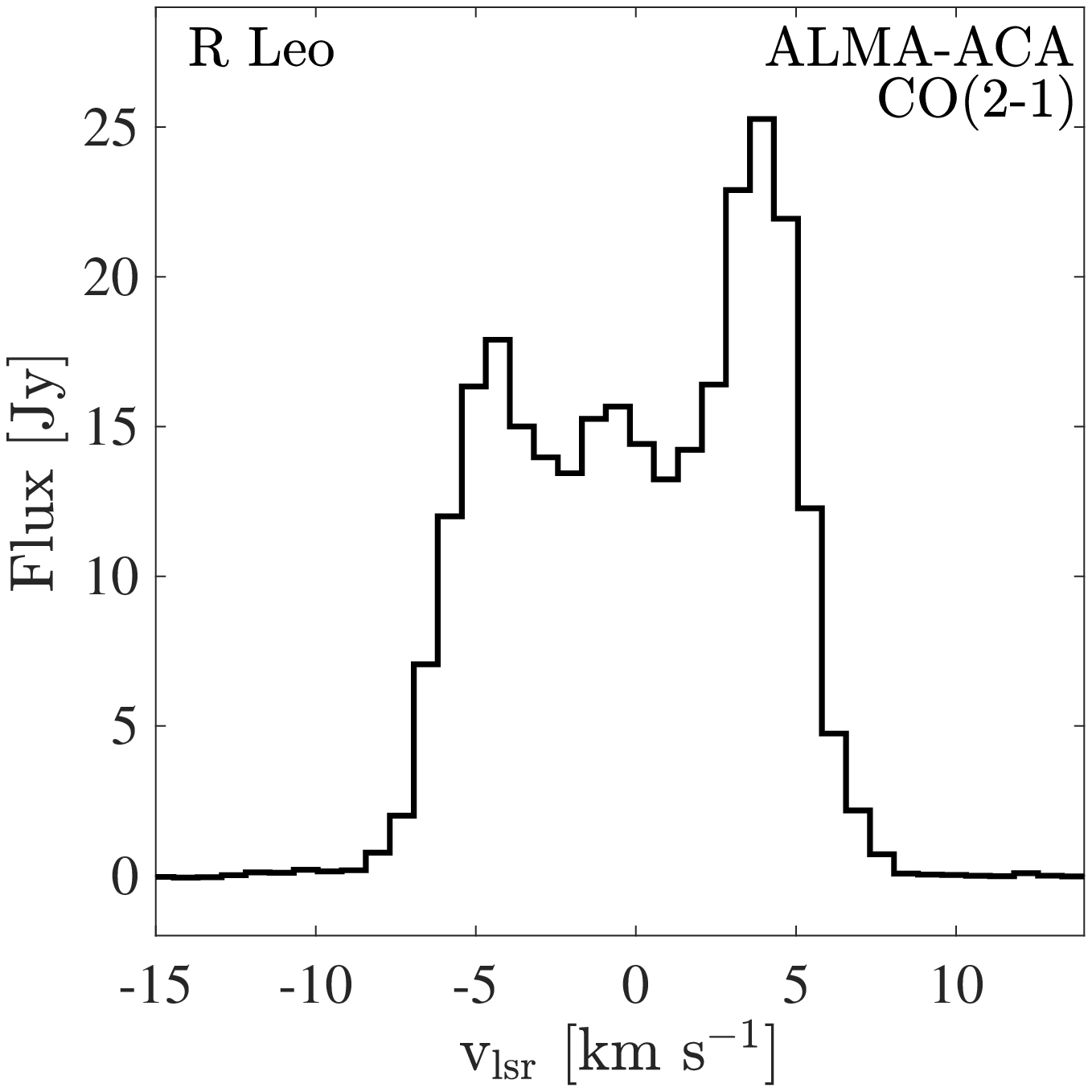}
\includegraphics[height=4.45cm]{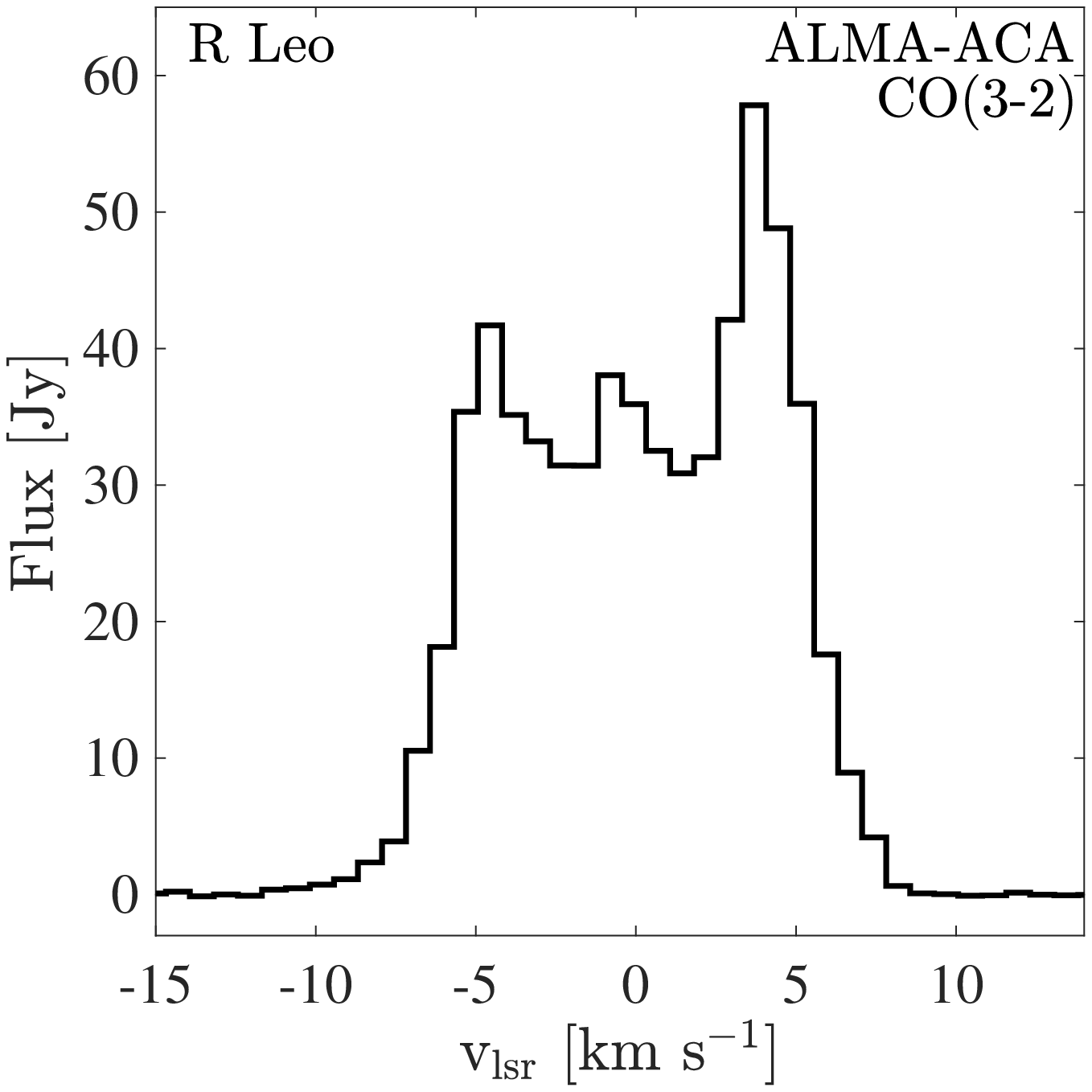}
\includegraphics[height=4.45cm]{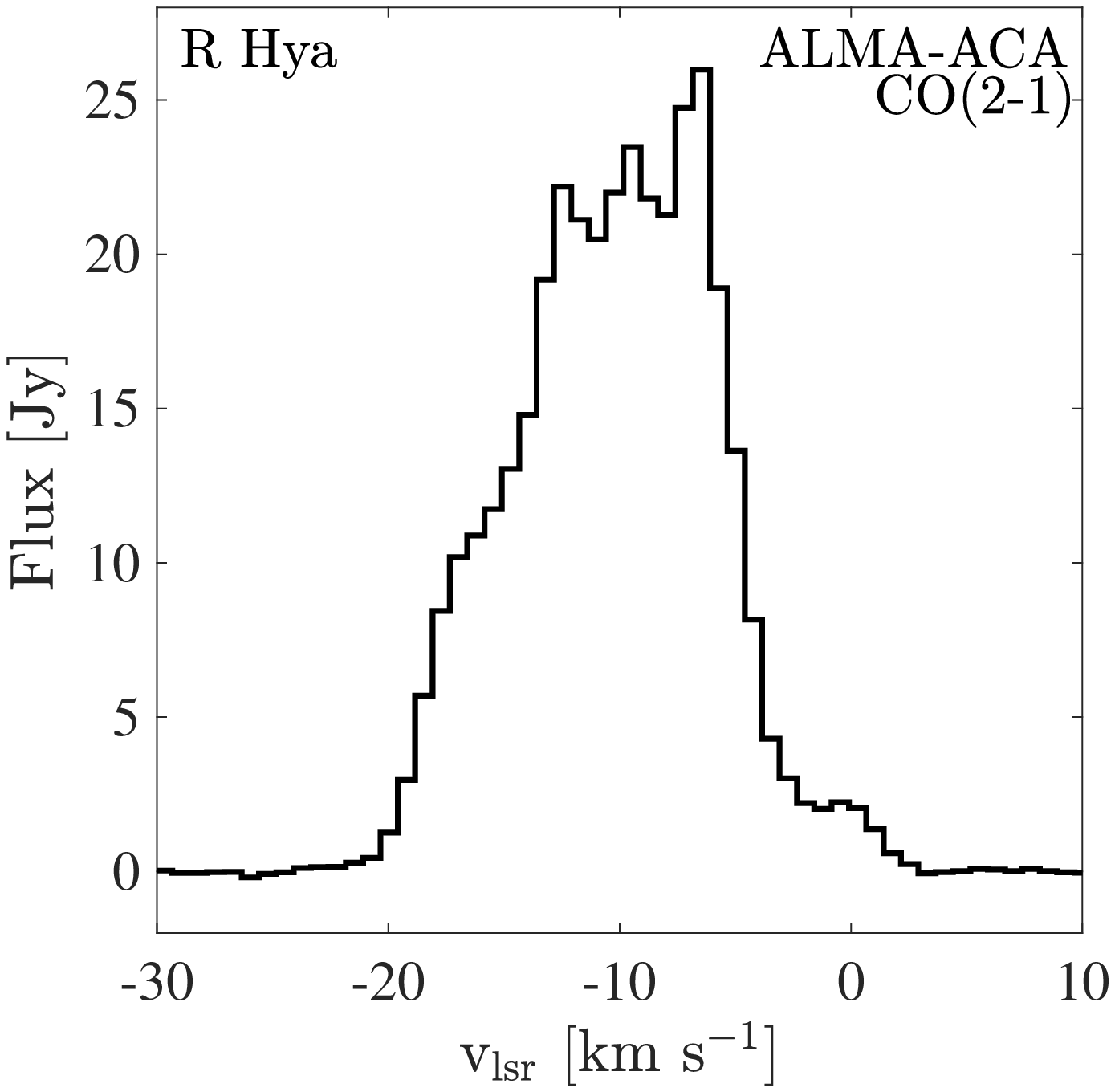}
\includegraphics[height=4.45cm]{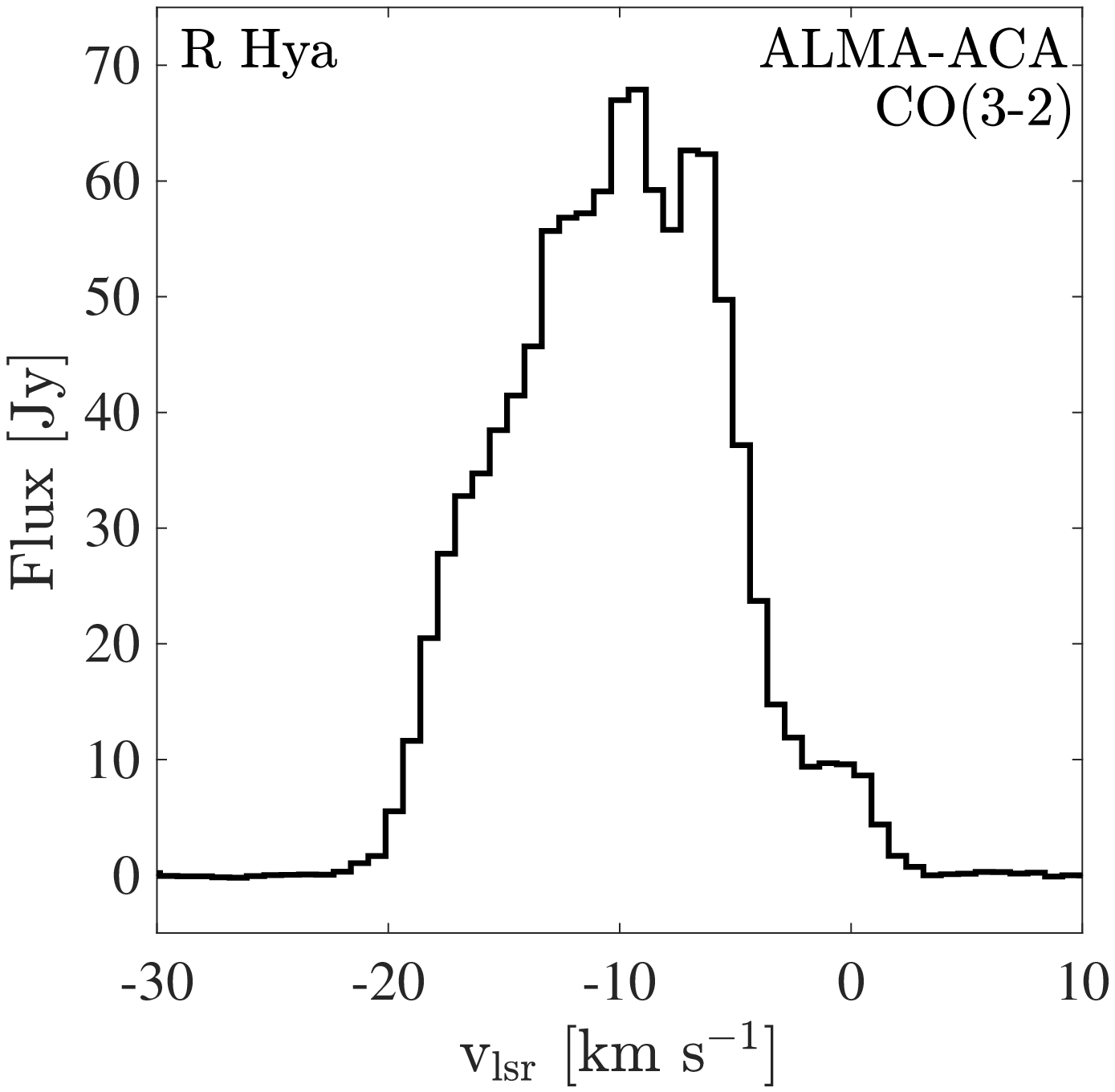}

\includegraphics[height=4.5cm]{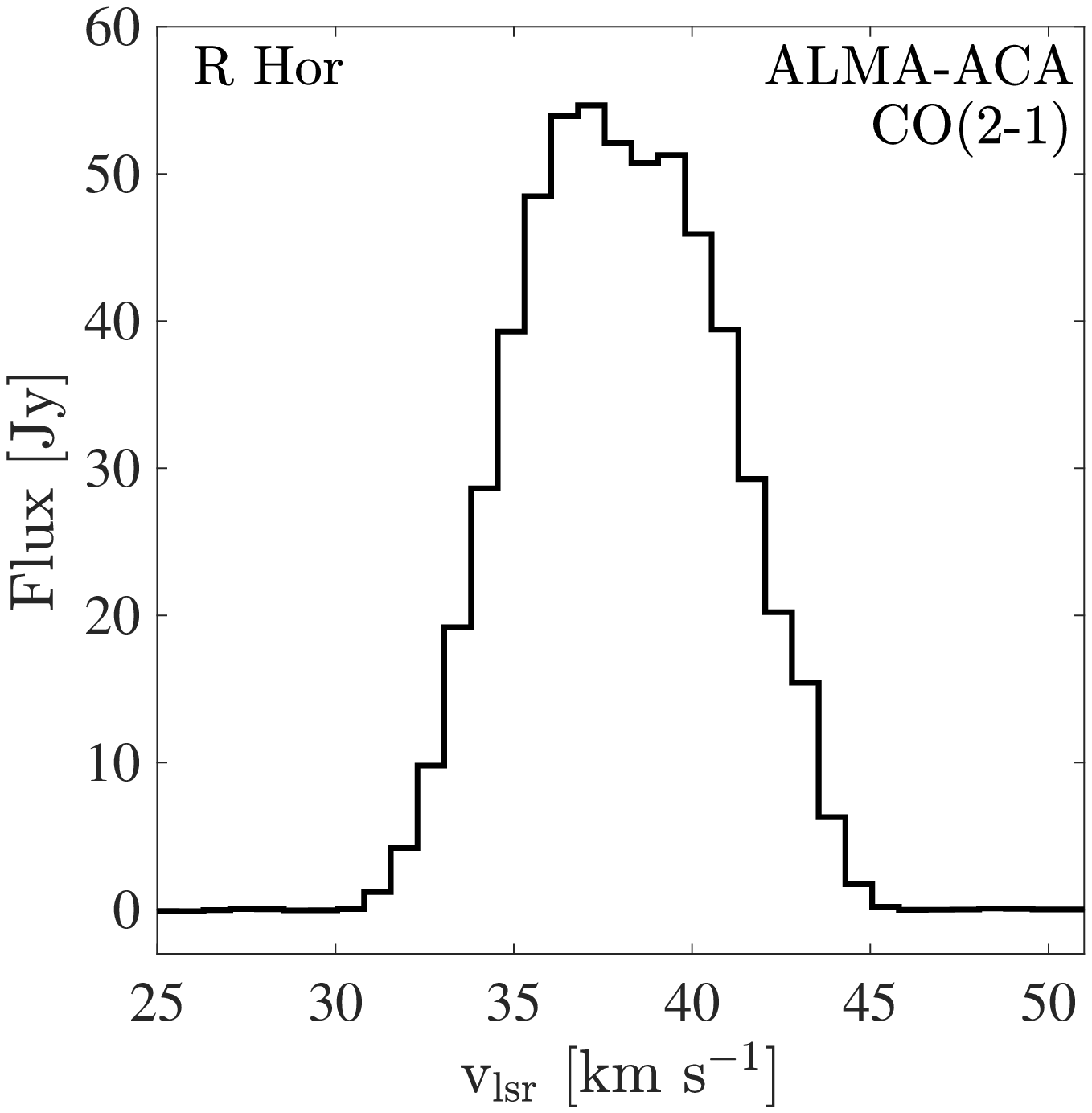}
\includegraphics[height=4.5cm]{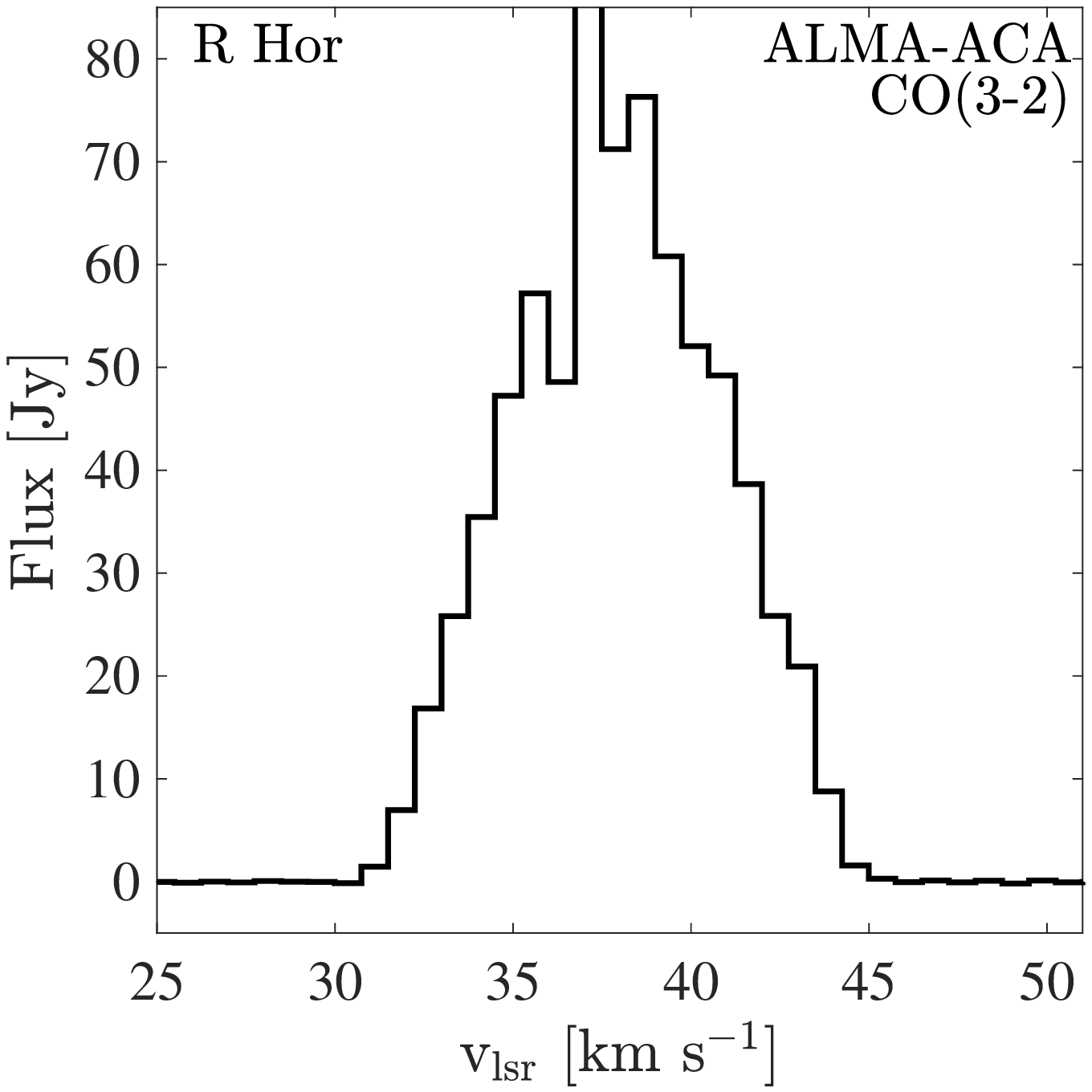}
\includegraphics[height=4.5cm]{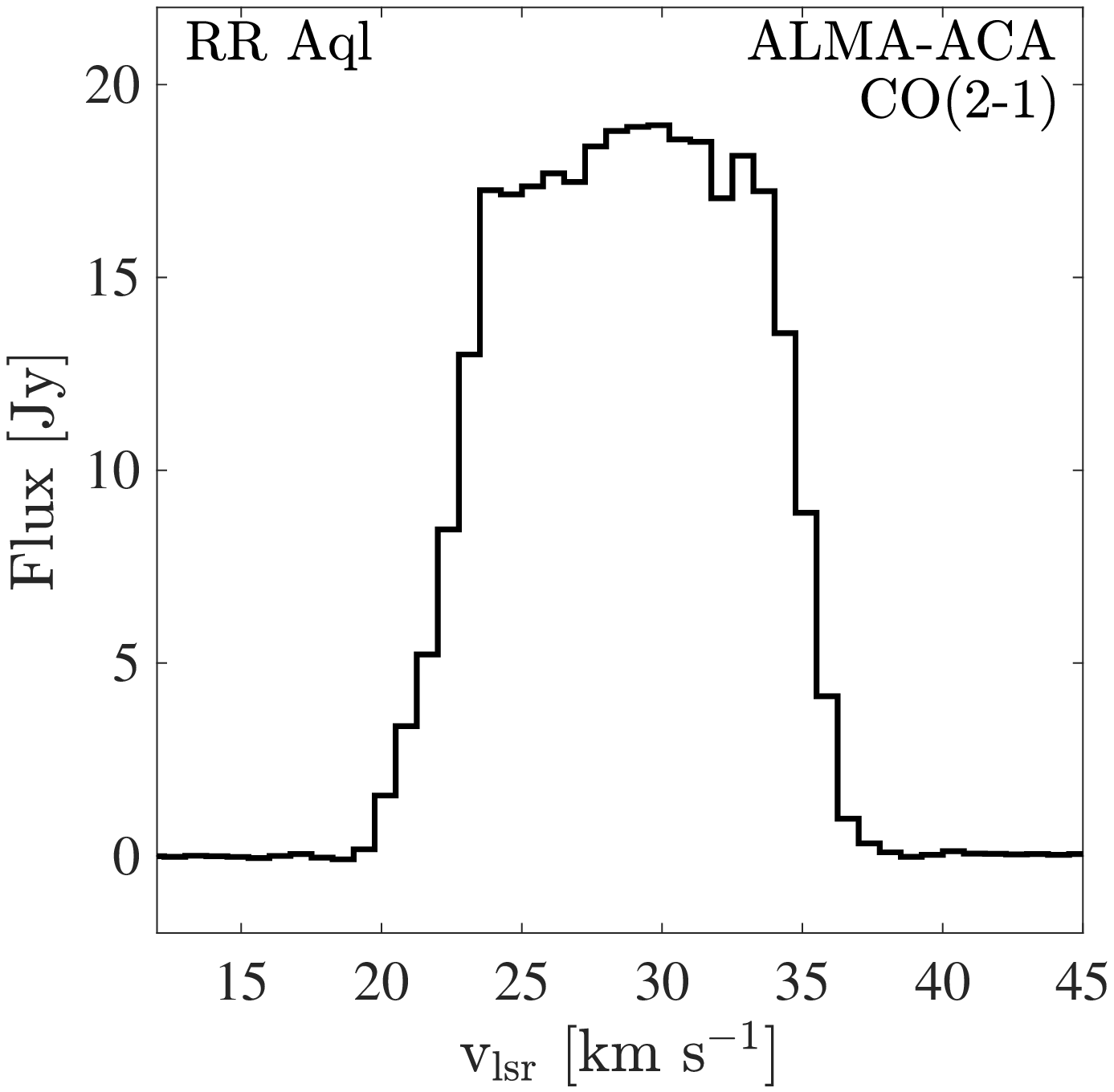}
\includegraphics[height=4.5cm]{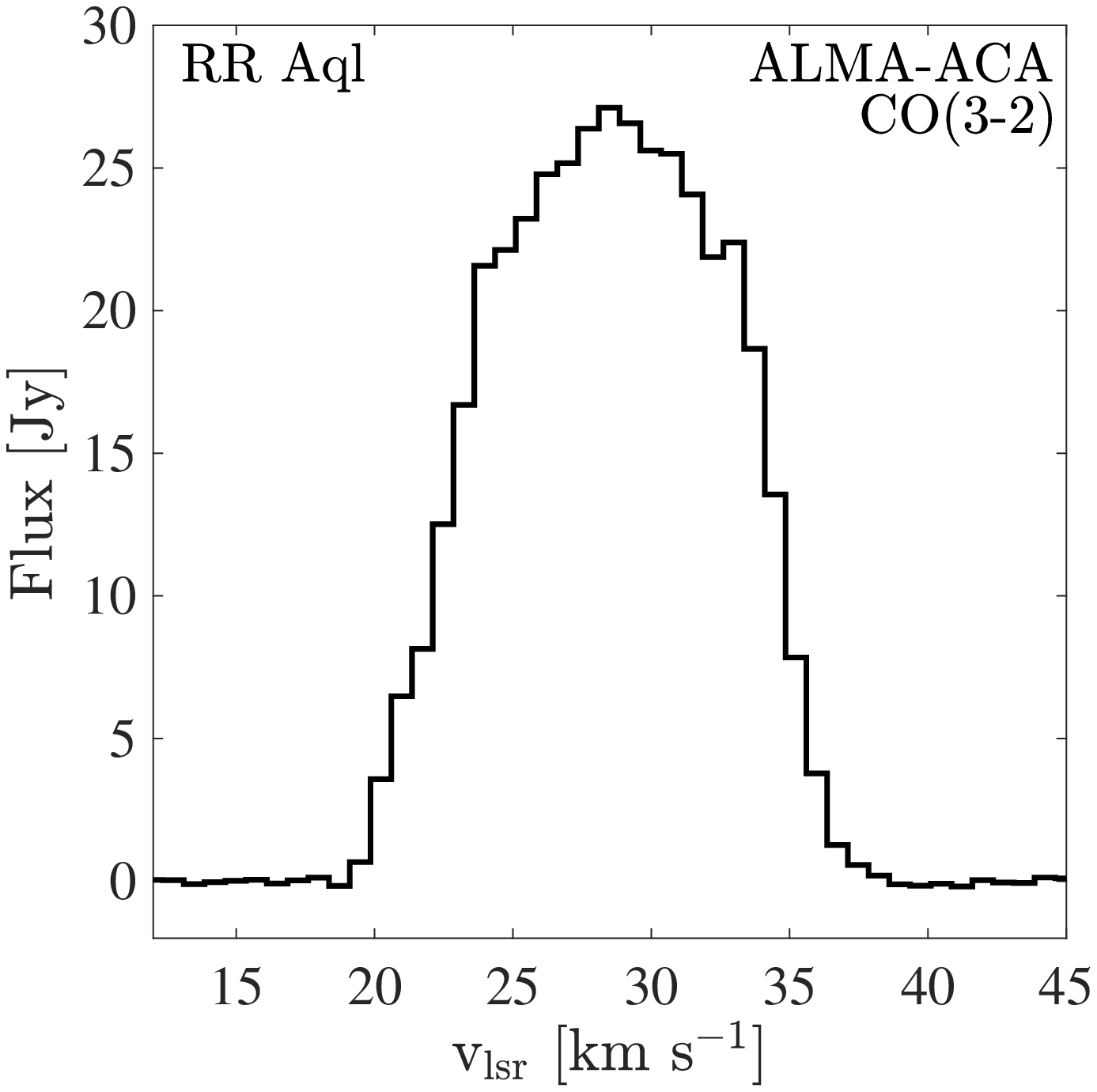}

\includegraphics[height=4.5cm]{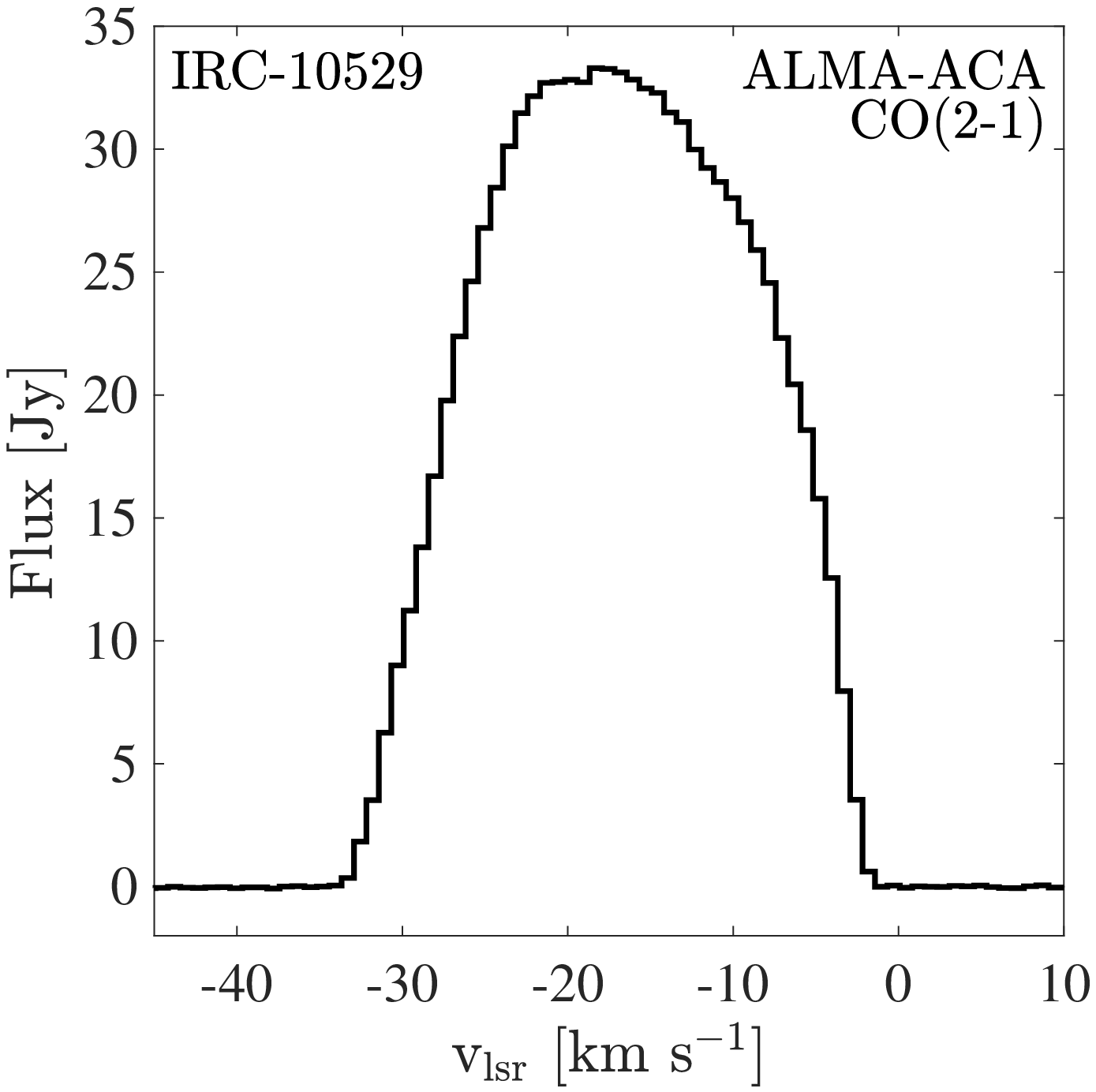}
\includegraphics[height=4.5cm]{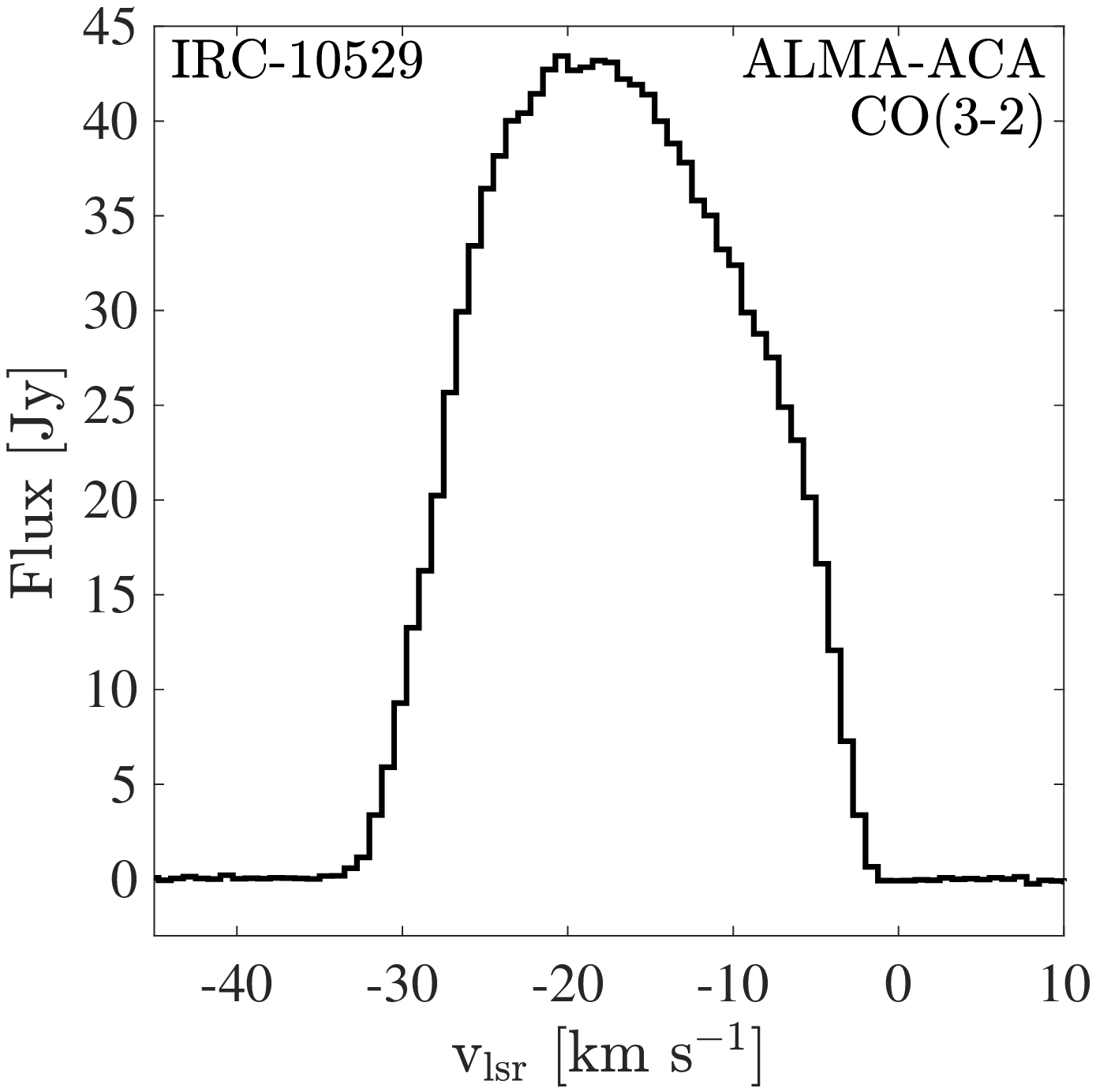}
\includegraphics[height=4.5cm]{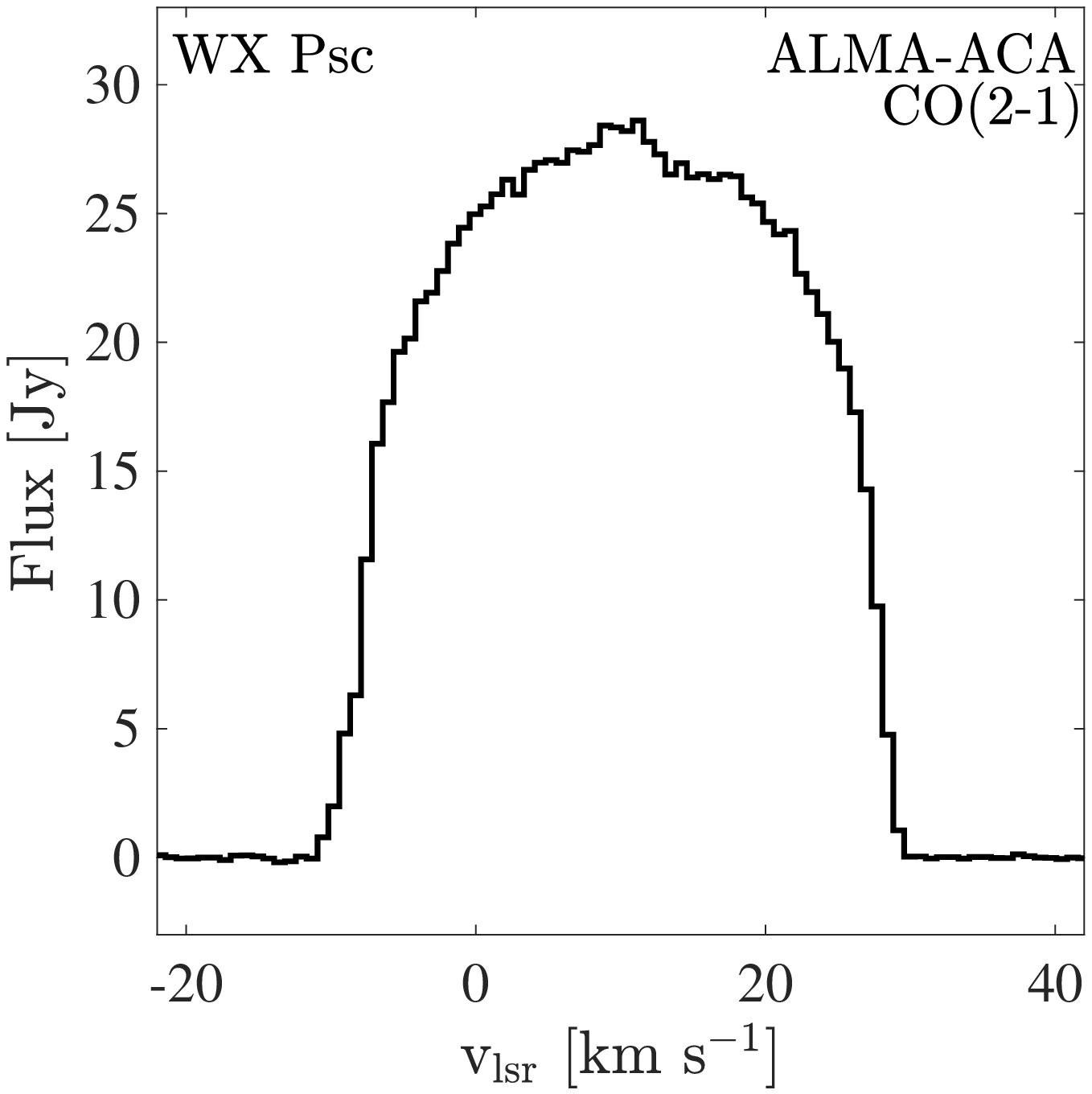}
\includegraphics[height=4.5cm]{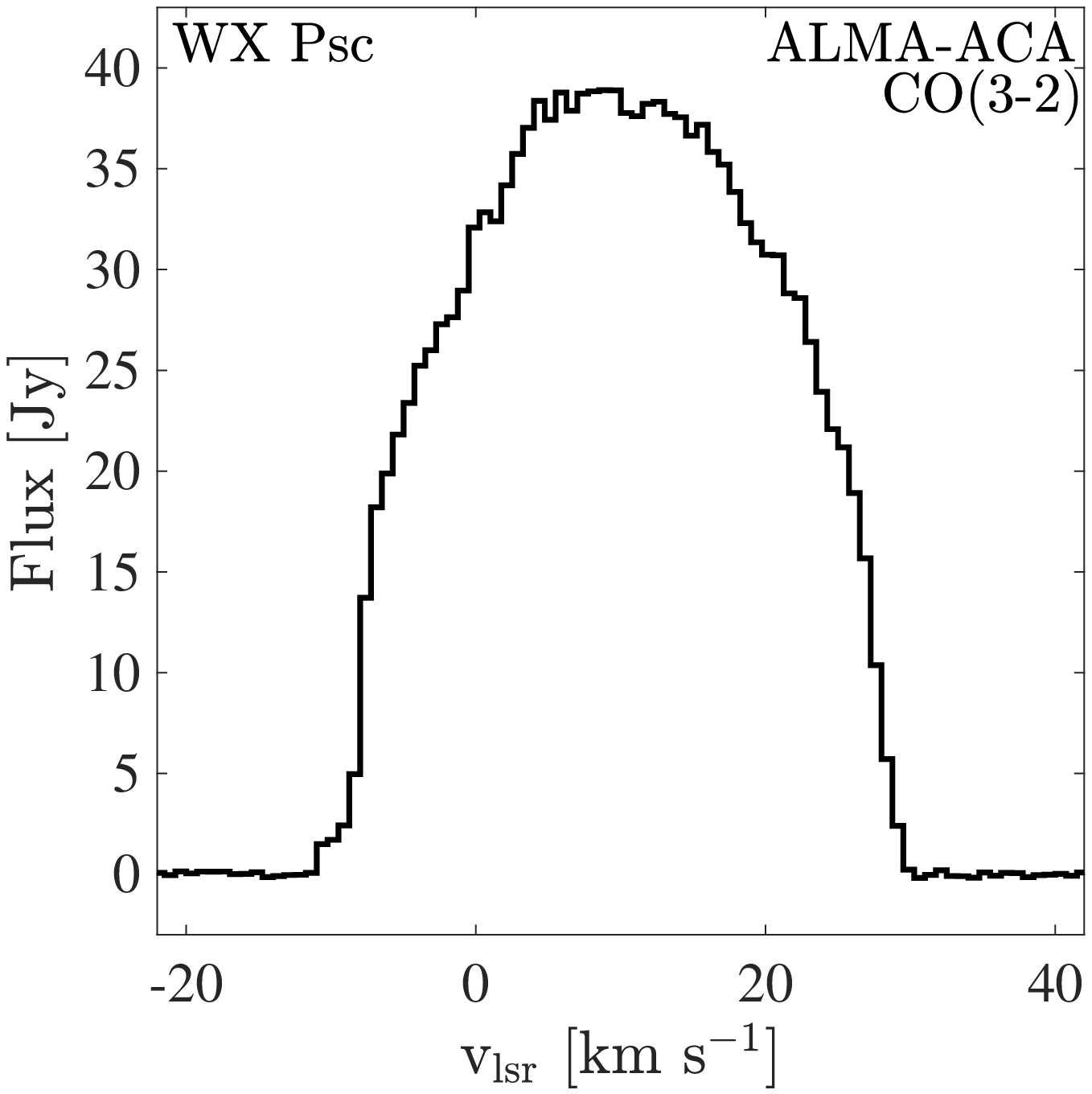}
\caption{CO $J$\,=\,2$\rightarrow$1 and 3$\rightarrow$2 line profiles measured toward the M-type AGB stars of the sample discussed in this paper. The source name is given in the upper left corner and the transition is in the upper right corner of each plot.}
\label{linesM_MSR}
\end{figure*}


\begin{figure*}[t]
\includegraphics[height=4.5cm]{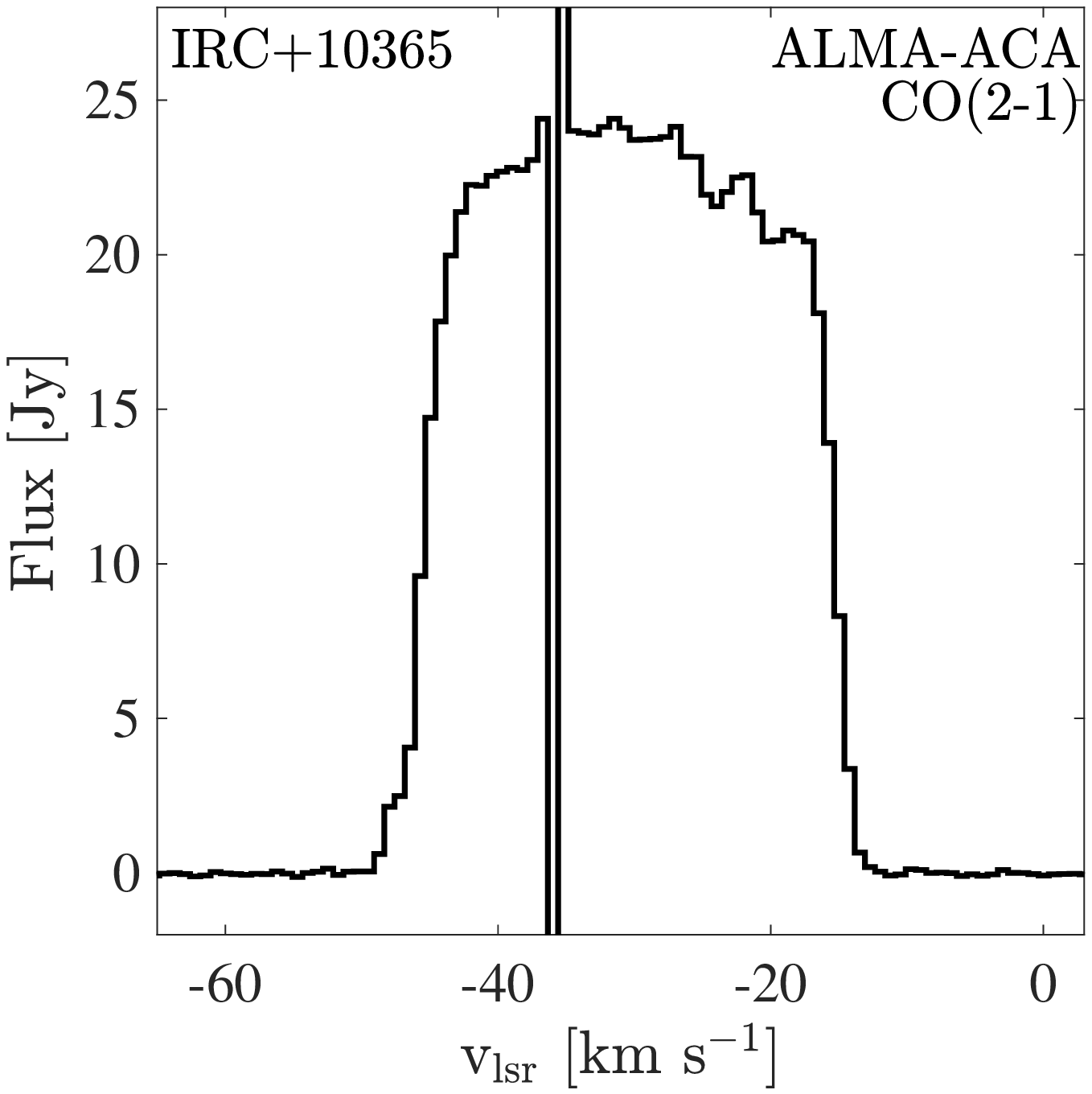}
\includegraphics[height=4.5cm]{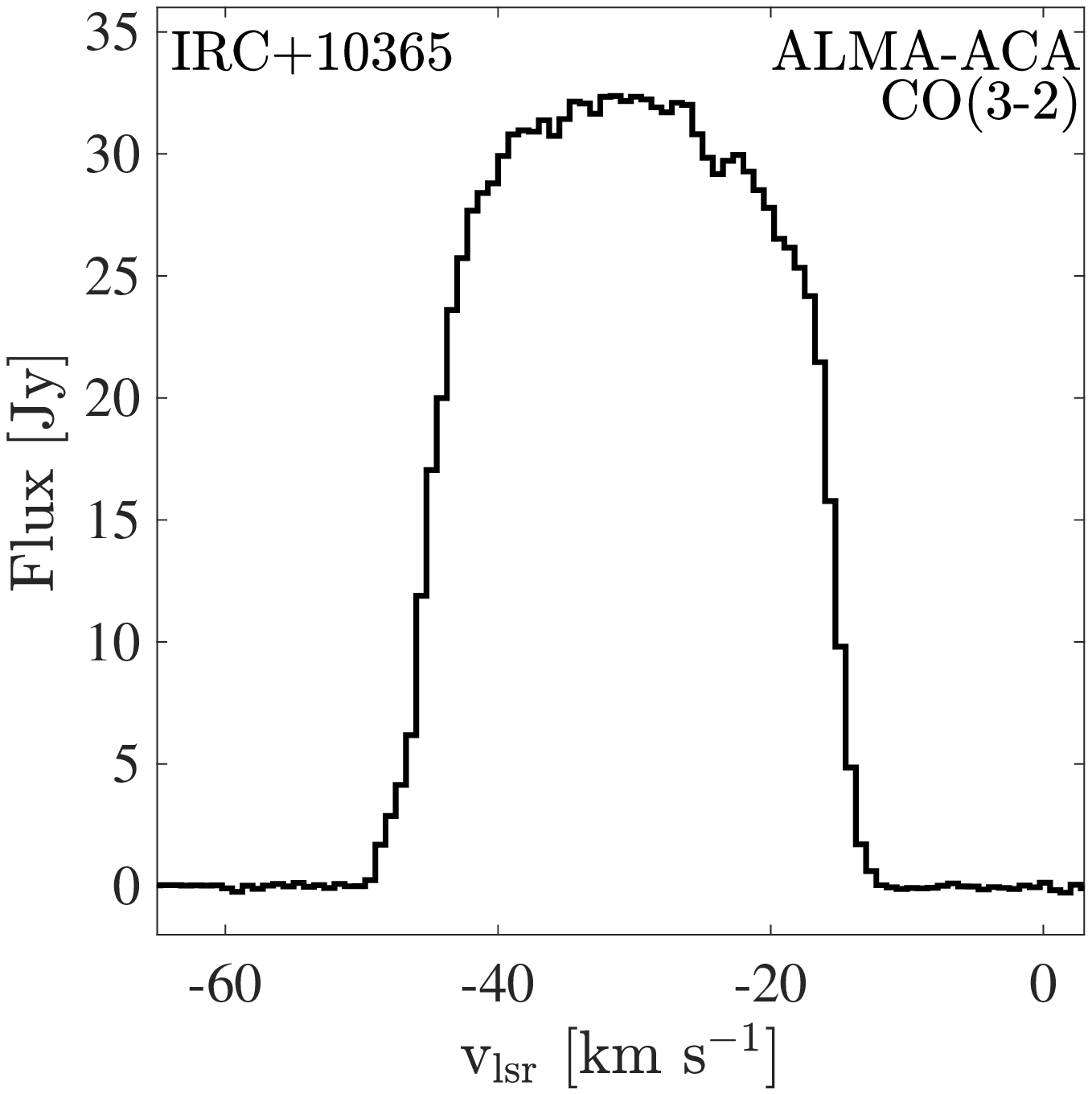}
\caption{CO $J$\,=\,2$\rightarrow$1 and 3$\rightarrow$2 line profiles measured toward the M-type AGB stars of the sample discussed in this paper. The source name is given in the upper left corner and the transition is in the upper right corner of each plot.}
\label{linesM_M}
\end{figure*}


\begin{figure*}[t]
\includegraphics[height=4.5cm]{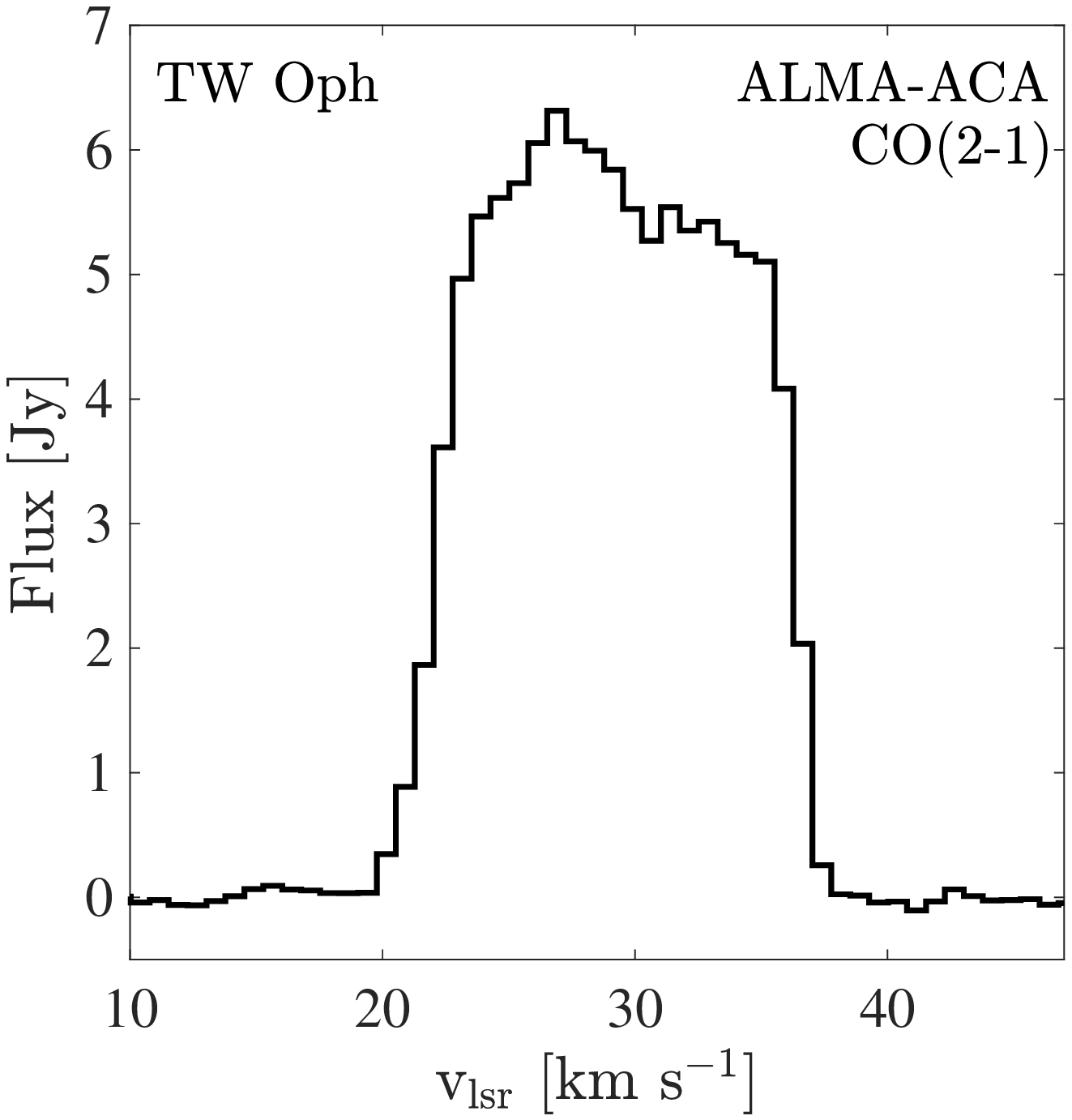}
\includegraphics[height=4.5cm]{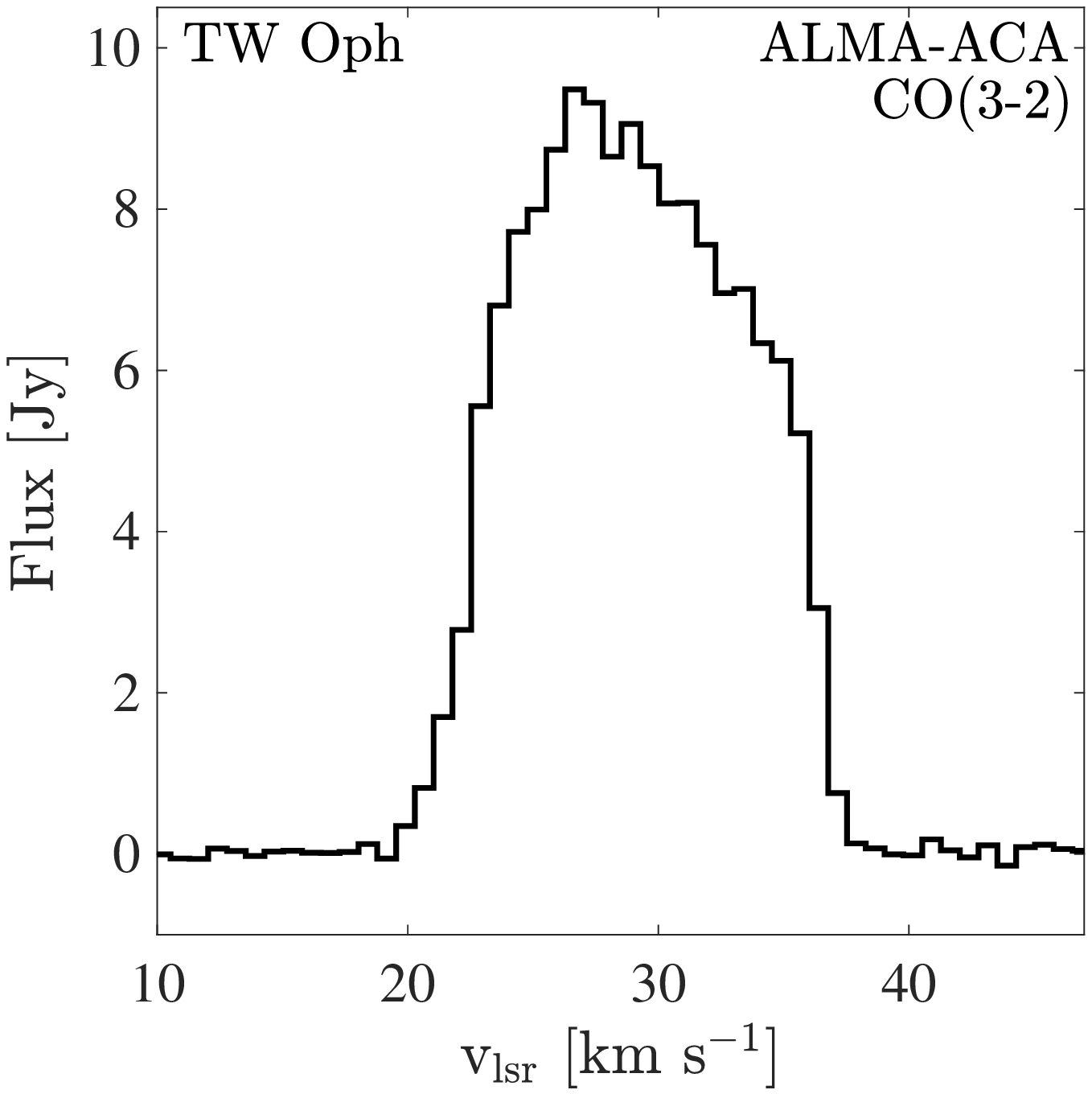}
\includegraphics[height=4.5cm]{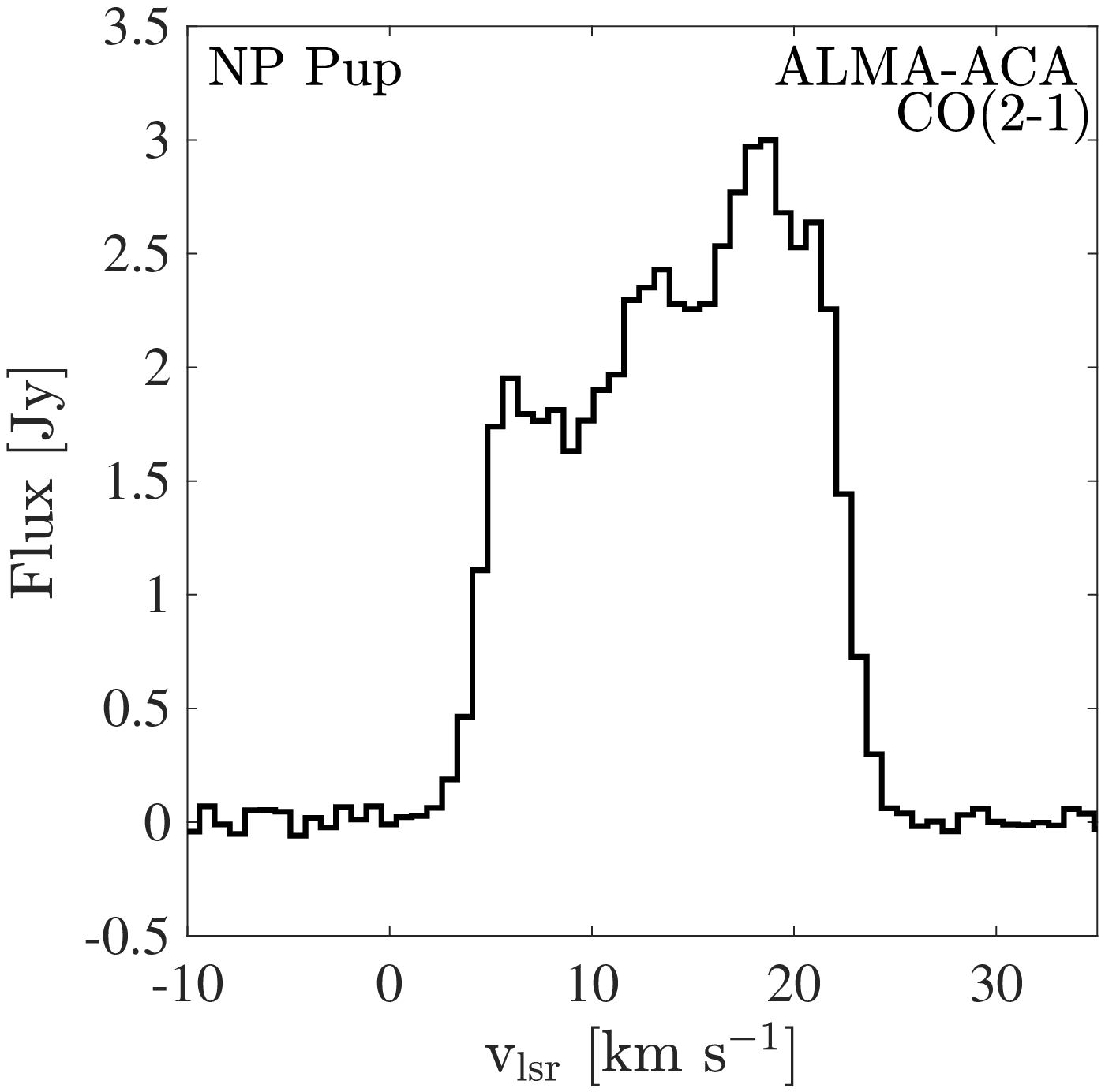}
\includegraphics[height=4.5cm]{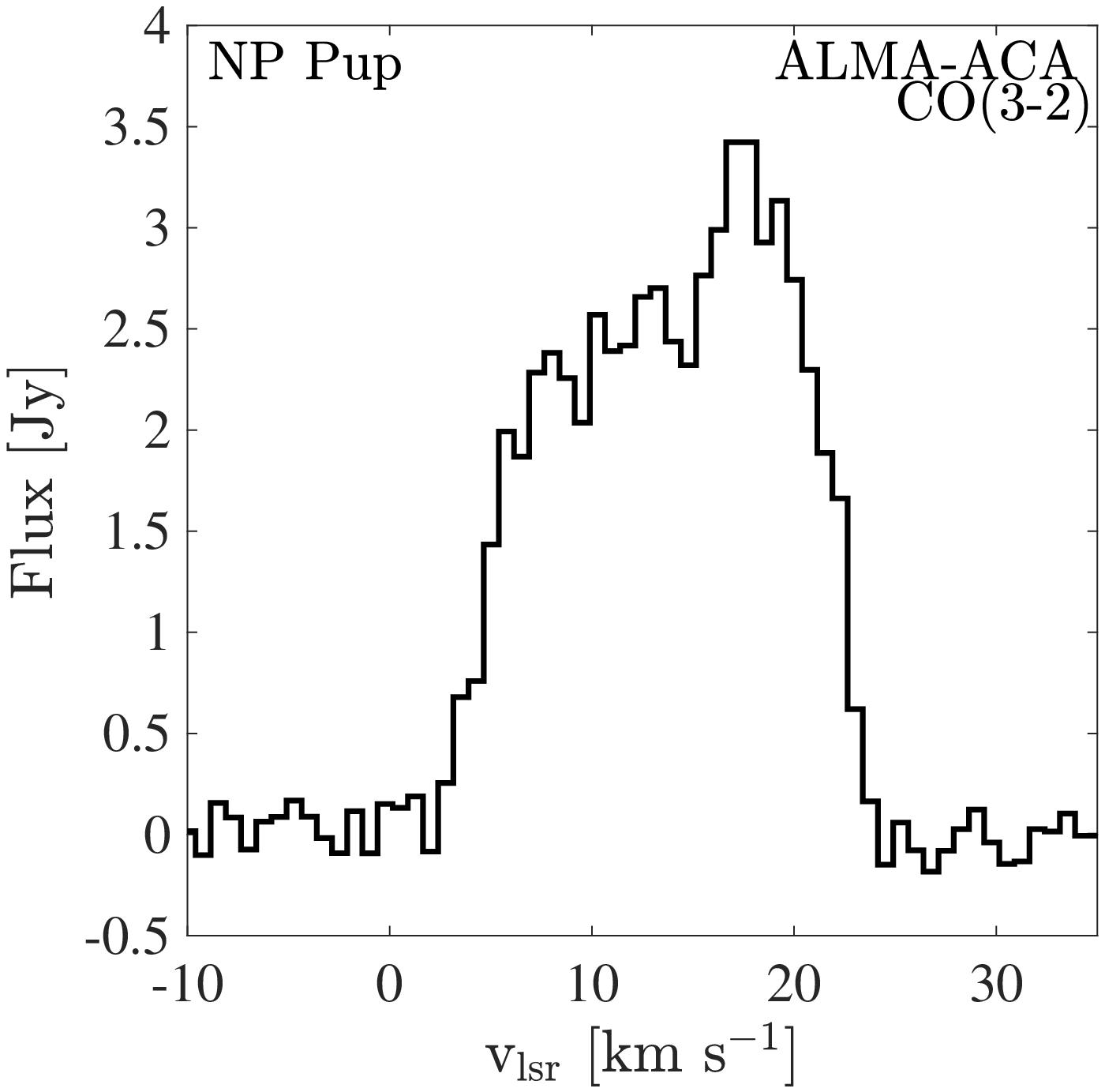}

\includegraphics[height=4.5cm]{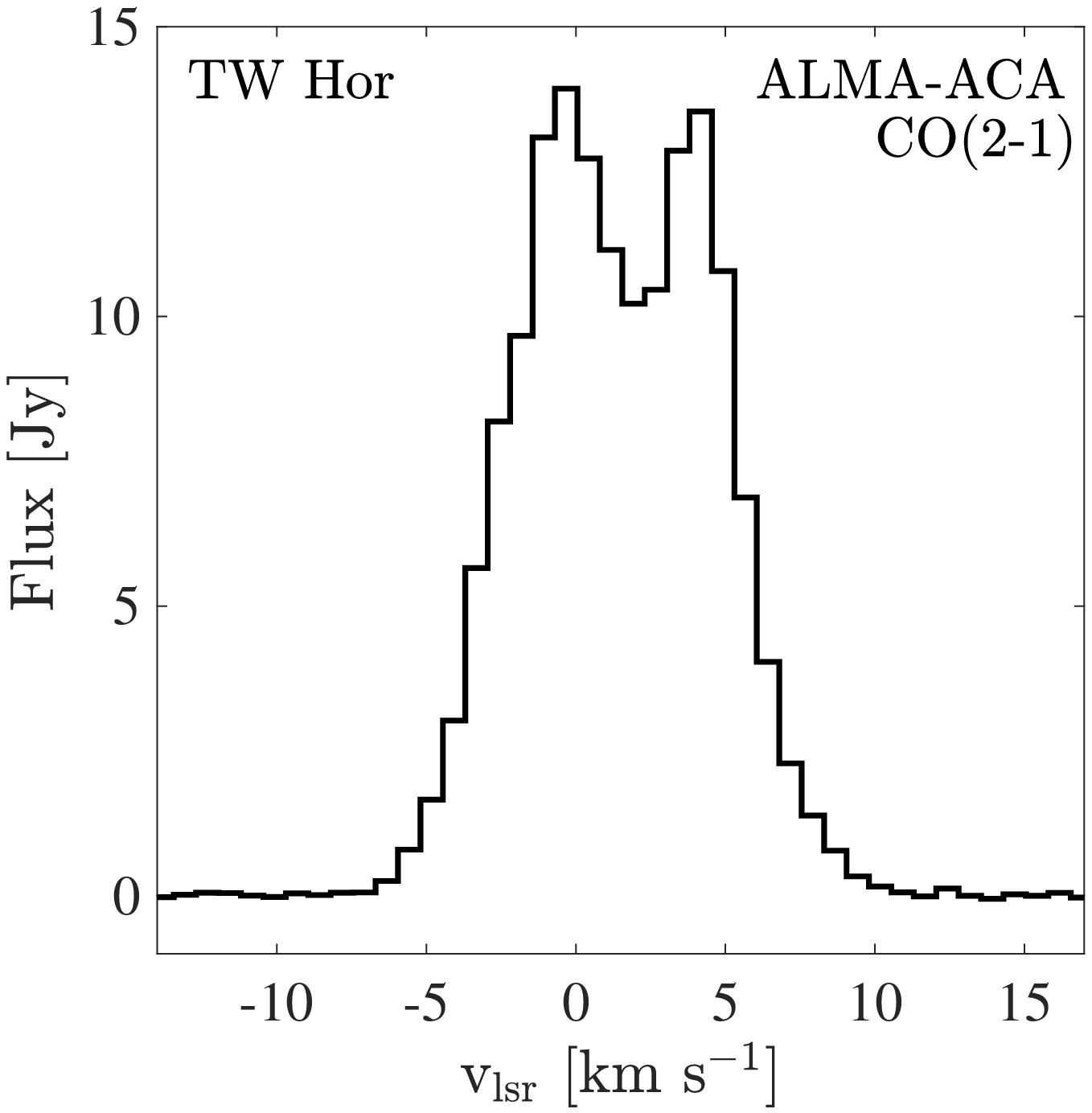}
\includegraphics[height=4.5cm]{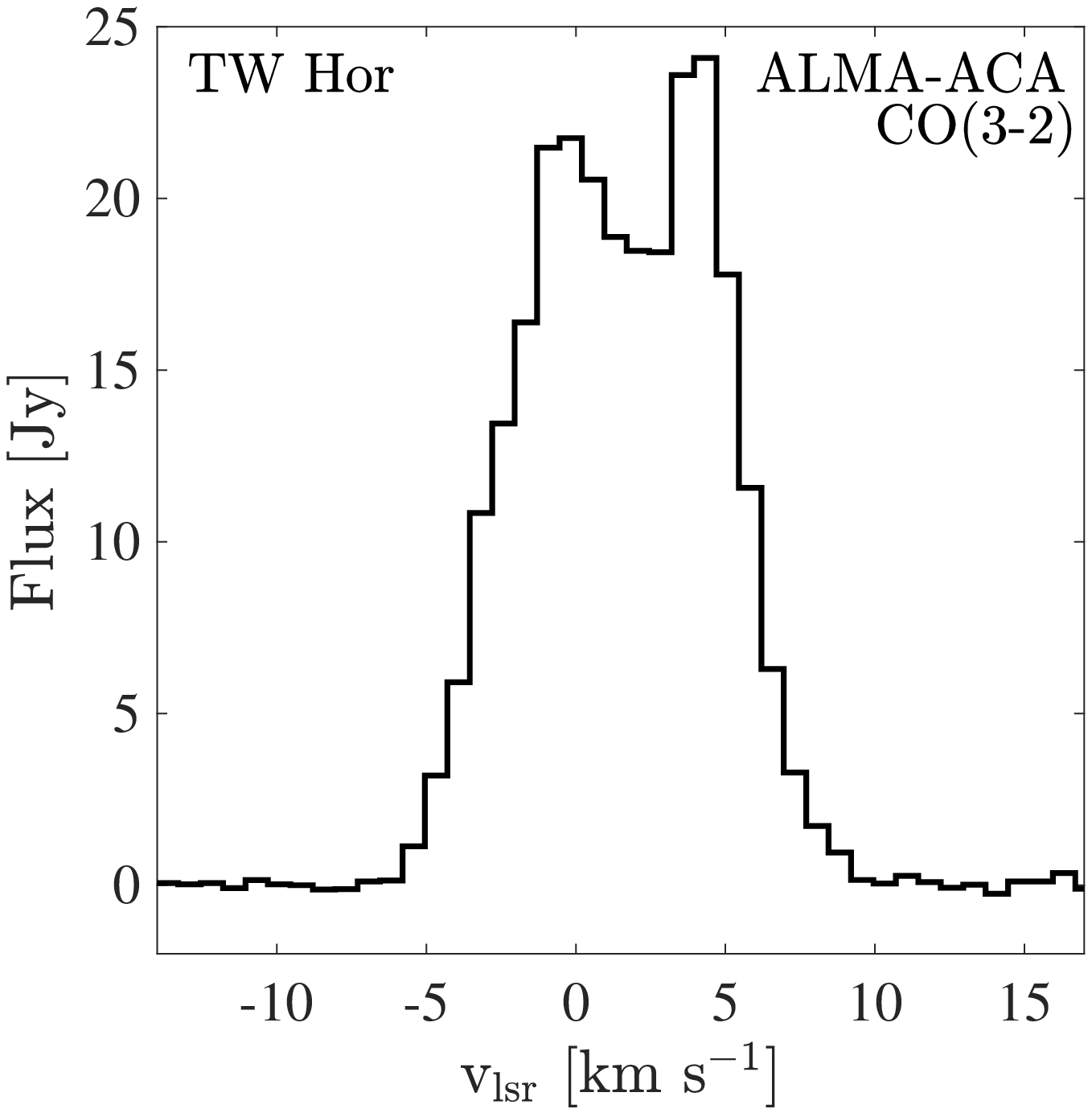}
\includegraphics[height=4.5cm]{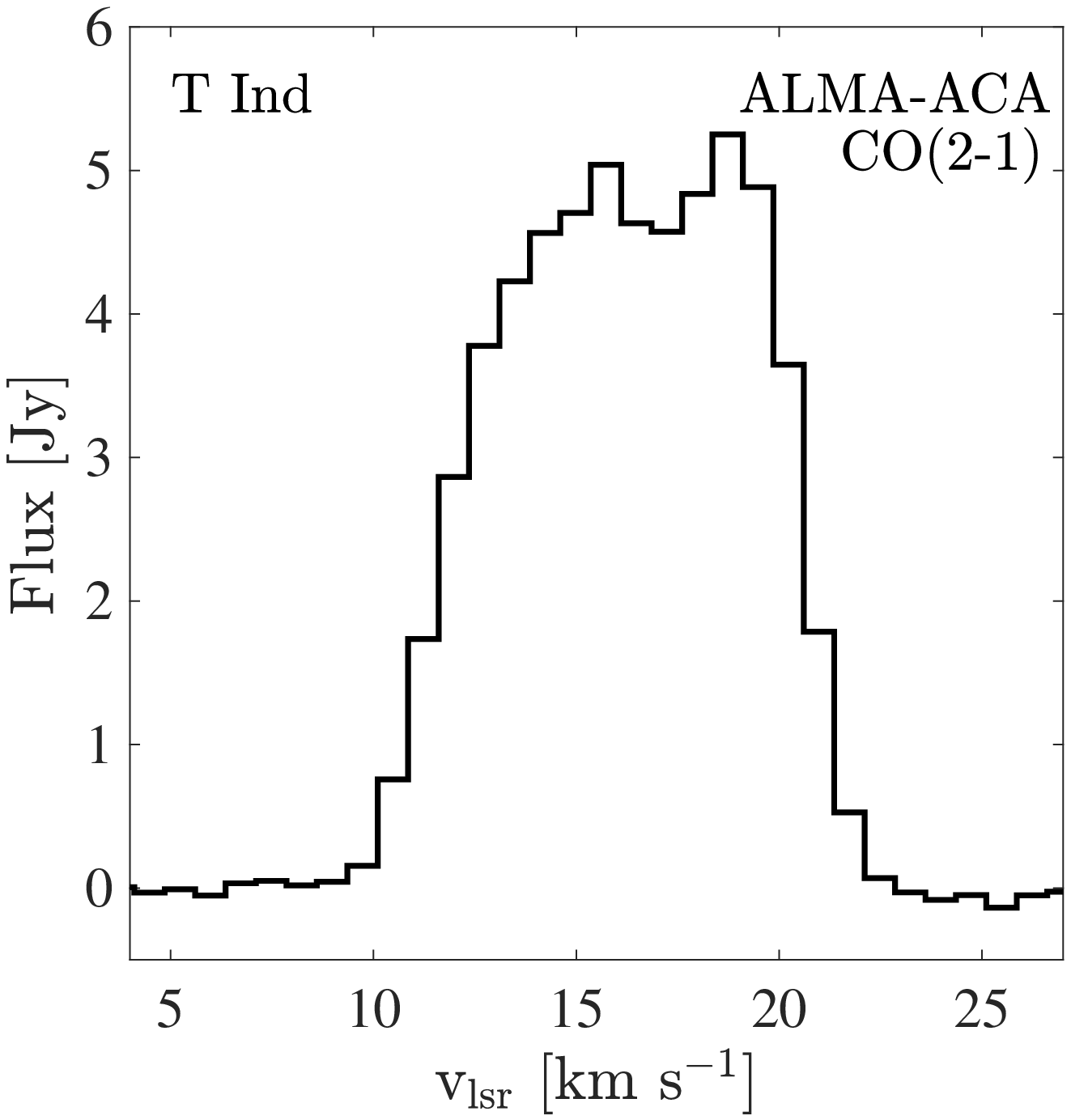}
\includegraphics[height=4.5cm]{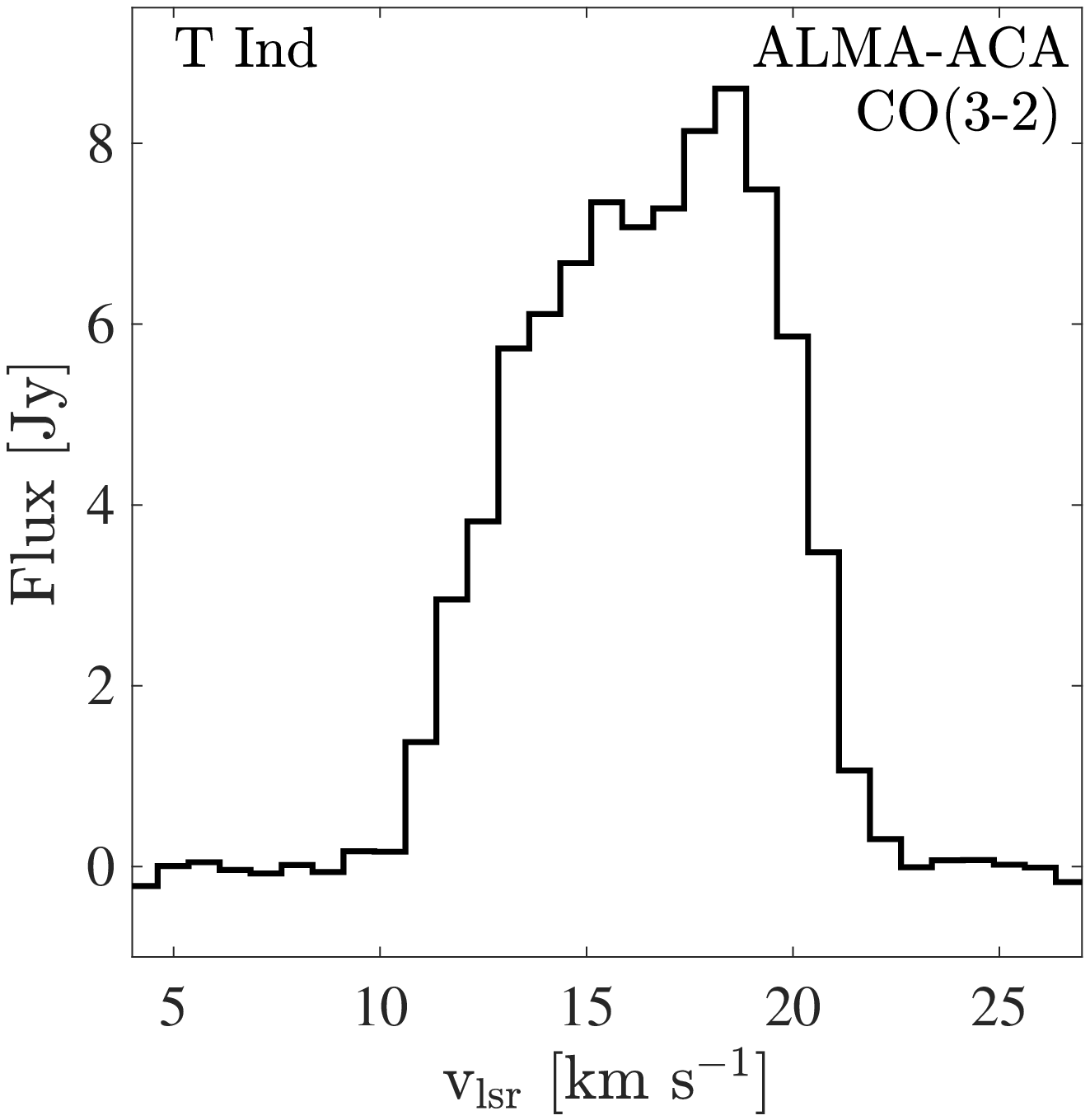}

\includegraphics[height=4.5cm]{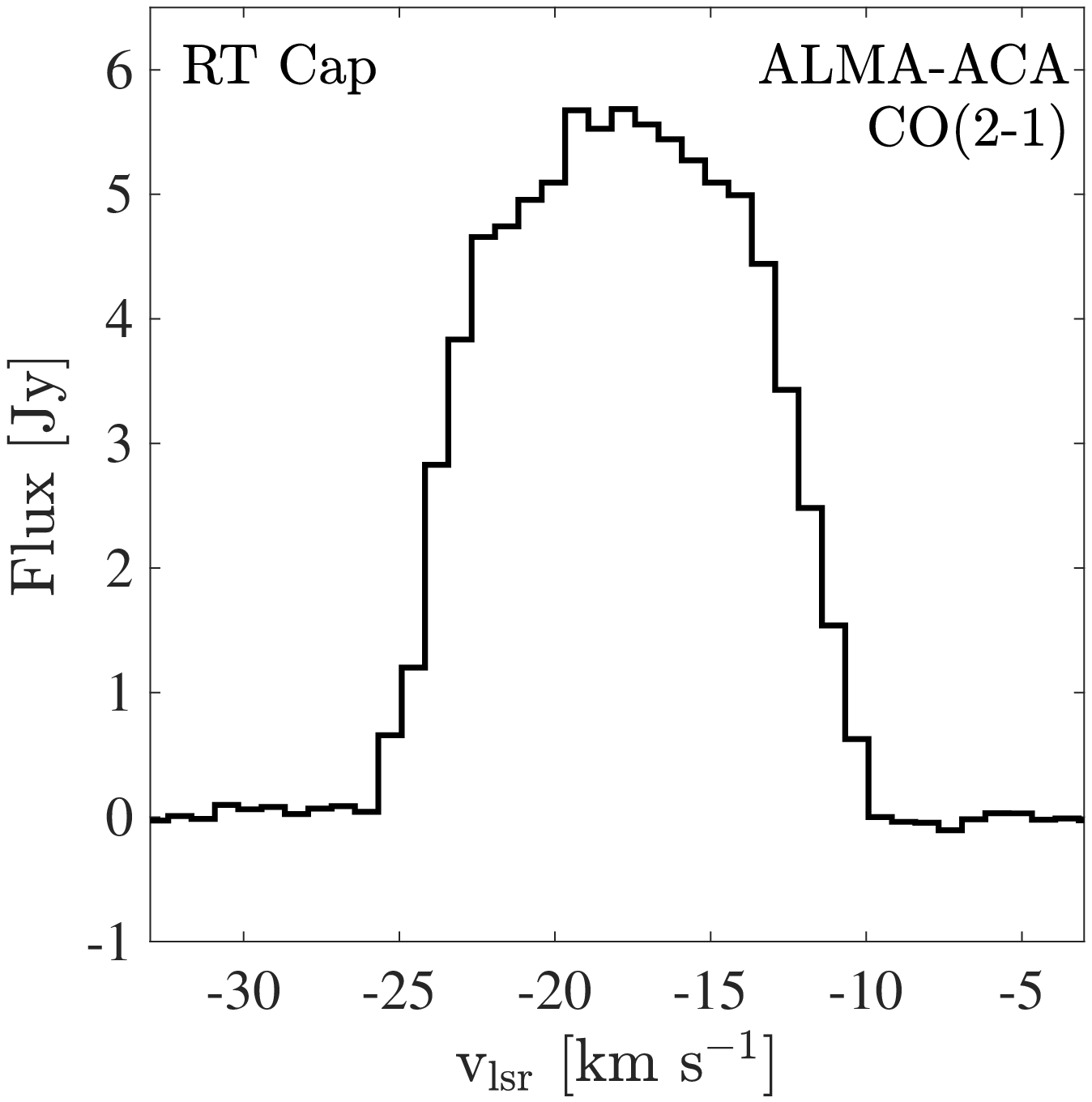}
\includegraphics[height=4.5cm]{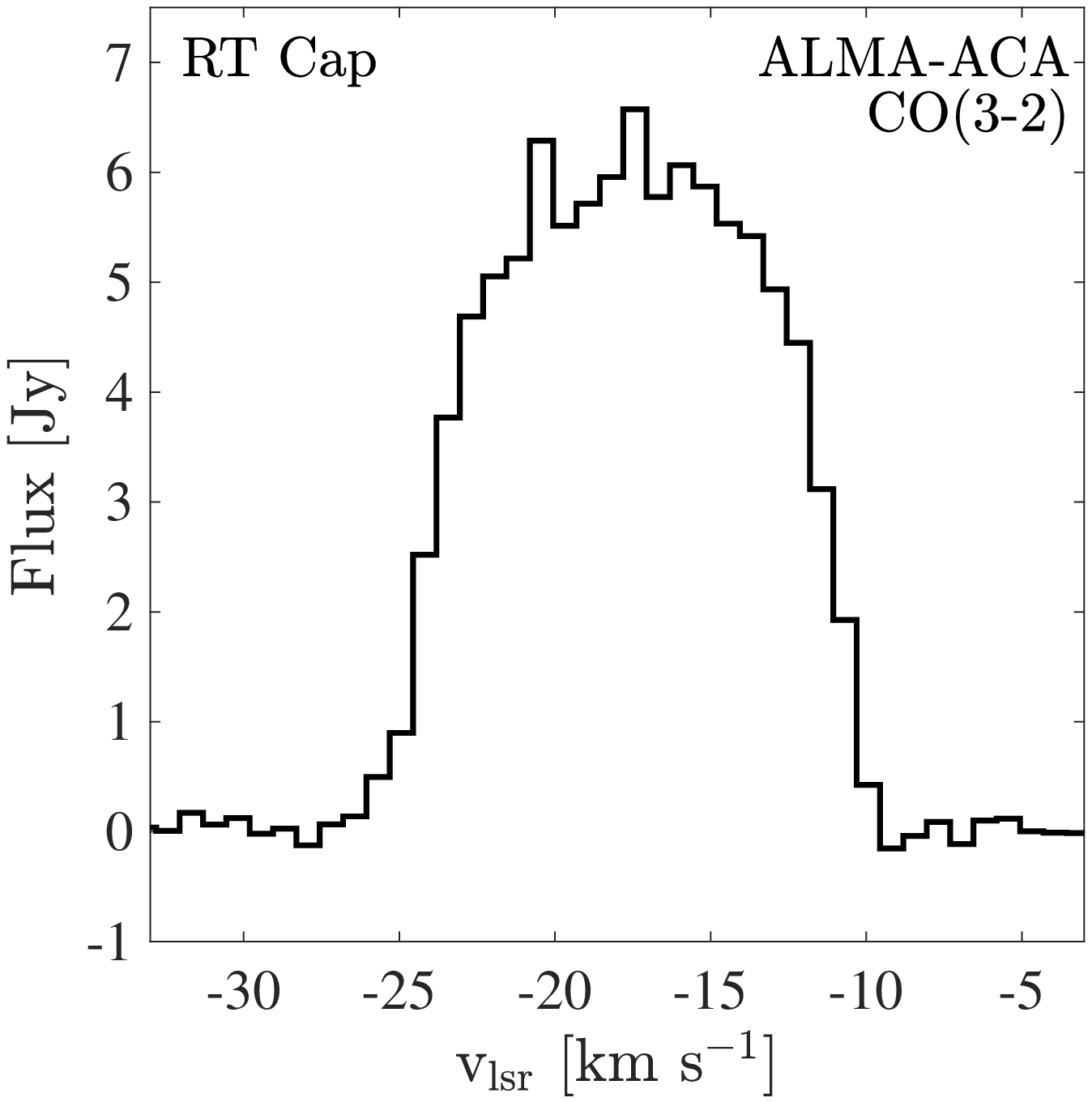}
\includegraphics[height=4.5cm]{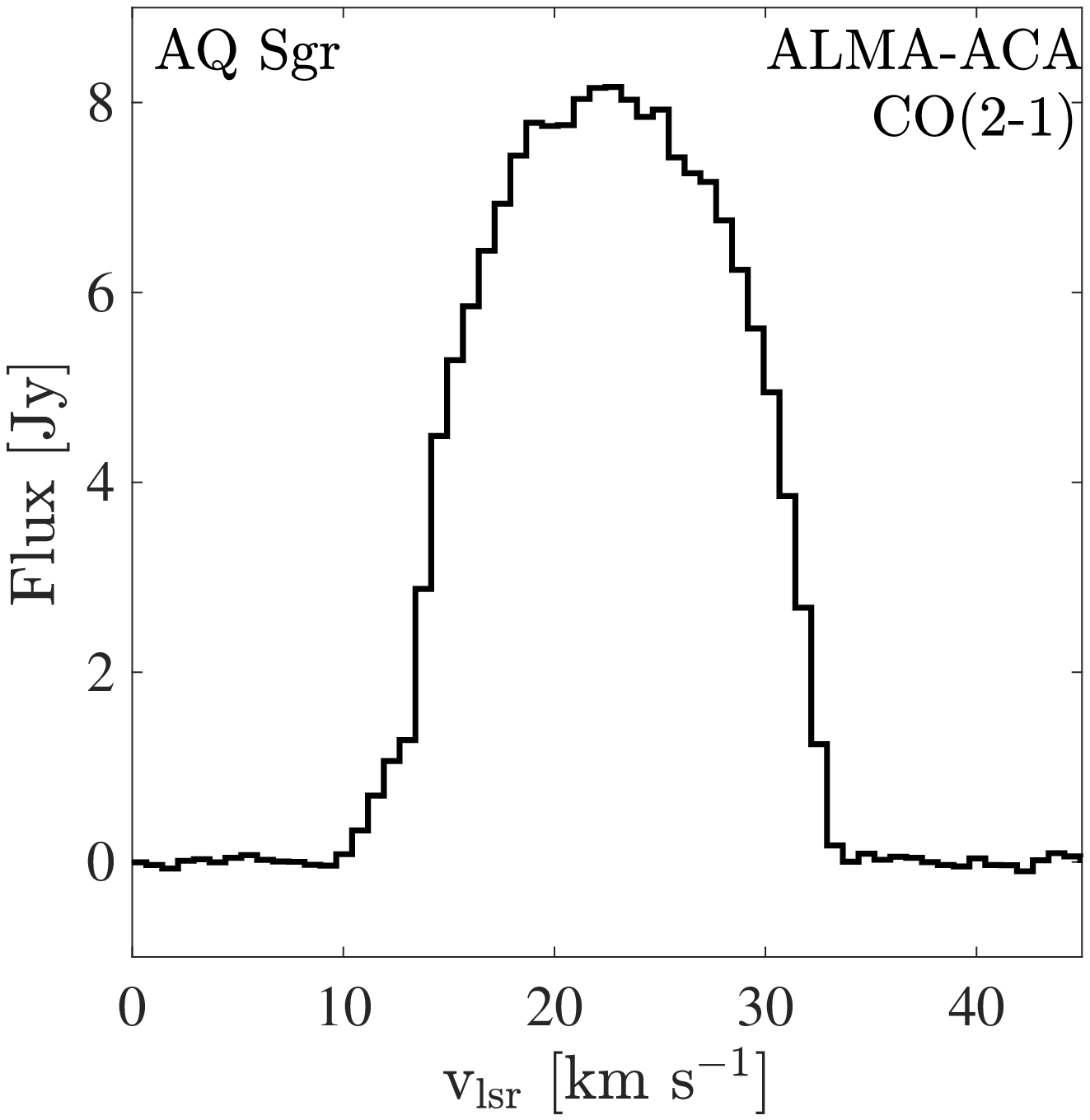}
\includegraphics[height=4.5cm]{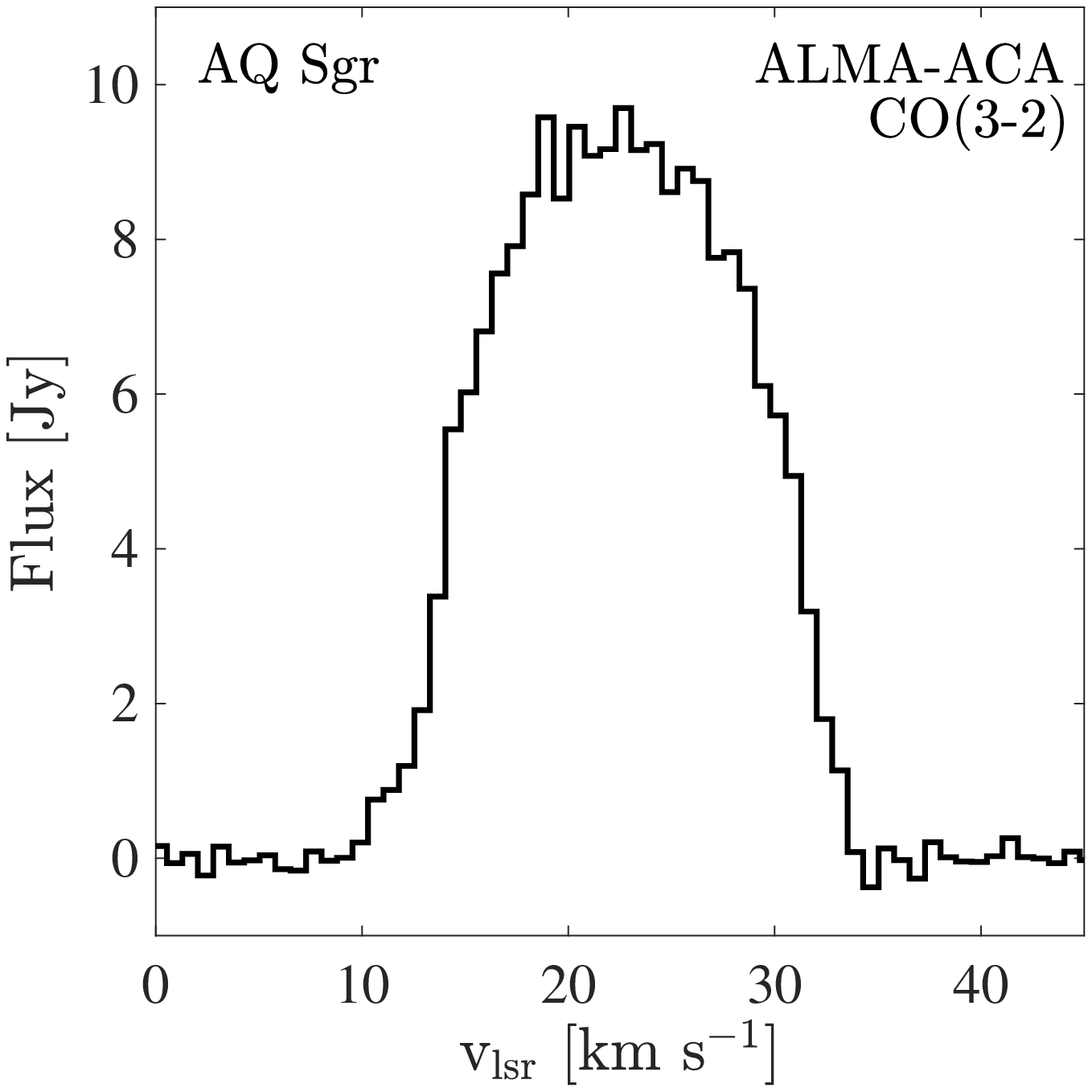}

\includegraphics[height=4.5cm]{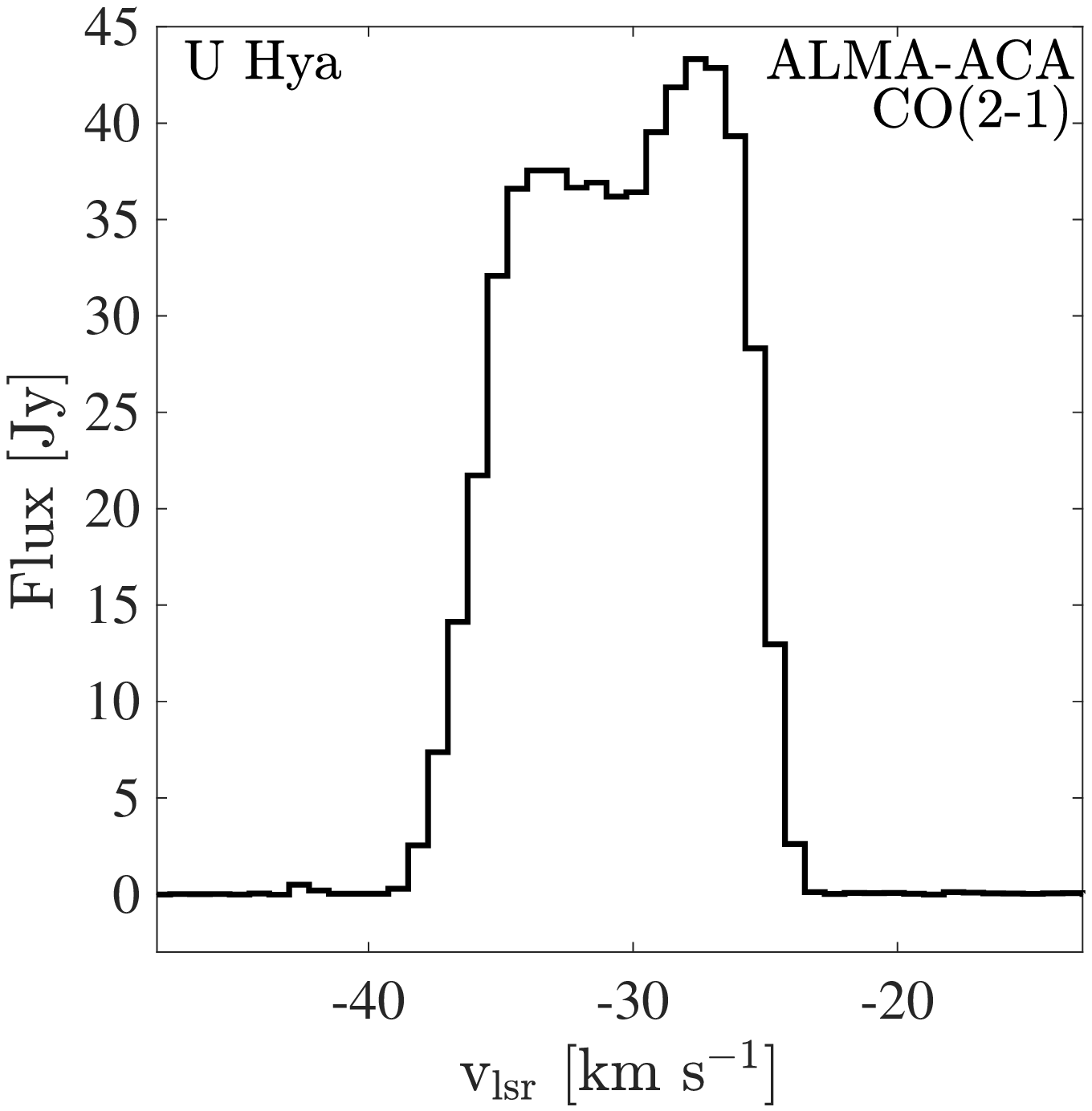}
\includegraphics[height=4.5cm]{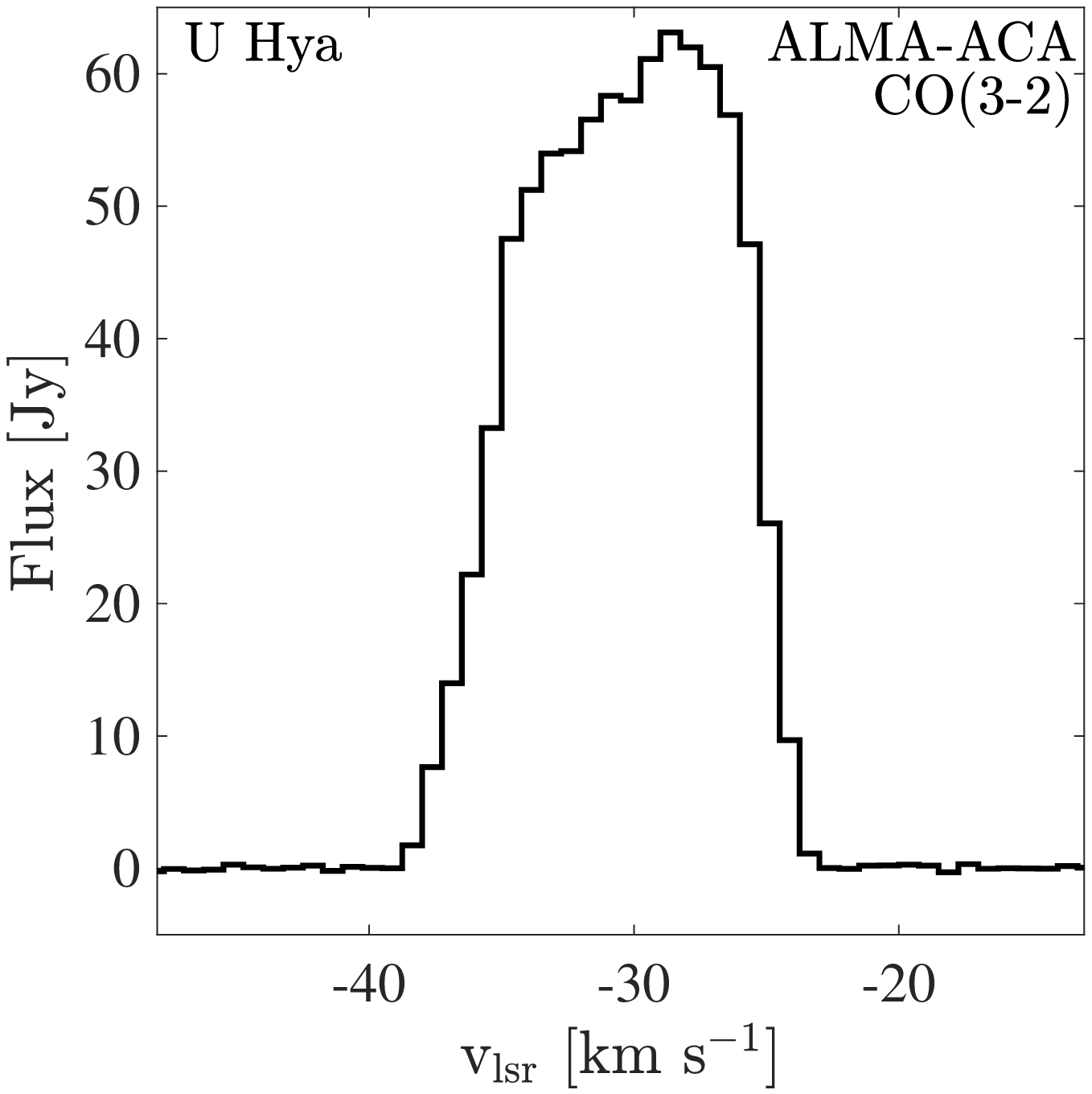}
\includegraphics[height=4.5cm]{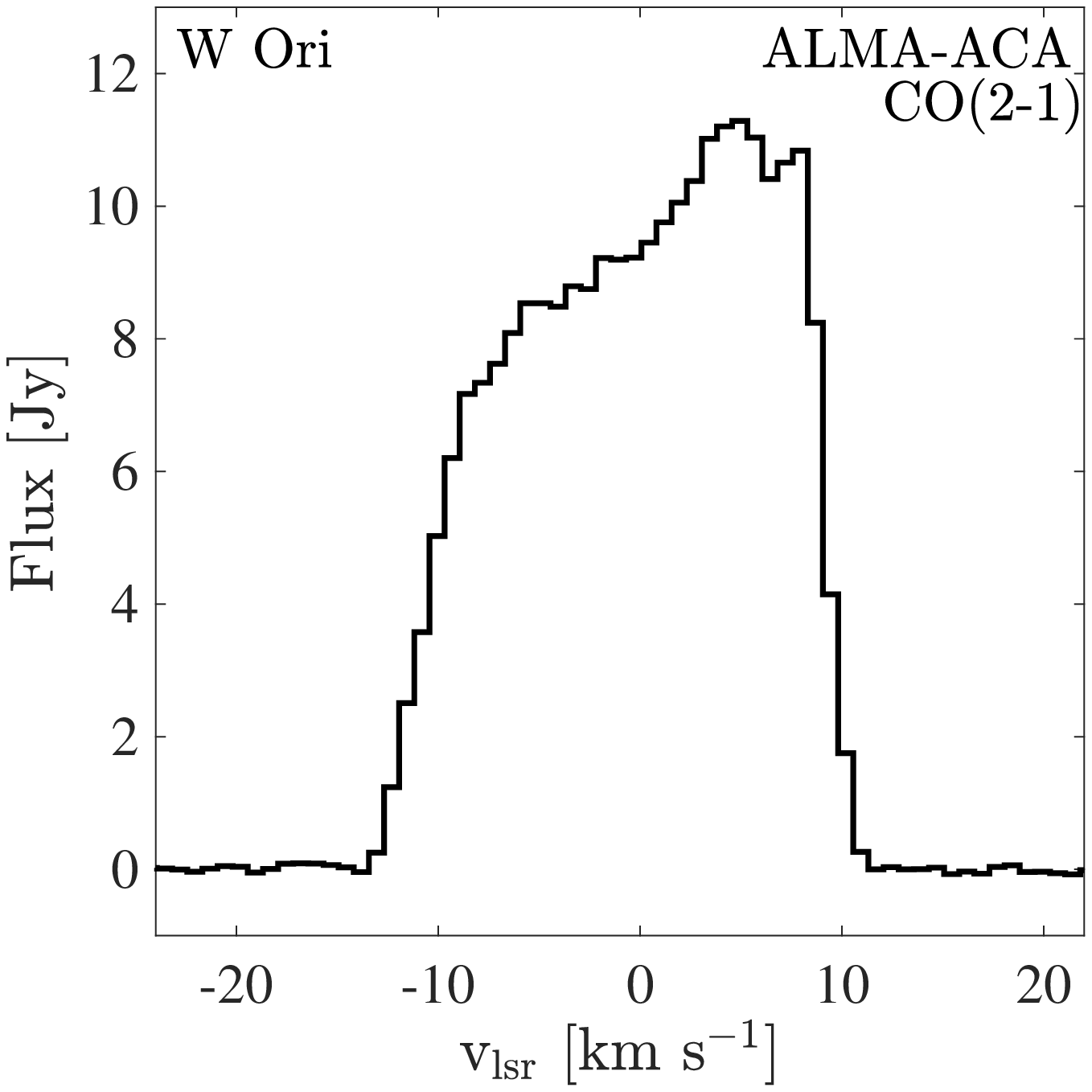}
\includegraphics[height=4.5cm]{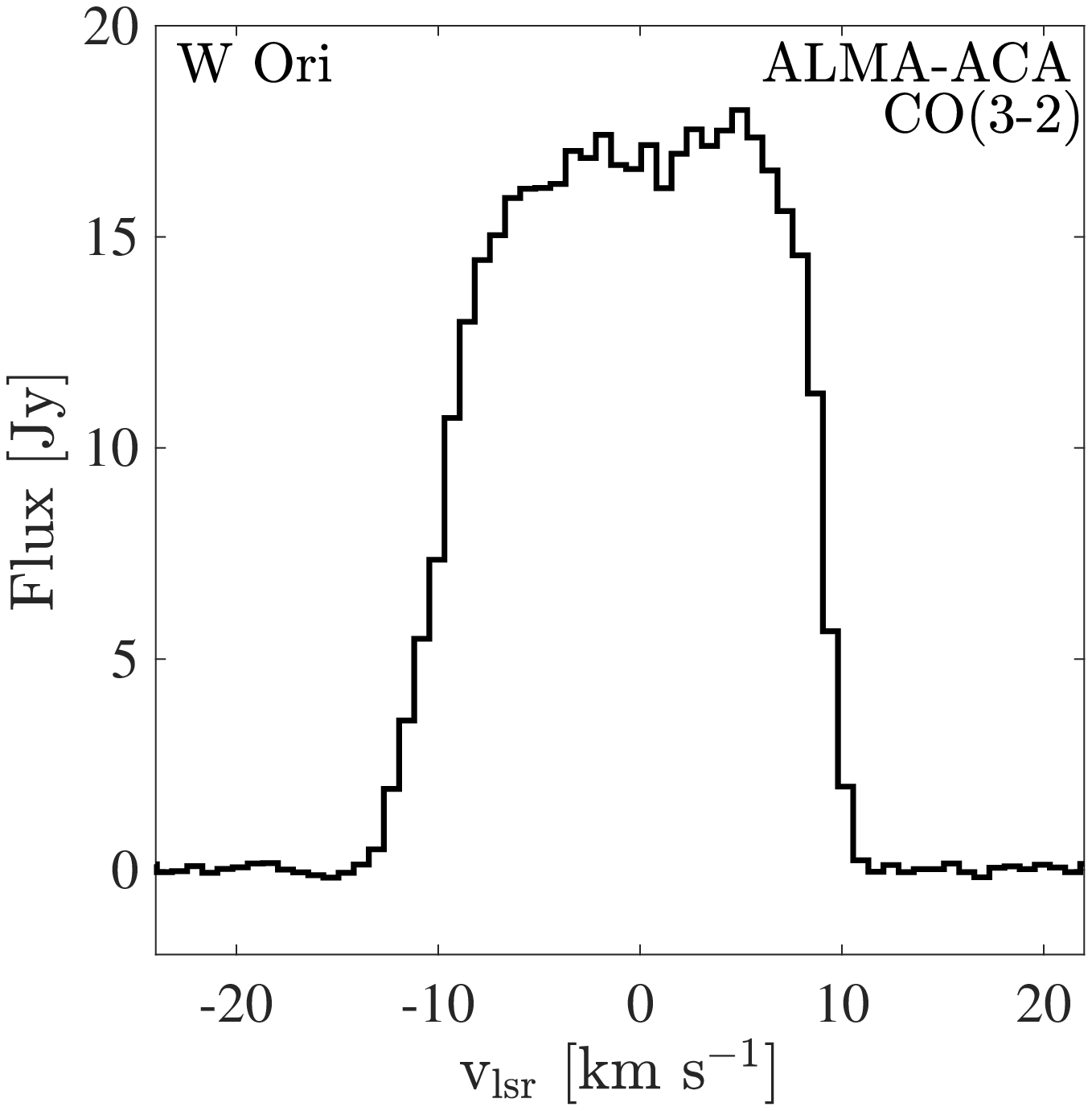}

\includegraphics[height=4.5cm]{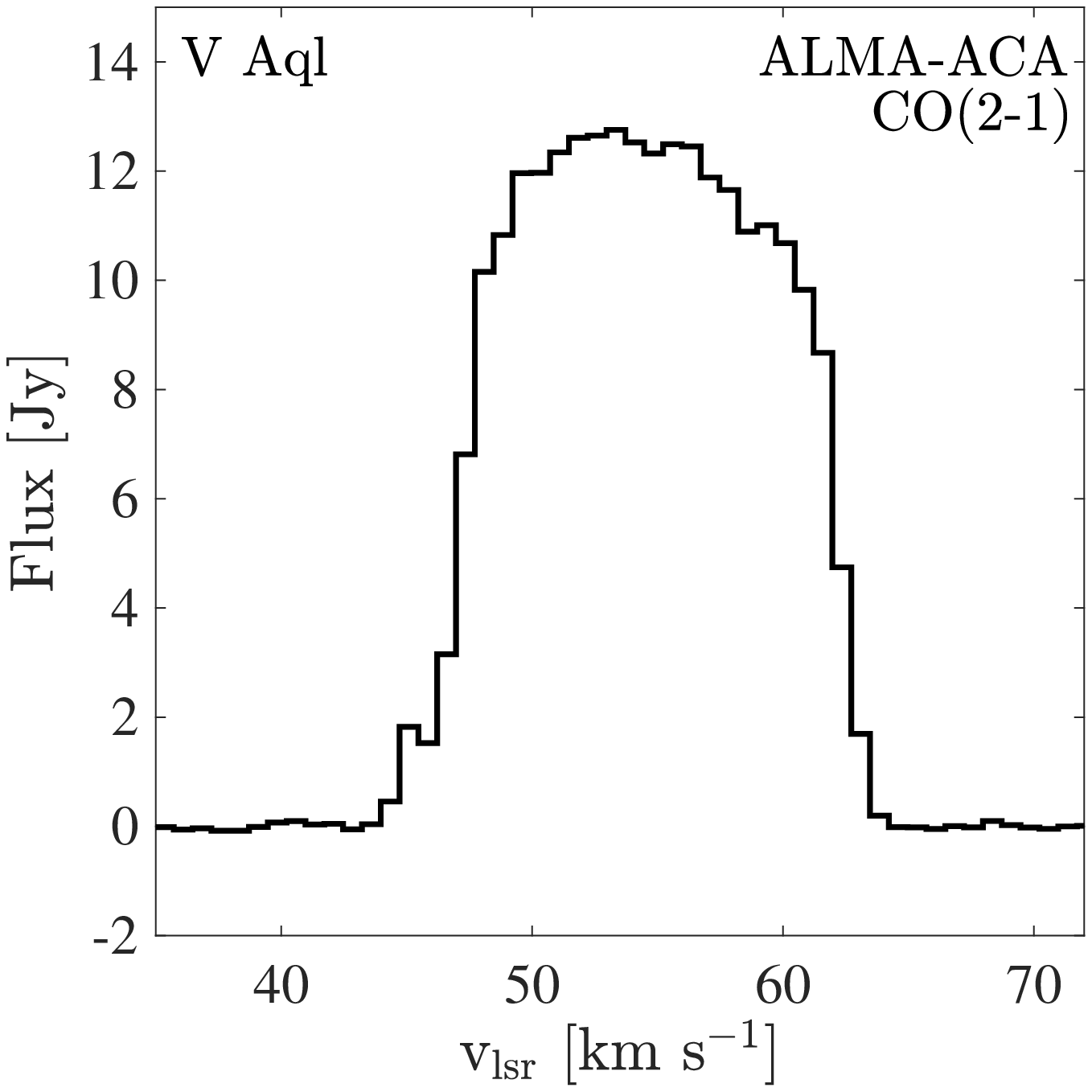}
\includegraphics[height=4.5cm]{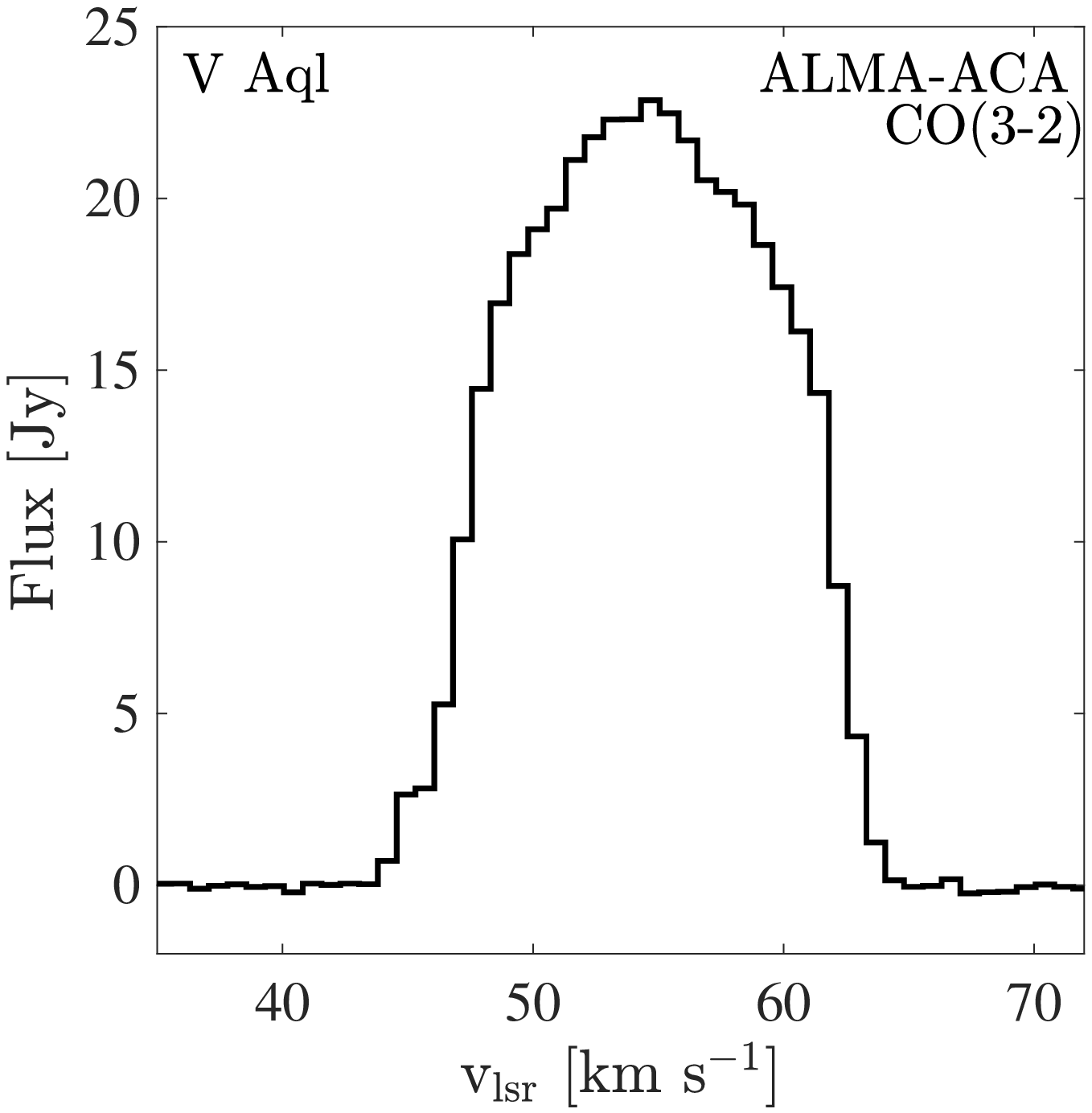}
\includegraphics[height=4.5cm]{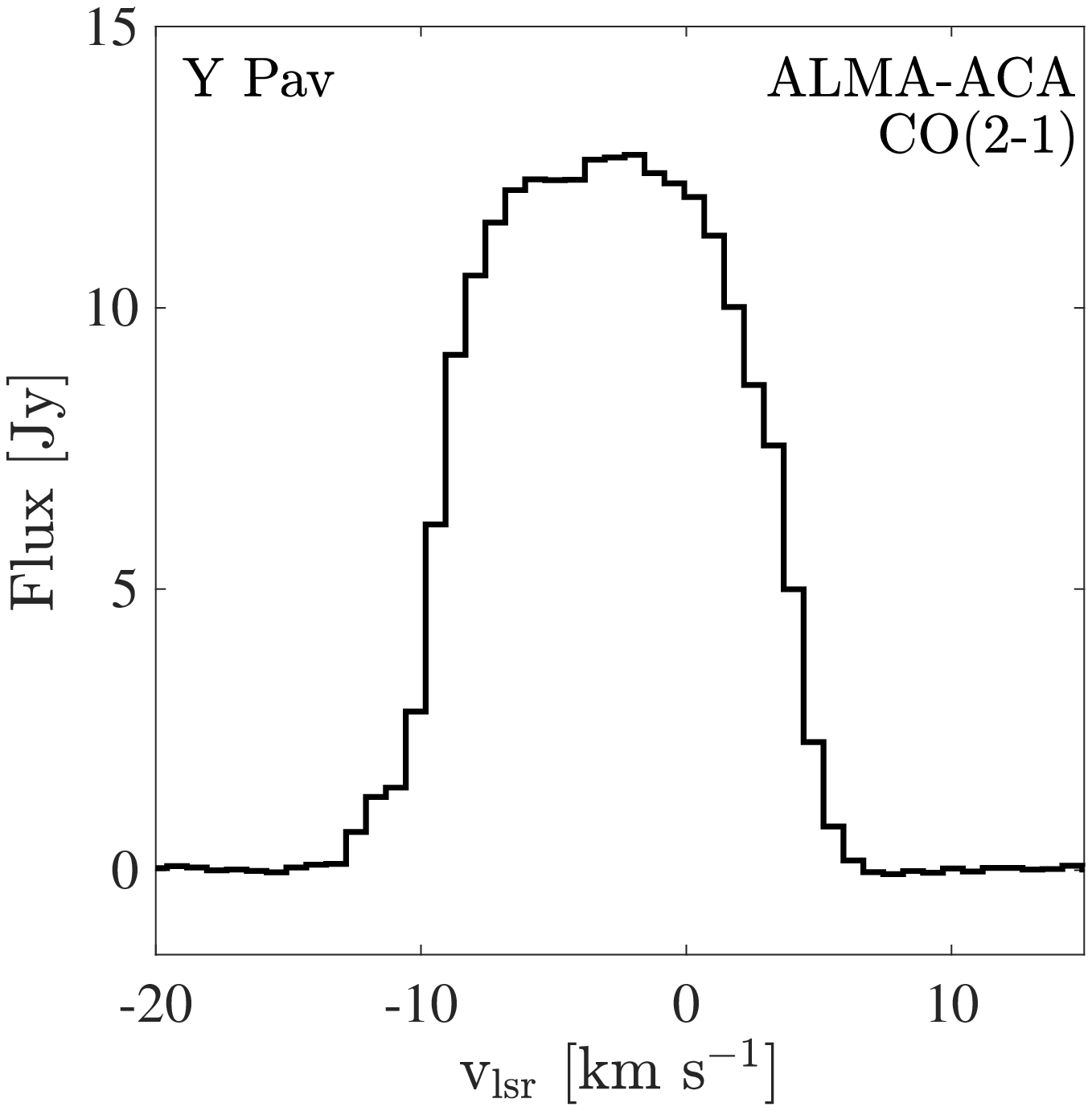}
\includegraphics[height=4.5cm]{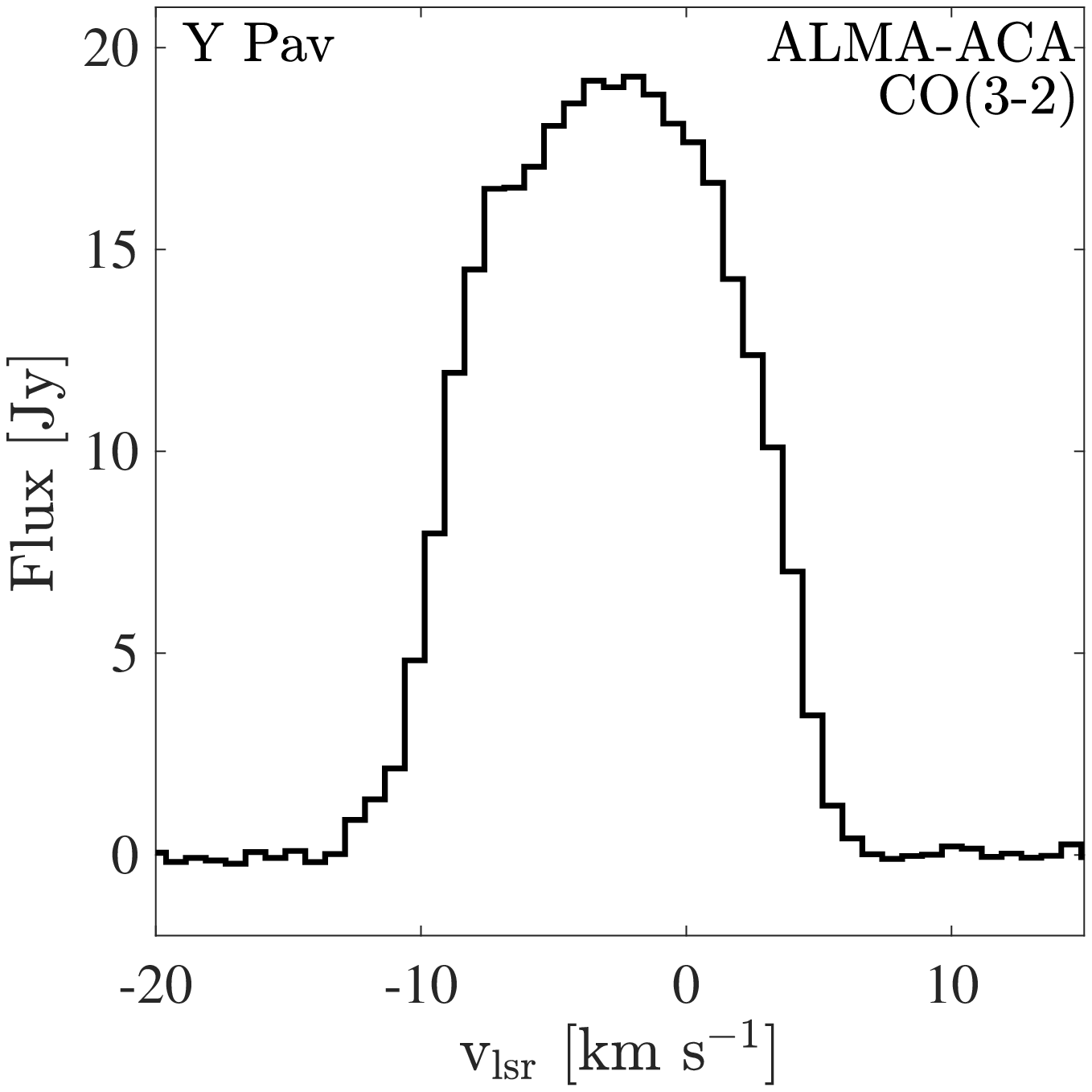}
\caption{CO $J$\,=\,2$\rightarrow$1 and 3$\rightarrow$2 line profiles measured toward the C-type AGB stars of the sample discussed in this paper. The source name is given in the upper left corner and the transition is in the upper right corner of each plot.}
\label{linesC_SR}
\end{figure*}


\begin{figure*}[t]
\includegraphics[height=4.5cm]{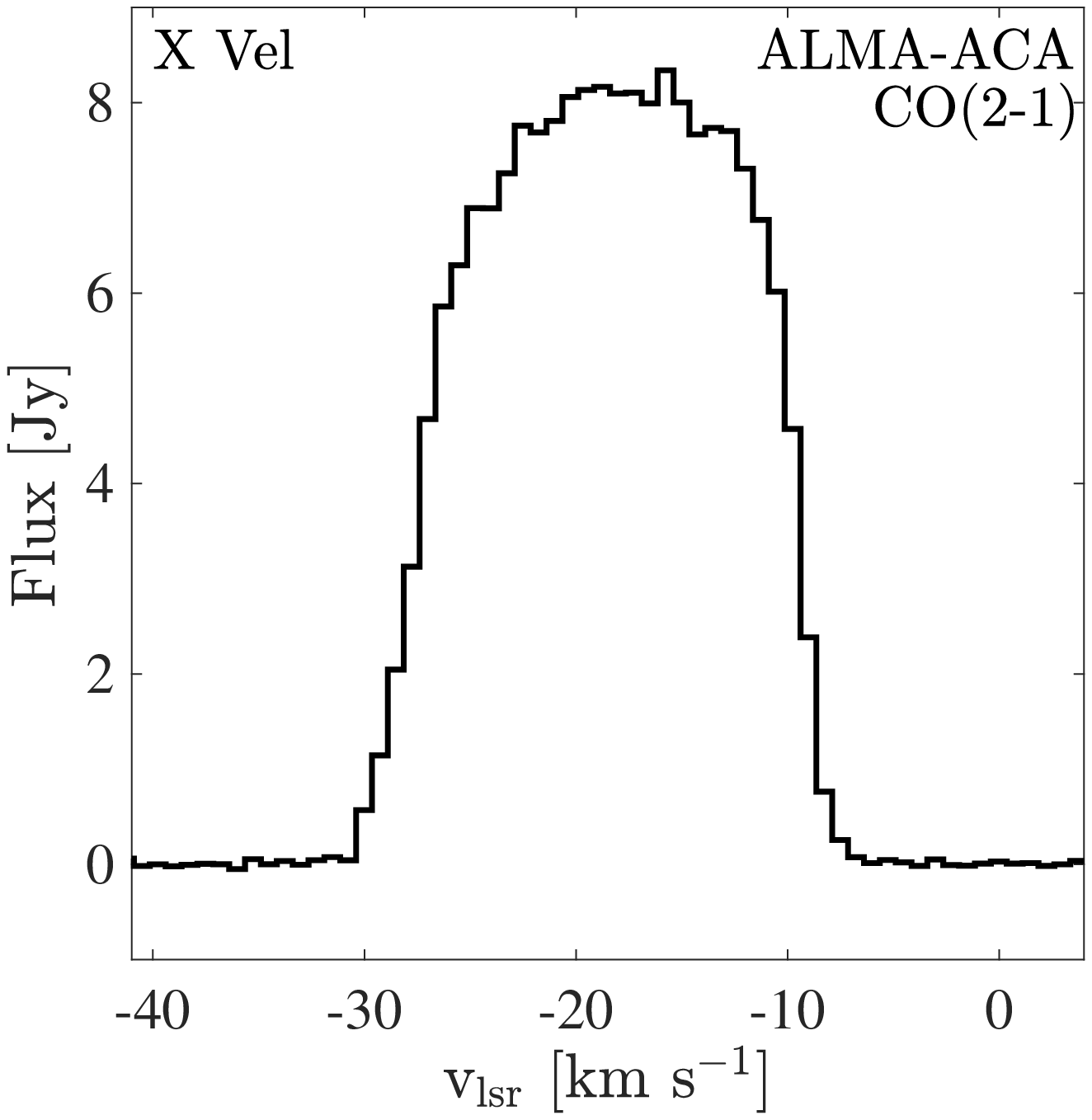}
\includegraphics[height=4.5cm]{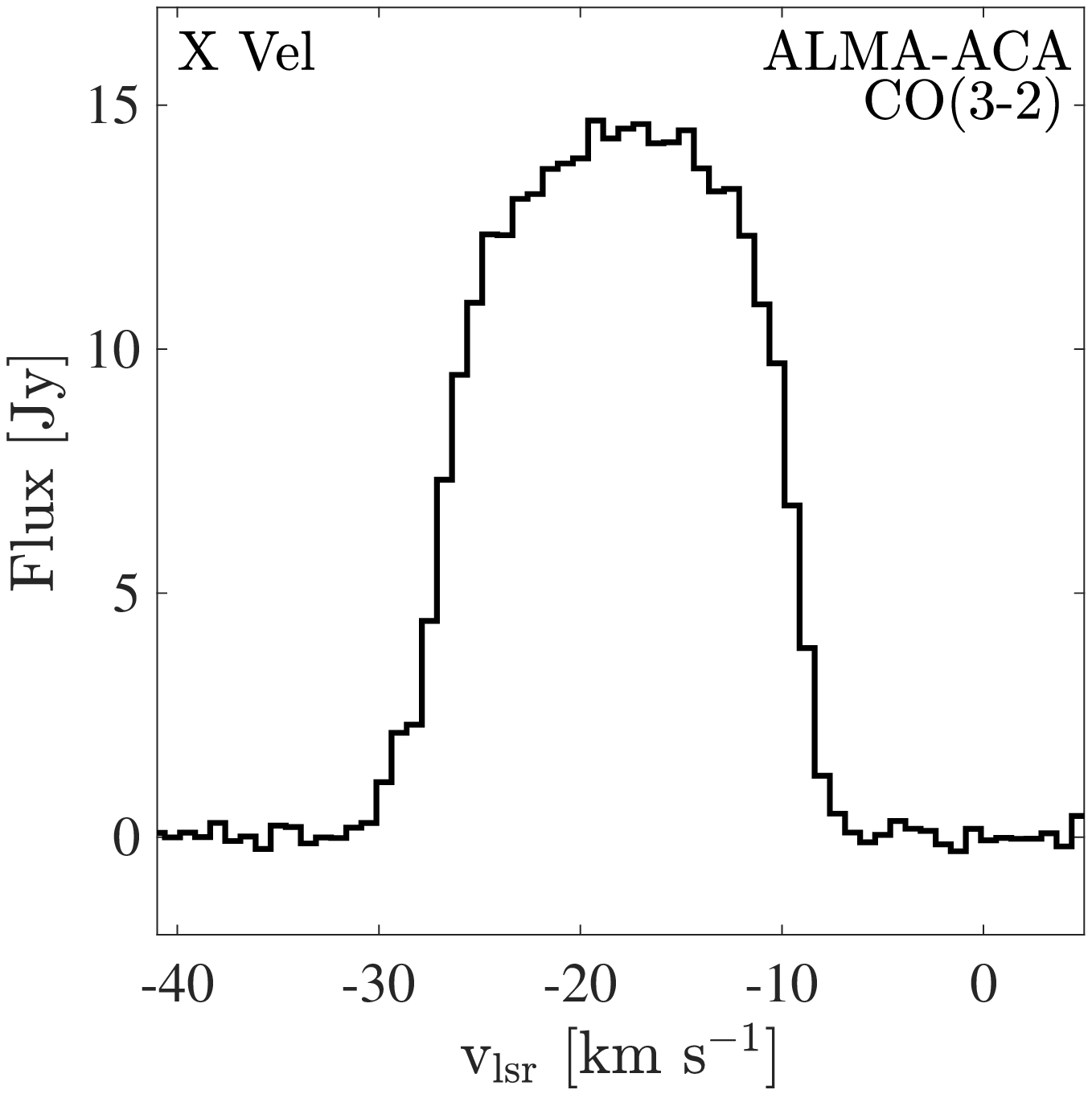}
\includegraphics[height=4.5cm]{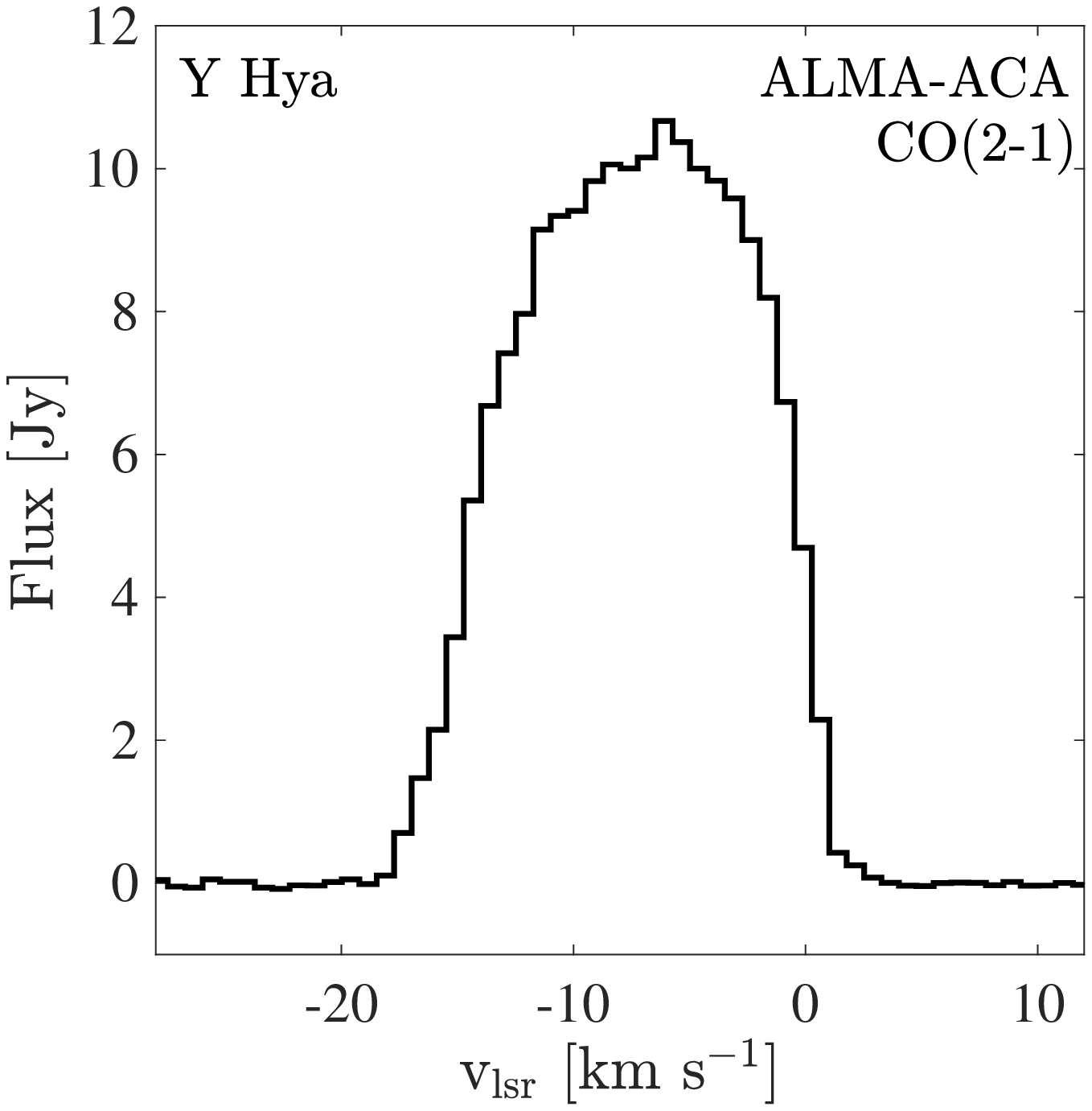}
\includegraphics[height=4.5cm]{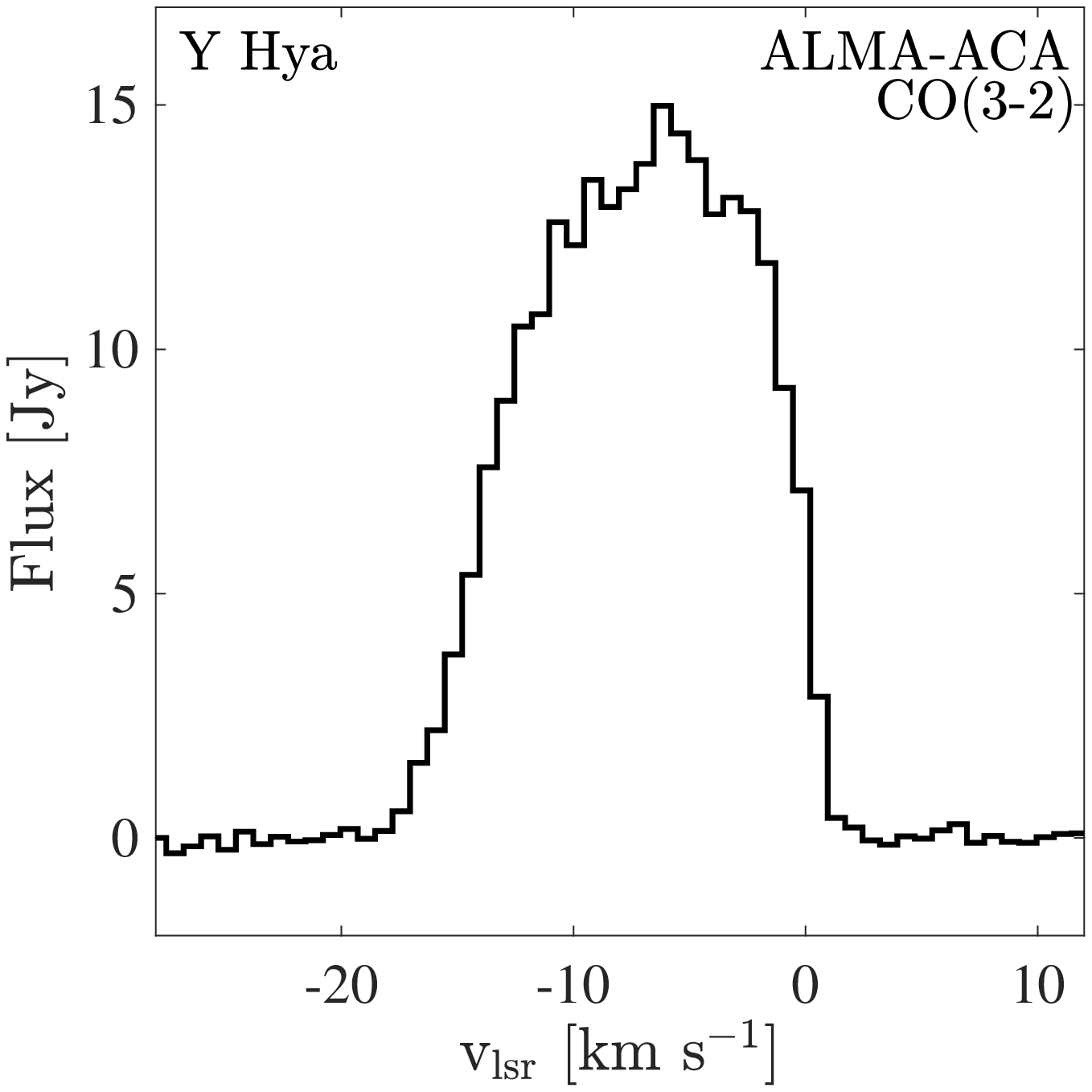}

\includegraphics[height=4.5cm]{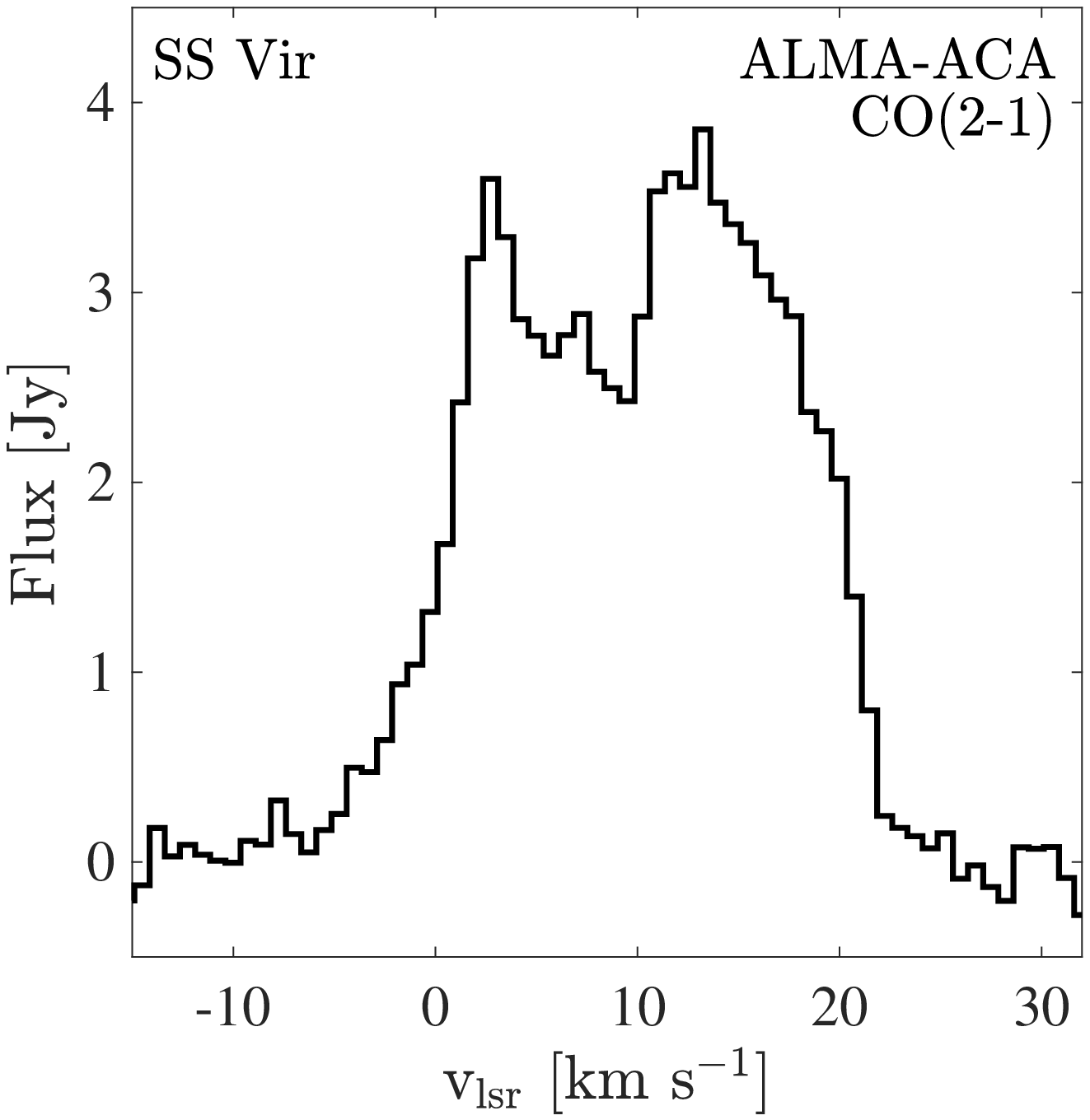}
\includegraphics[height=4.5cm]{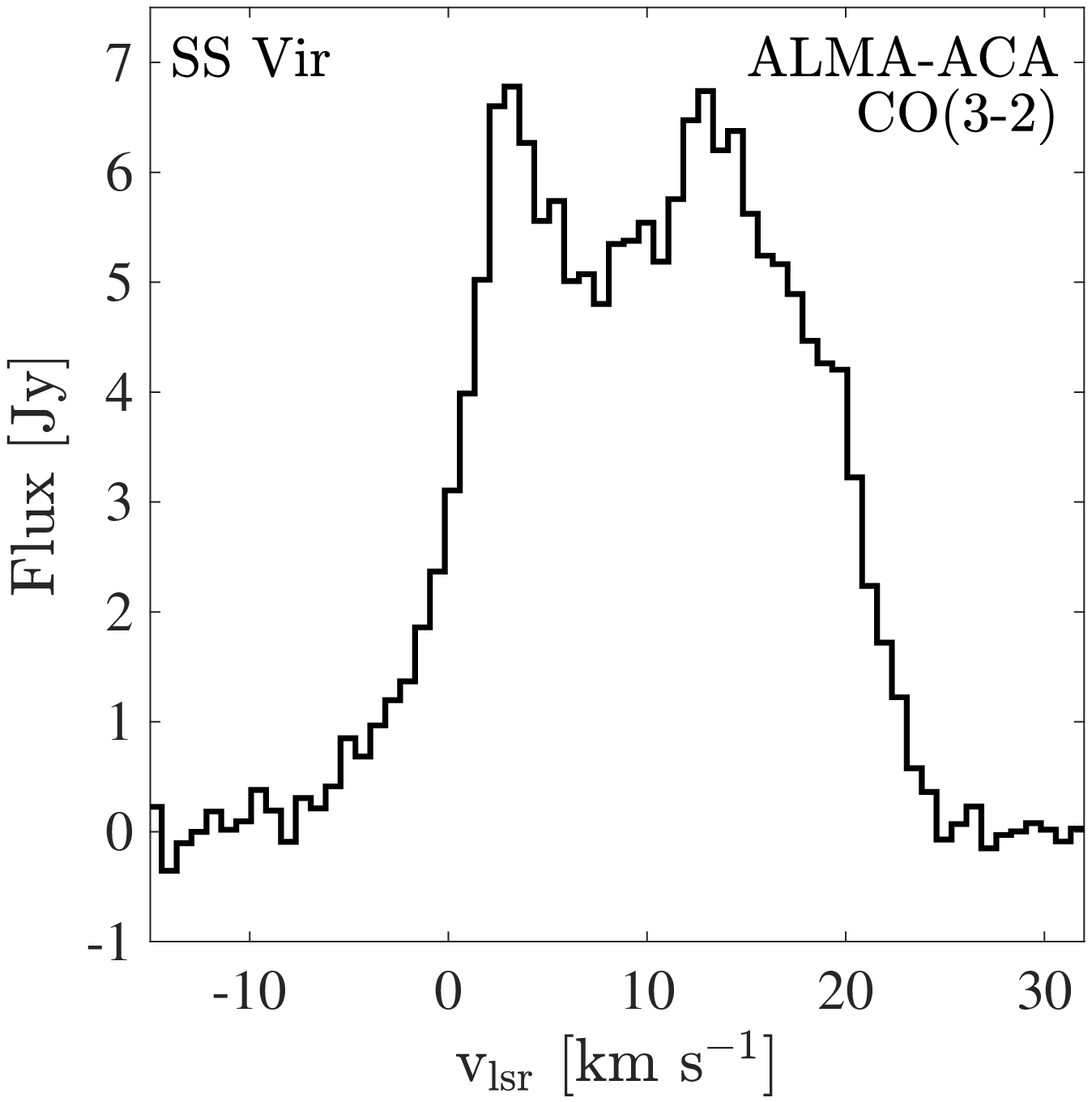}
\includegraphics[height=4.5cm]{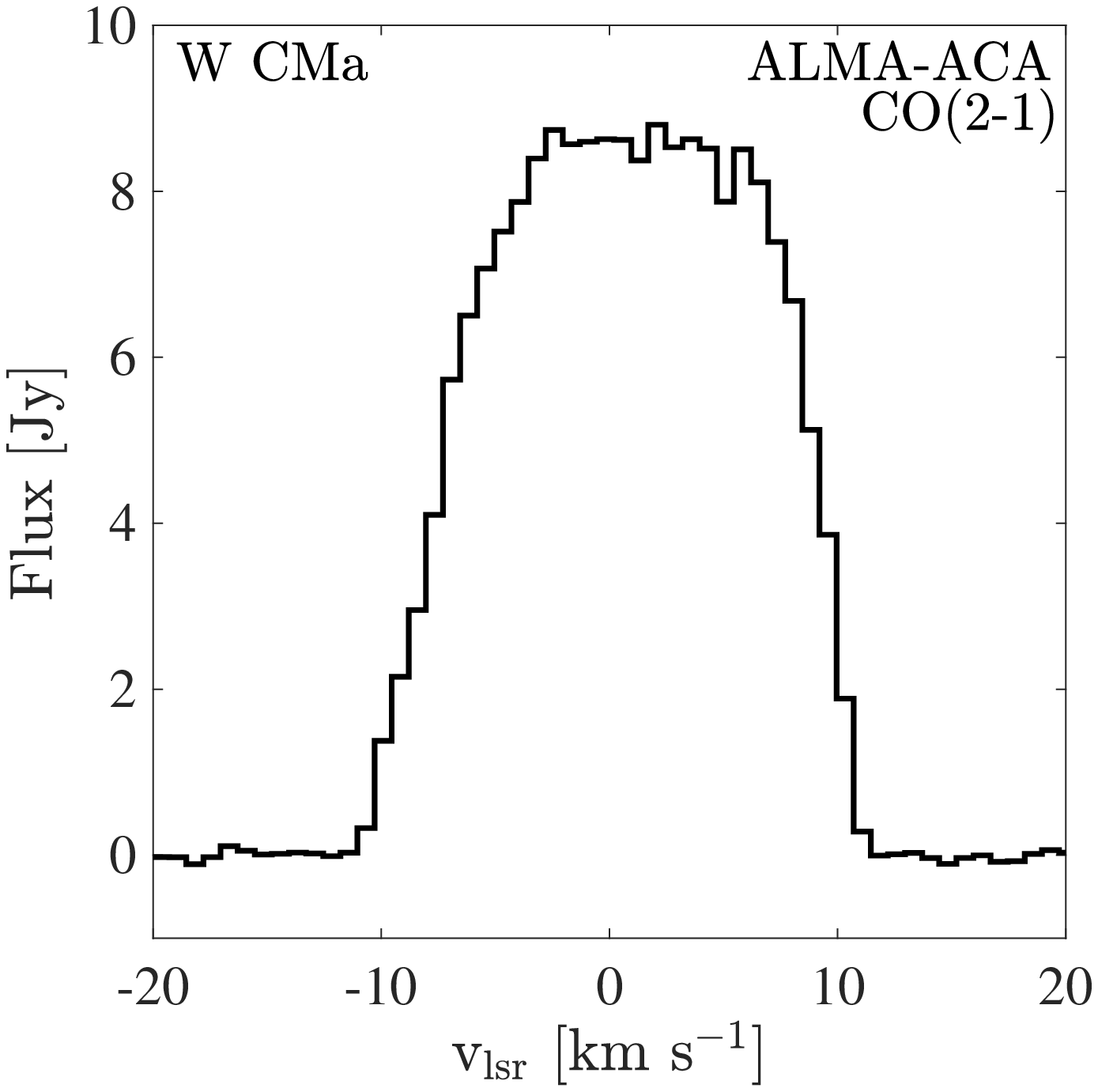}
\includegraphics[height=4.5cm]{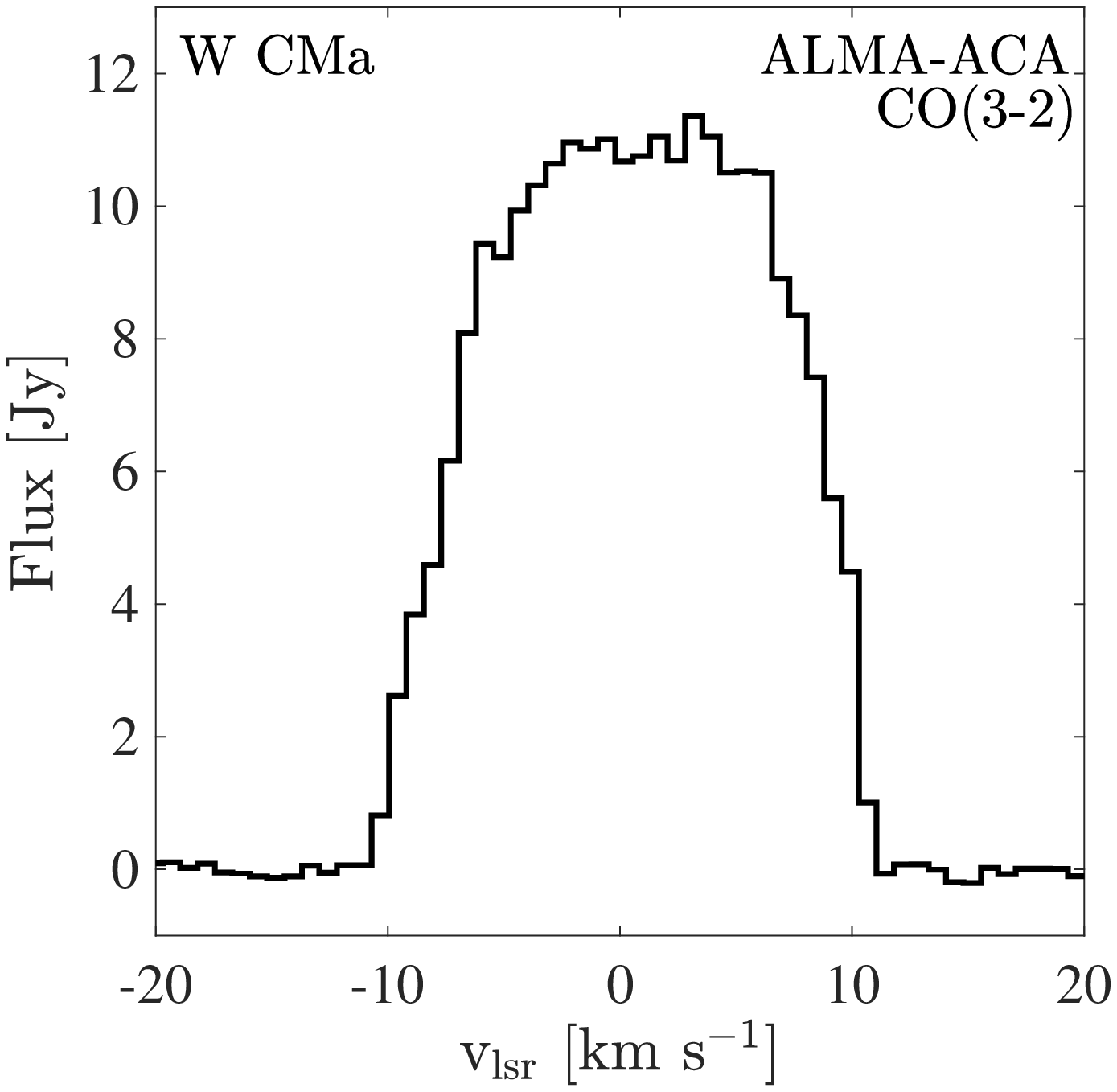}

\includegraphics[height=4.5cm]{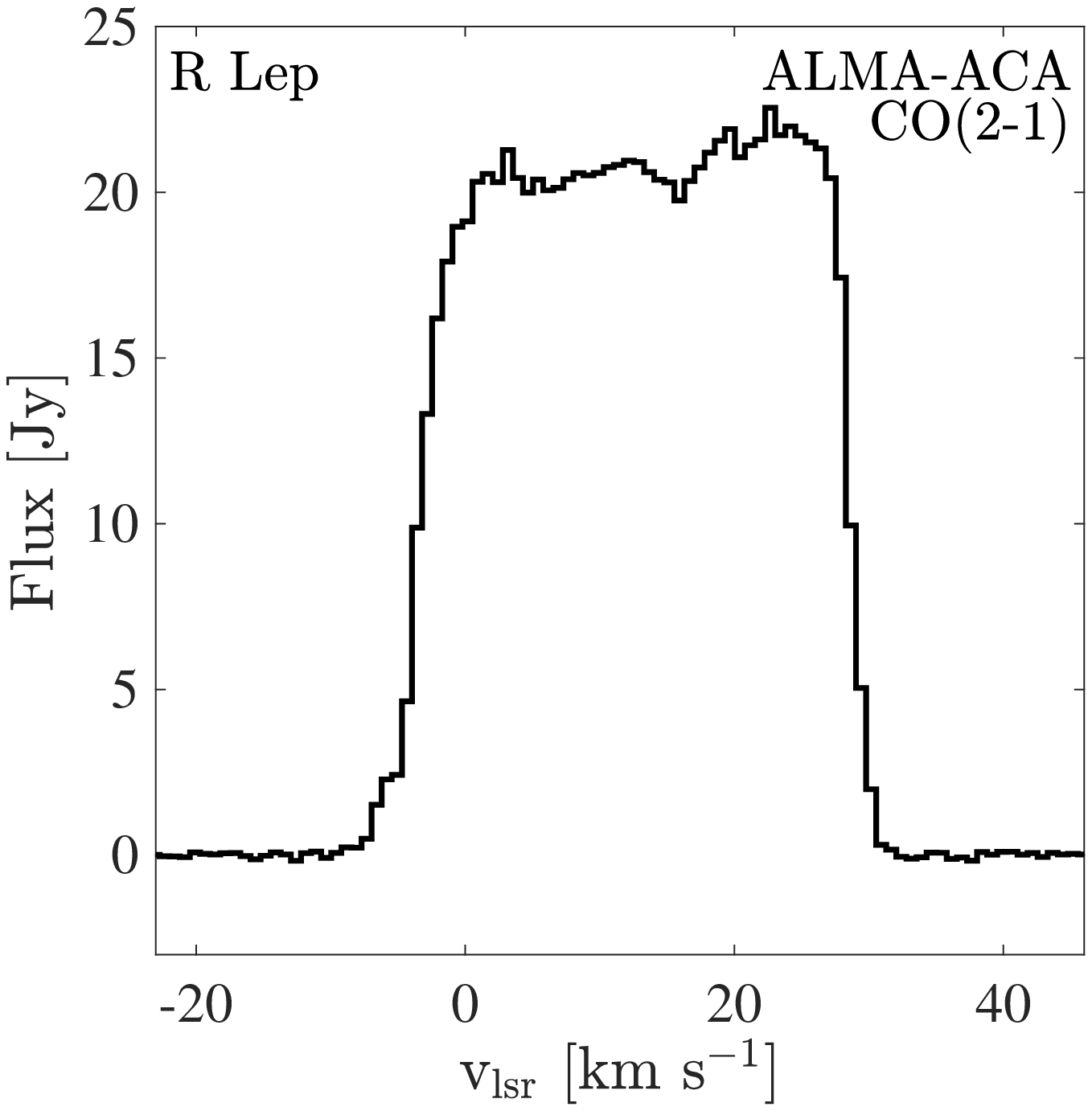}
\includegraphics[height=4.5cm]{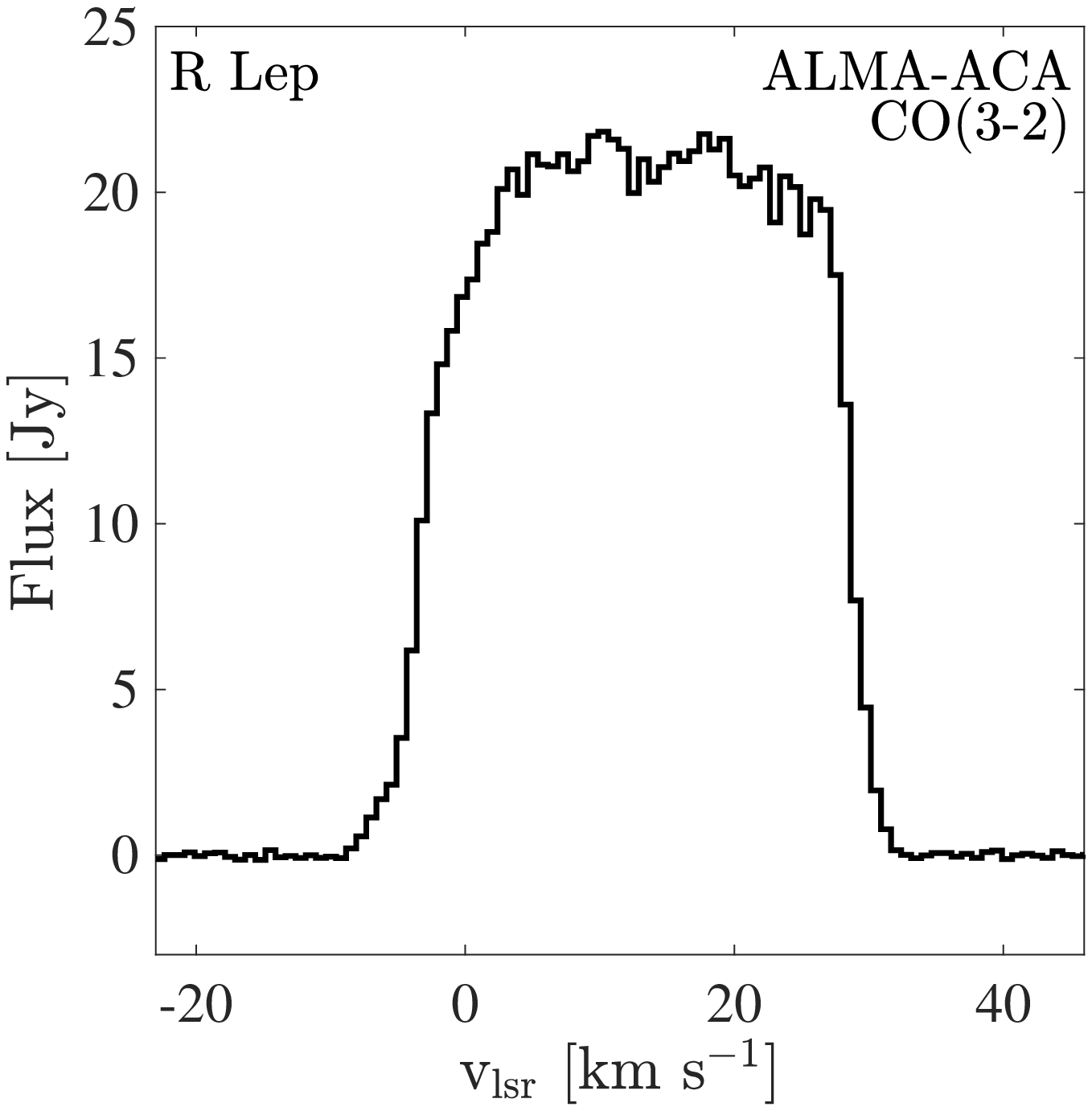}
\includegraphics[height=4.5cm]{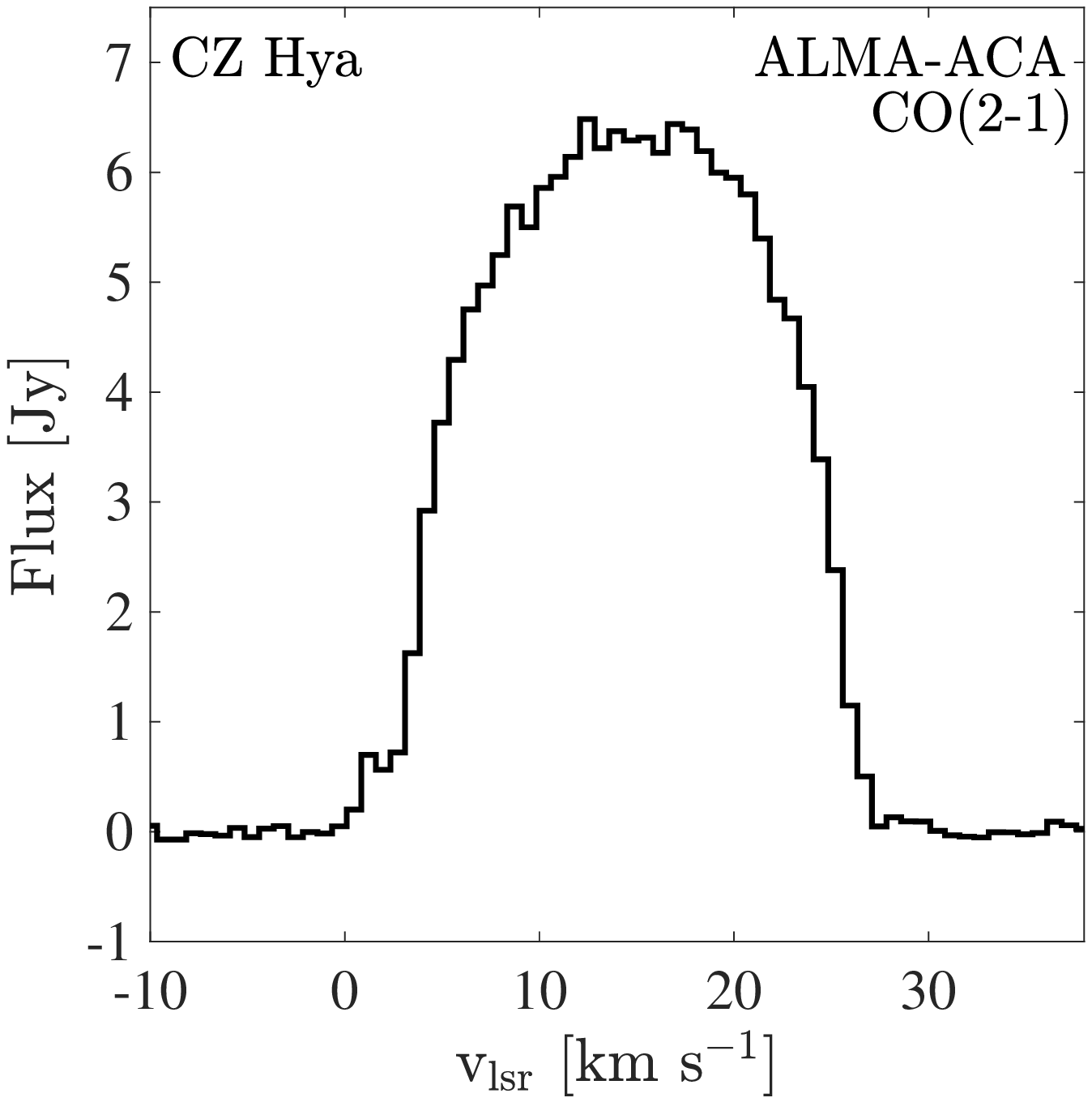}
\includegraphics[height=4.5cm]{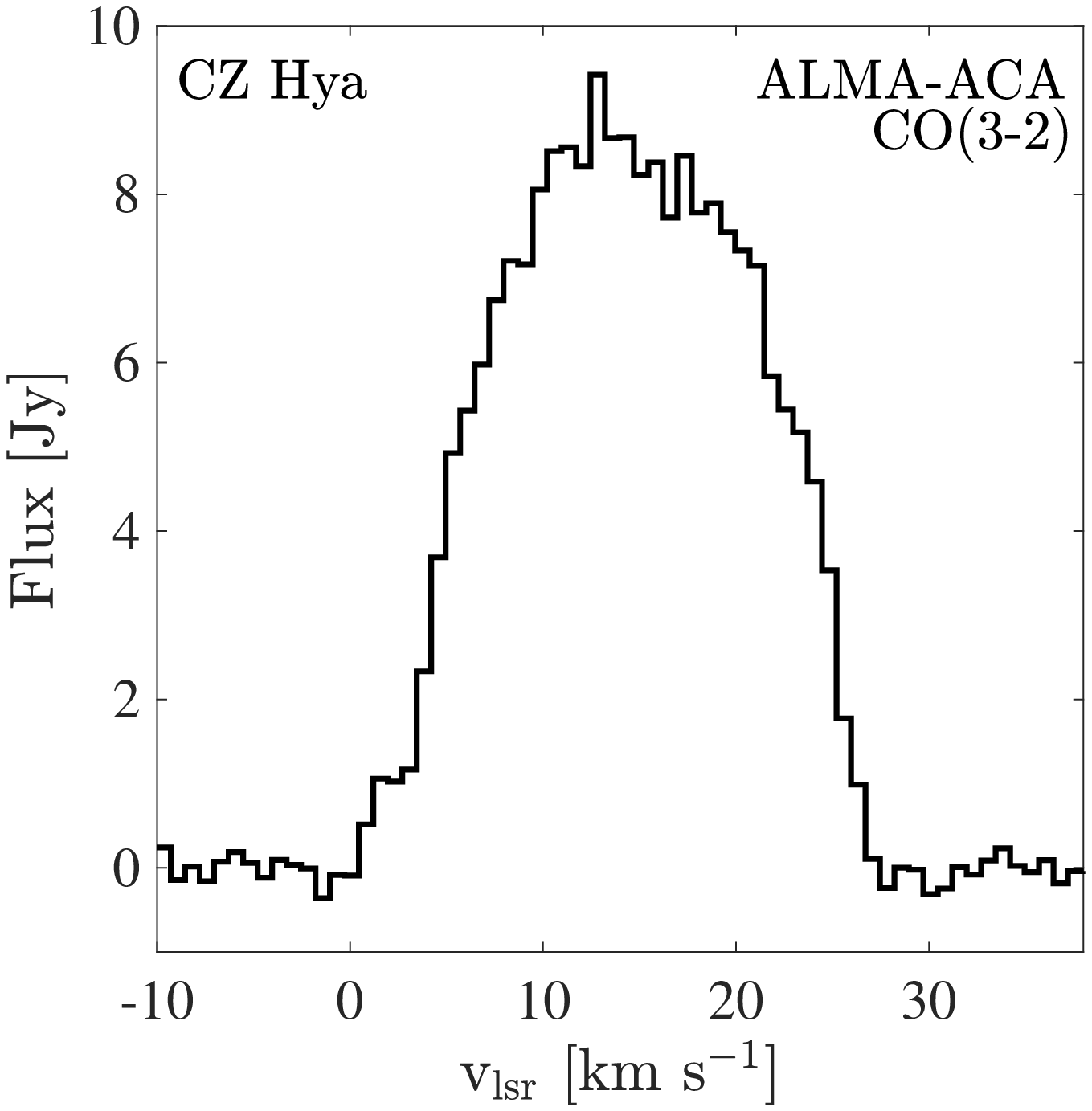}

\includegraphics[height=4.5cm]{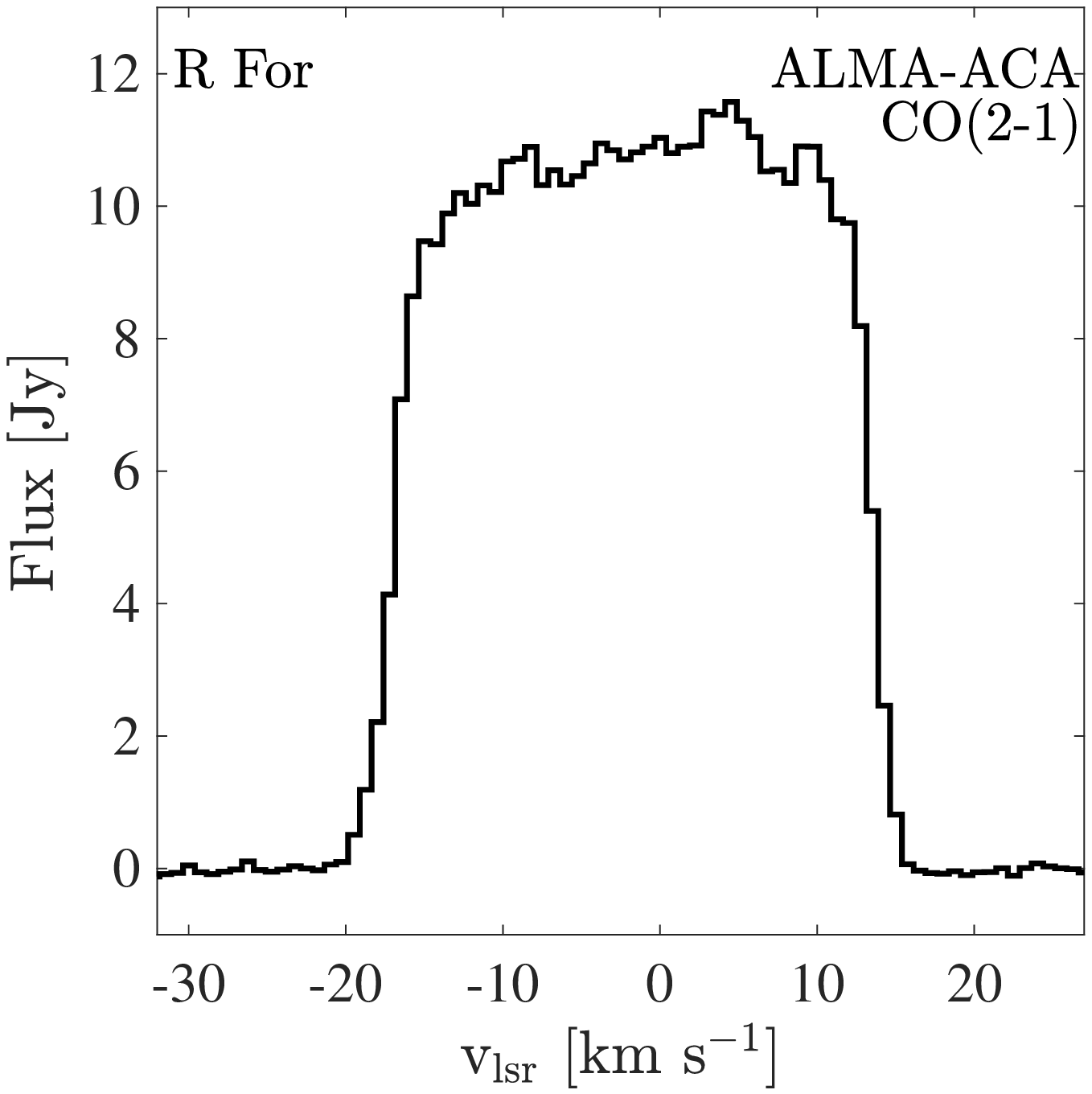}
\includegraphics[height=4.5cm]{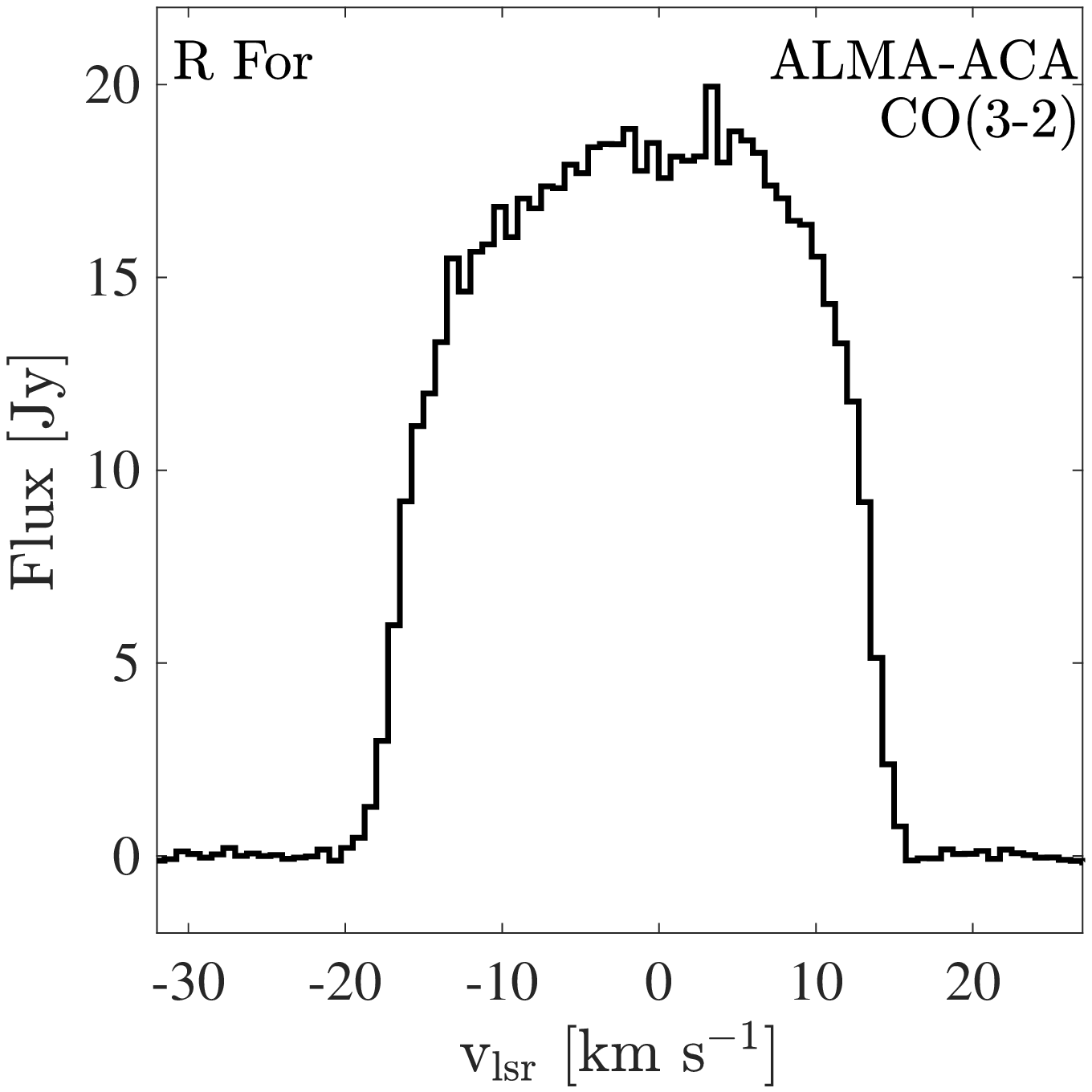}
\includegraphics[height=4.5cm]{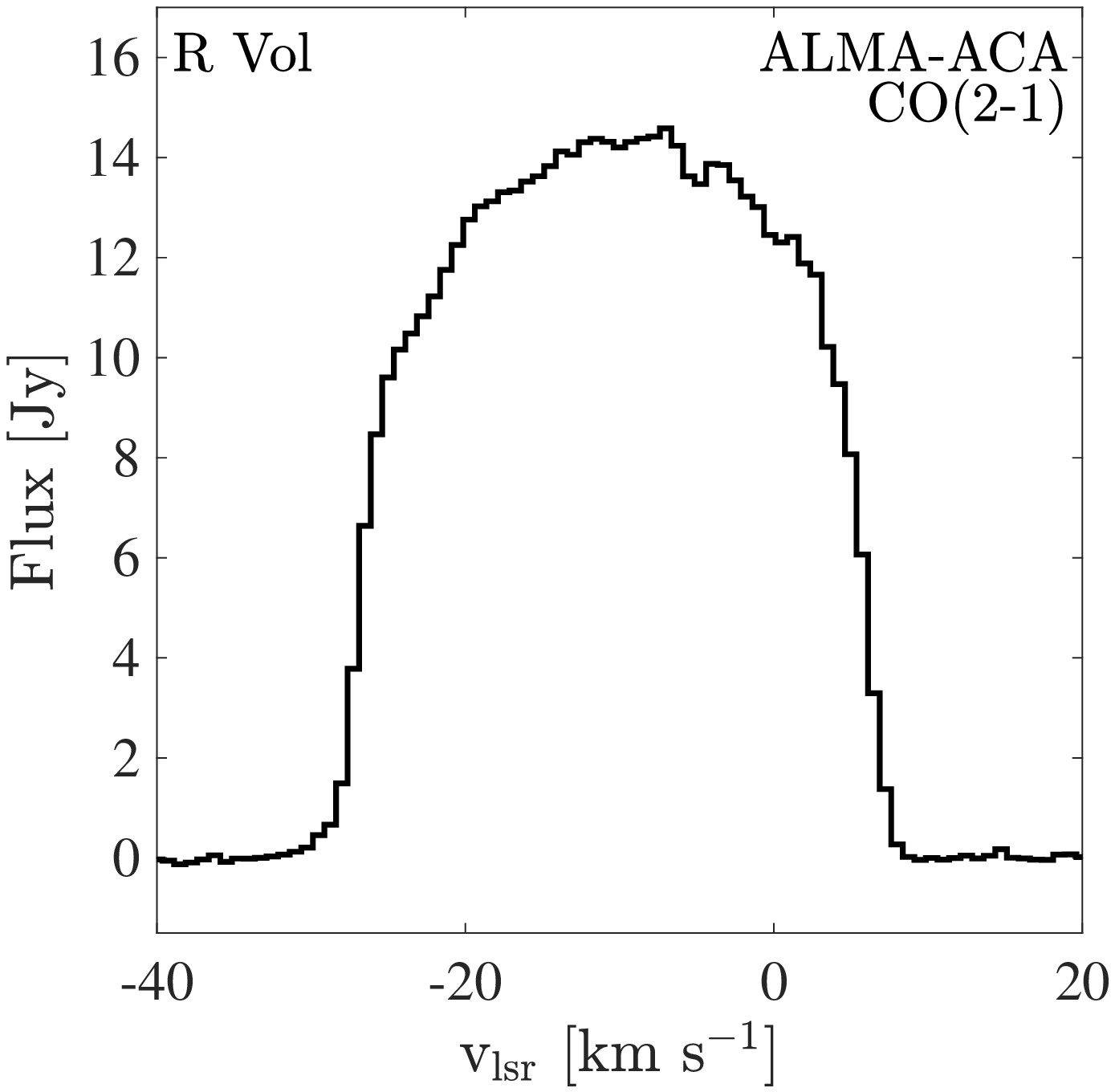}
\includegraphics[height=4.5cm]{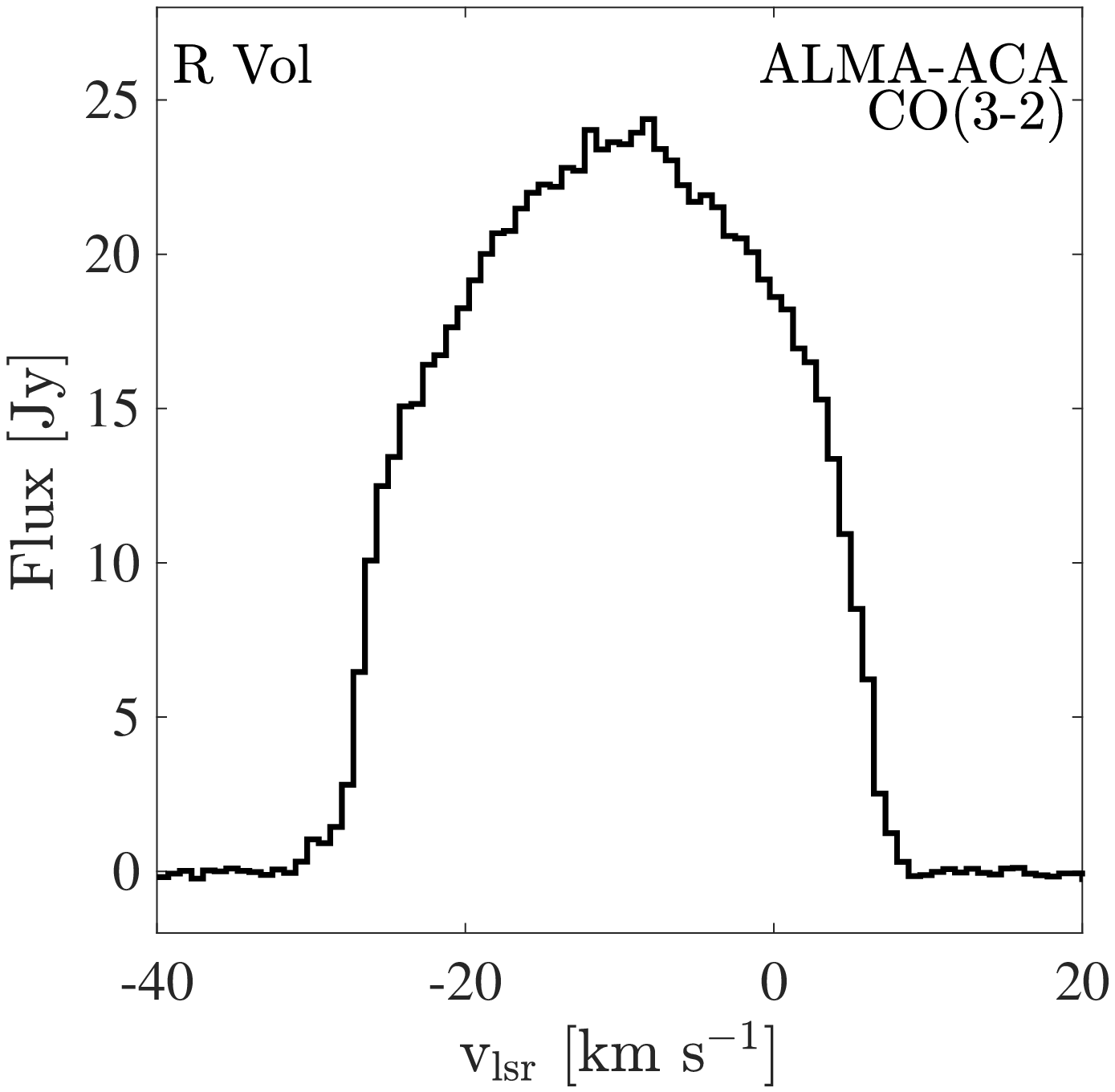}

\includegraphics[height=4.5cm]{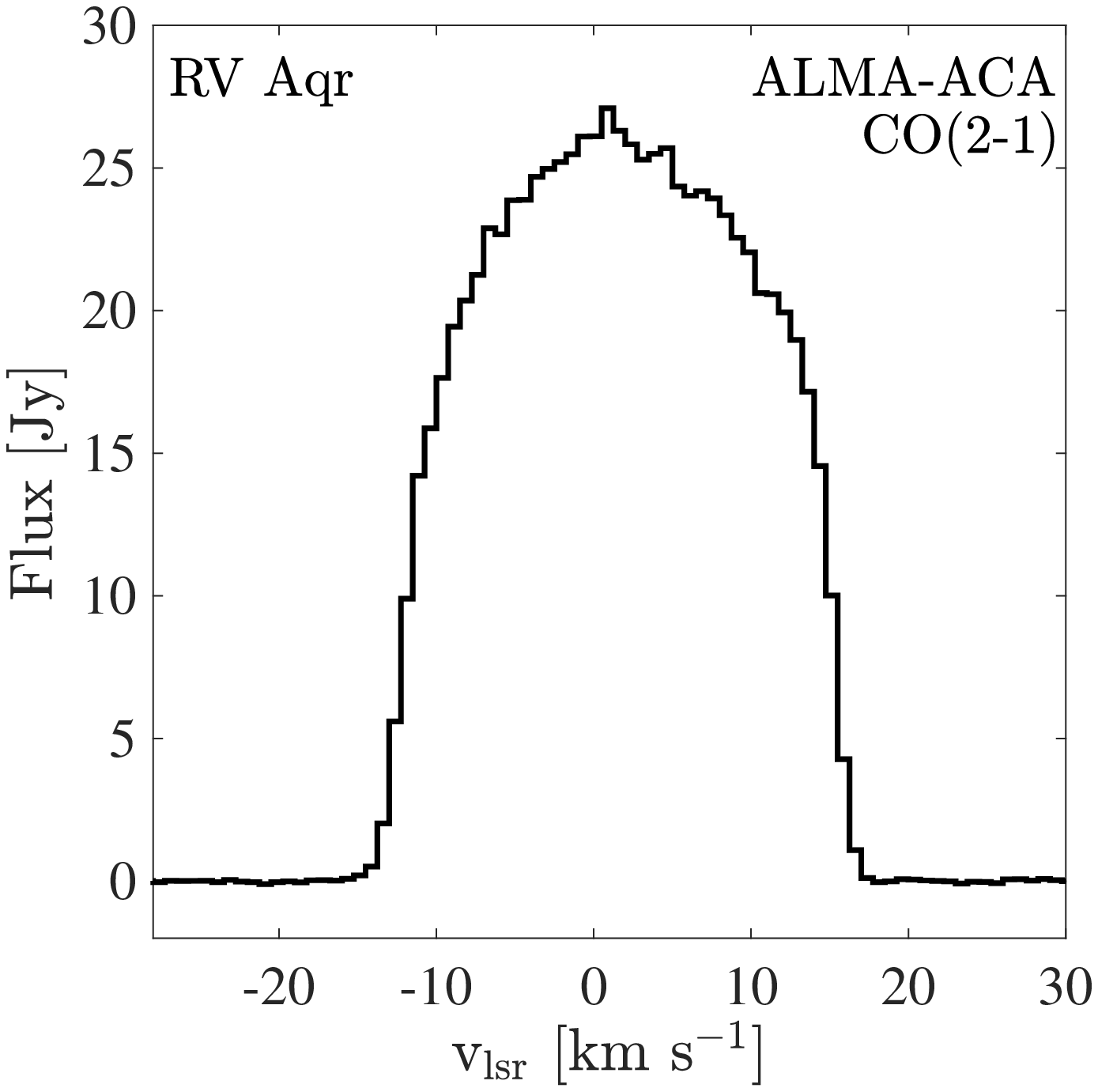}
\includegraphics[height=4.5cm]{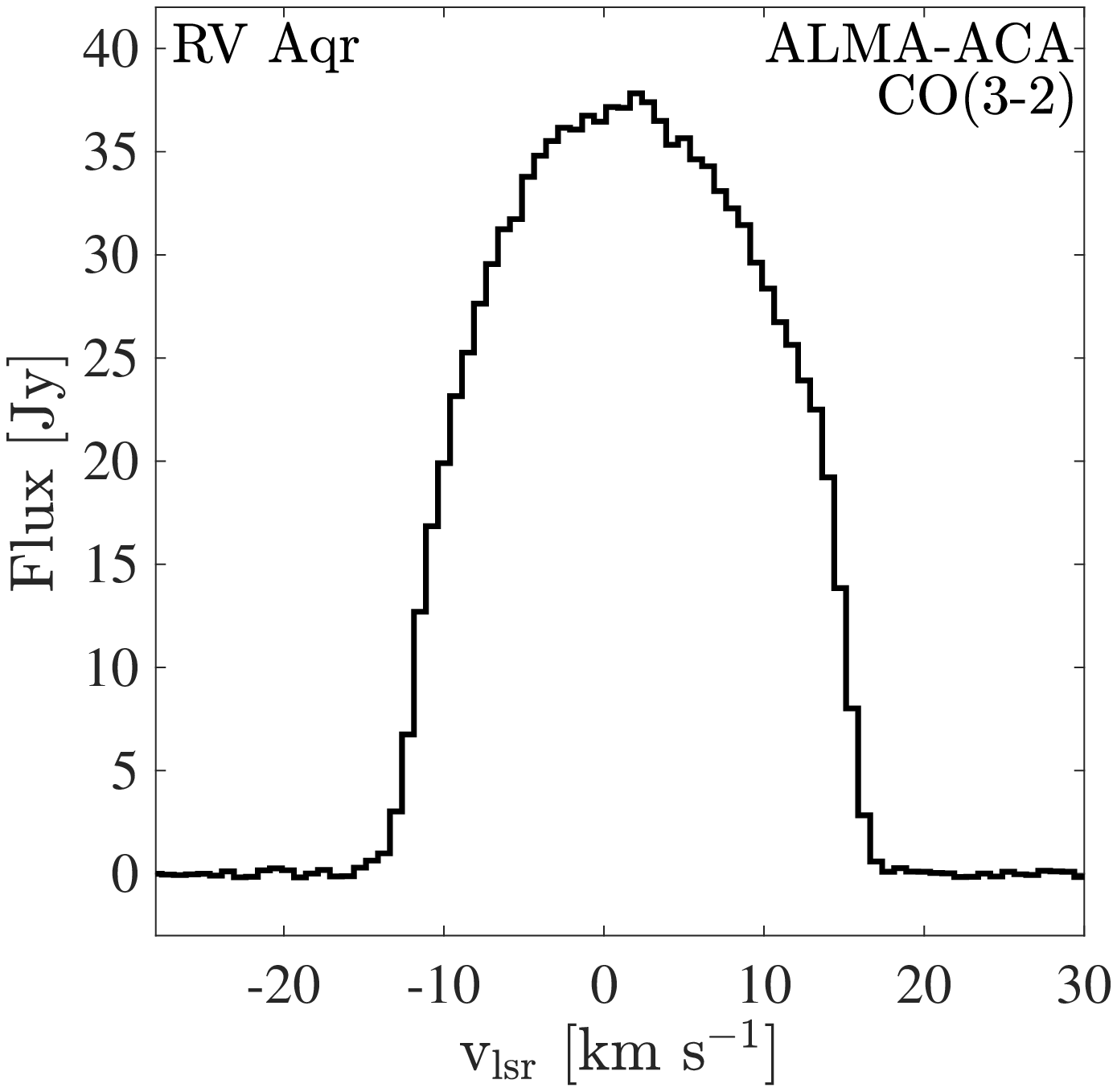}
\includegraphics[height=4.5cm]{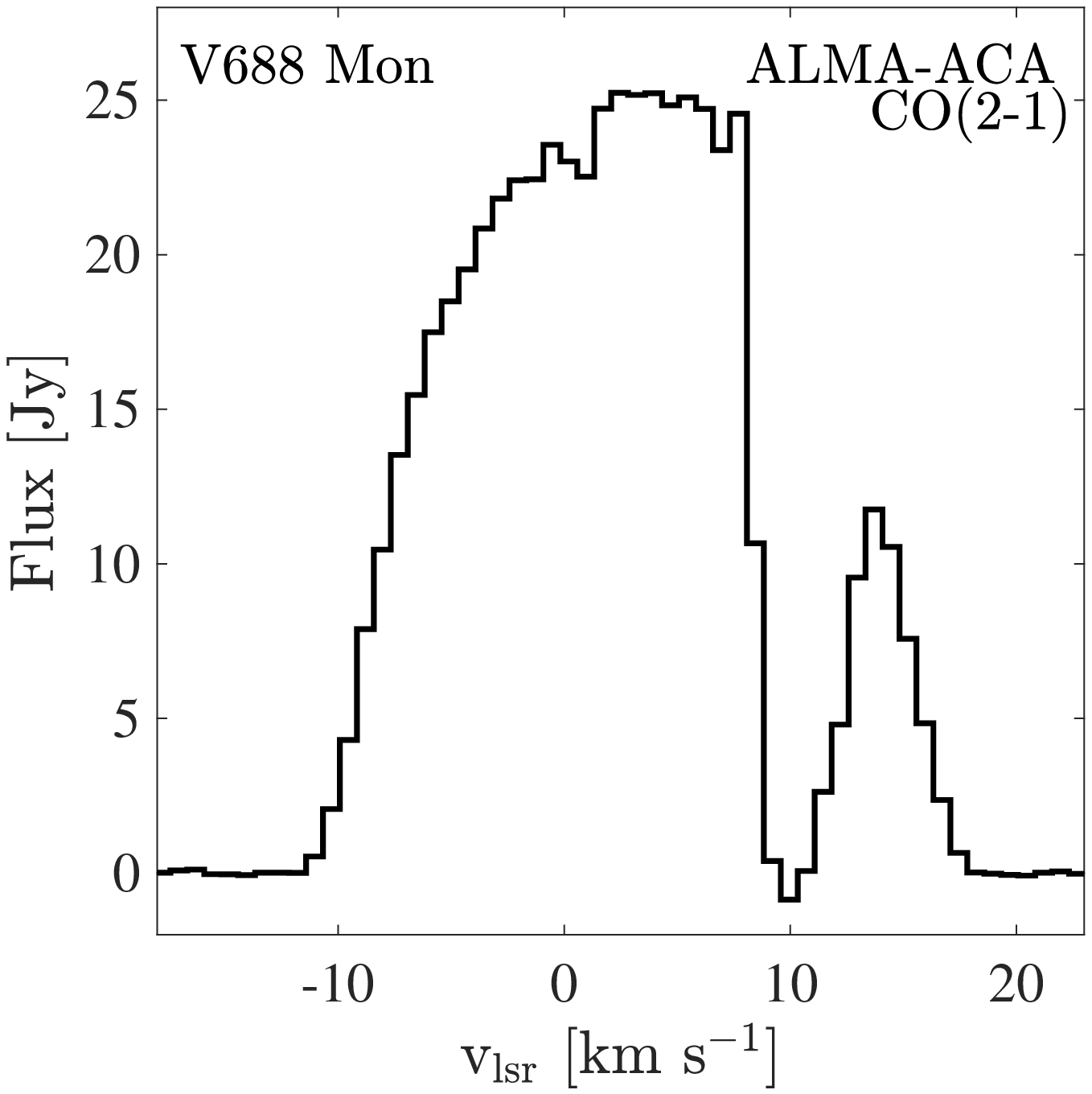}
\includegraphics[height=4.5cm]{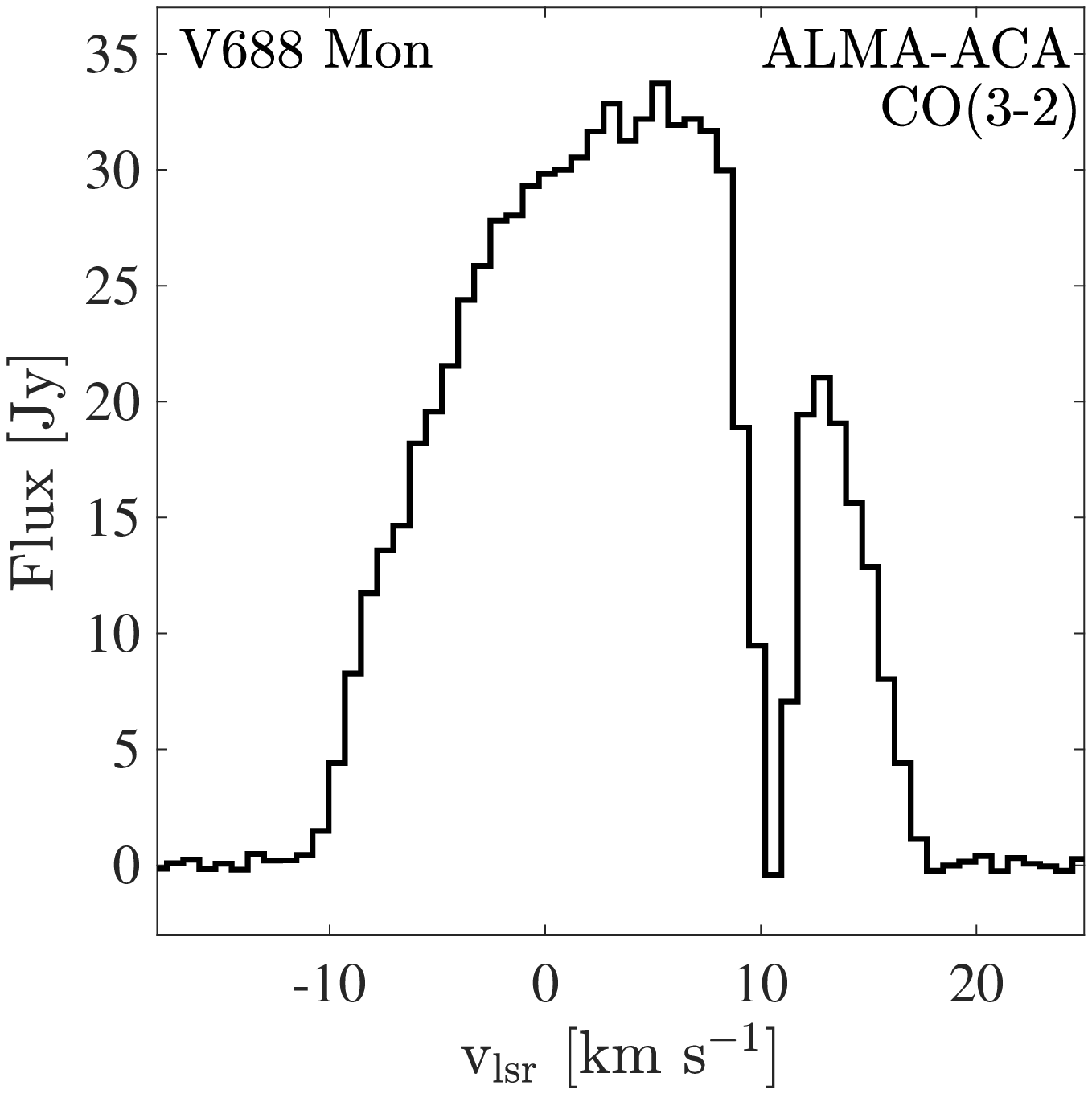}
\caption{CO $J$\,=\,2$\rightarrow$1 and 3$\rightarrow$2 line profiles measured toward the C-type AGB stars of the sample discussed in this paper. The source name is given in the upper left corner and the transition is in the upper right corner of each plot.}
\label{linesC_MSR}
\end{figure*}


\begin{figure*}[t]
\includegraphics[height=4.5cm]{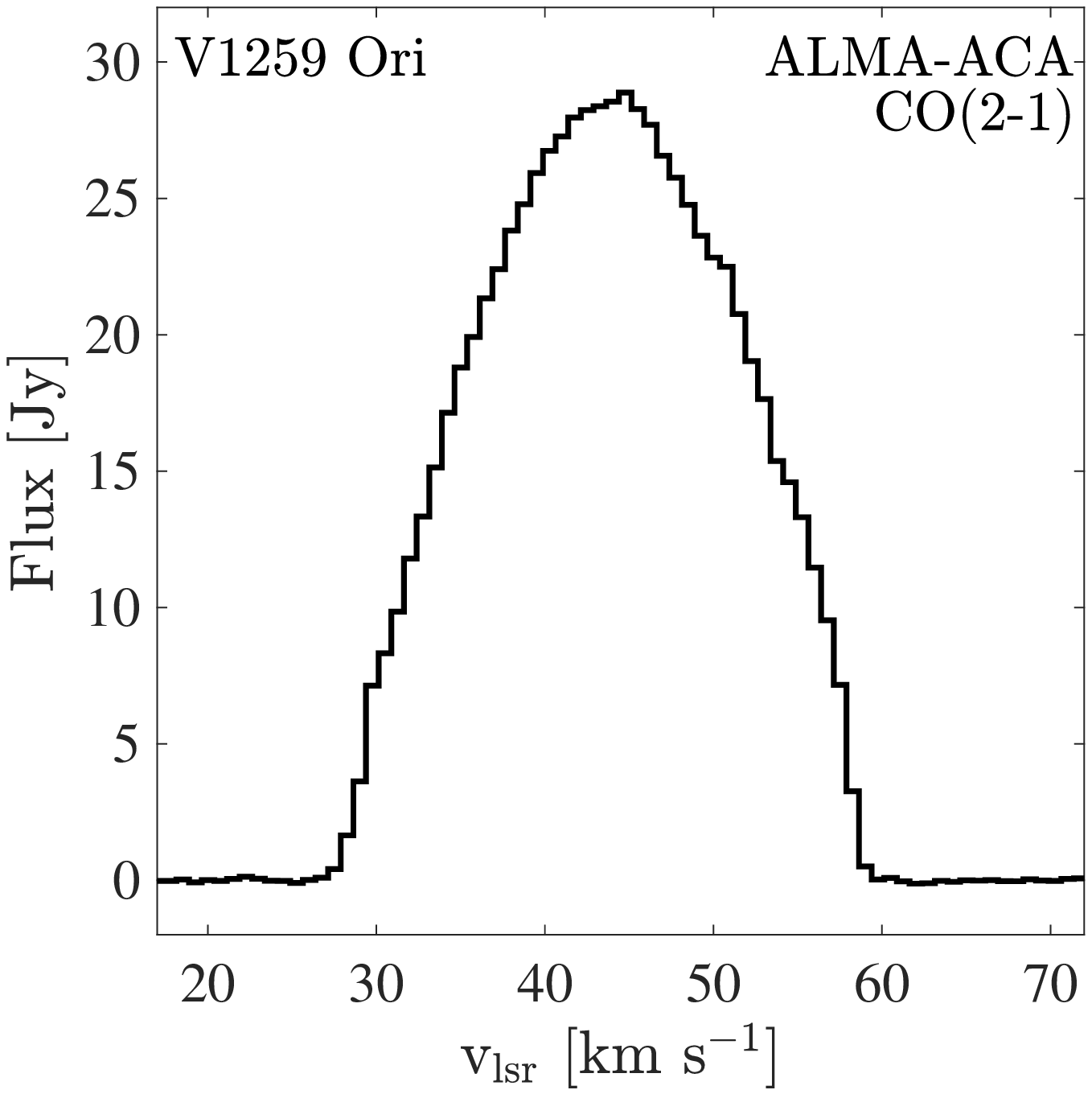}
\includegraphics[height=4.5cm]{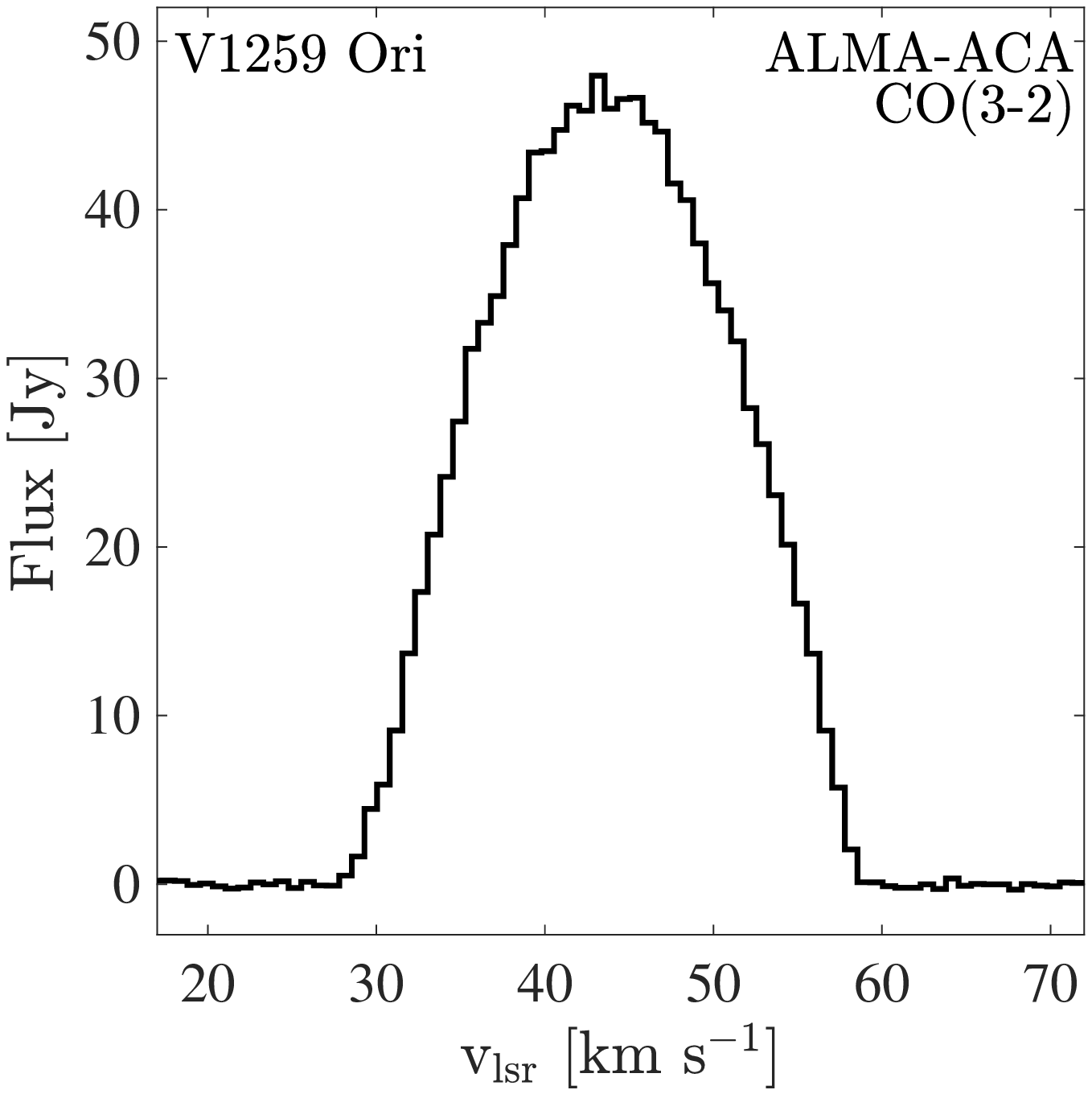}
\caption{CO $J$\,=\,2$\rightarrow$1 and 3$\rightarrow$2 line profiles measured toward the C-type AGB stars of the sample discussed in this paper. The source name is given in the upper left corner and the transition is in the upper right corner of each plot.}
\label{linesC_M}
\end{figure*}


\section{Results from fitting to Gaussian emission distributions}

\begin{figure*}[t]
\includegraphics[height=4.5cm]{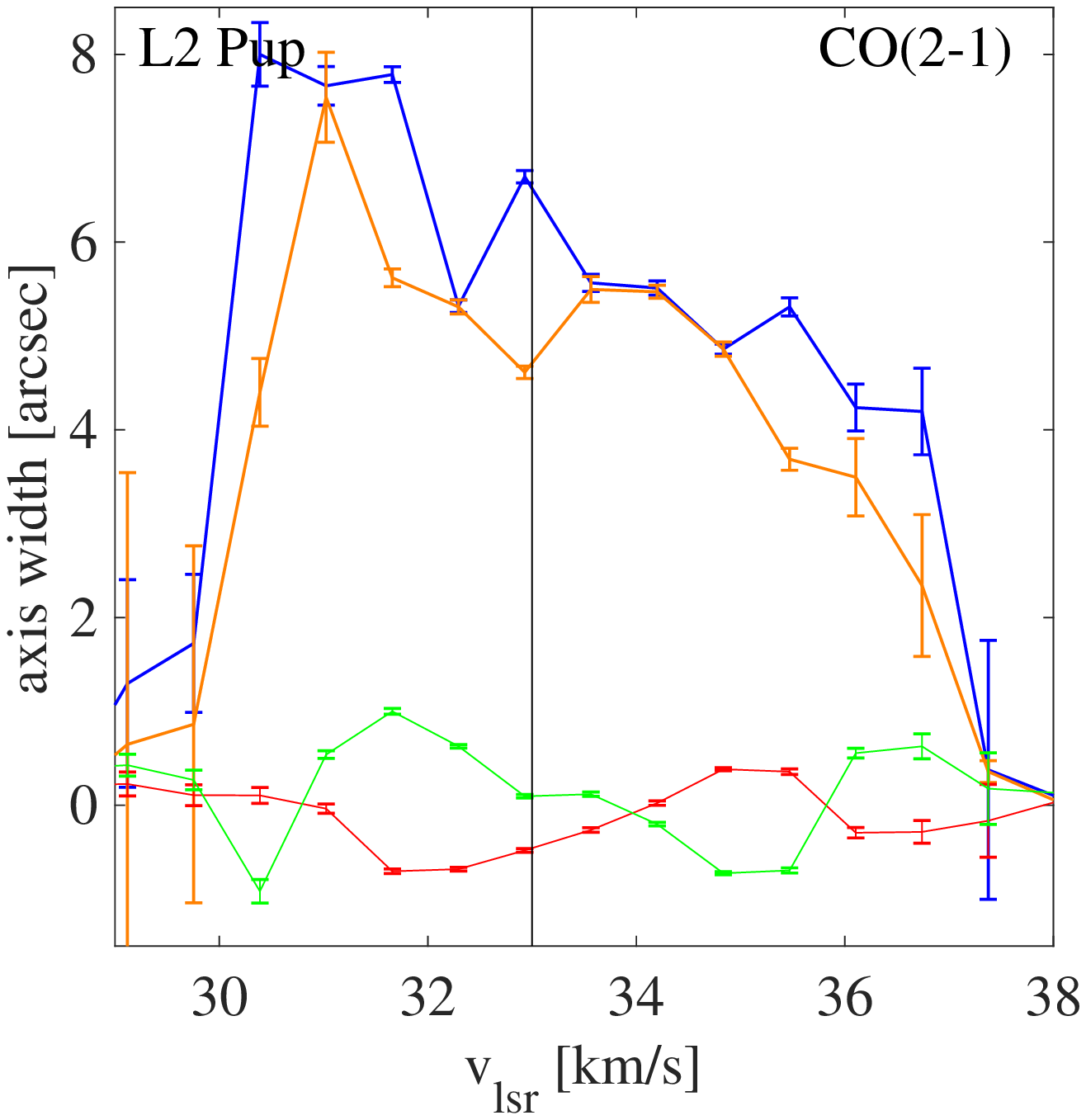}
\includegraphics[height=4.5cm]{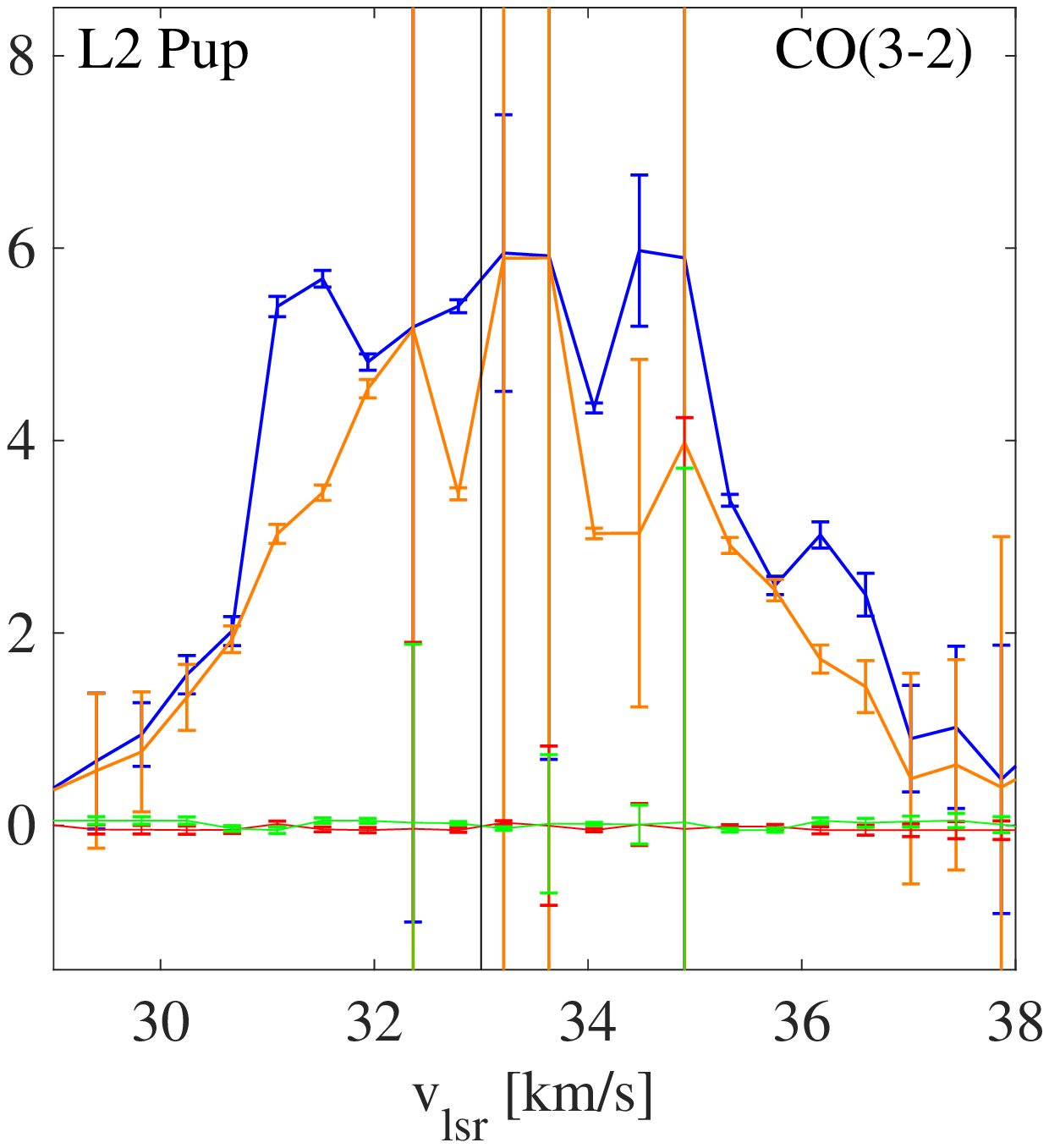}
\hspace{0.2cm}
\includegraphics[height=4.5cm]{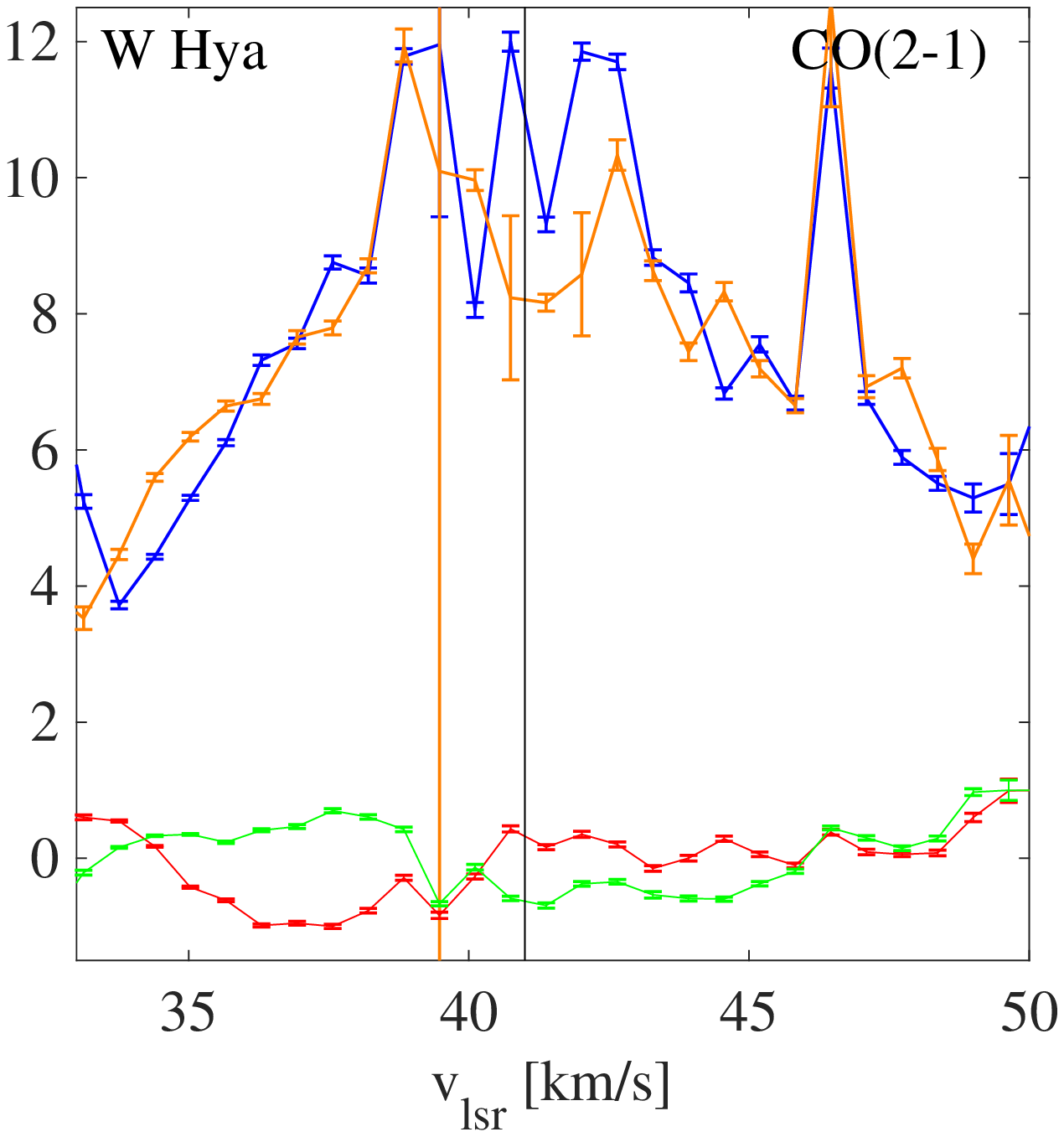}
\includegraphics[height=4.5cm]{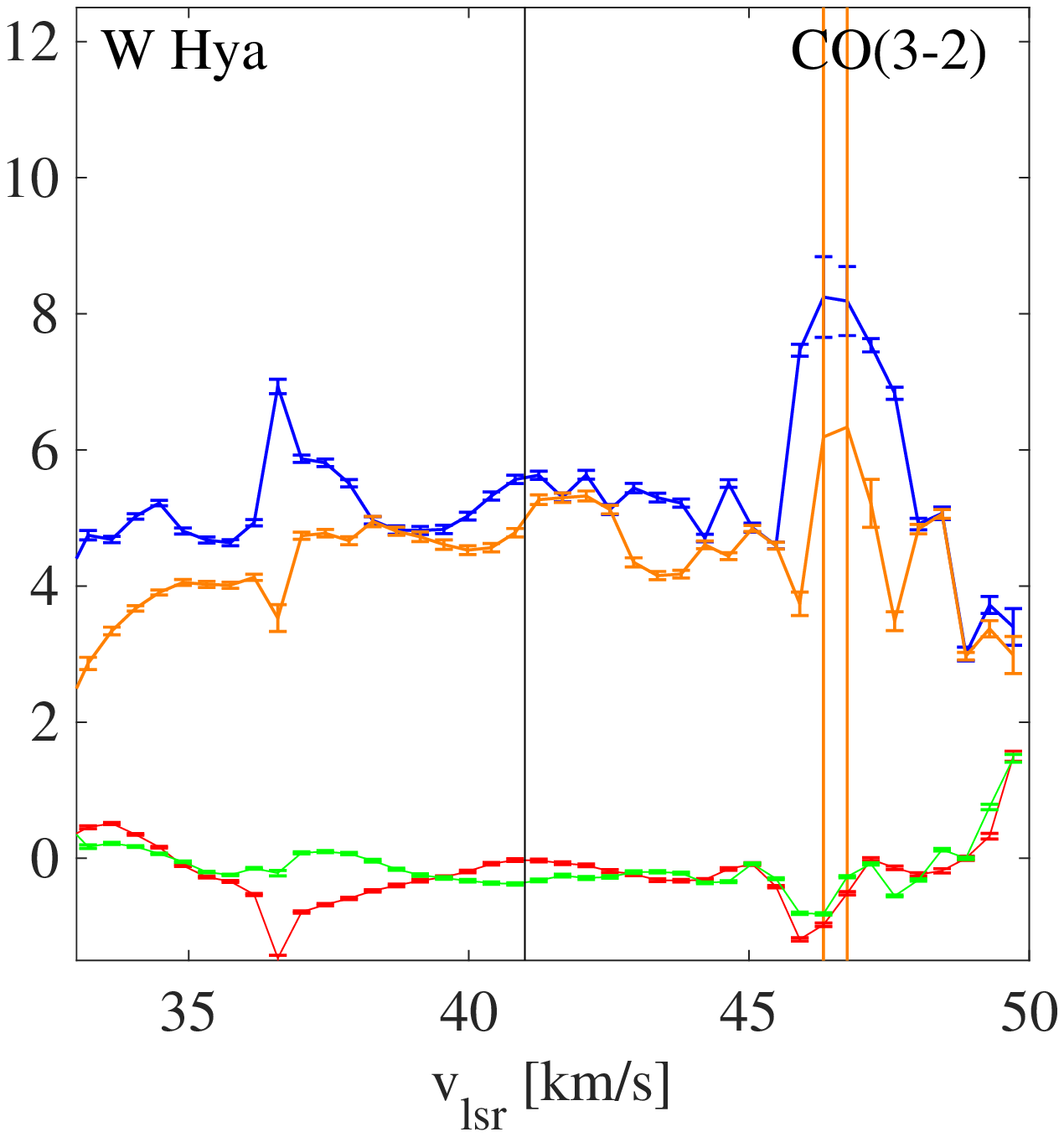}

\includegraphics[height=4.5cm]{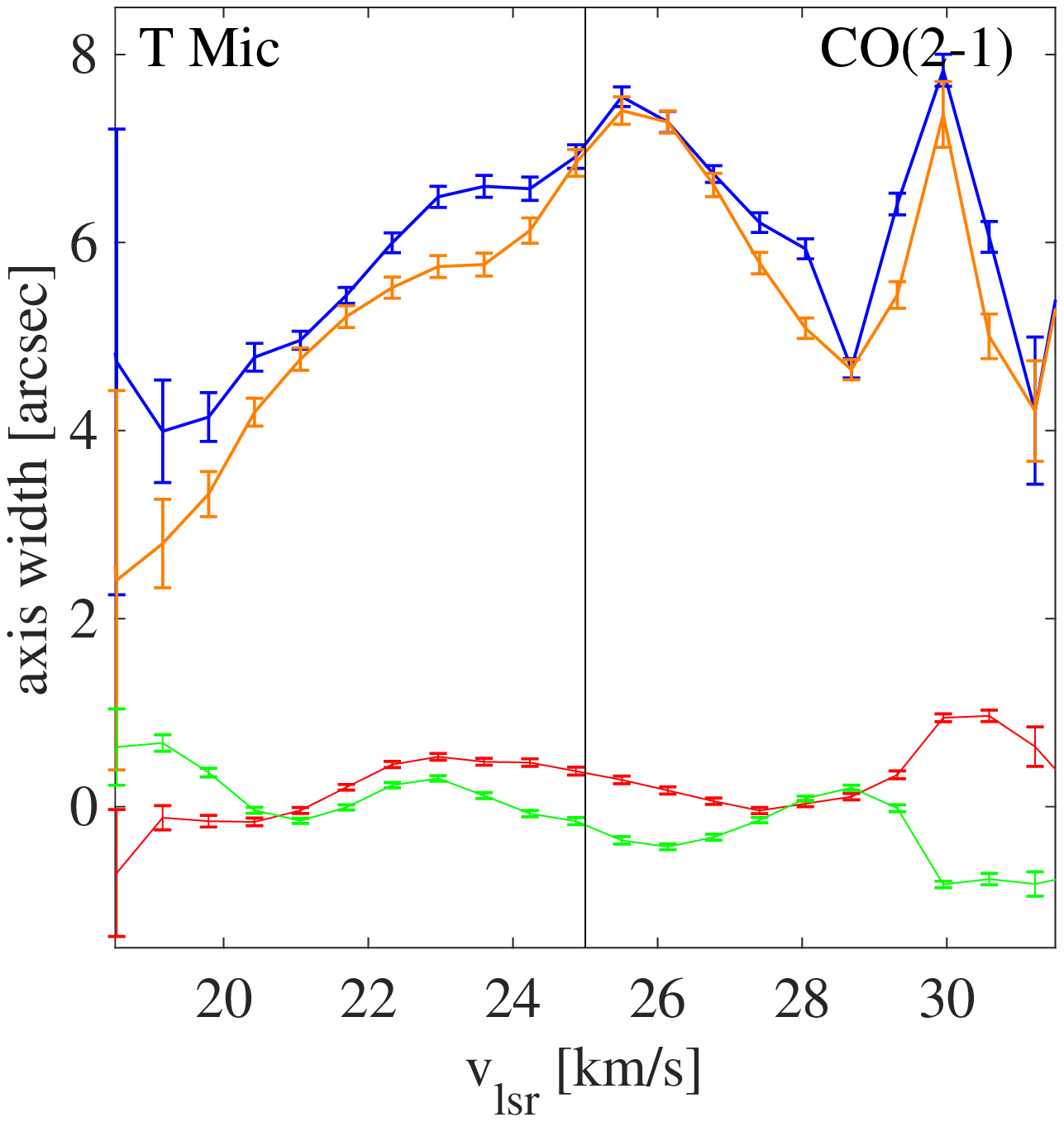}
\hspace{0.03cm}
\includegraphics[height=4.5cm]{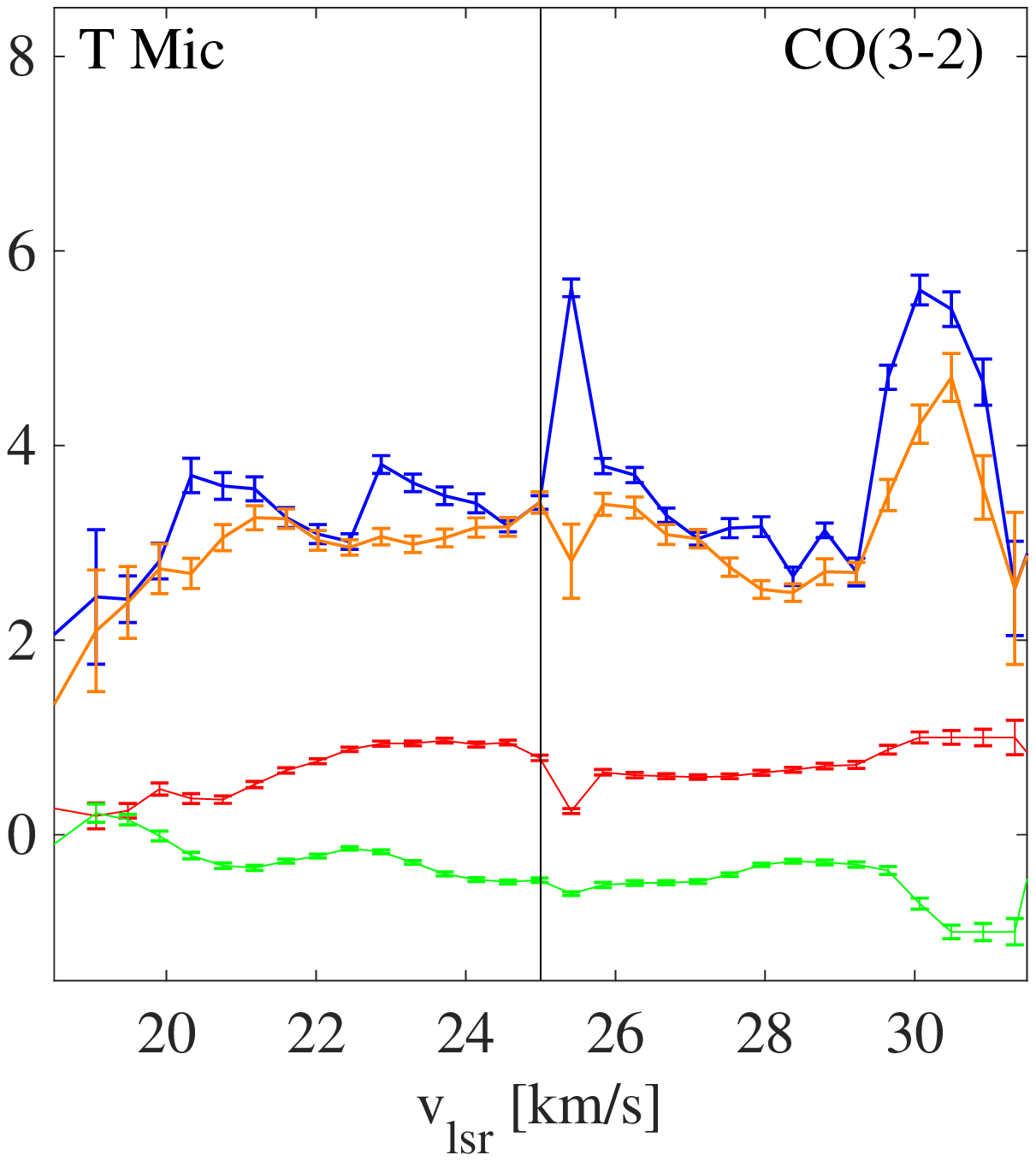}
\hspace{0.27cm}
\includegraphics[height=4.5cm]{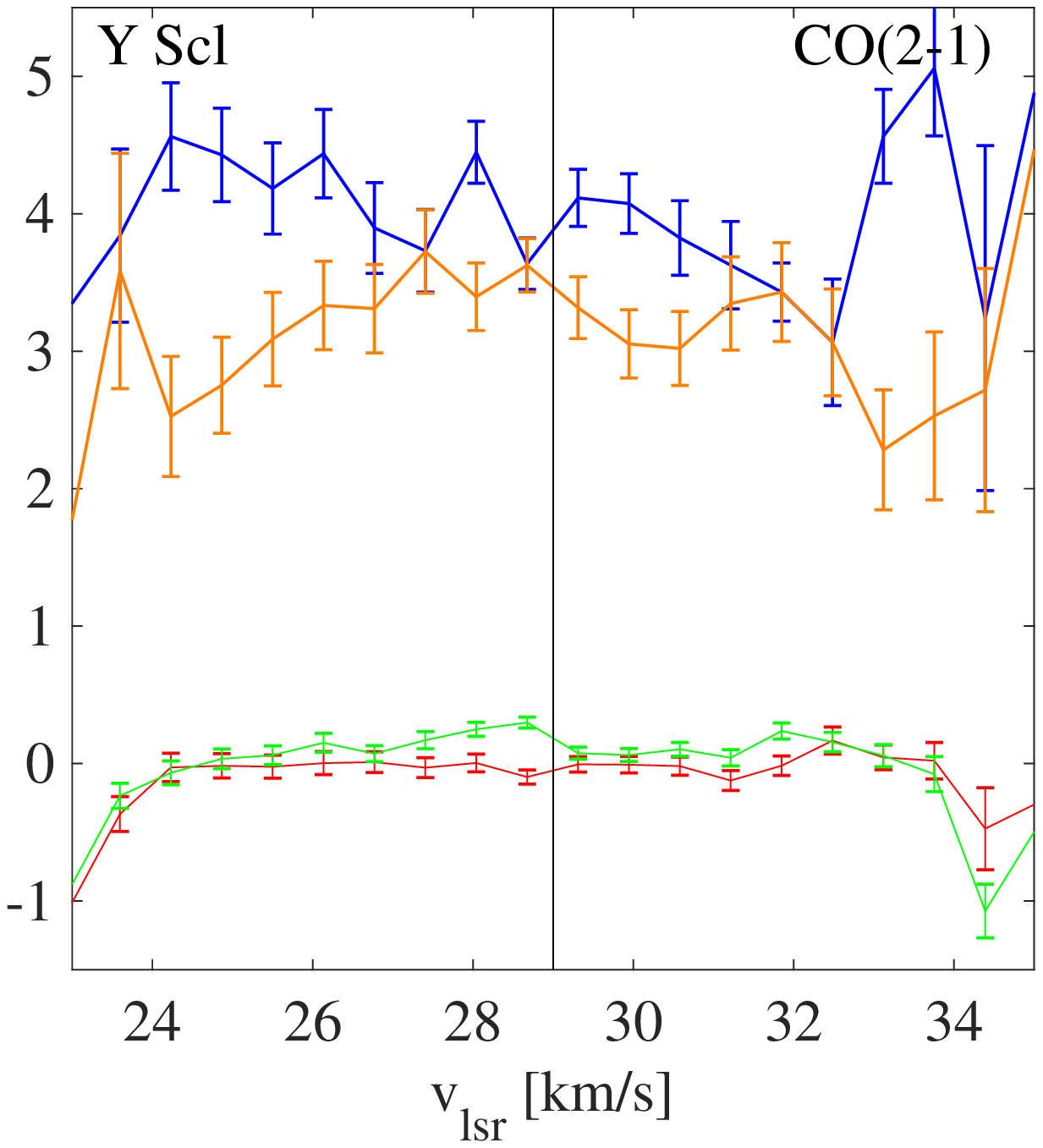}
\hspace{0.05cm}
\includegraphics[height=4.5cm]{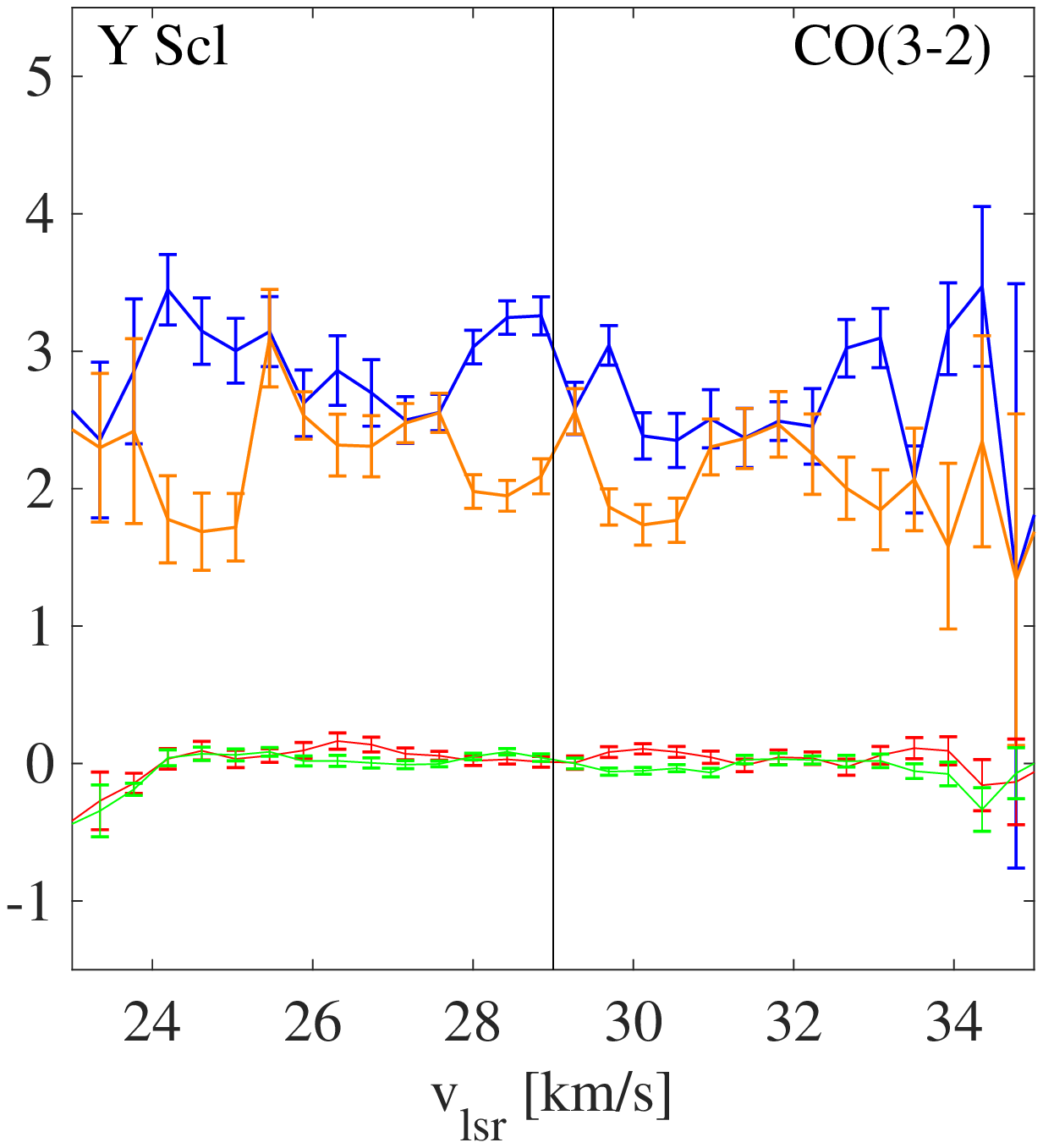}

\includegraphics[height=4.5cm]{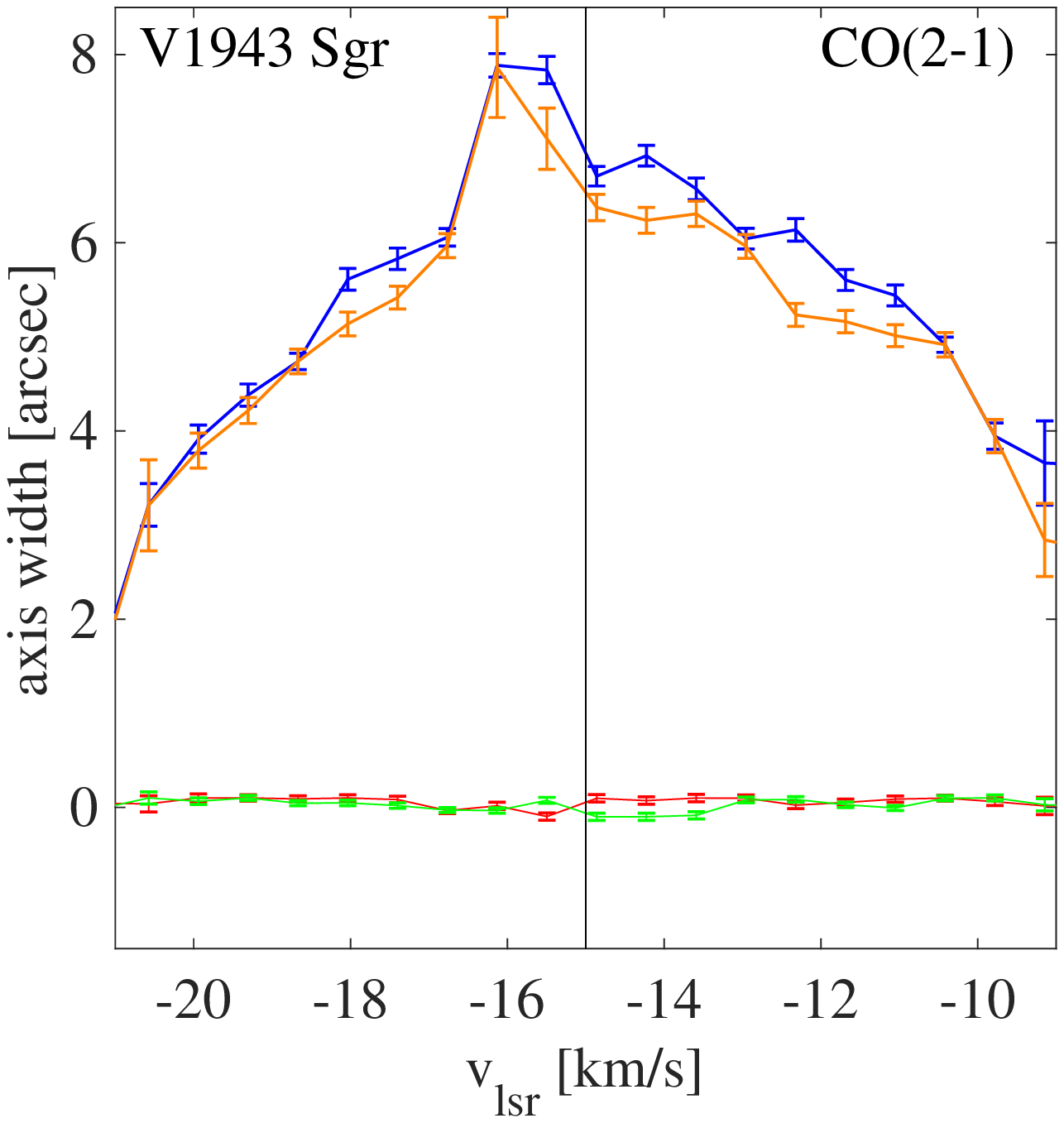}
\hspace{0.03cm}
\includegraphics[height=4.5cm]{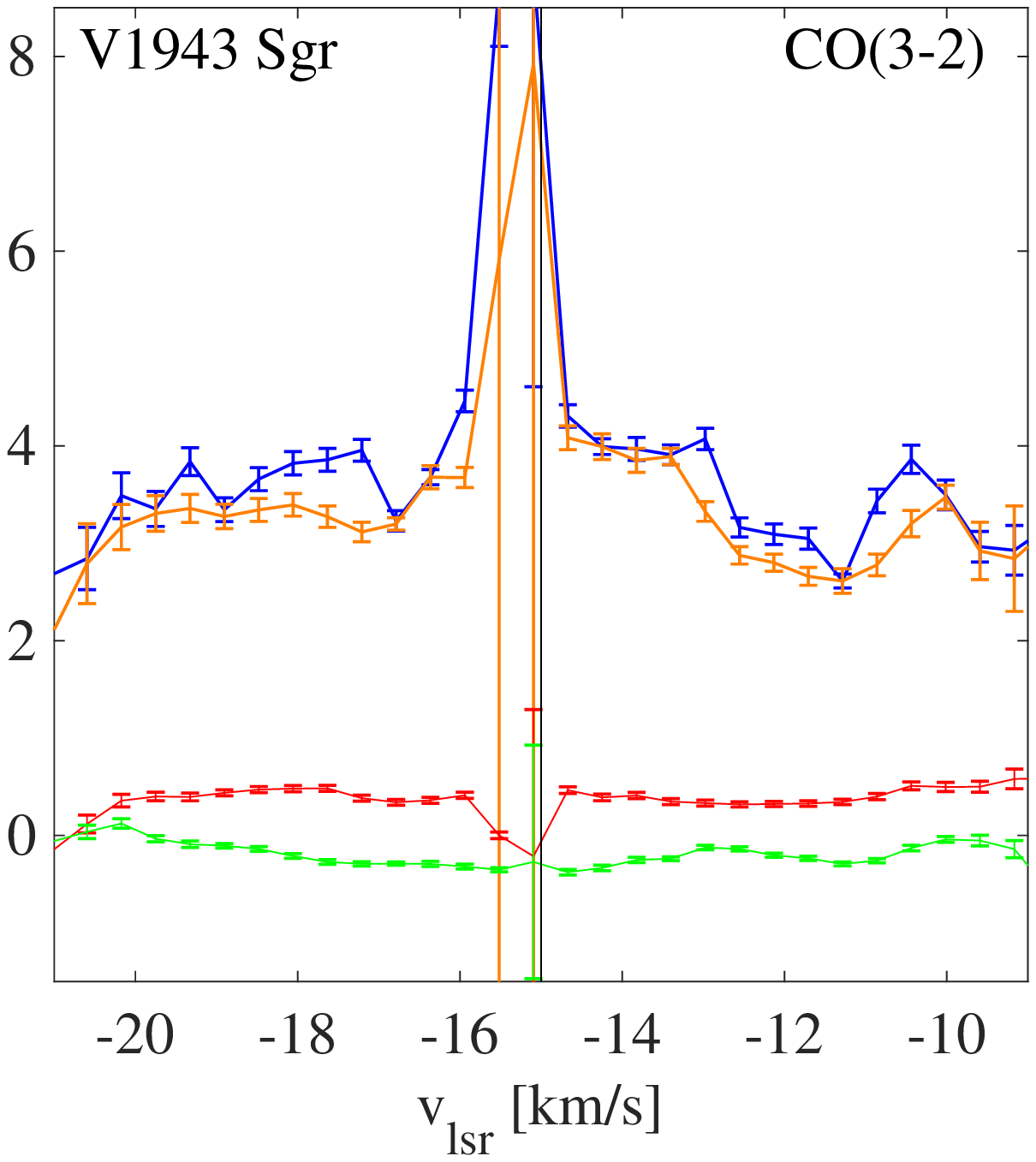}
\hspace{0.27cm}
\includegraphics[height=4.5cm]{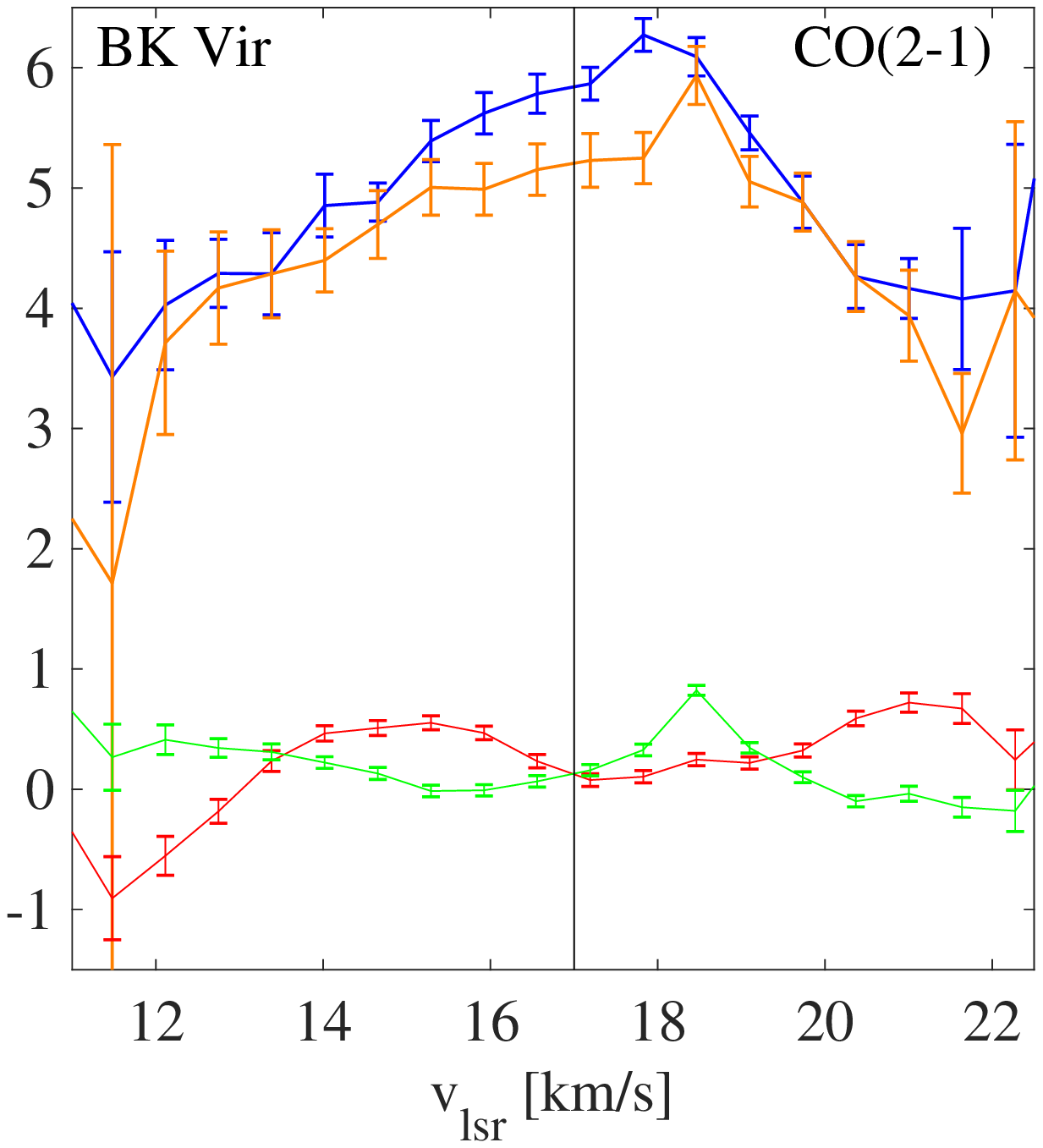}
\hspace{0.05cm}
\includegraphics[height=4.5cm]{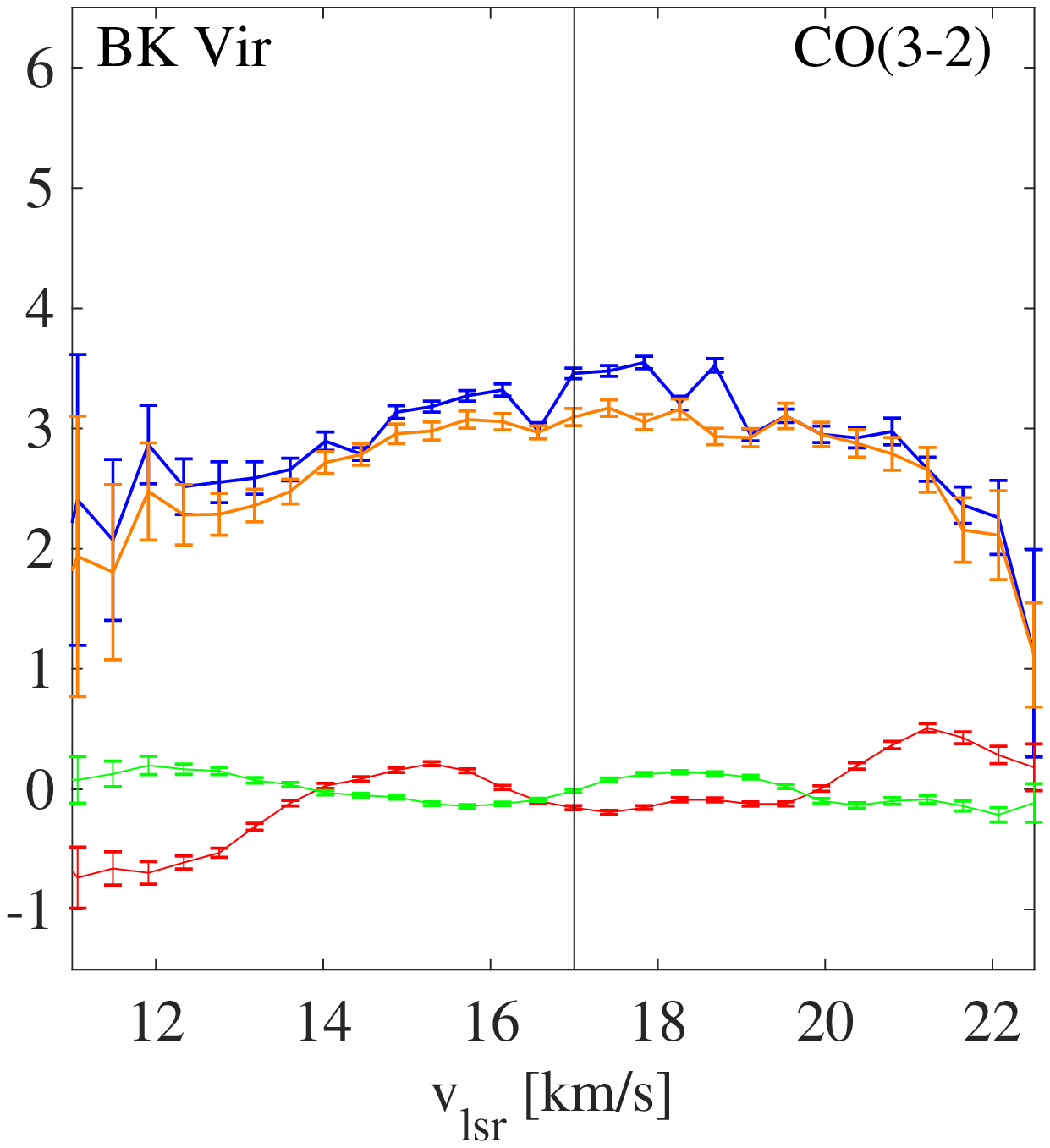}

\includegraphics[height=4.5cm]{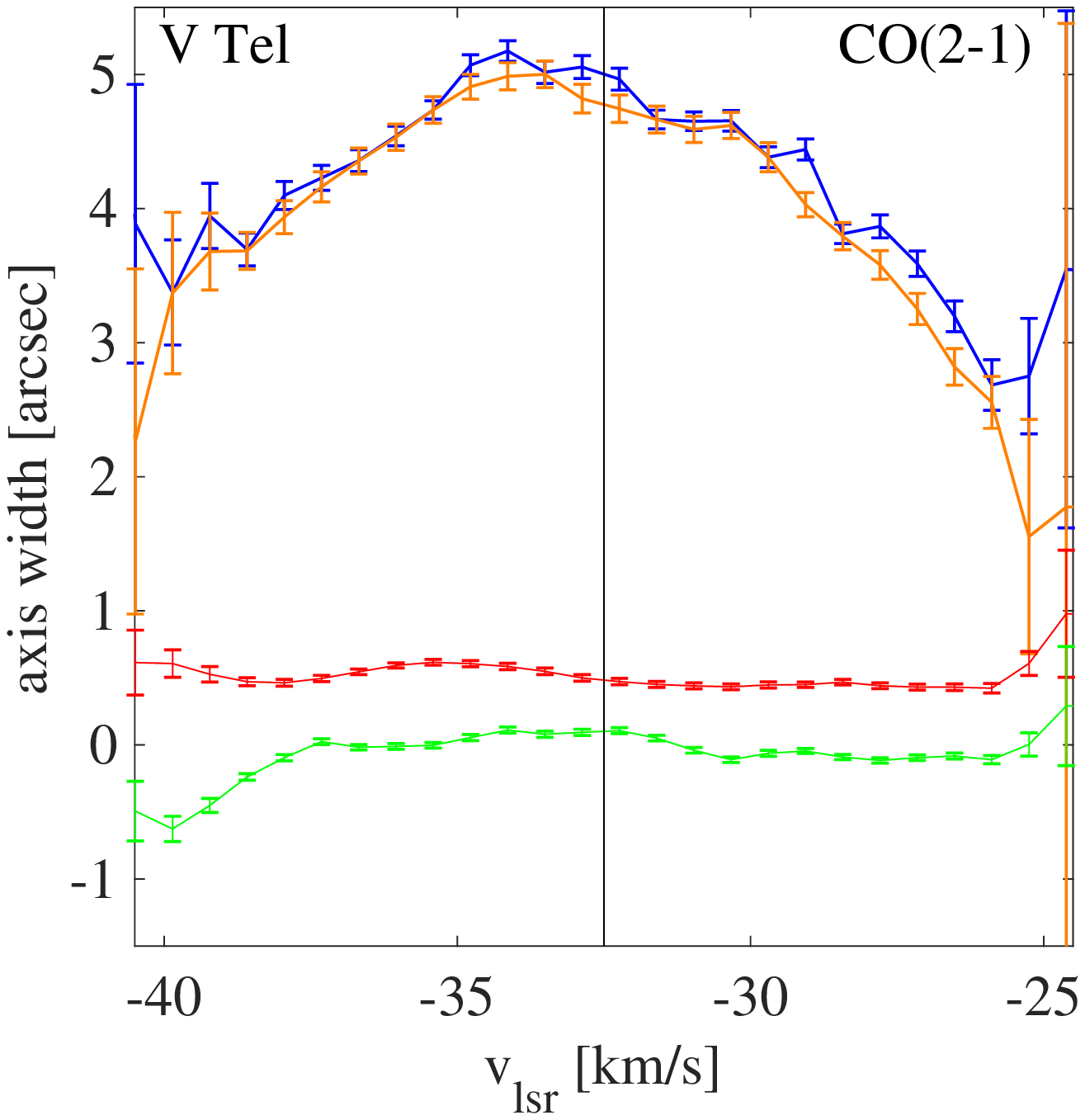}
\includegraphics[height=4.5cm]{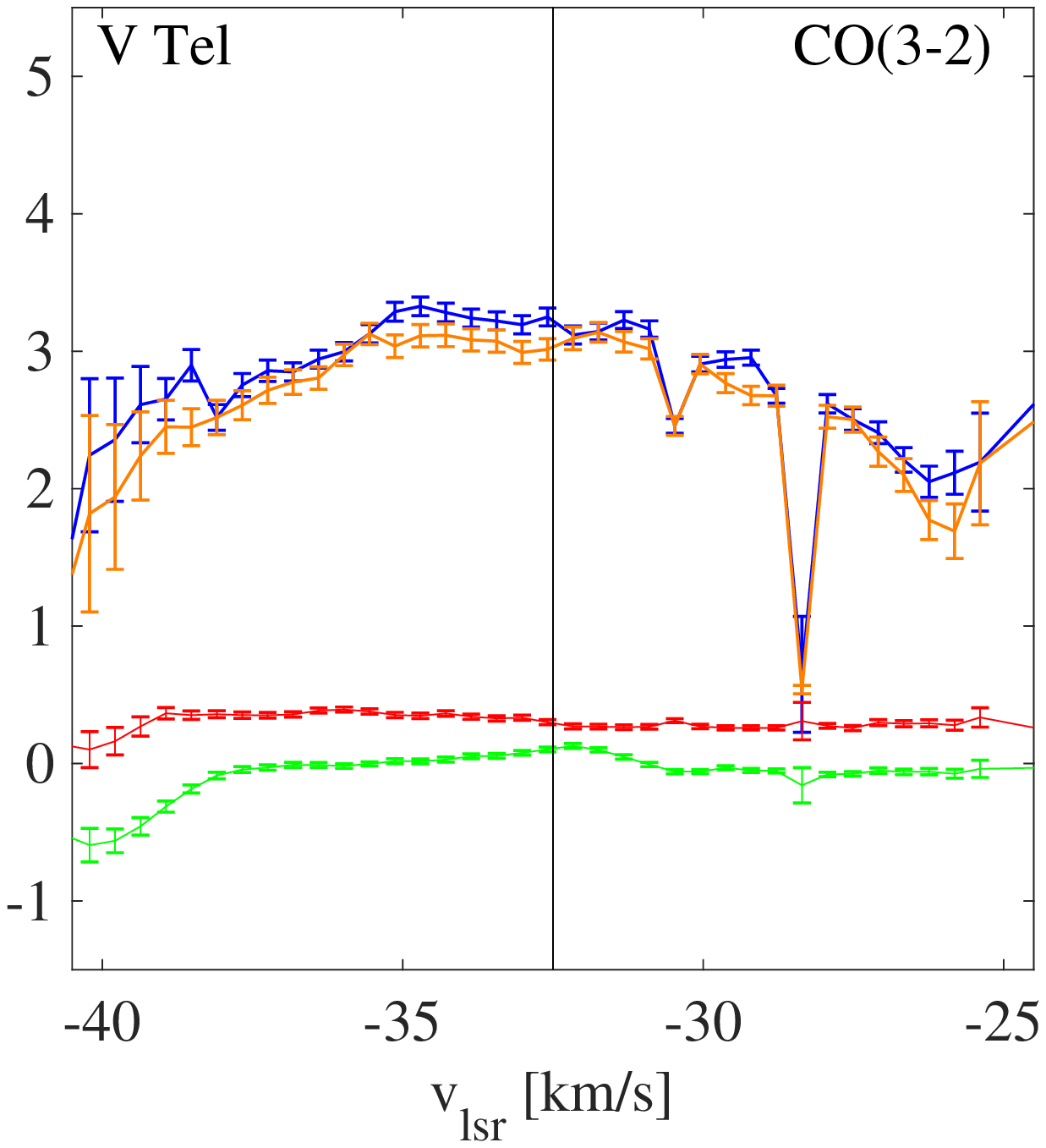}
\hspace{0.27cm}
\includegraphics[height=4.5cm]{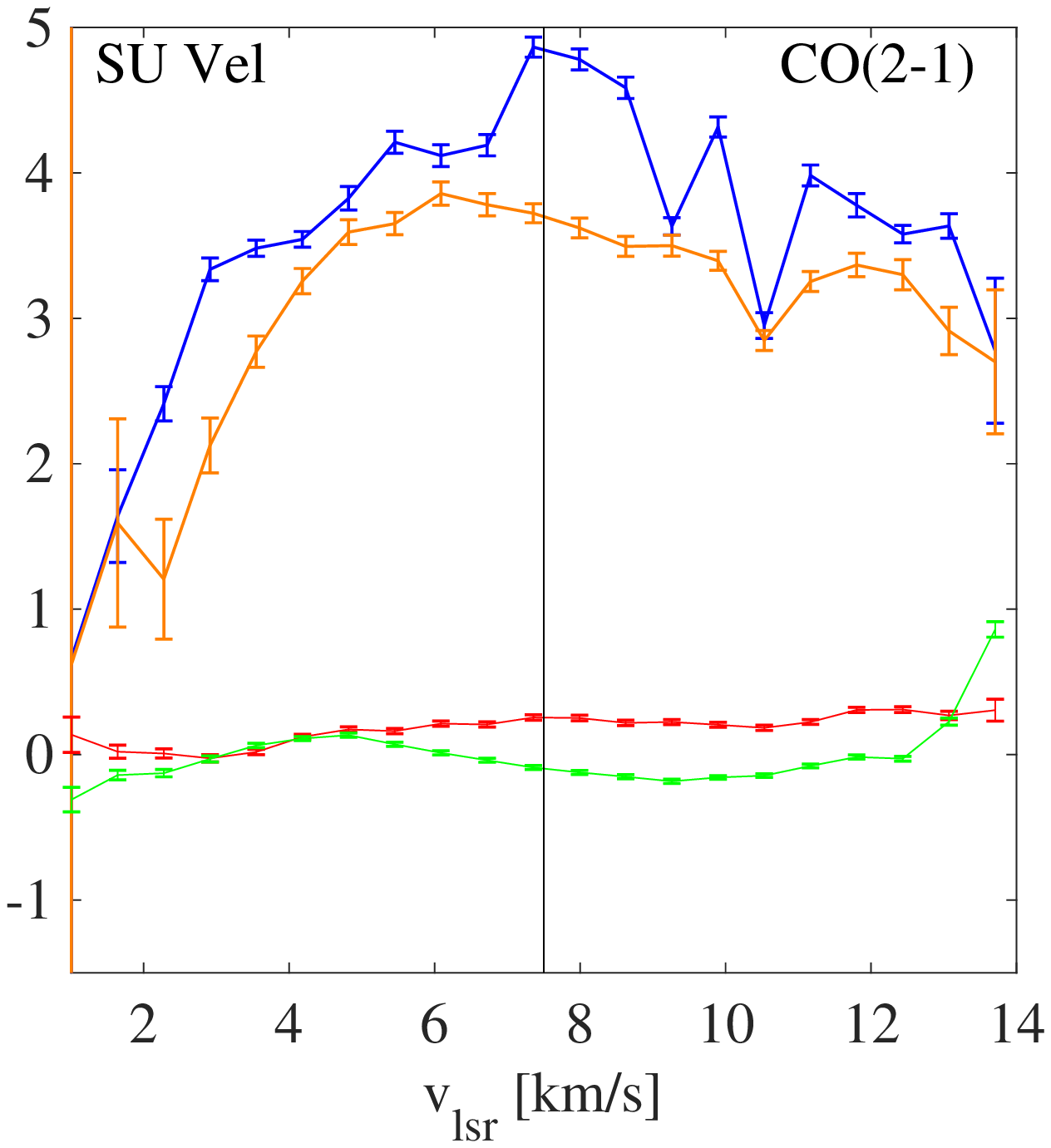}
\hspace{0.05cm}
\includegraphics[height=4.5cm]{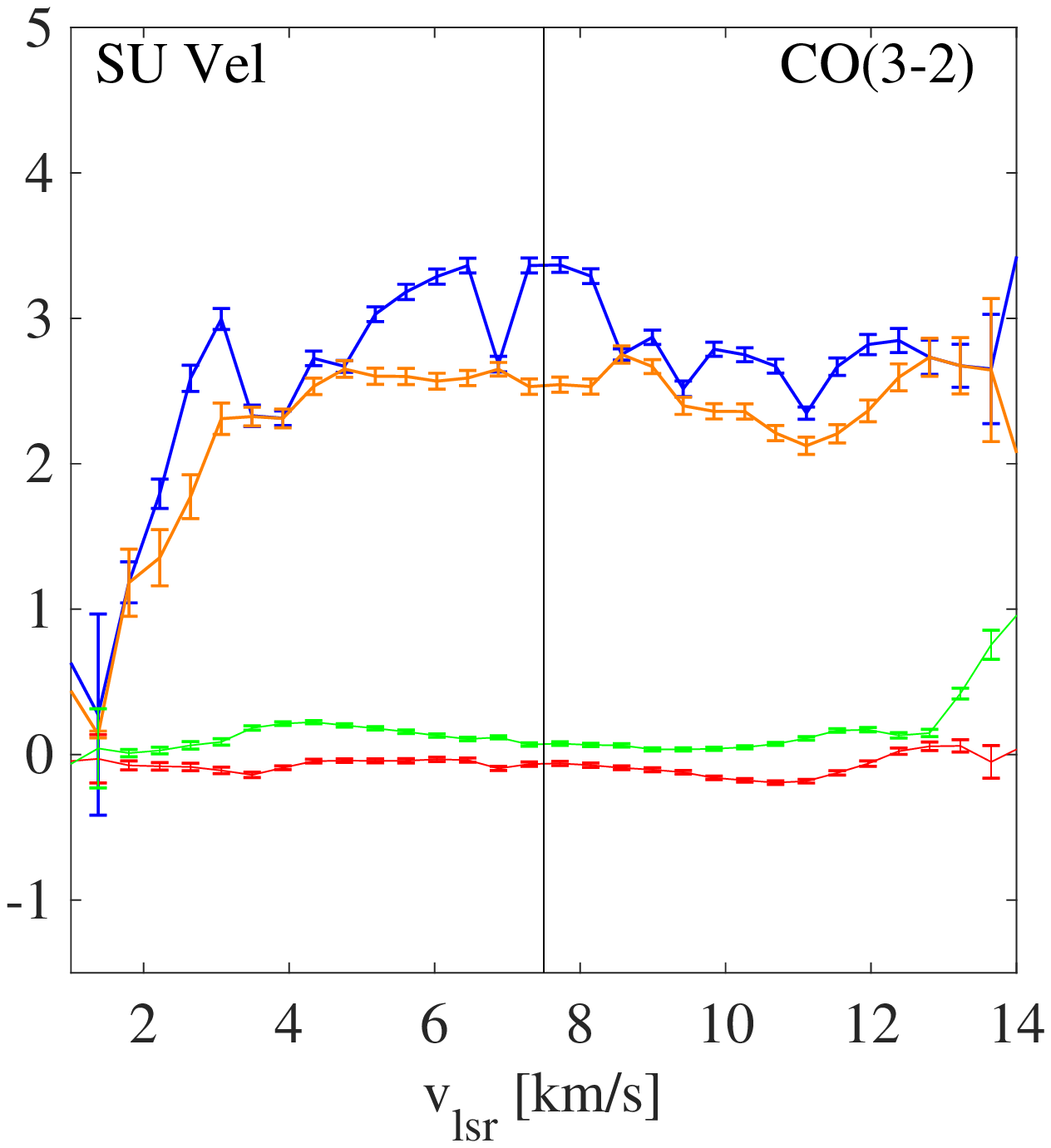}

\includegraphics[height=4.5cm]{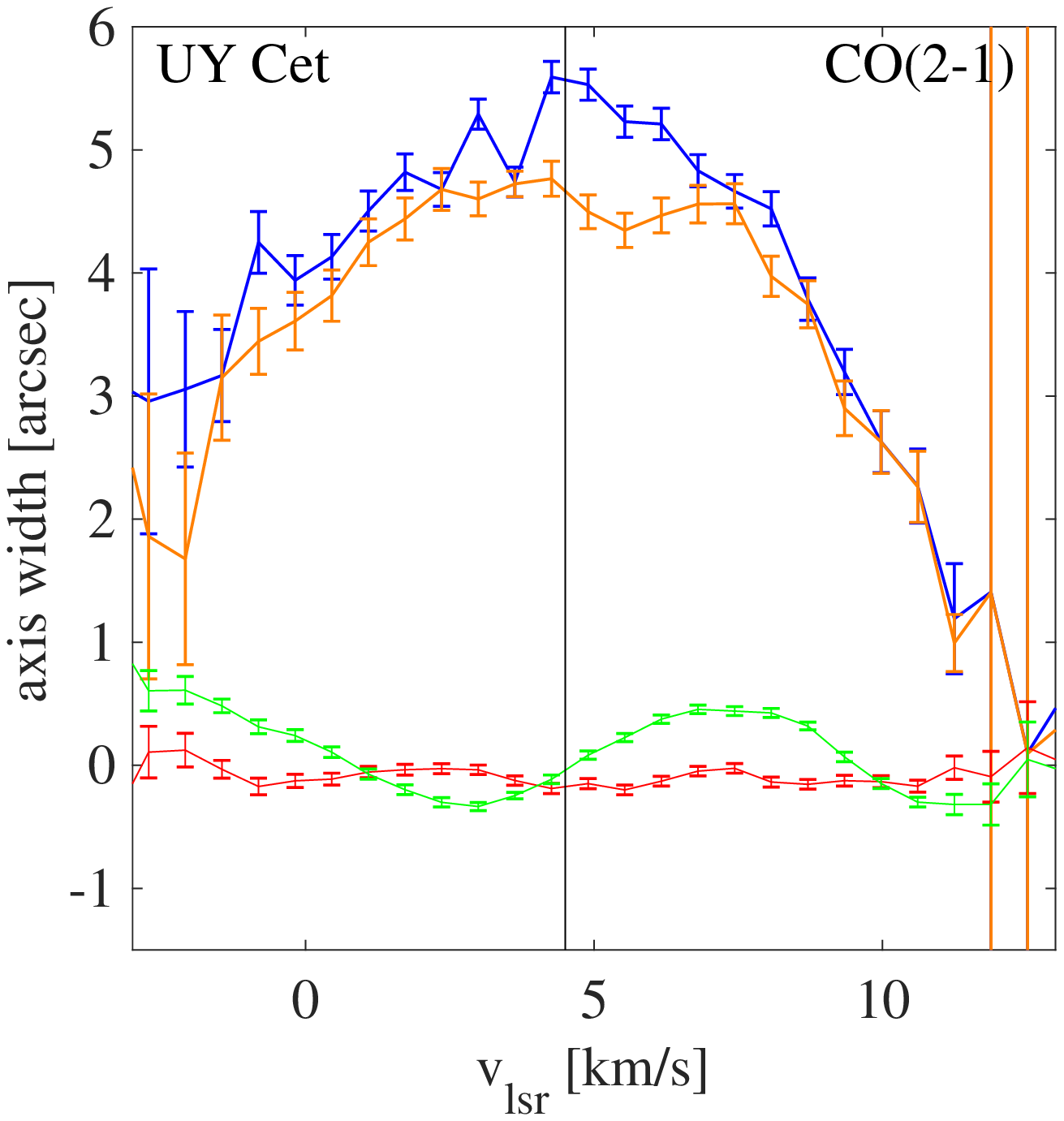}
\hspace{0.05cm}
\includegraphics[height=4.5cm]{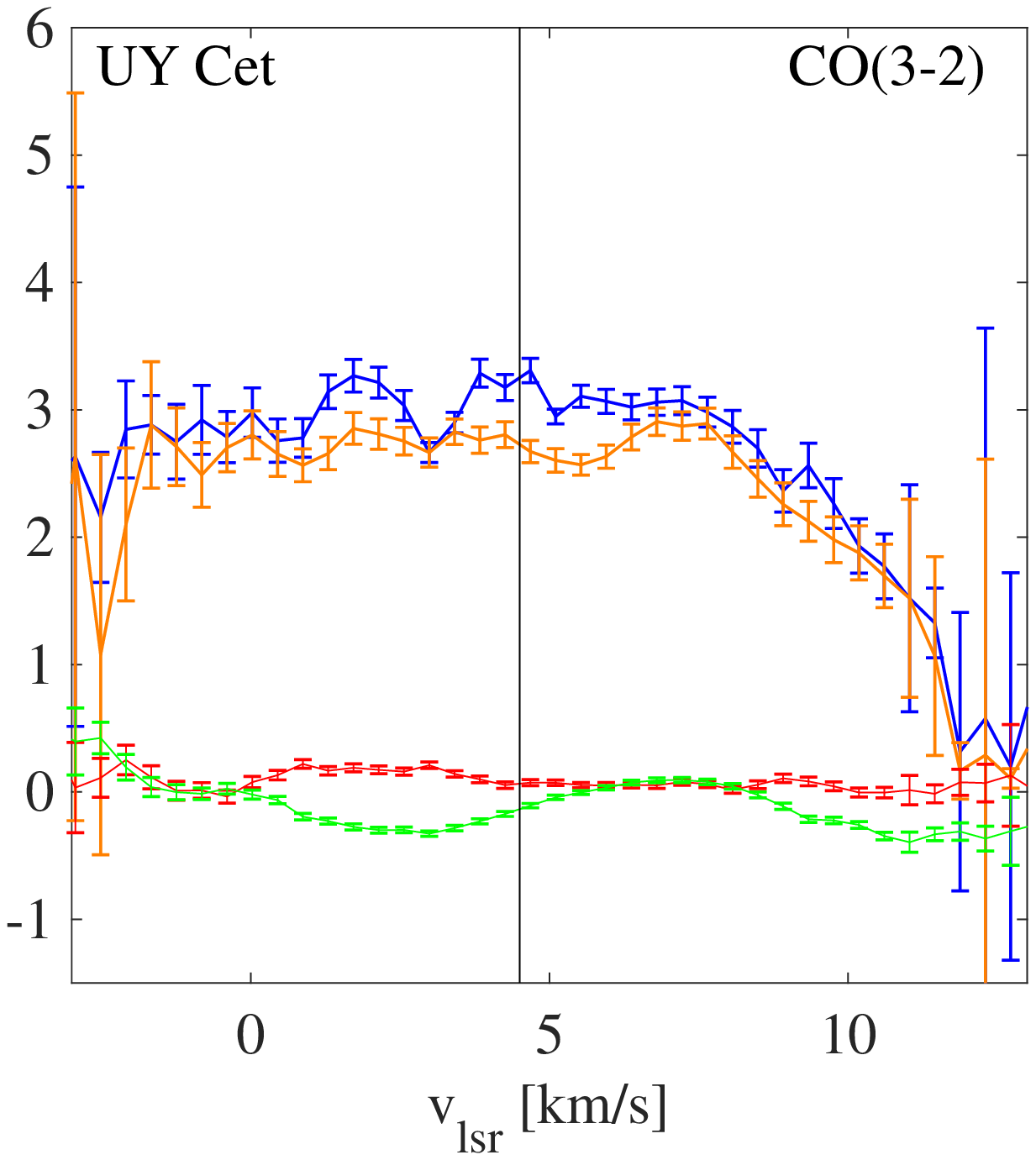}
\hspace{0.3cm}
\includegraphics[height=4.5cm]{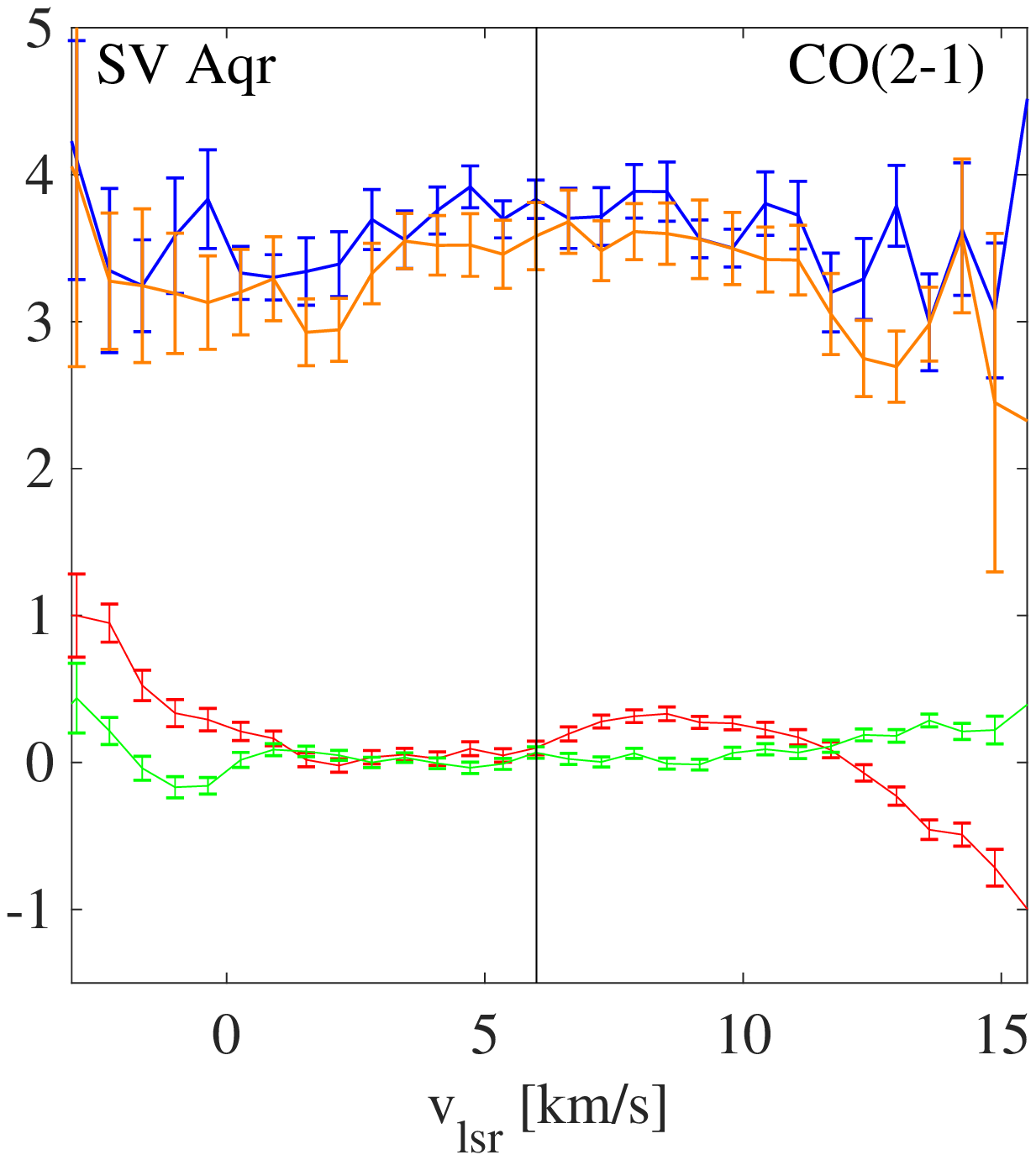}
\hspace{0.18cm}
\includegraphics[height=4.5cm]{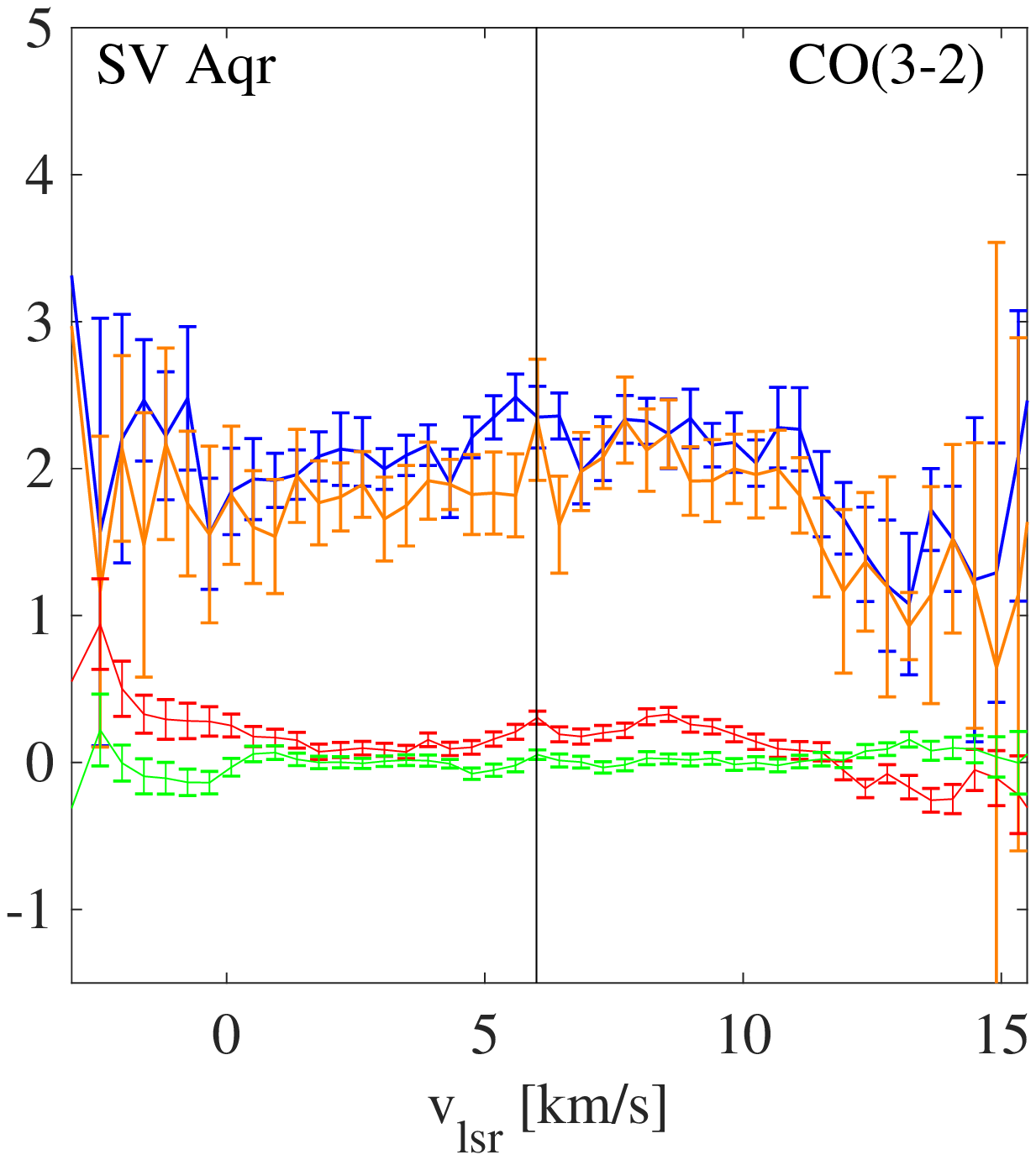}
\caption{Results from the visibility fitting to the data measured toward the M-type AGB stars of the sample discussed in this paper. The source name is given in the upper left corner and the transition is in the upper right corner of each plot. The upper blue and orange lines show the major and minor axis of the best-fit Gaussian in each channel, respectively. The lower red and green lines show the RA and Dec offset relative to the center position, respectively.}
\label{uvM_SR}
\end{figure*}


\begin{figure*}[t]
\includegraphics[height=4.5cm]{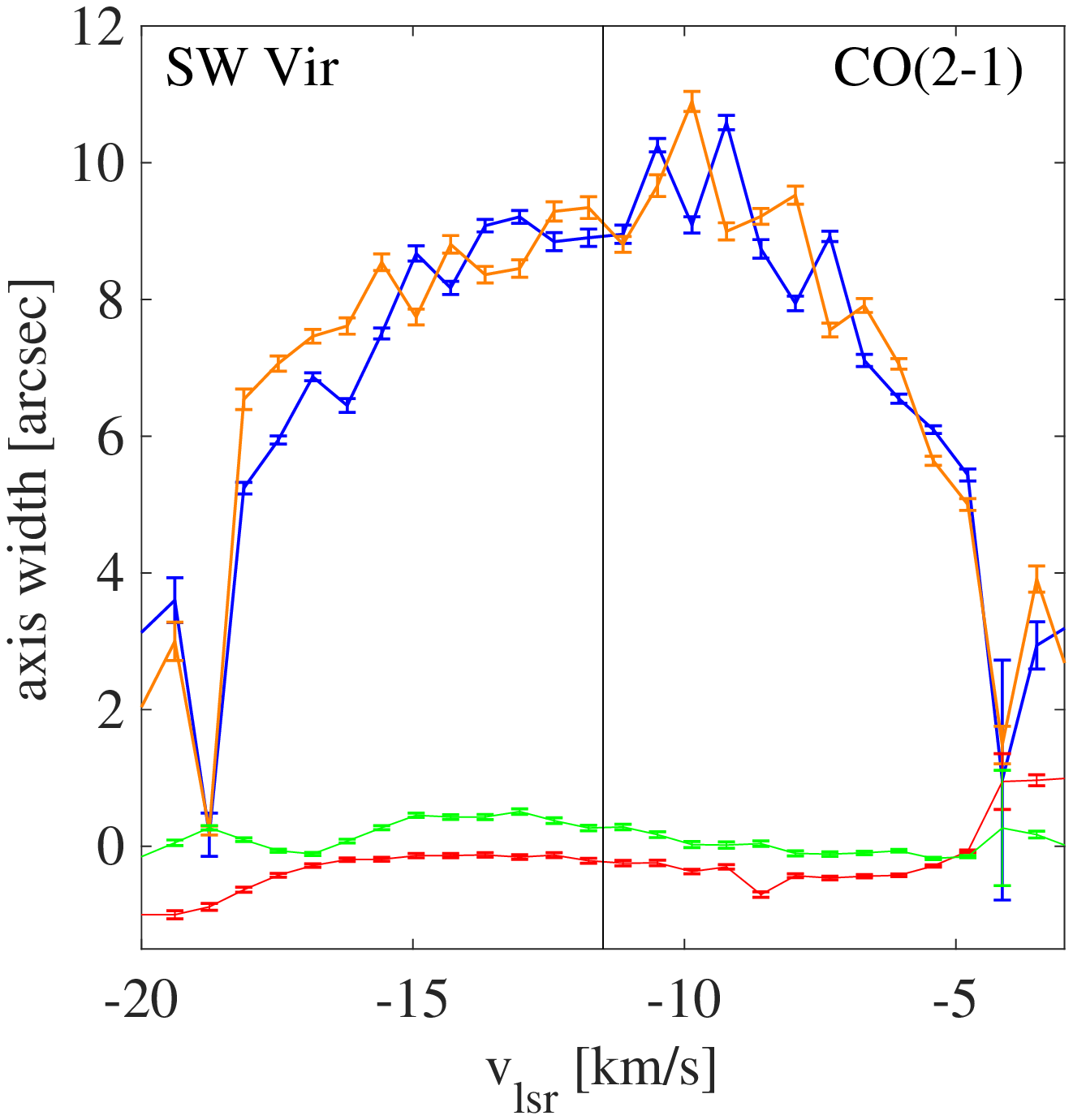}
\hspace{0.05cm}
\includegraphics[height=4.5cm]{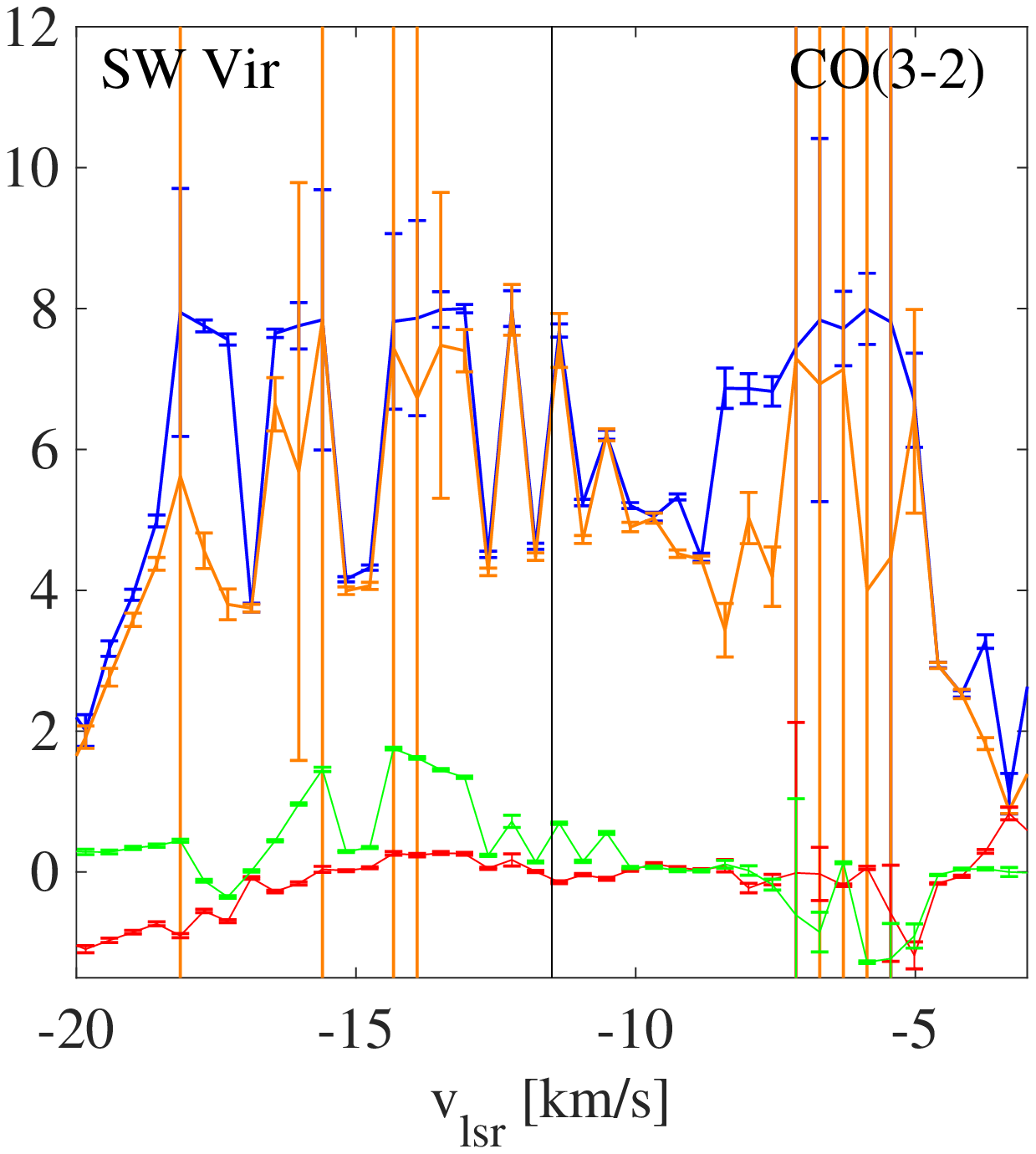}
\hspace{0.2cm}
\includegraphics[height=4.5cm]{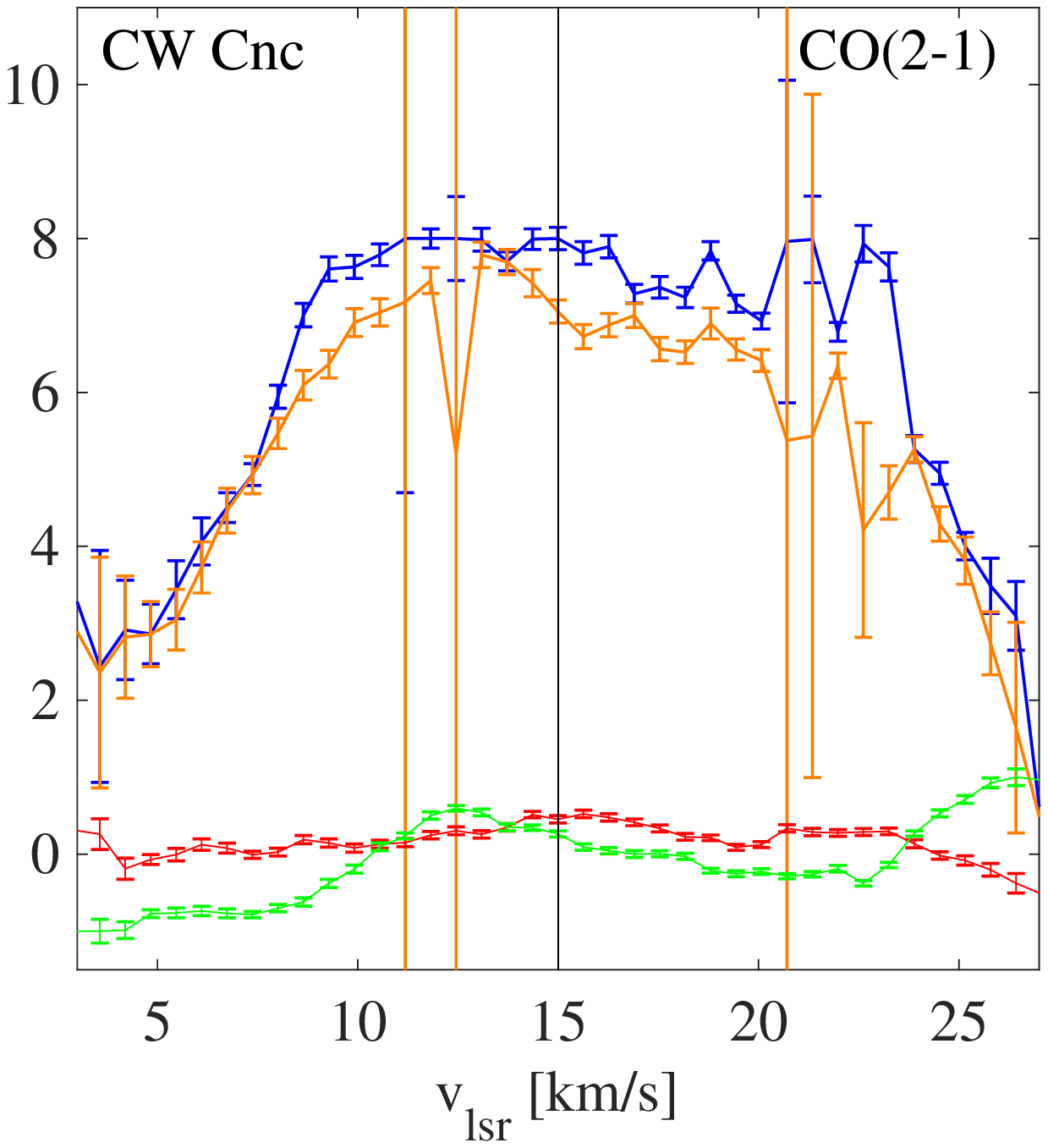}
\hspace{0.05cm}
\includegraphics[height=4.5cm]{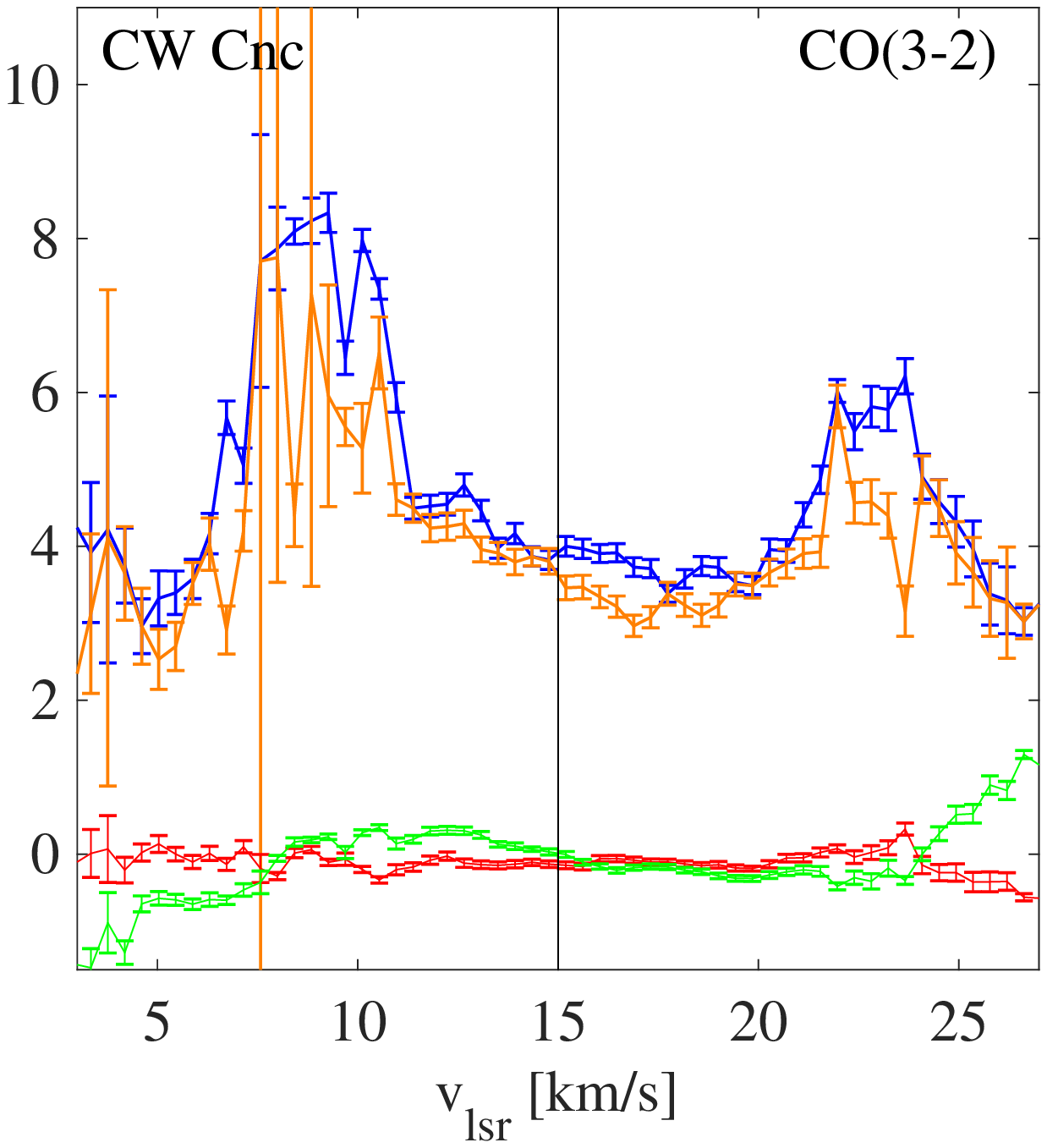}

\includegraphics[height=4.5cm]{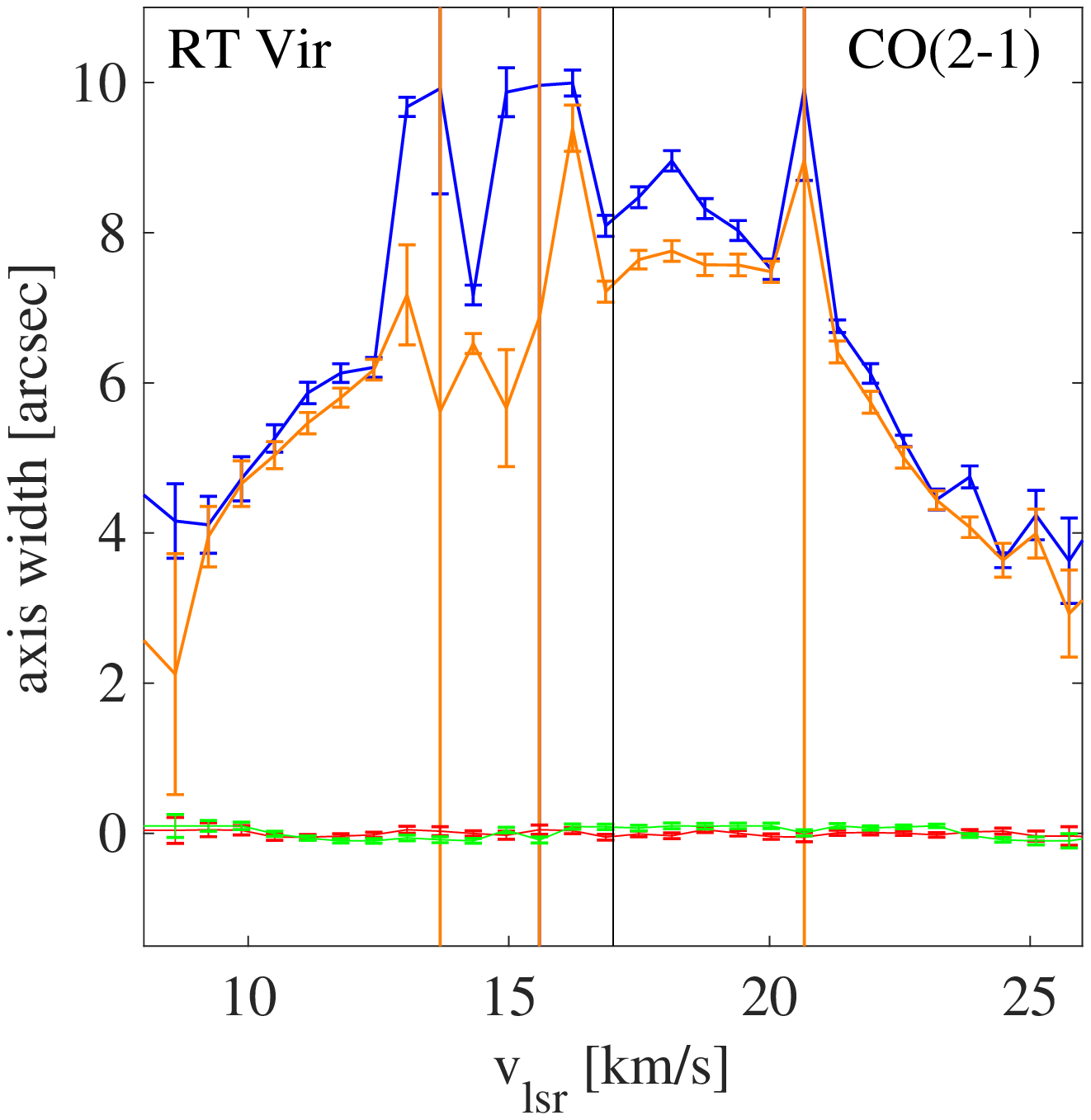}
\includegraphics[height=4.5cm]{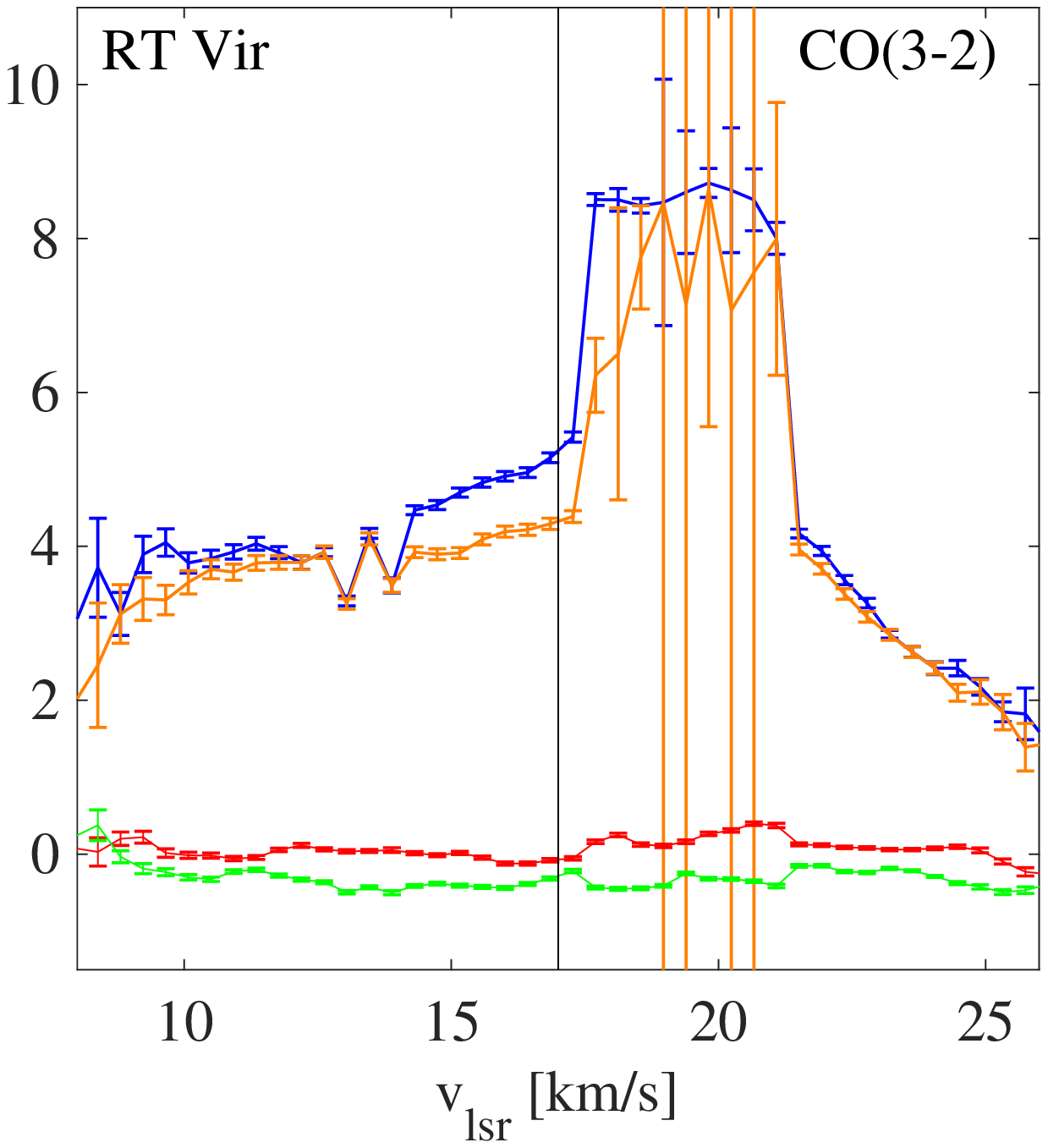}
\hspace{0.2cm}
\includegraphics[height=4.5cm]{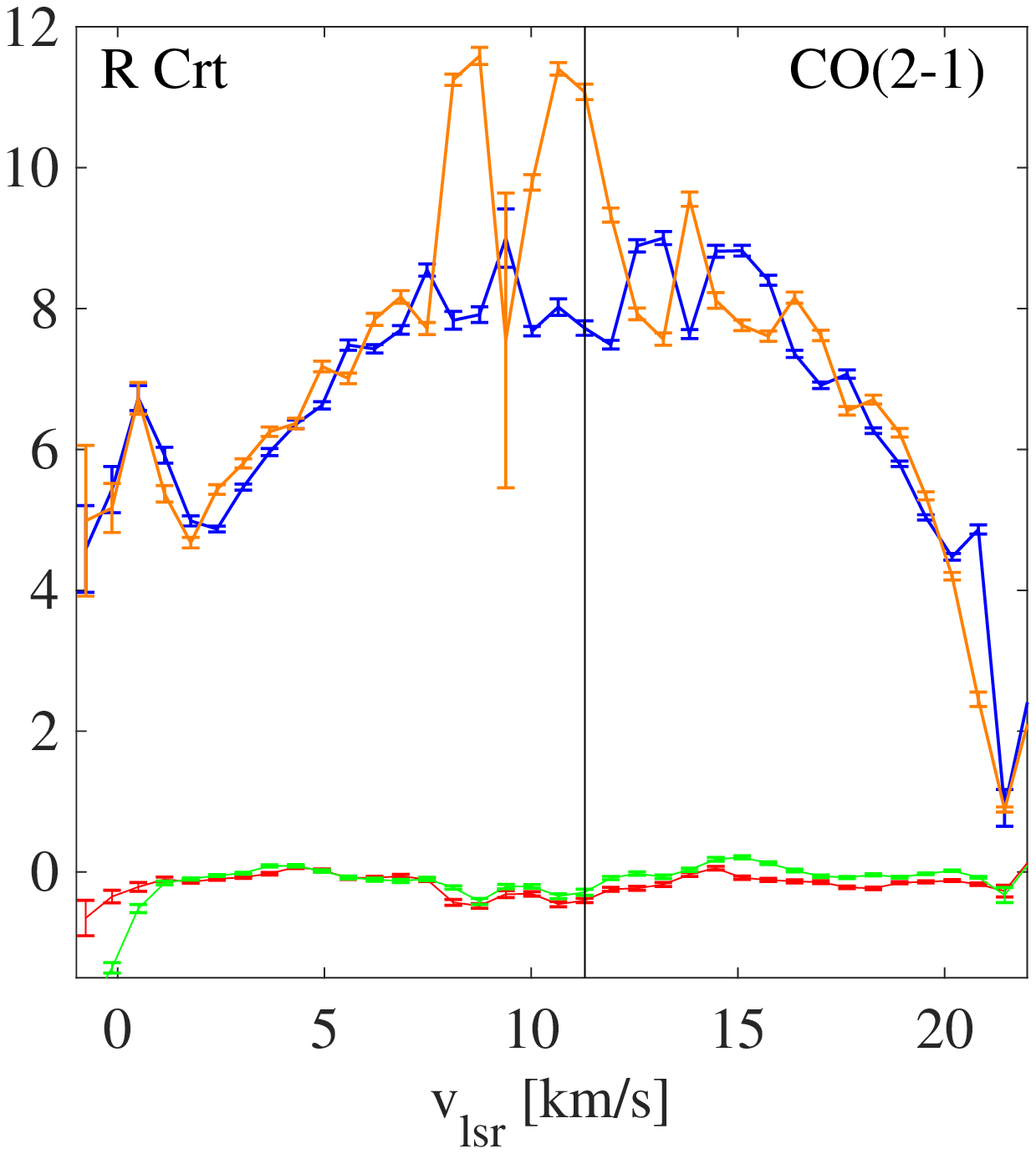}
\hspace{0.15cm}
\includegraphics[height=4.5cm]{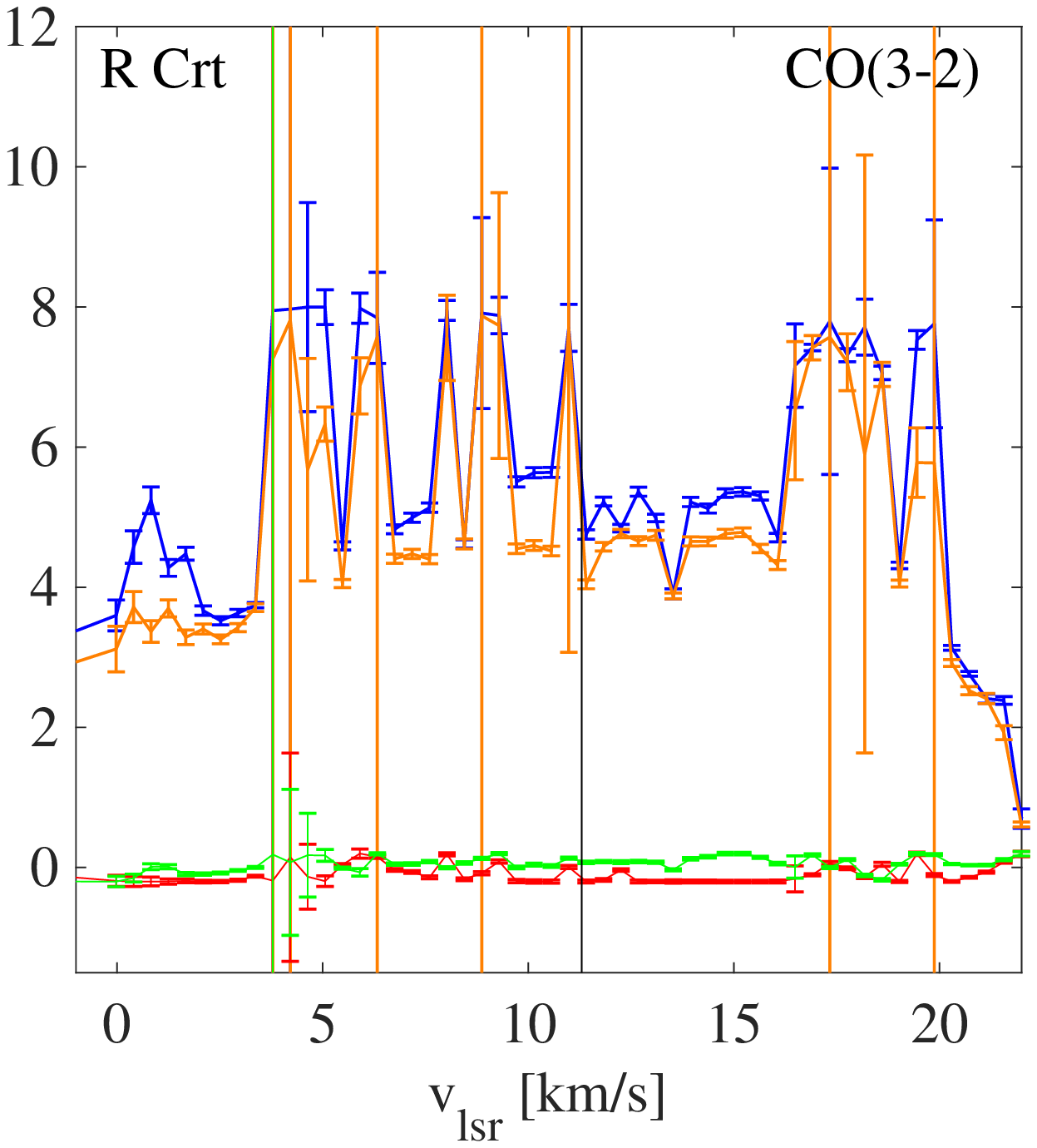}

\includegraphics[height=4.5cm]{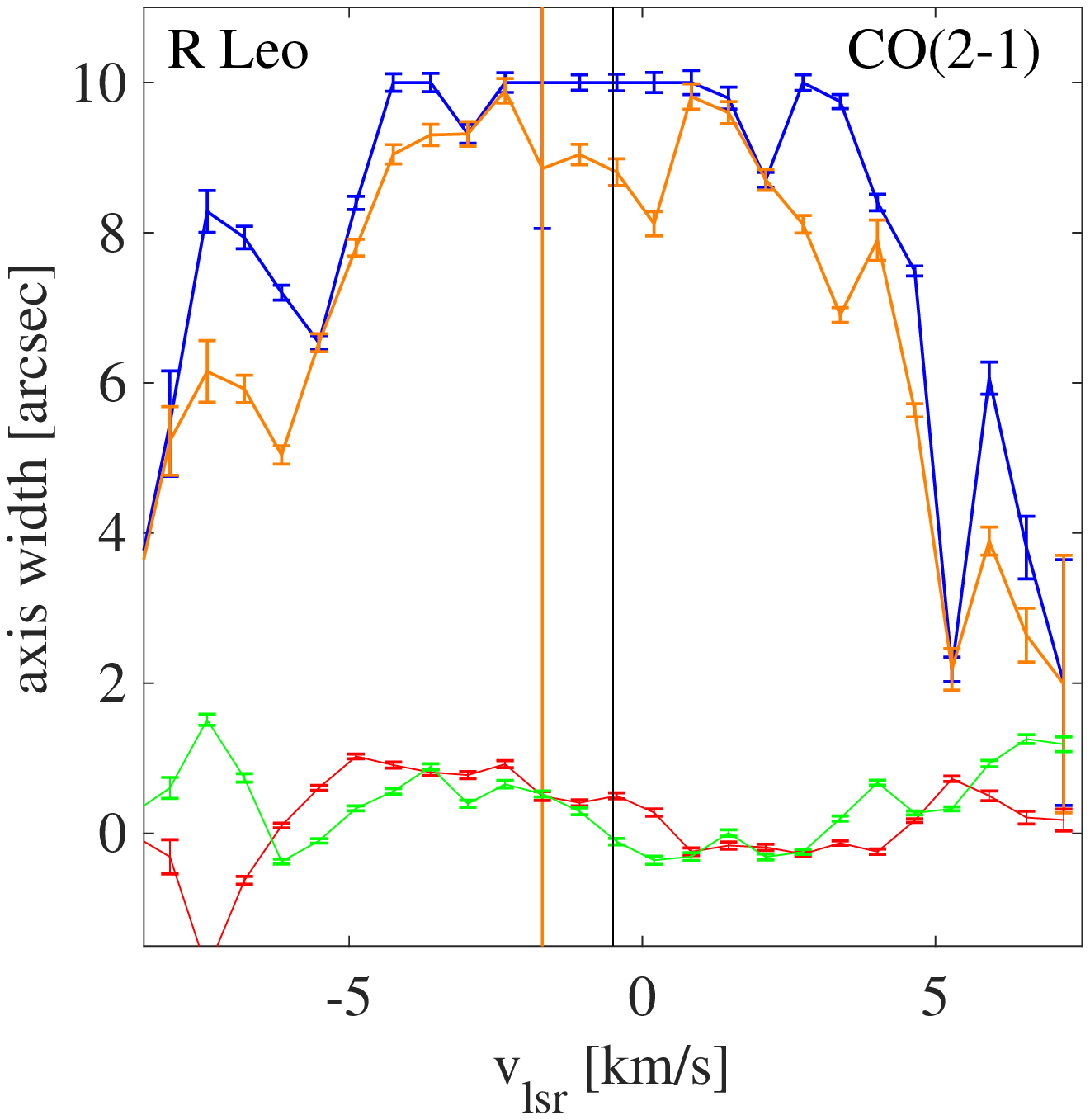}
\includegraphics[height=4.5cm]{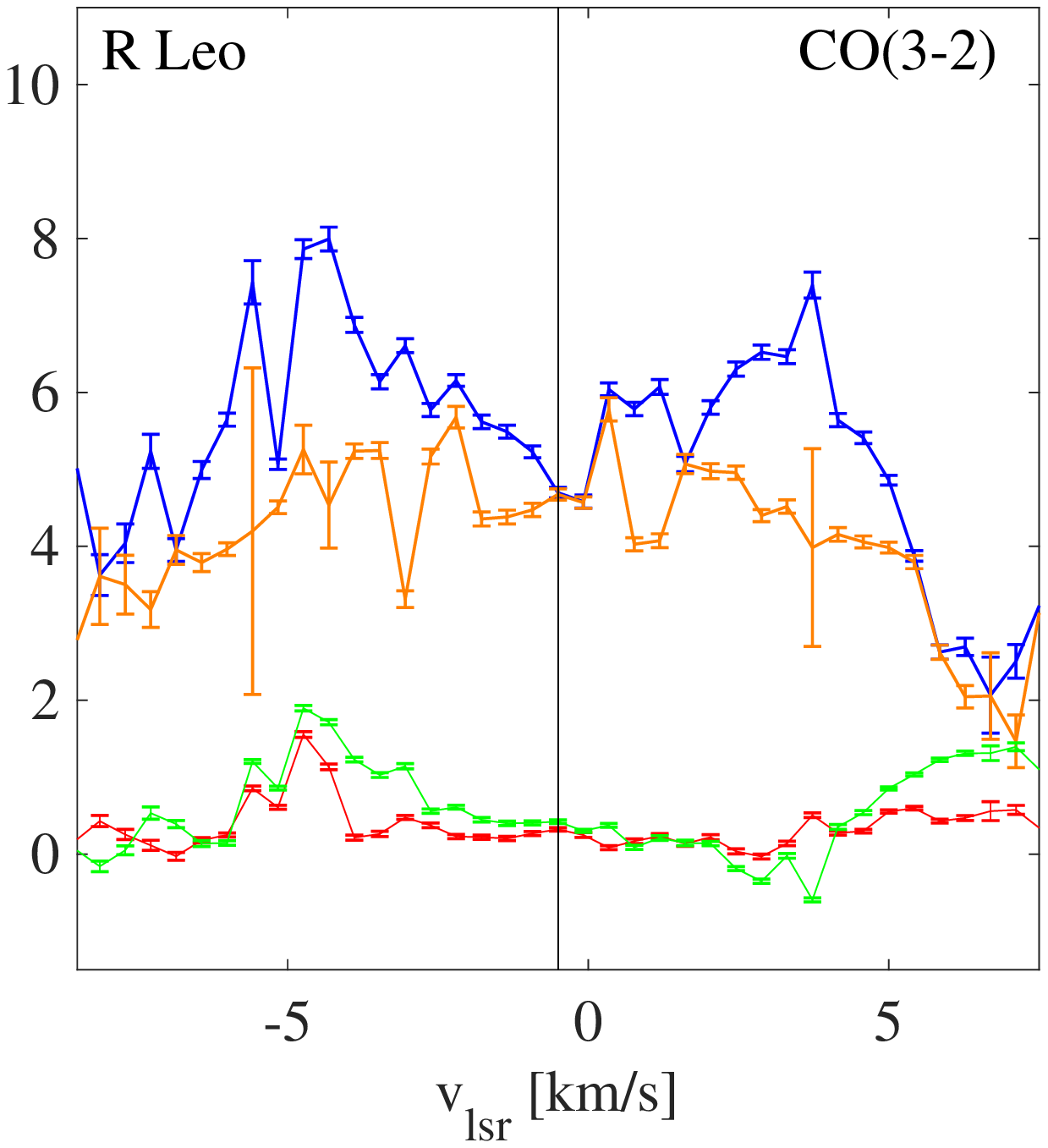}
\hspace{0.2cm}
\includegraphics[height=4.5cm]{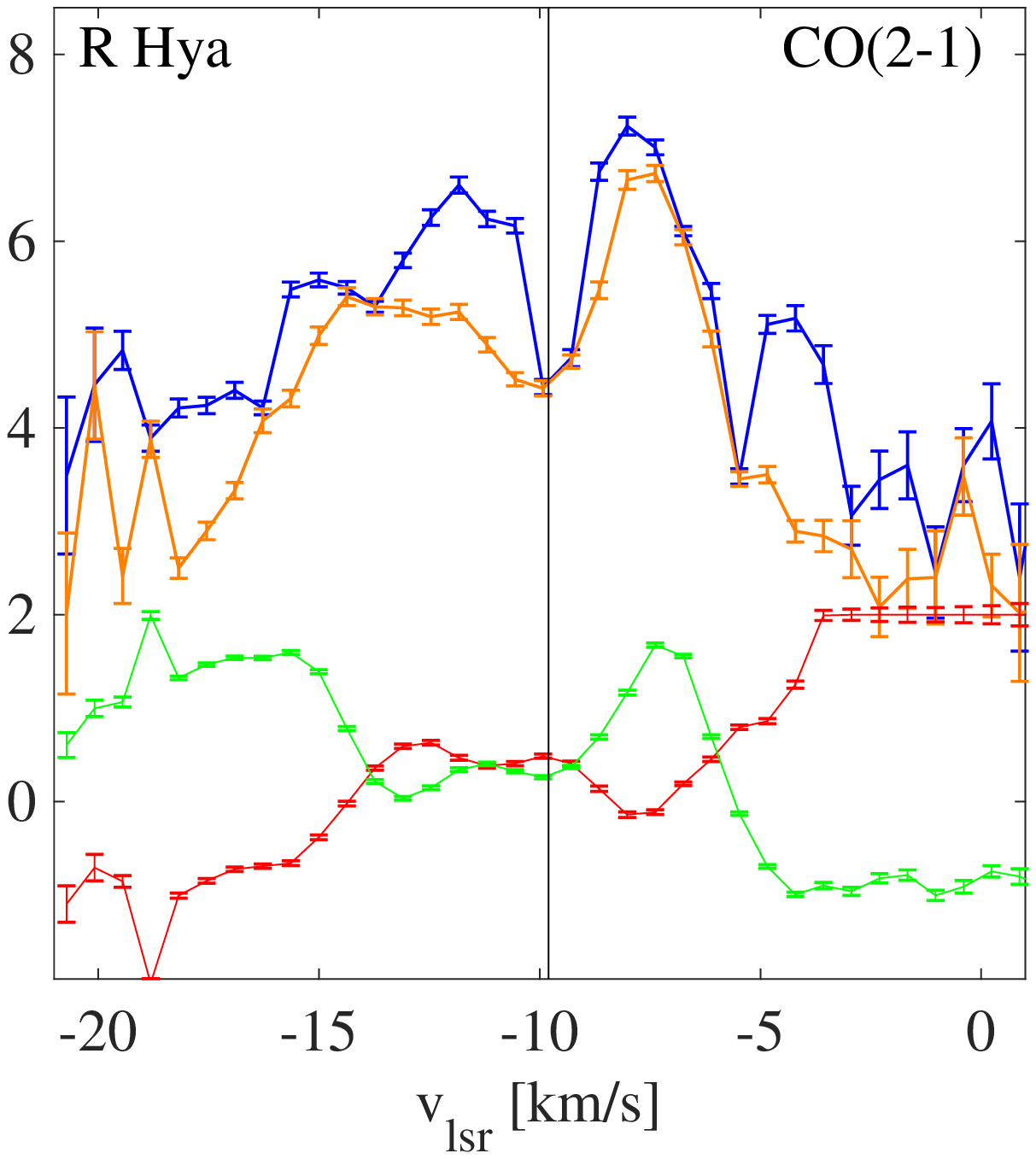}
\hspace{0.15cm}
\includegraphics[height=4.5cm]{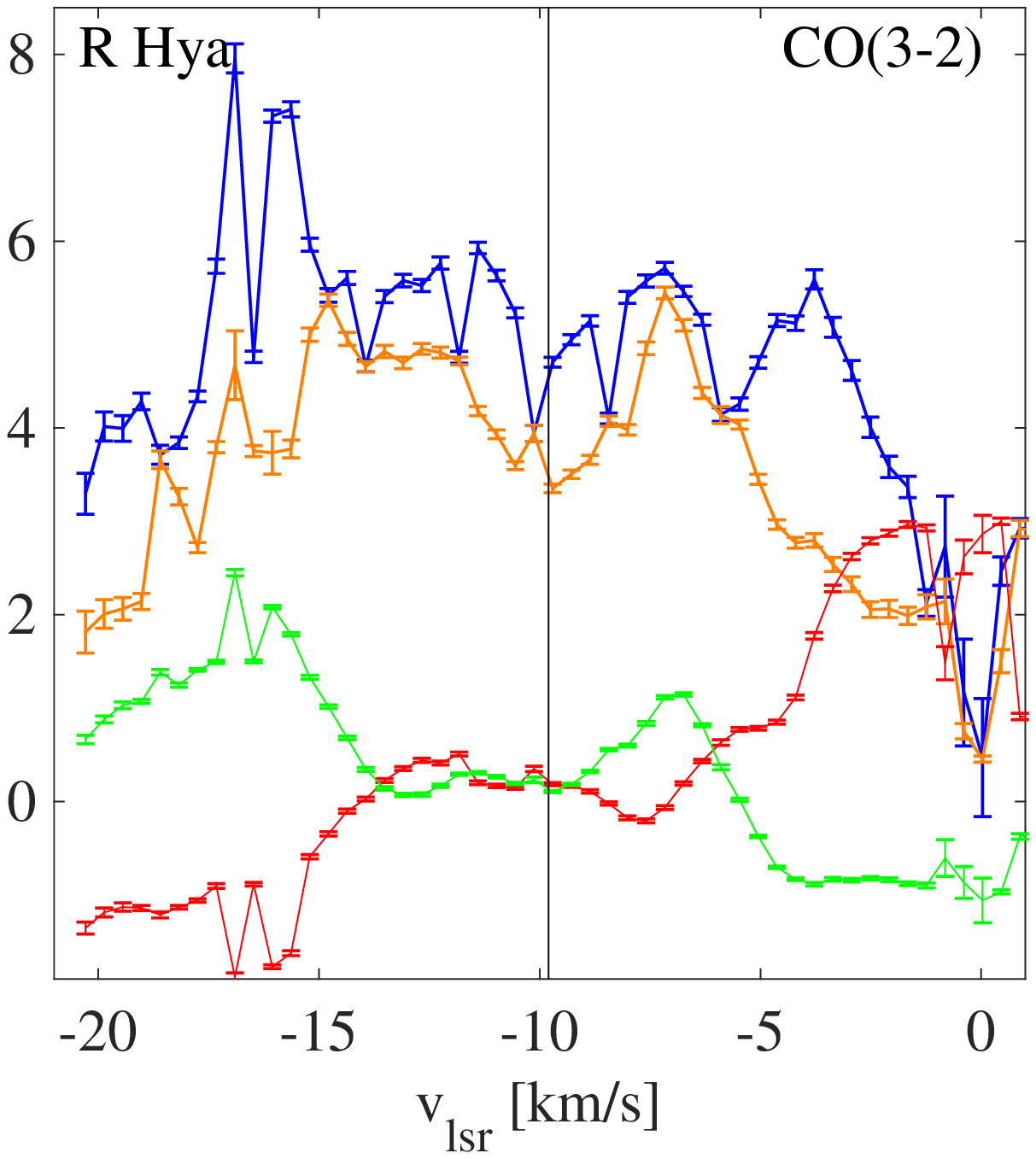}

\includegraphics[height=4.5cm]{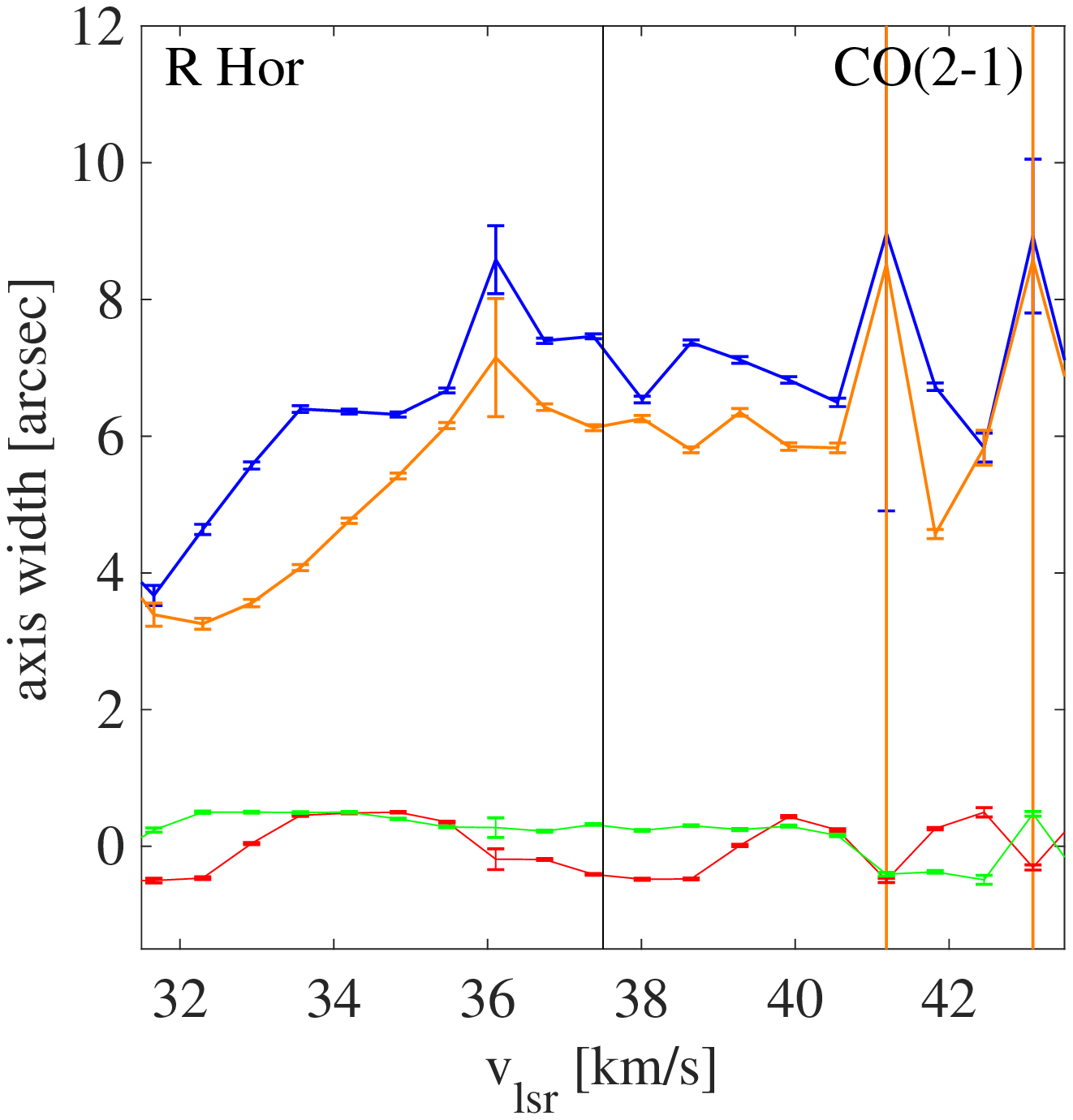}
\hspace{0.03cm}
\includegraphics[height=4.5cm]{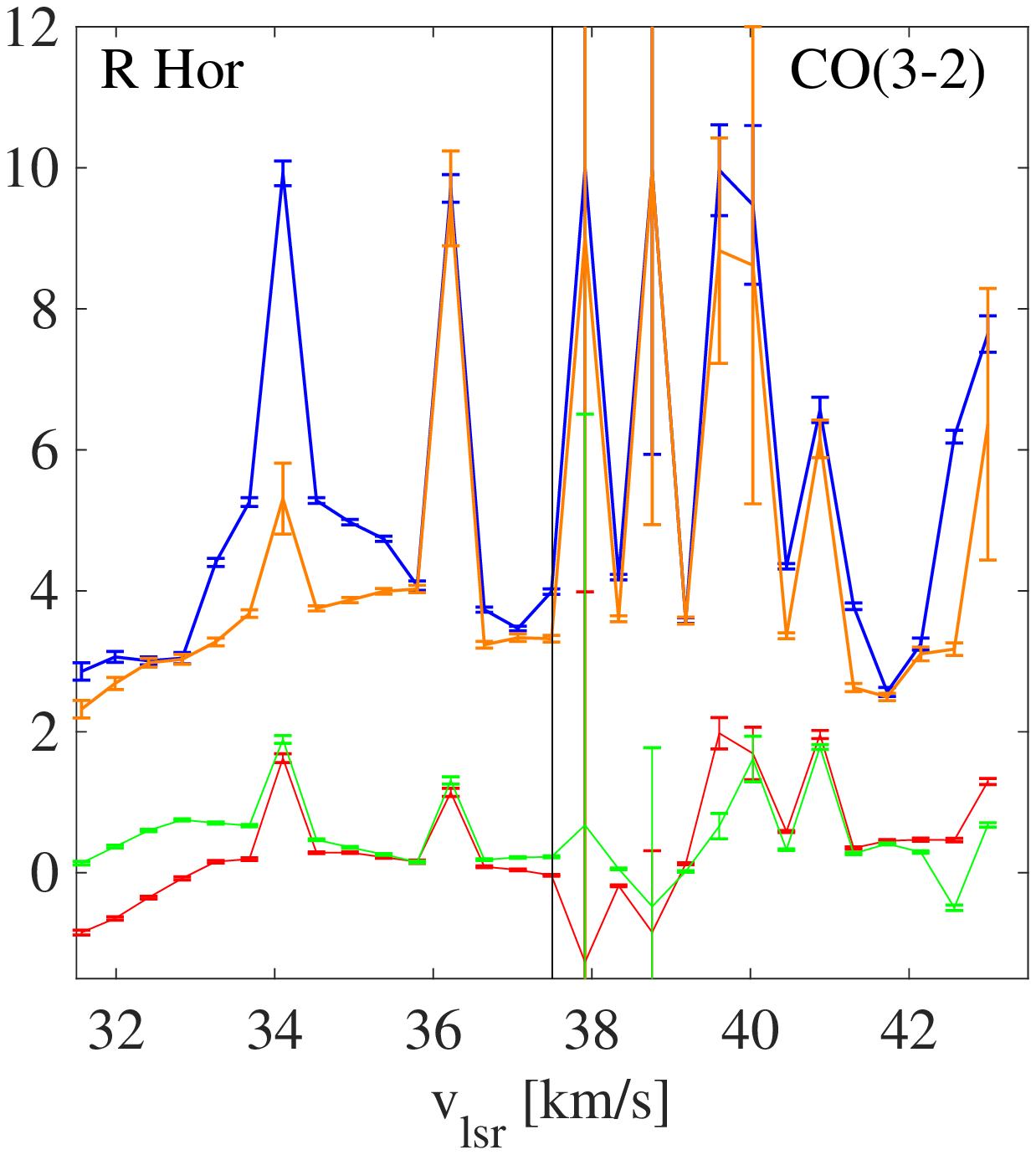}
\hspace{0.2cm}
\includegraphics[height=4.5cm]{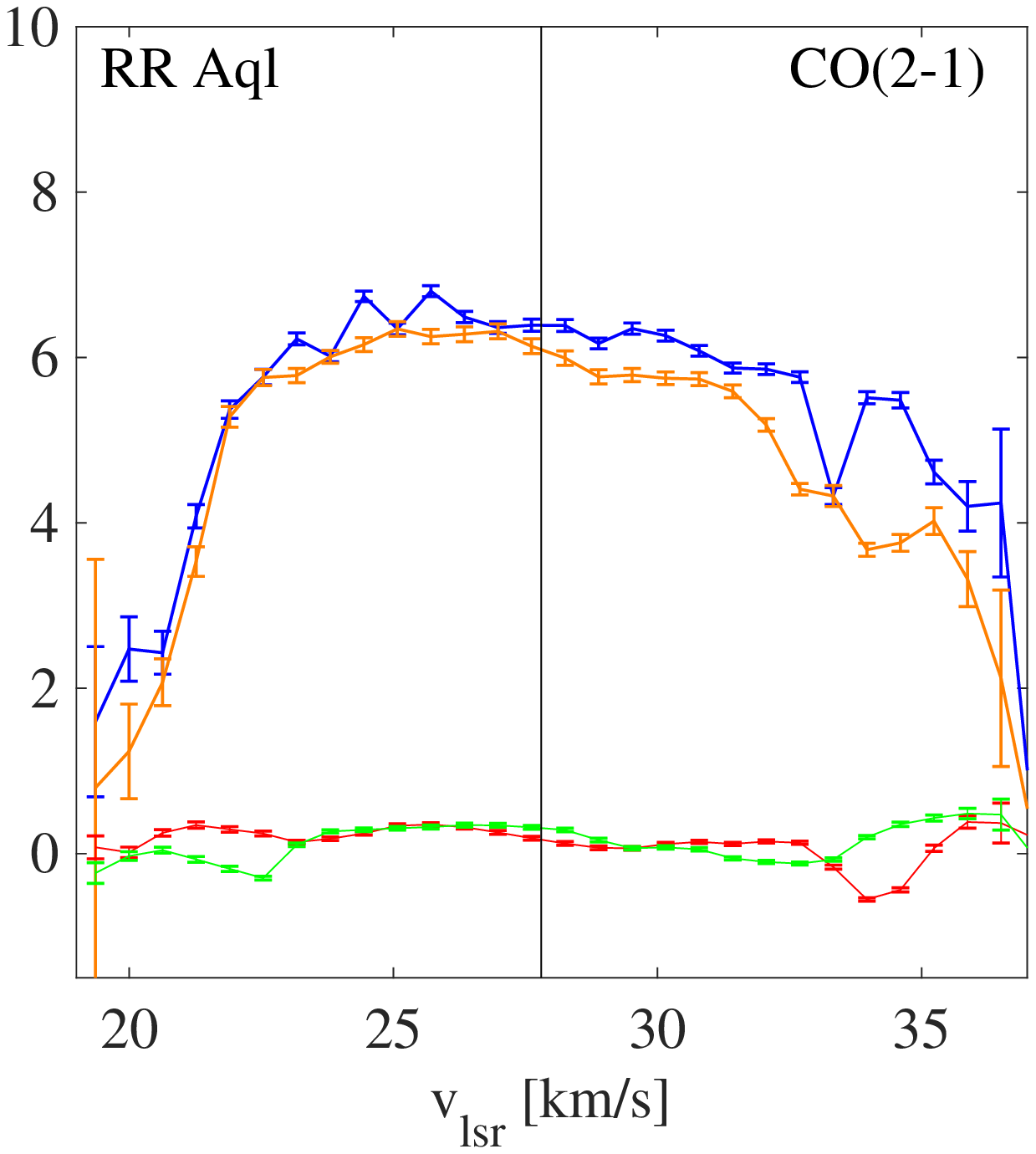}
\hspace{0.1cm}
\includegraphics[height=4.5cm]{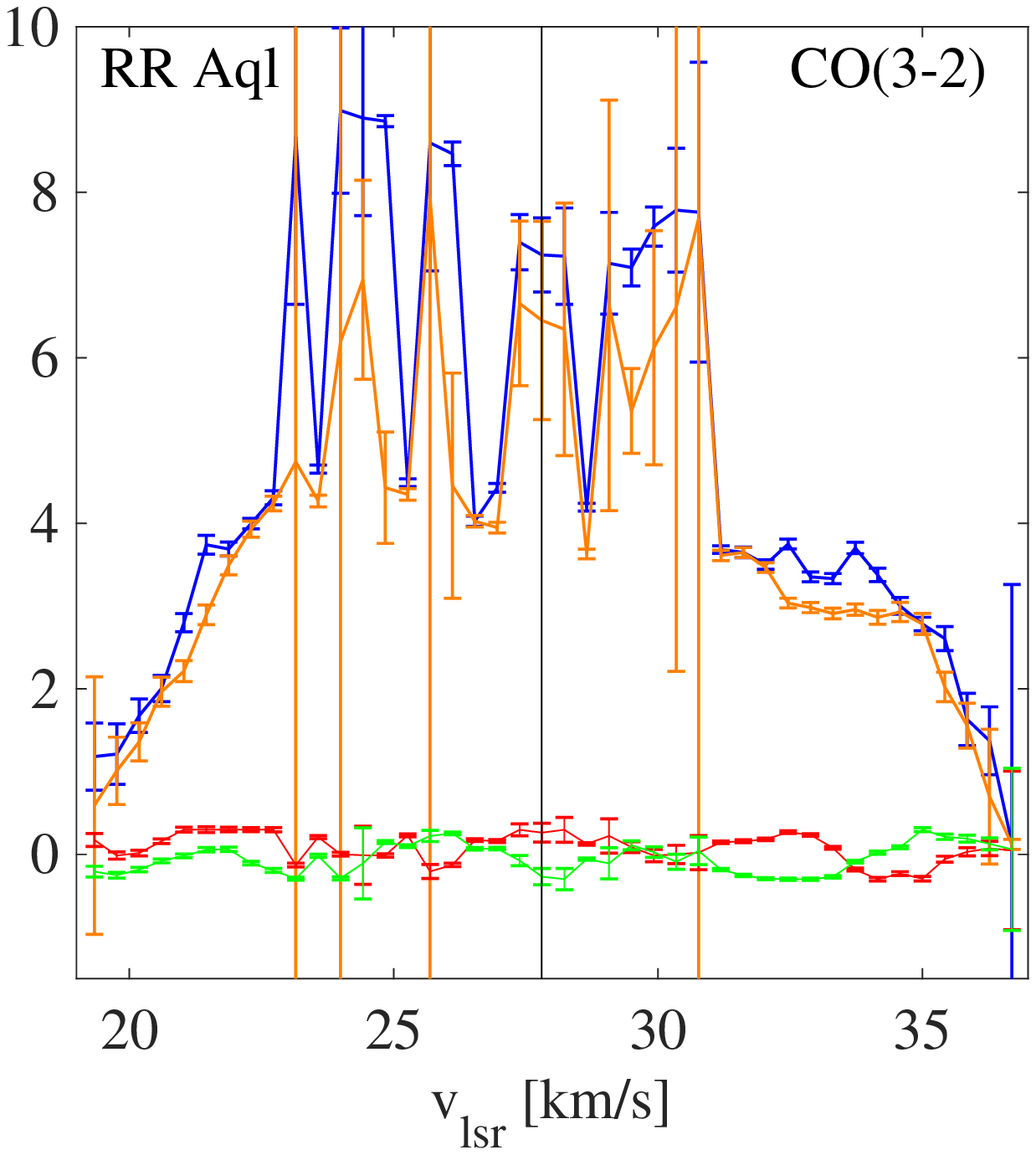}

\includegraphics[height=4.5cm]{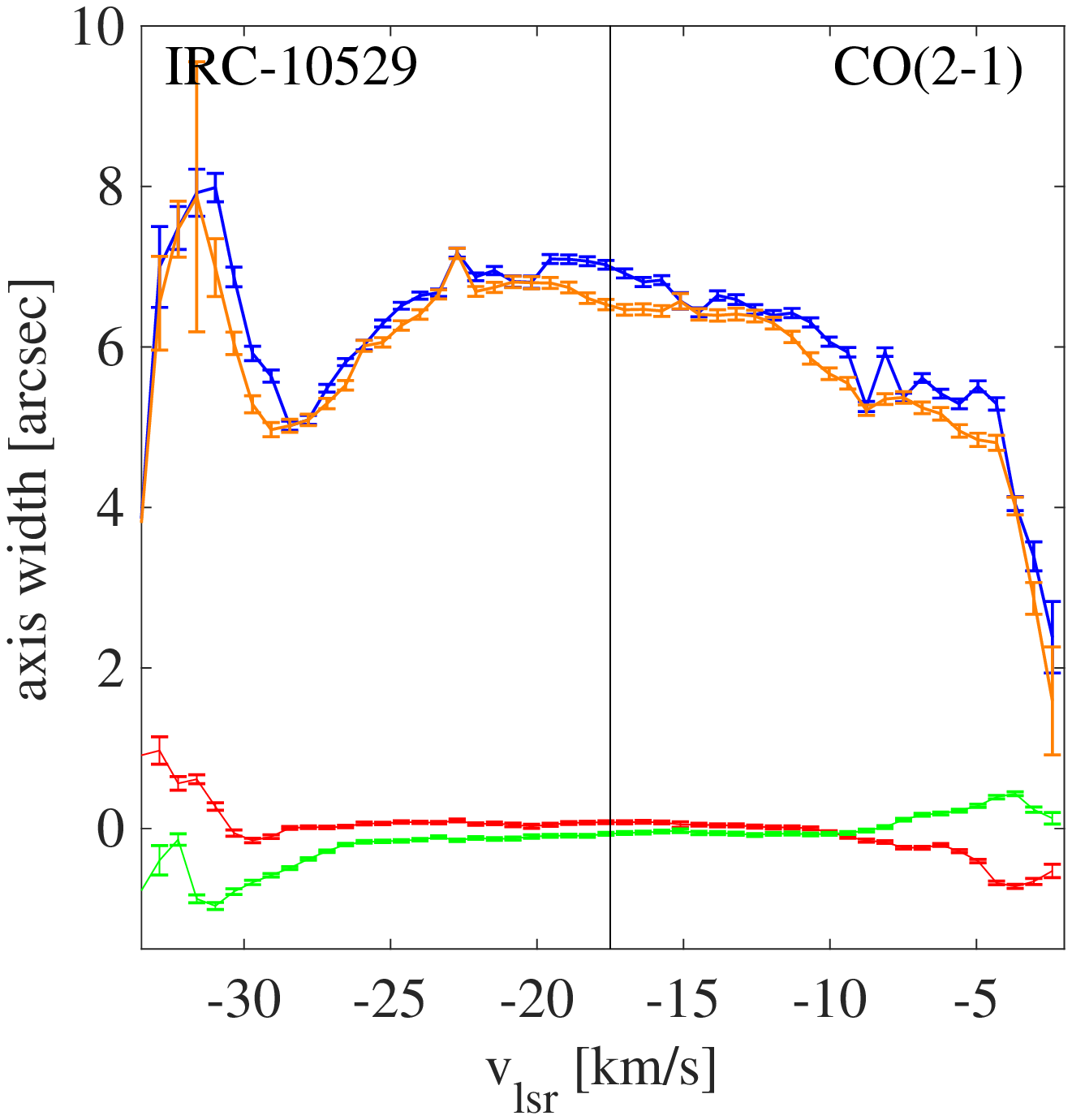}
\hspace{0.05cm}
\includegraphics[height=4.5cm]{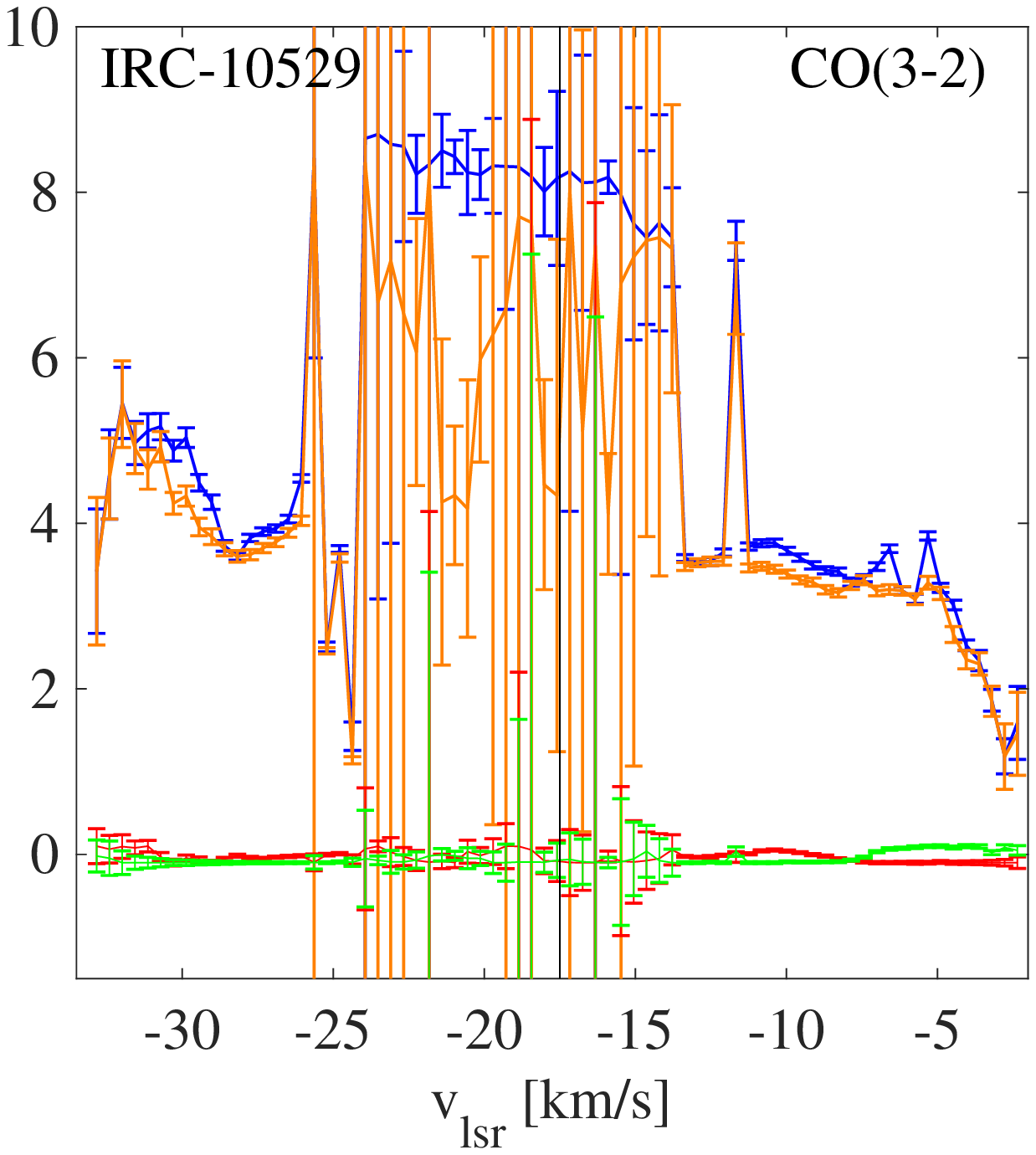}
\hspace{0.18cm}
\includegraphics[height=4.5cm]{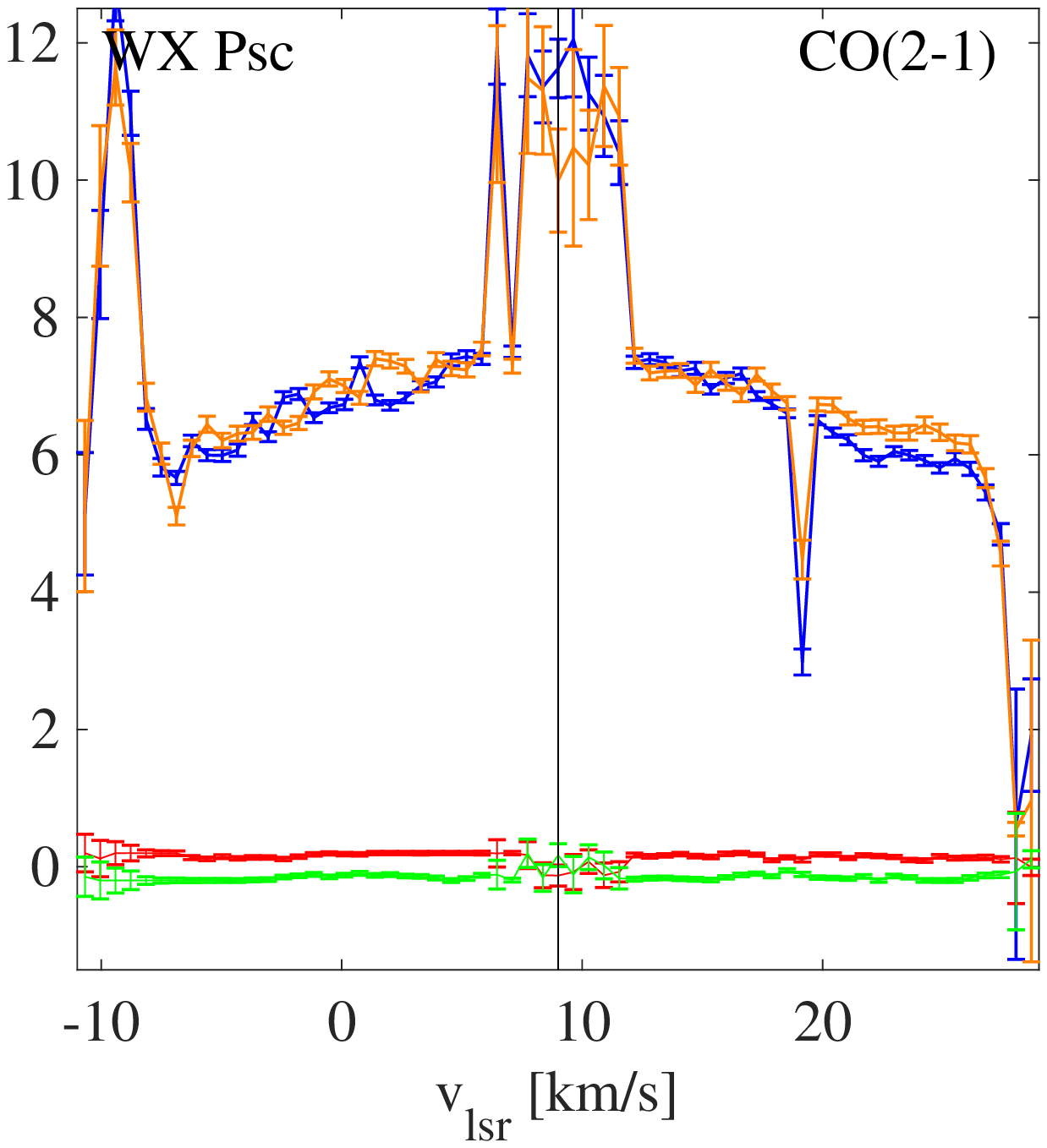}
\hspace{0.01cm}
\includegraphics[height=4.5cm]{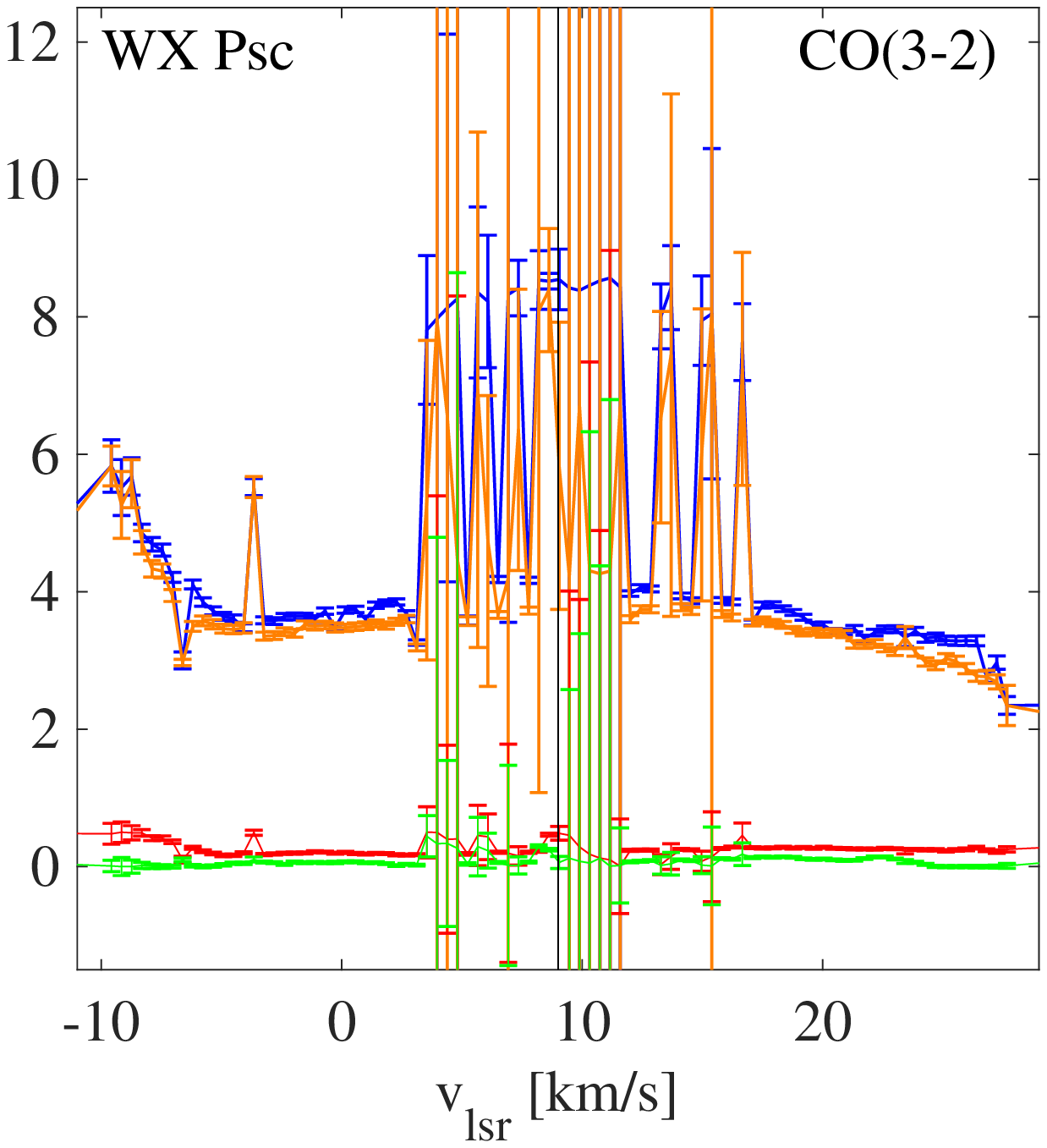}
\caption{Results from the visibility fitting to the data measured toward the M-type AGB stars of the sample discussed in this paper. The source name is given in the upper left corner and the transition is in the upper right corner of each plot. The upper blue and orange lines show the major and minor axis of the best-fit Gaussian in each channel, respectively. The lower red and green lines show the RA and Dec offset relative to the center position, respectively.}
\label{uvM_SRM}
\end{figure*}


\begin{figure*}[t]
\includegraphics[height=4.5cm]{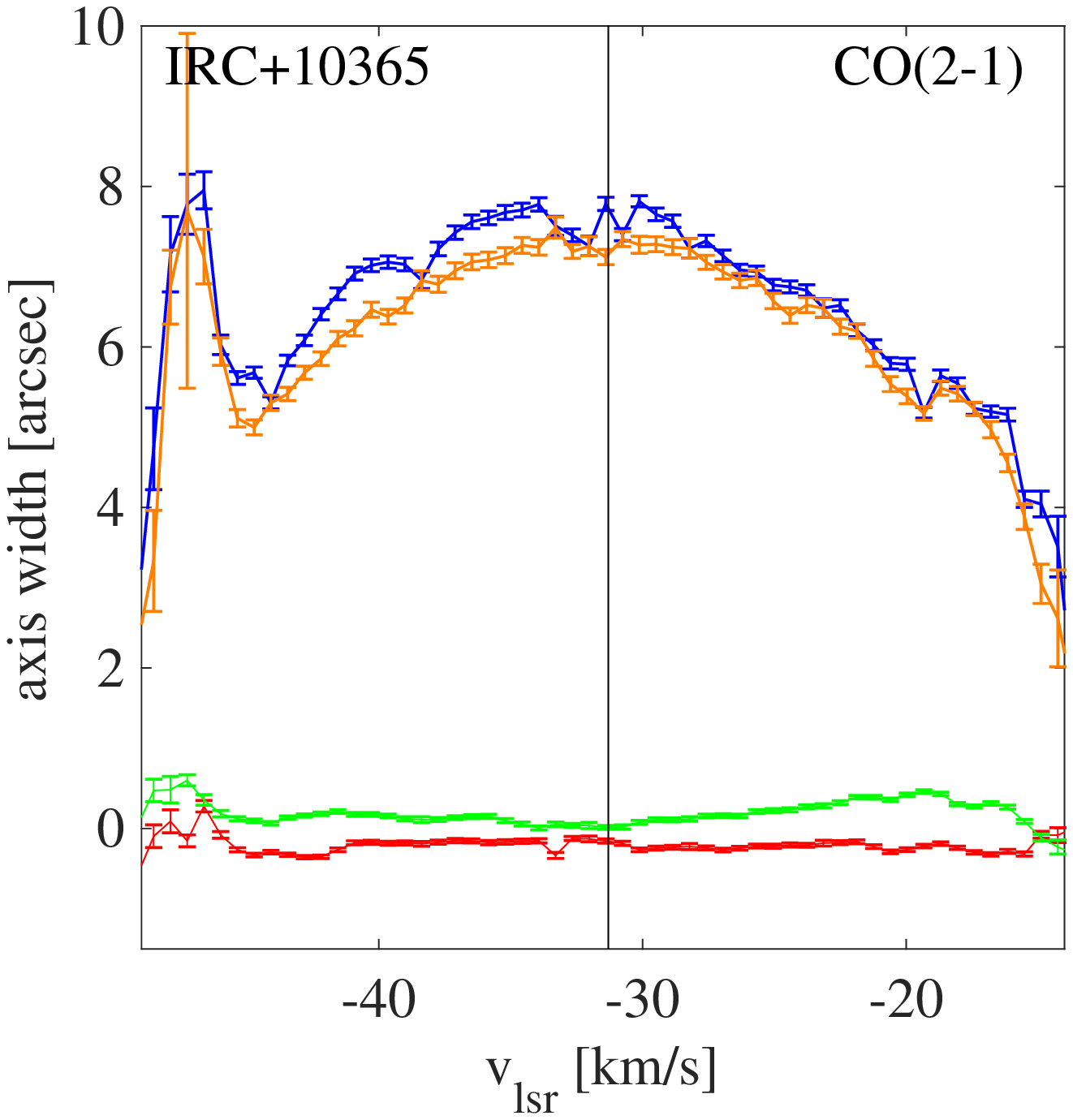}
\hspace{0.05cm}
\includegraphics[height=4.5cm]{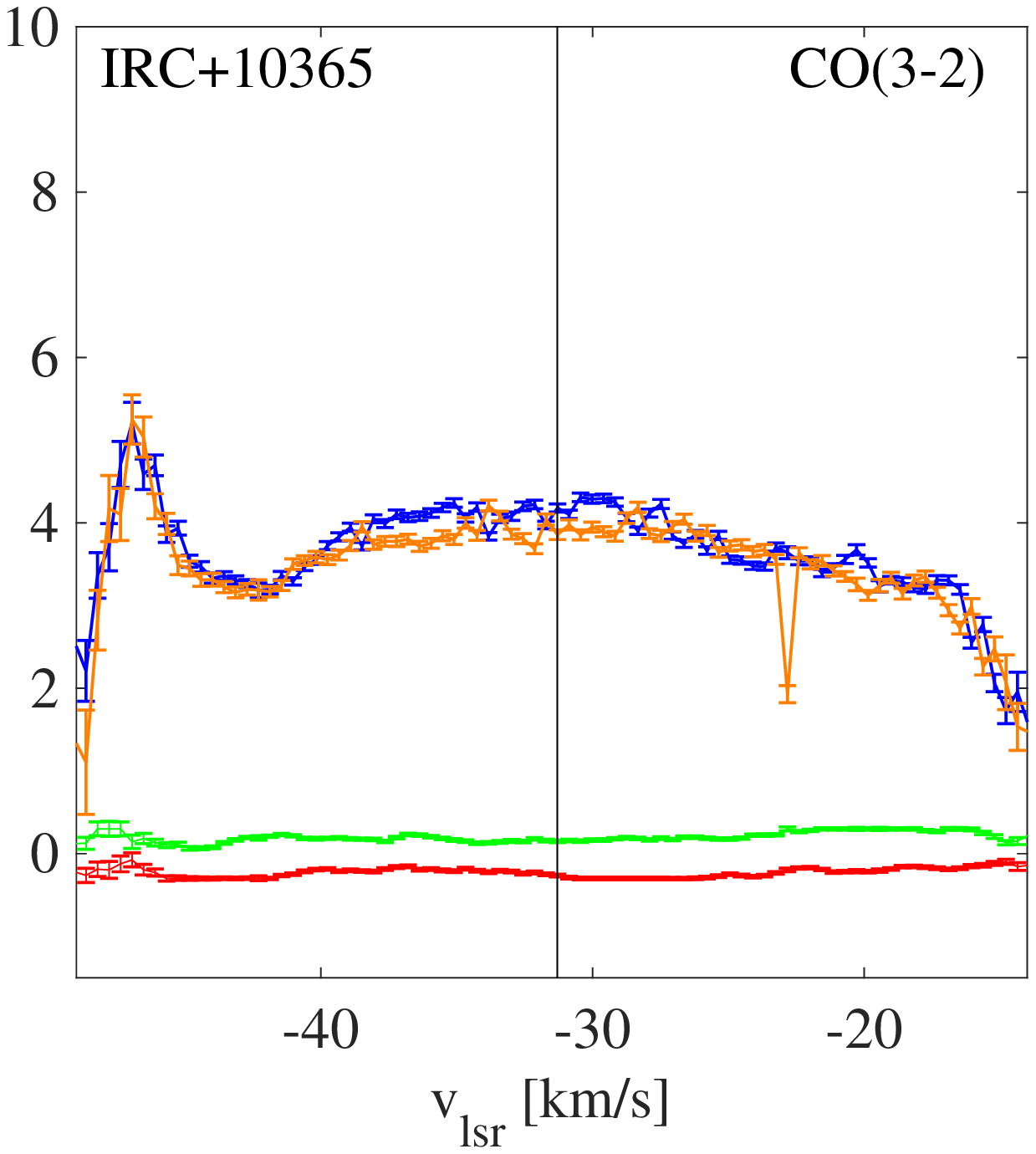}
\caption{Results from the visibility fitting to the data measured toward the M-type AGB stars of the sample discussed in this paper. The source name is given in the upper left corner and the transition is in the upper right corner of each plot. The upper blue and orange lines show the major and minor axis of the best-fit Gaussian in each channel, respectively. The lower red and green lines show the RA and Dec offset relative to the center position, respectively.}
\label{uvM_M}
\end{figure*}


\begin{figure*}[t]
\includegraphics[height=4.5cm]{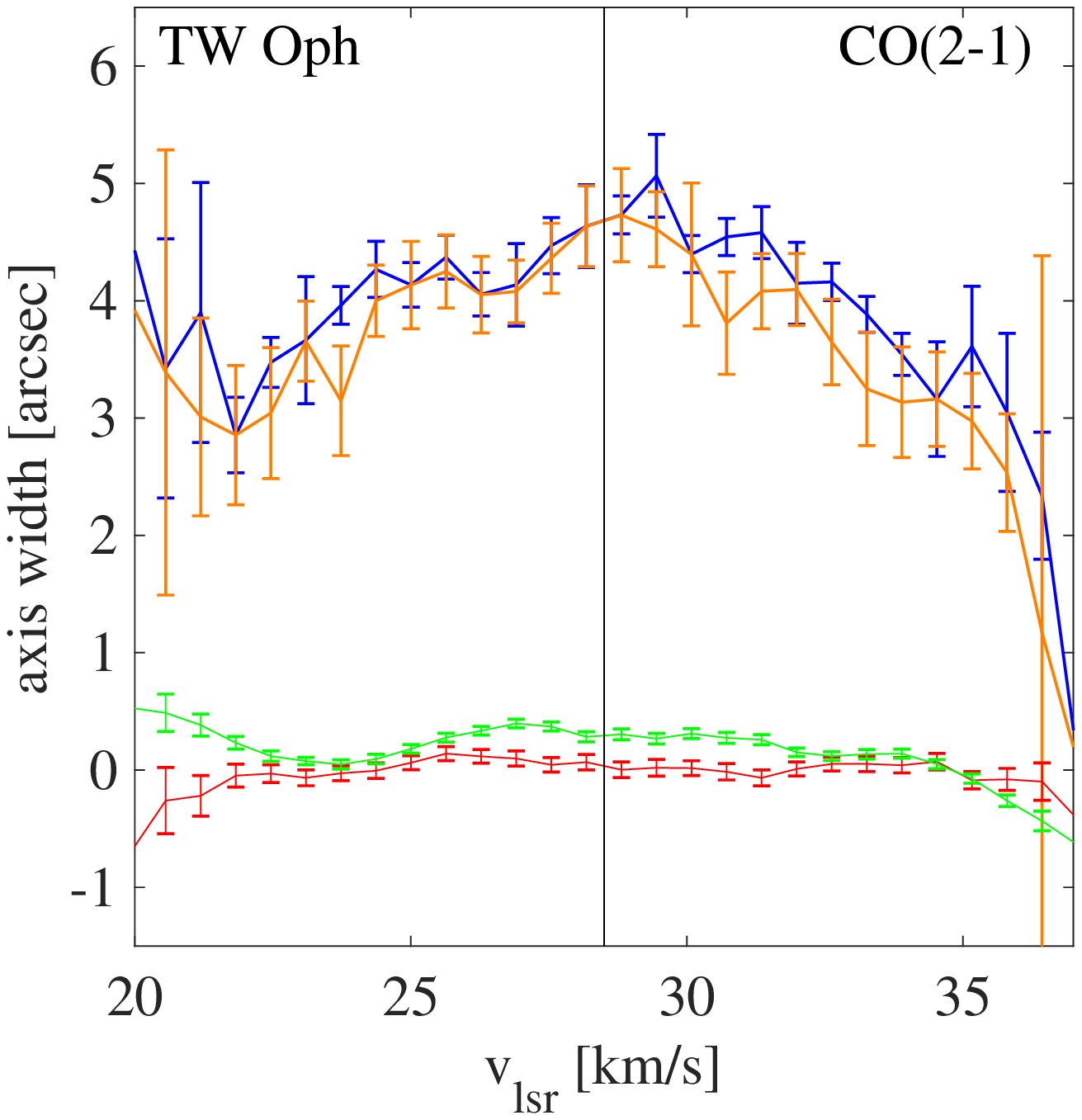}
\hspace{0.03cm}
\includegraphics[height=4.5cm]{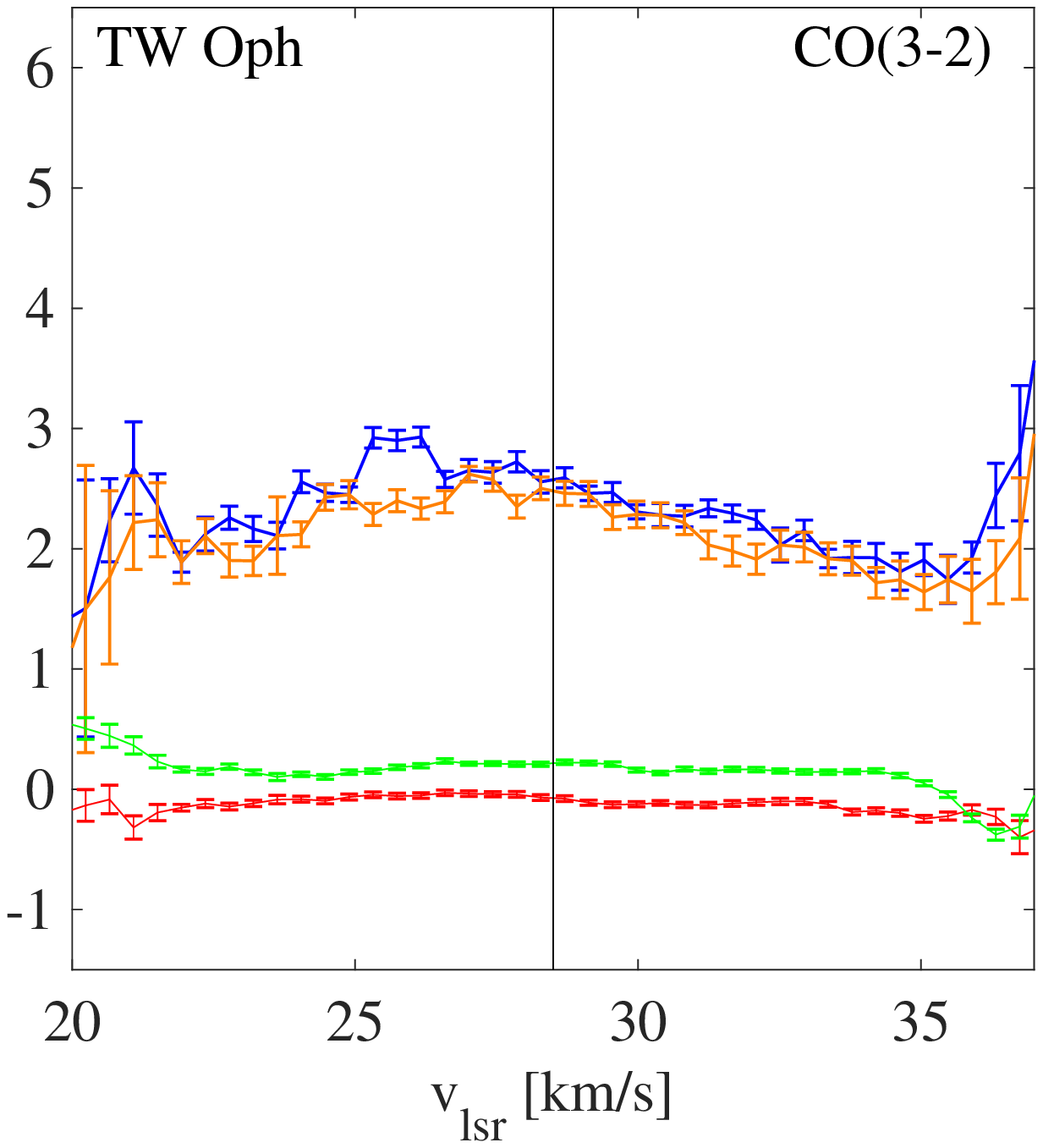}
\hspace{0.2cm}
\includegraphics[height=4.5cm]{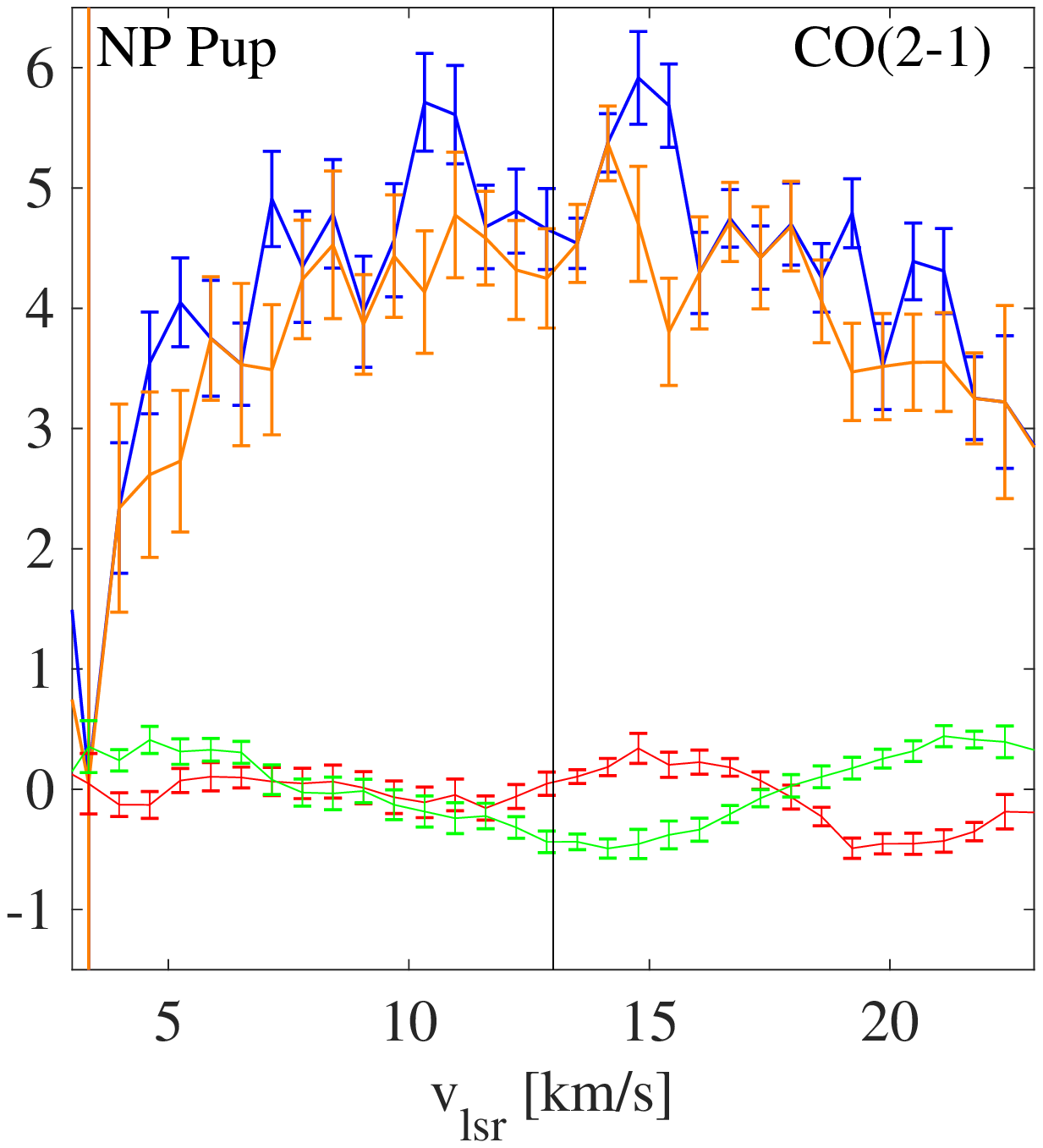}
\hspace{0.03cm}
\includegraphics[height=4.5cm]{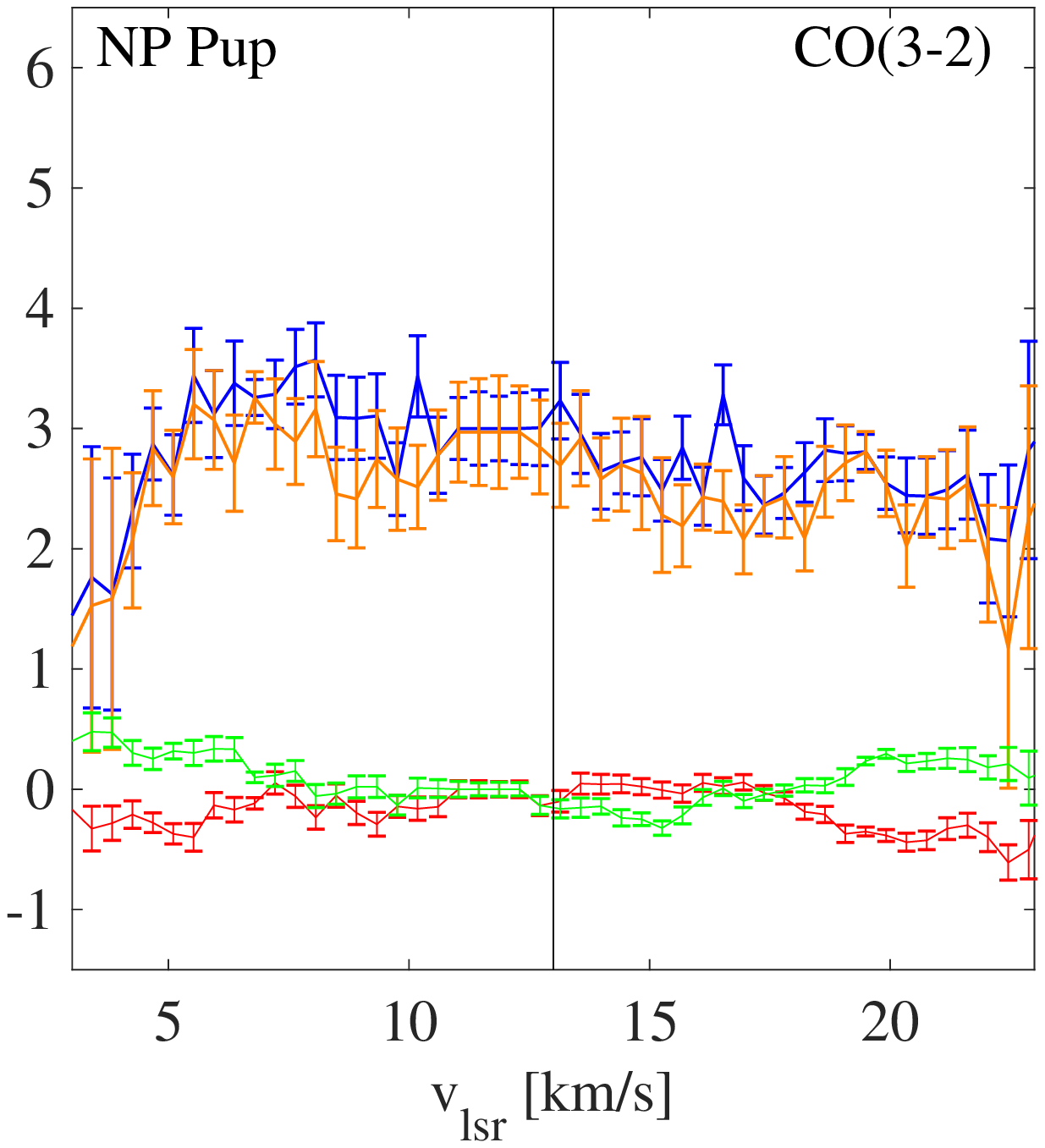}

\includegraphics[height=4.5cm]{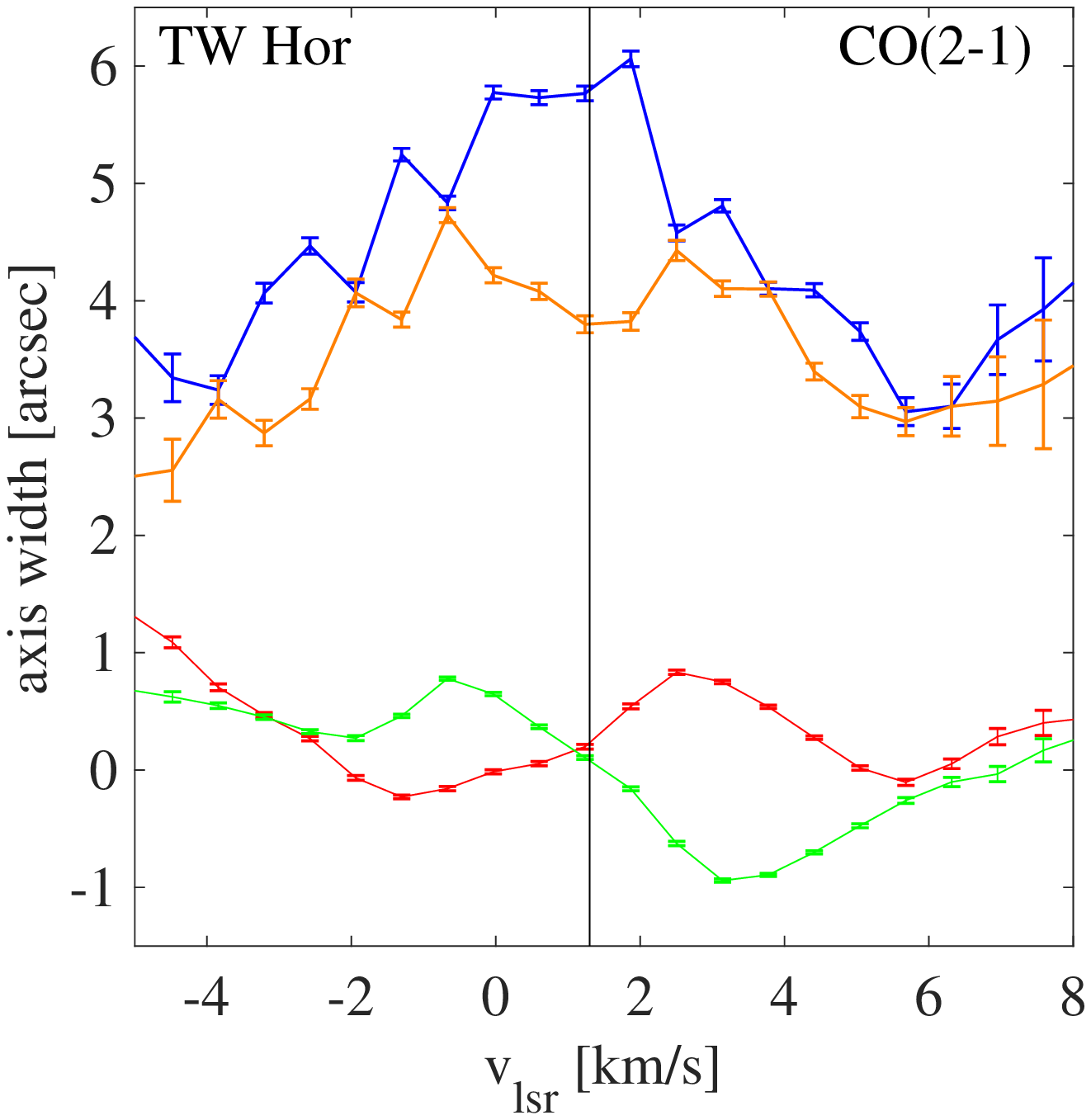}
\hspace{0.01cm}
\includegraphics[height=4.5cm]{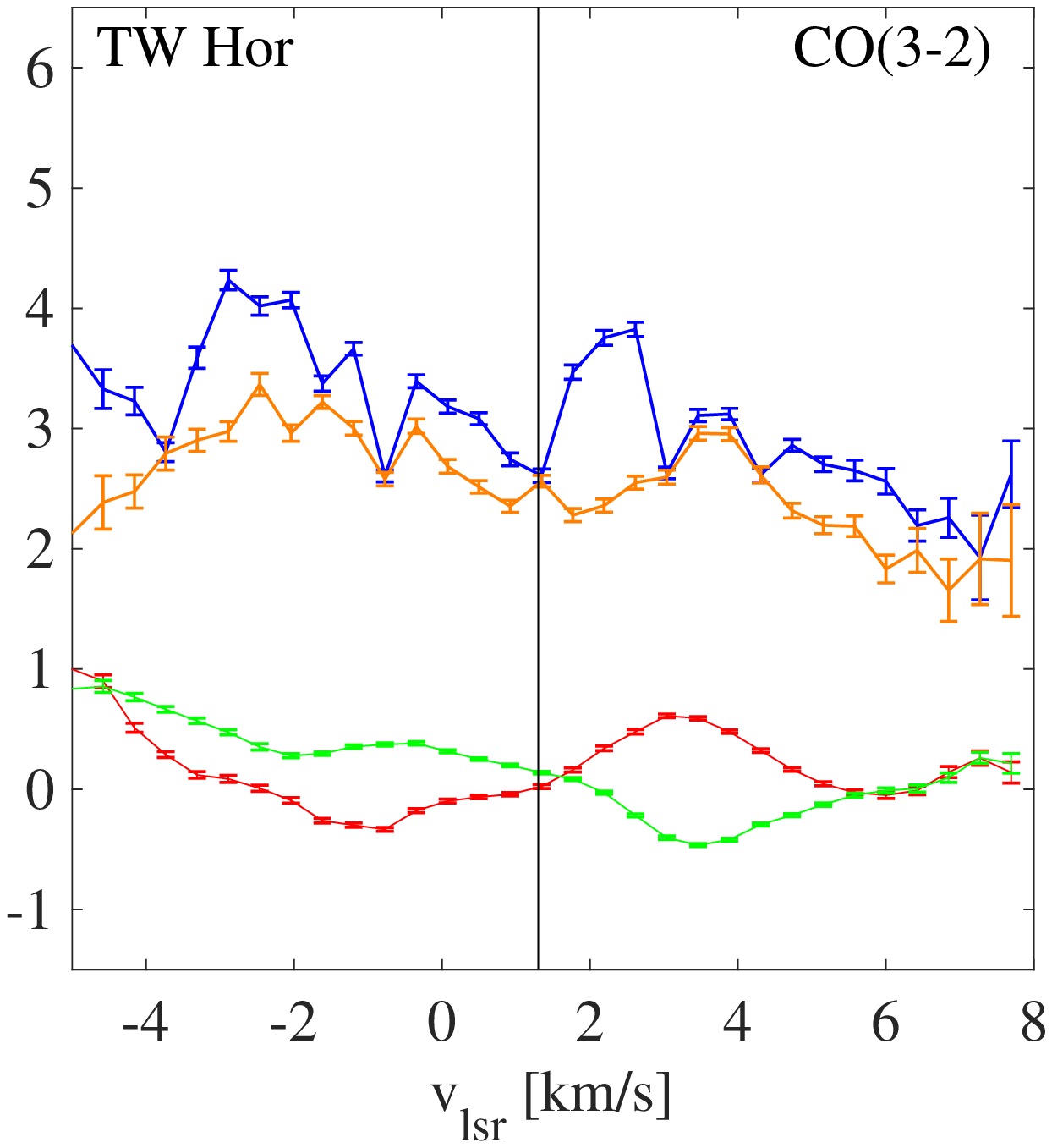}
\hspace{0.2cm}
\includegraphics[height=4.5cm]{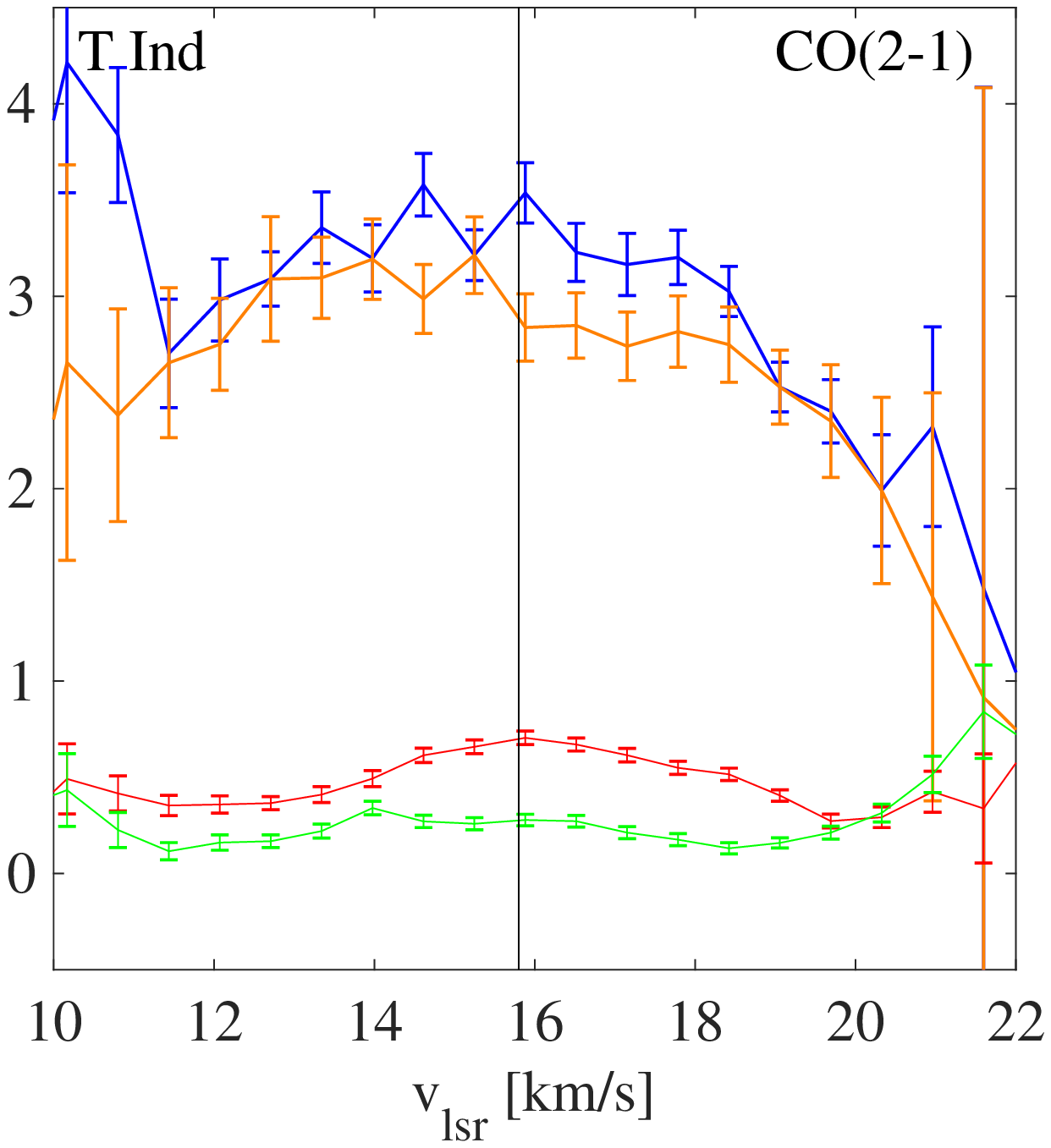}
\hspace{0.03cm}
\includegraphics[height=4.5cm]{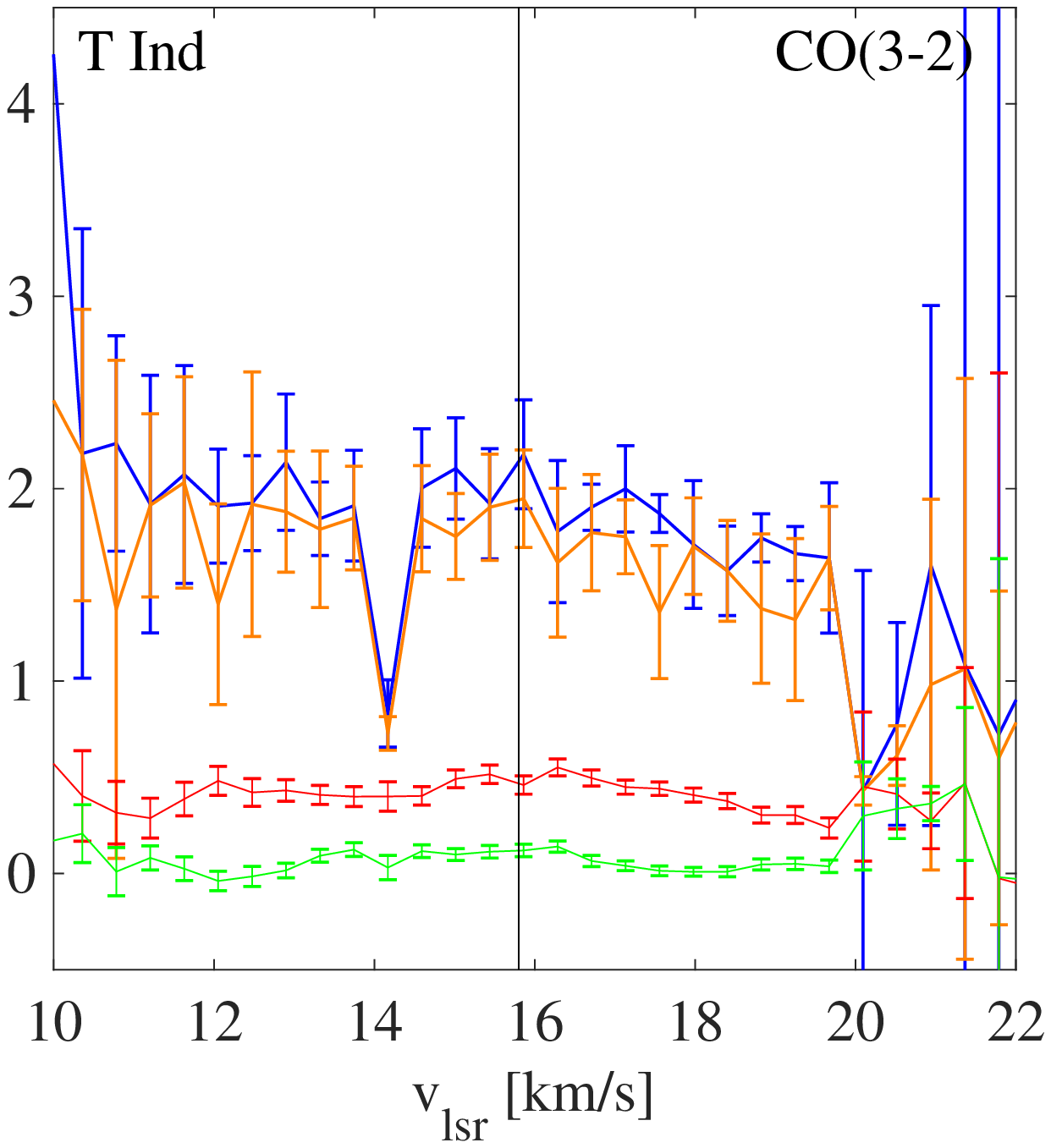}

\includegraphics[height=4.5cm]{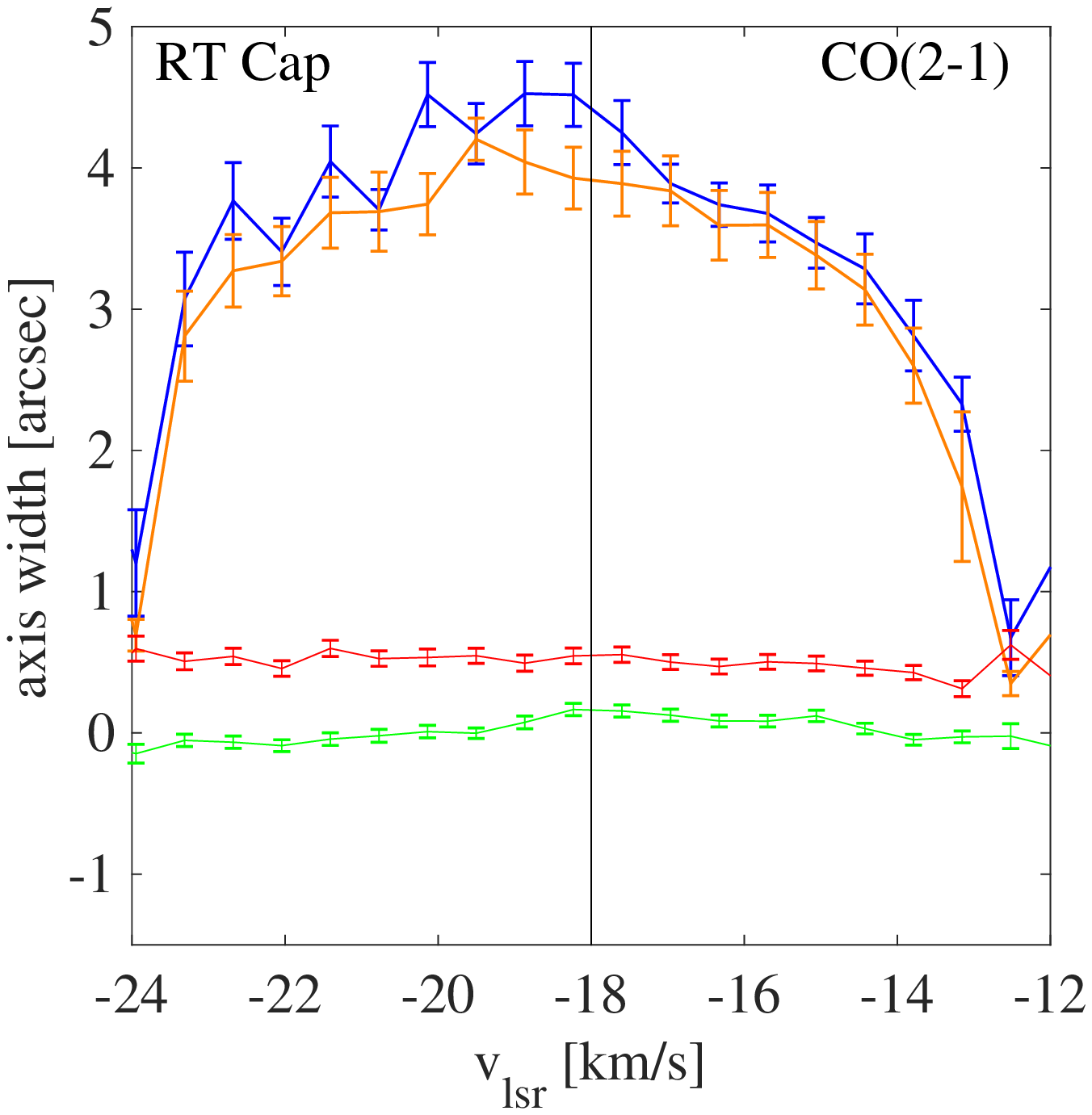}
\hspace{0.04cm}
\includegraphics[height=4.5cm]{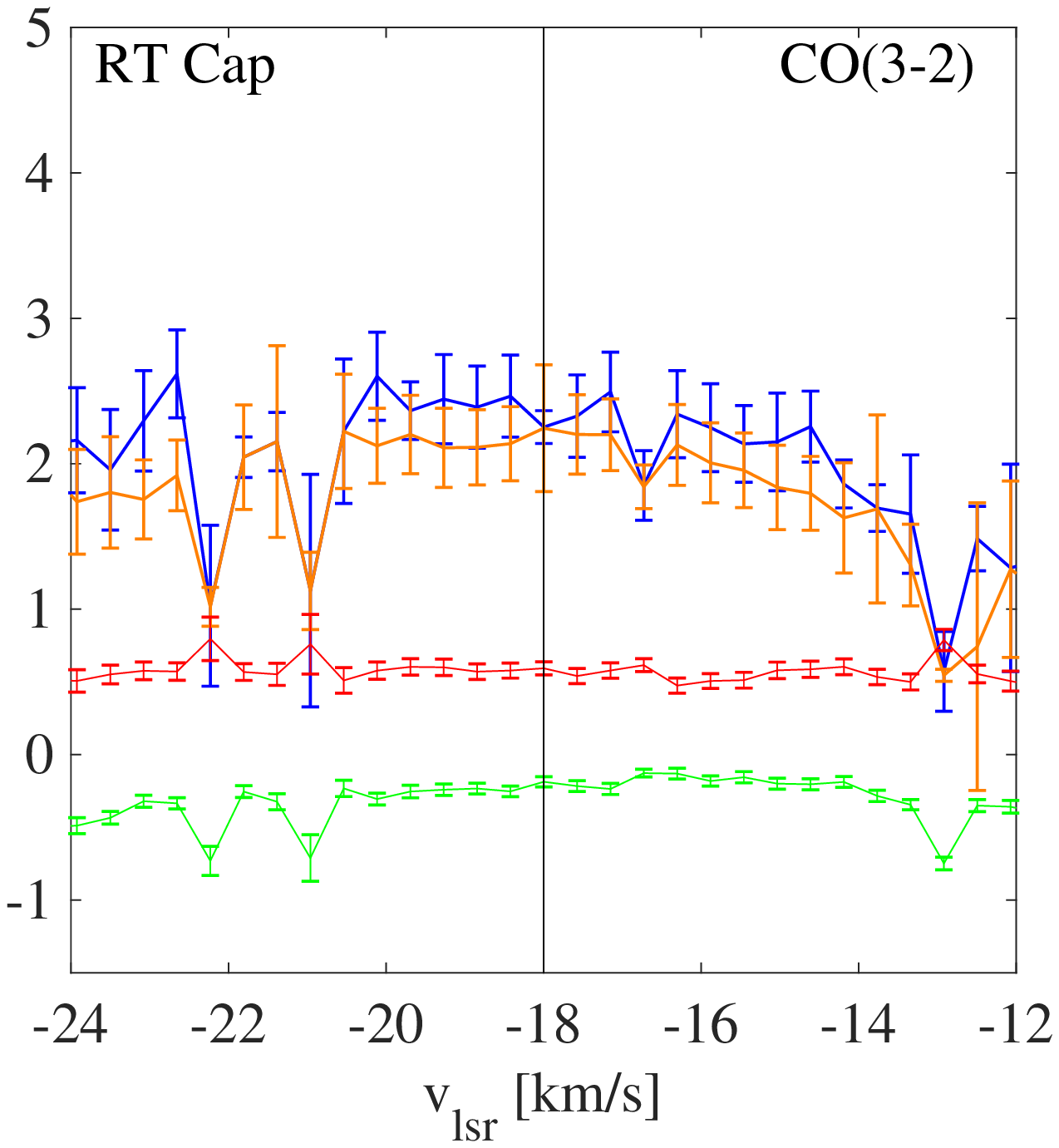}
\hspace{0.15cm}
\includegraphics[height=4.5cm]{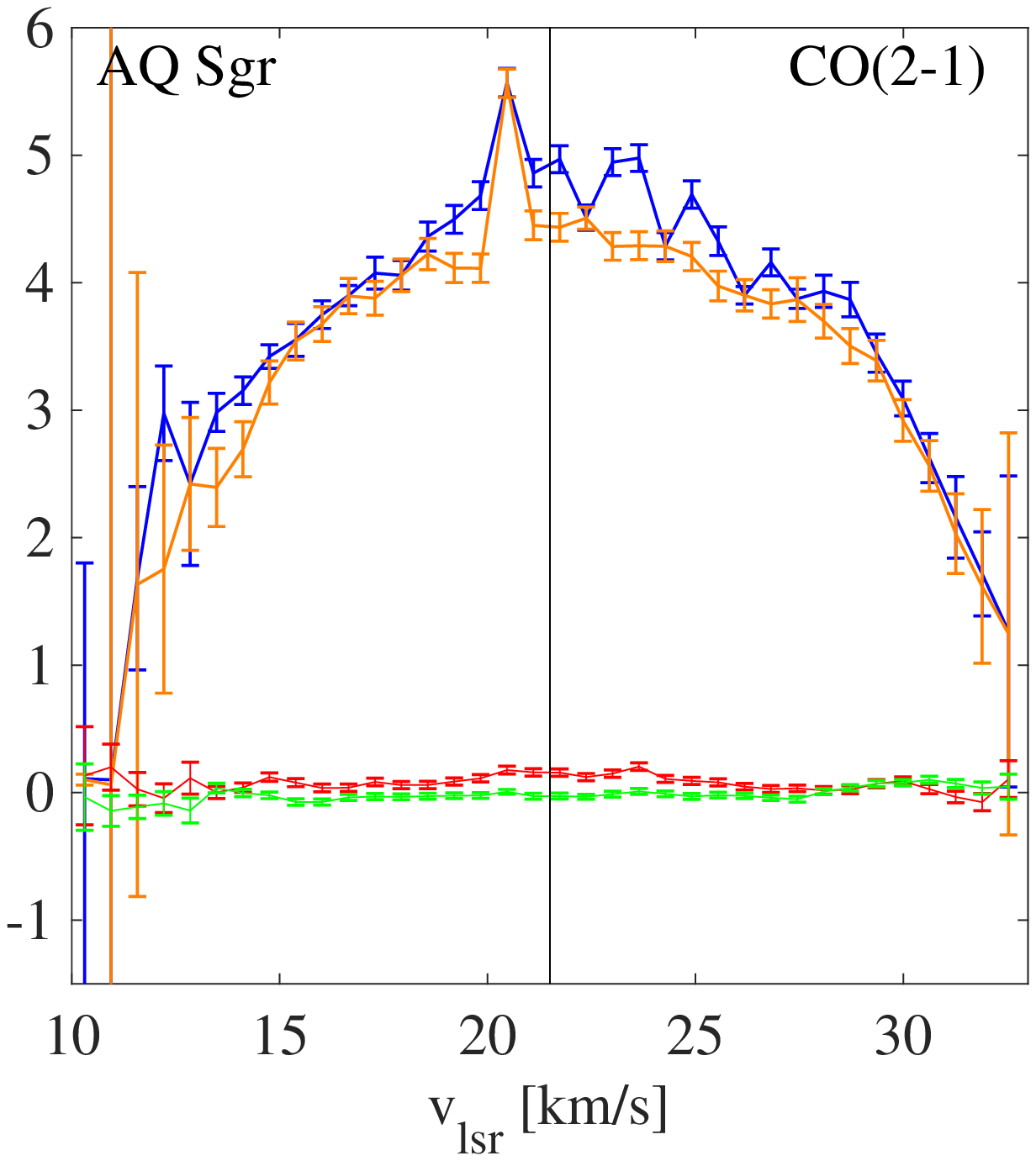}
\hspace{0.04cm}
\includegraphics[height=4.5cm]{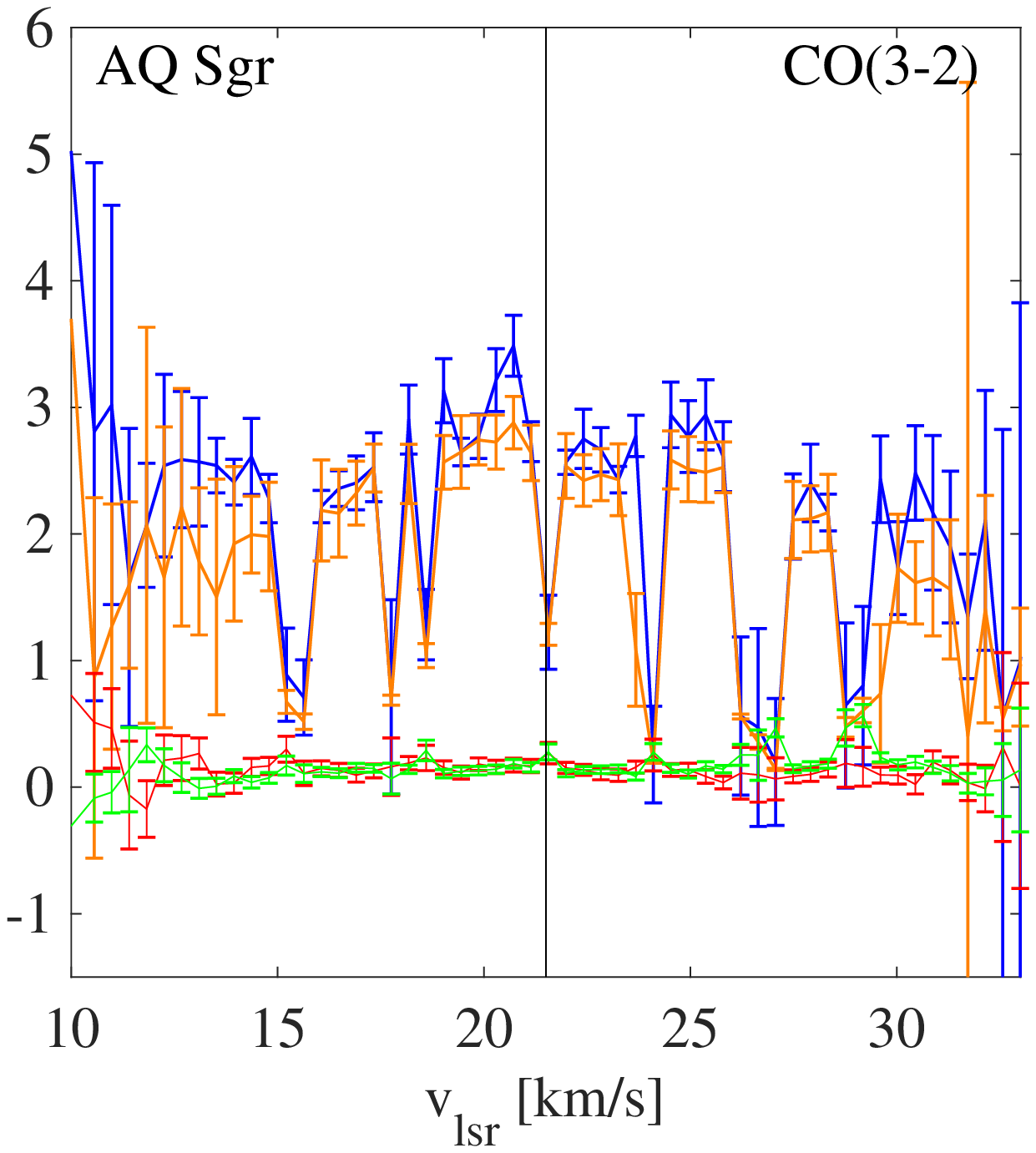}

\includegraphics[height=4.5cm]{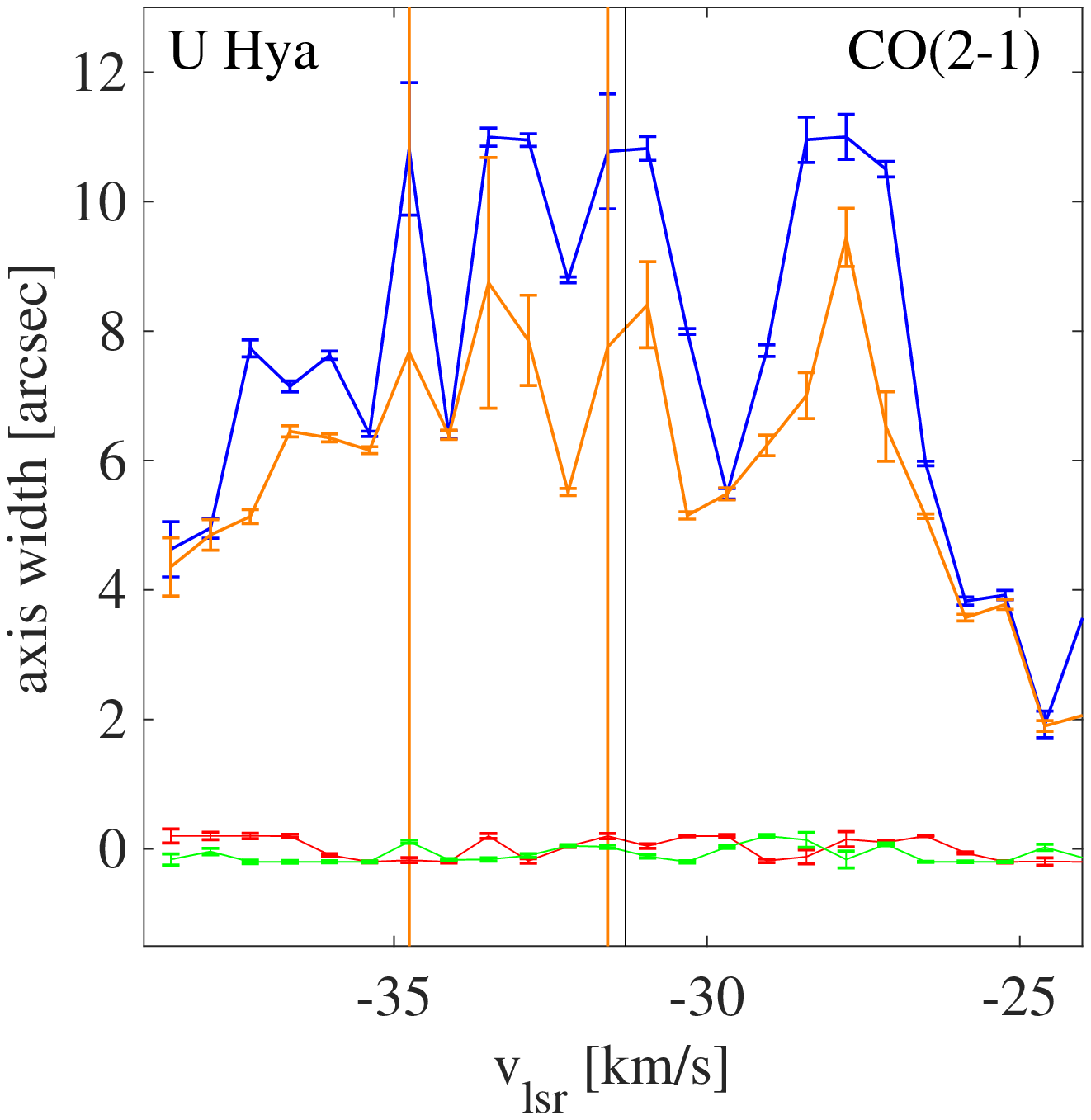}
\hspace{0.02cm}
\includegraphics[height=4.5cm]{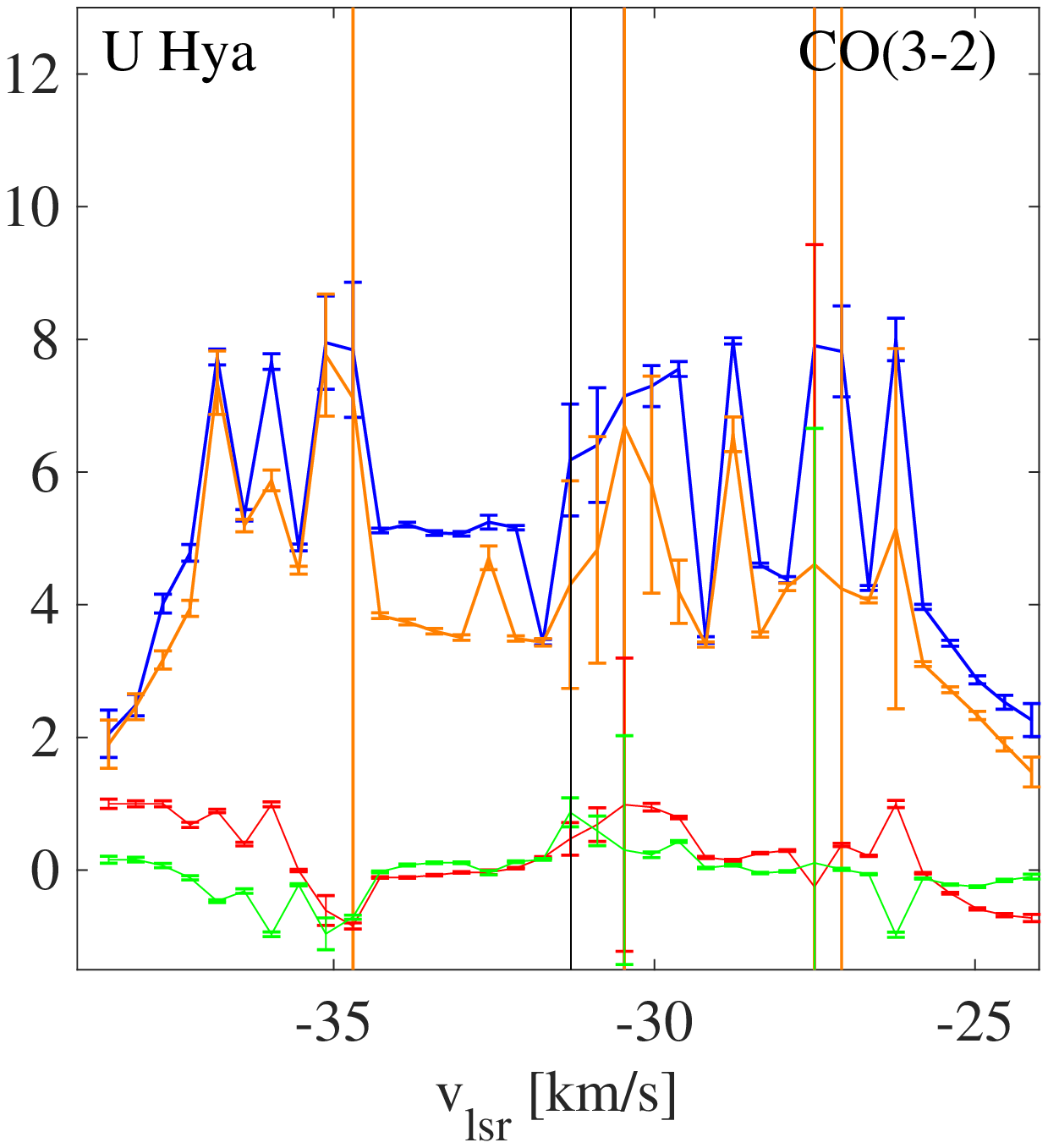}
\hspace{0.2cm}
\includegraphics[height=4.5cm]{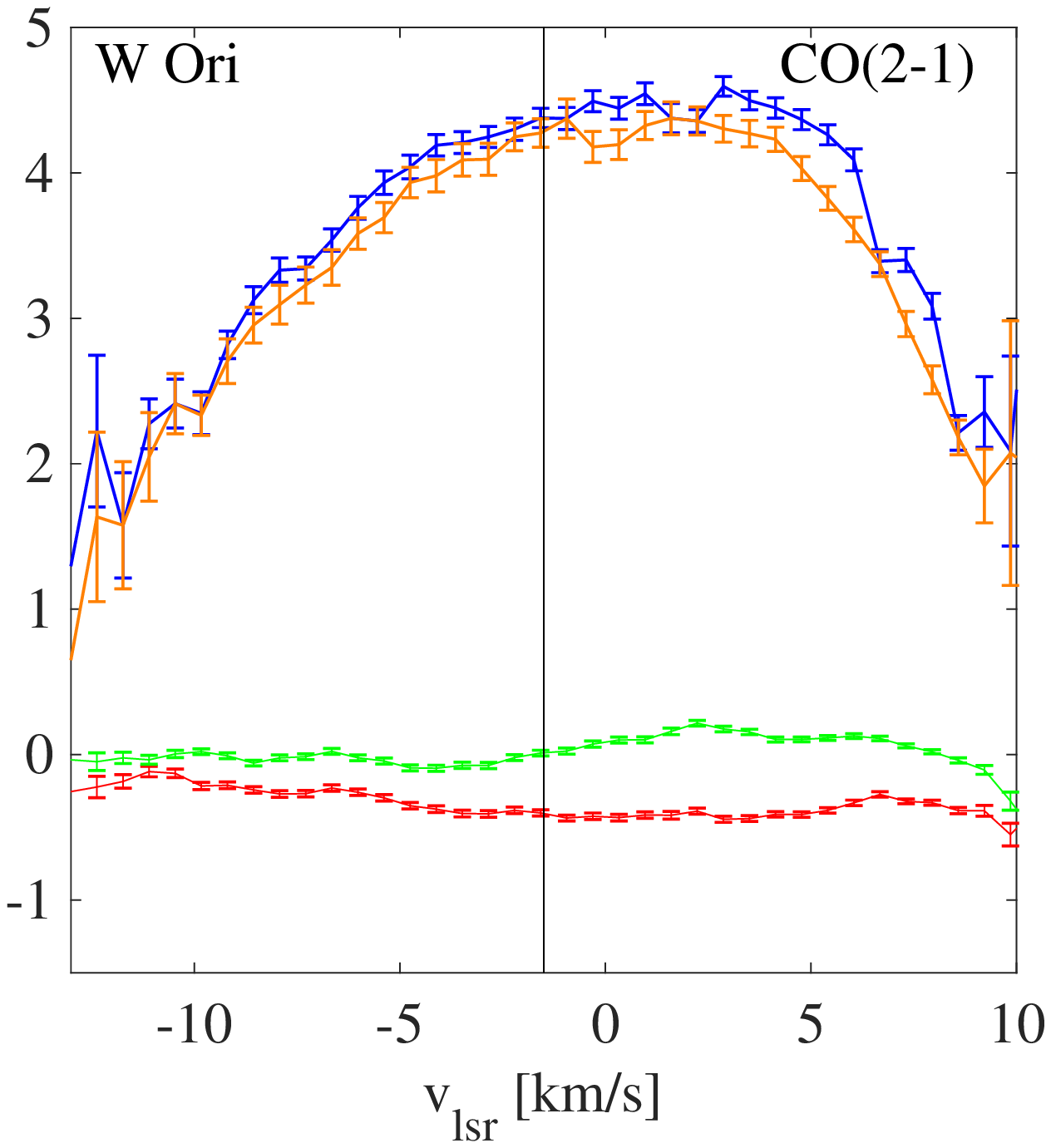}
\hspace{0.02cm}
\includegraphics[height=4.5cm]{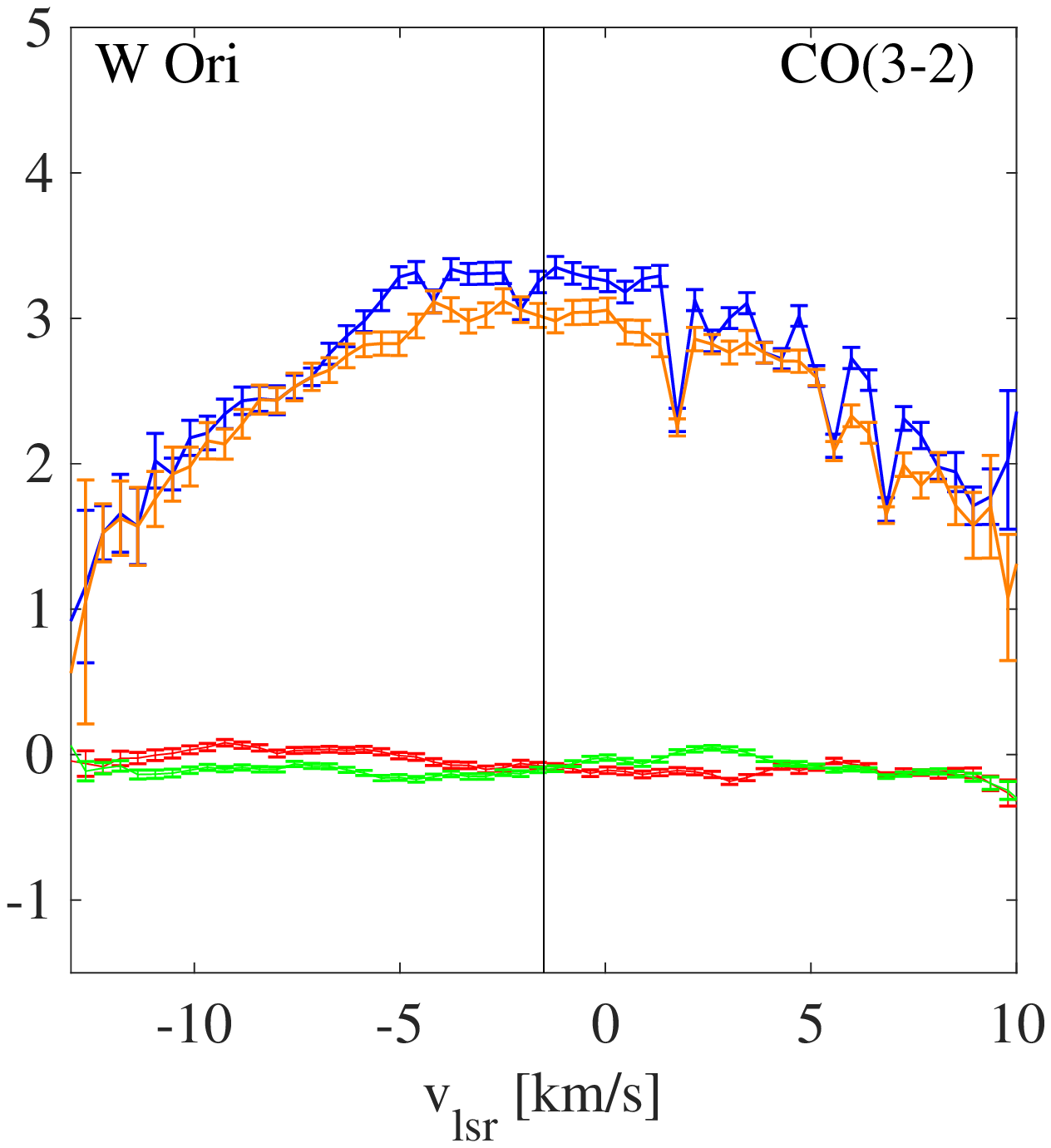}

\includegraphics[height=4.5cm]{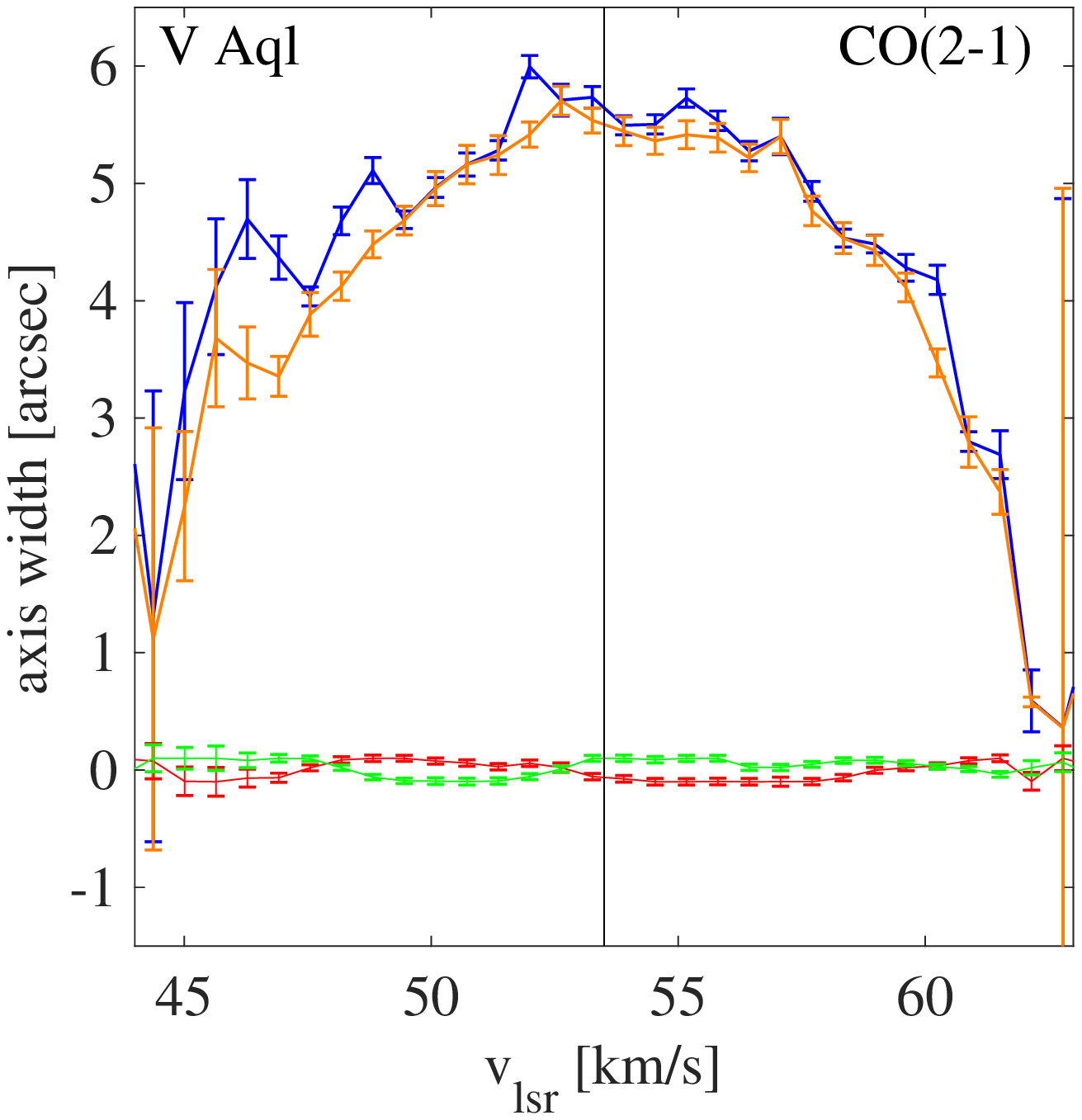}
\hspace{0.05cm}
\includegraphics[height=4.5cm]{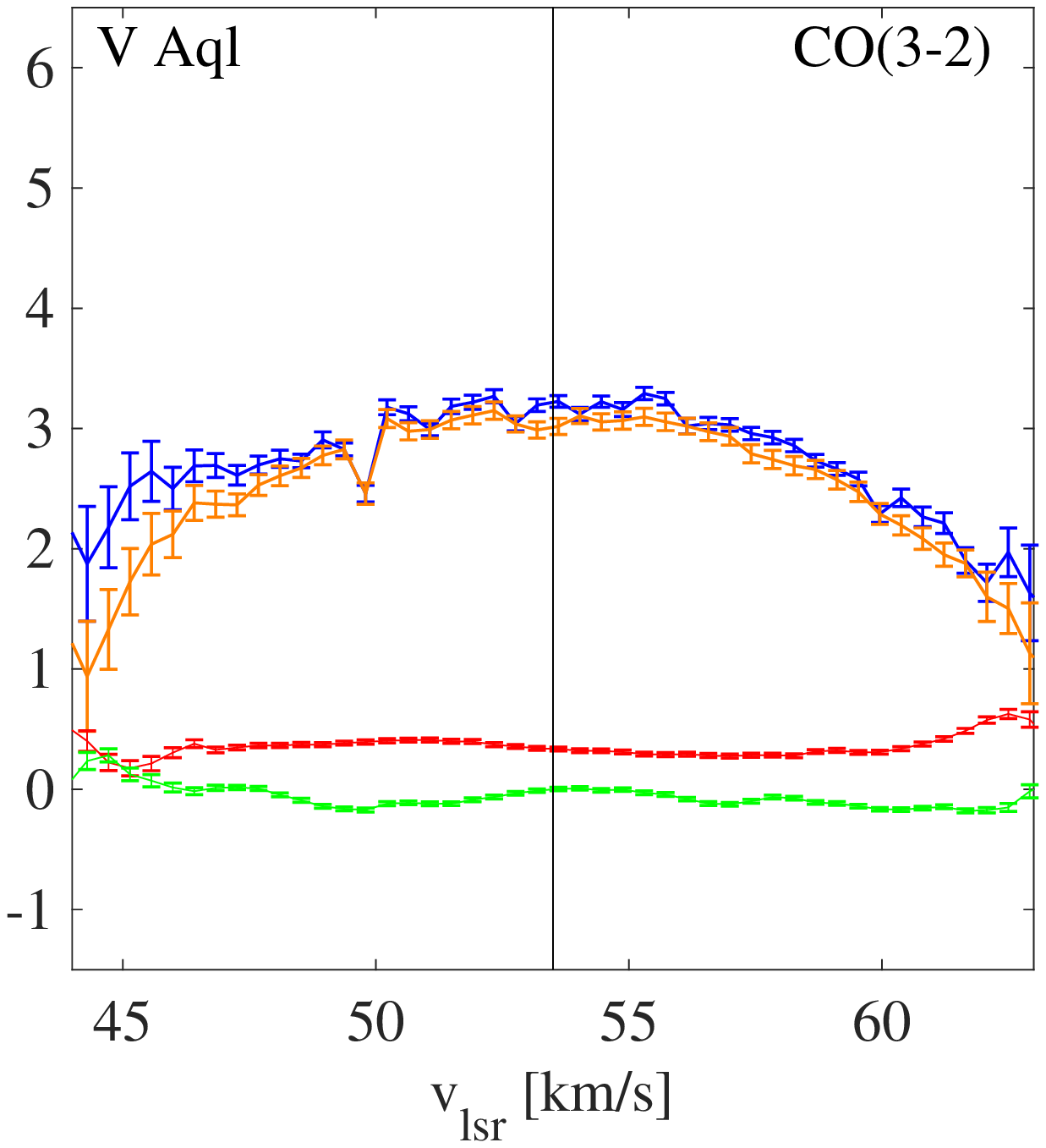}
\hspace{0.2cm}
\includegraphics[height=4.5cm]{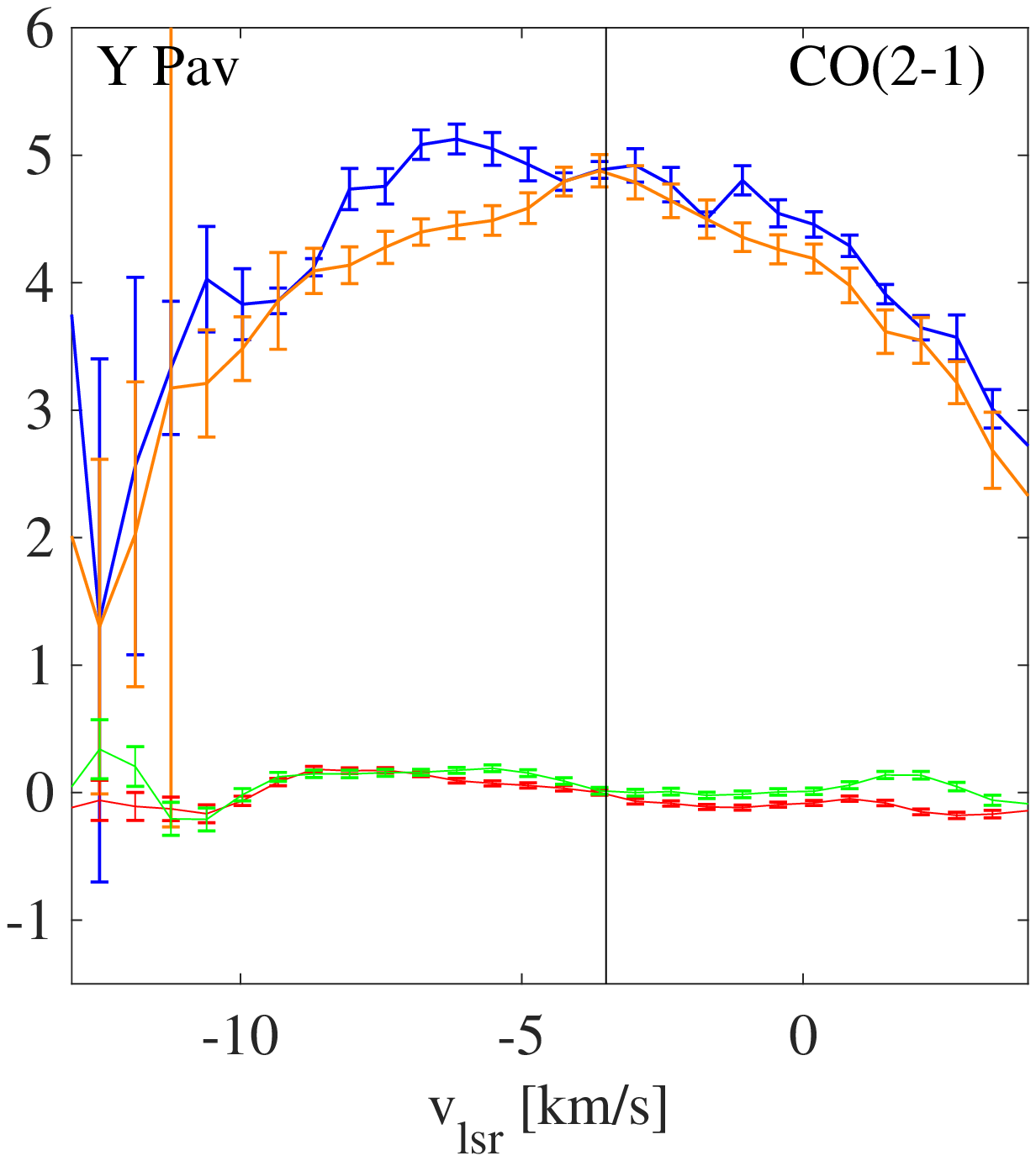}
\hspace{0.05cm}
\includegraphics[height=4.5cm]{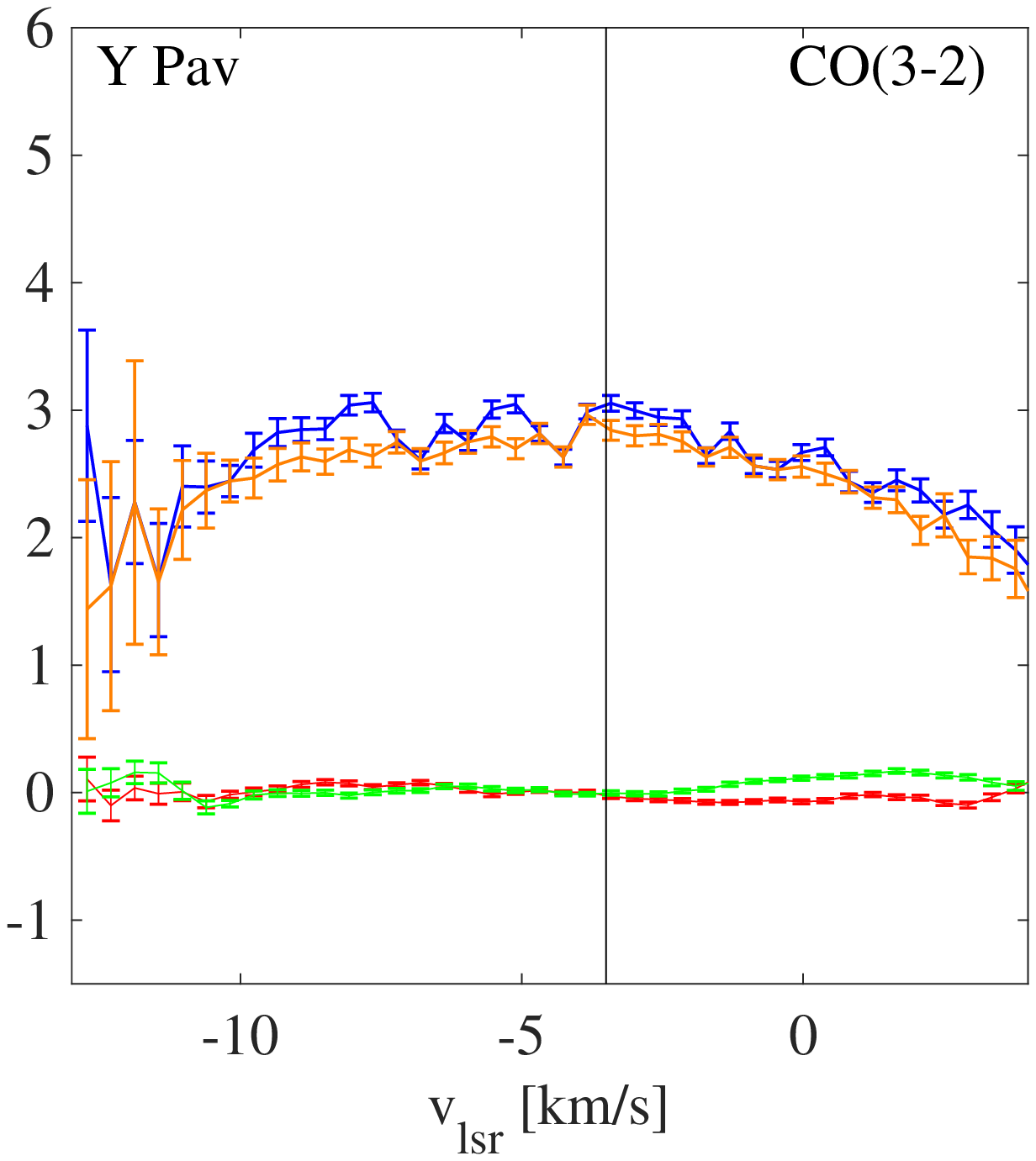}
\caption{Results from the visibility fitting to the data measured toward the C-type AGB stars of the sample discussed in this paper. The source name is given in the upper left corner and the transition is in the upper right corner of each plot. The upper blue and orange lines show the major and minor axis of the best-fit Gaussian in each channel, respectively. The lower red and green lines show the RA and Dec offset relative to the center position, respectively.}
\label{uvC_SR}
\end{figure*}


\begin{figure*}[t]
\includegraphics[height=4.5cm]{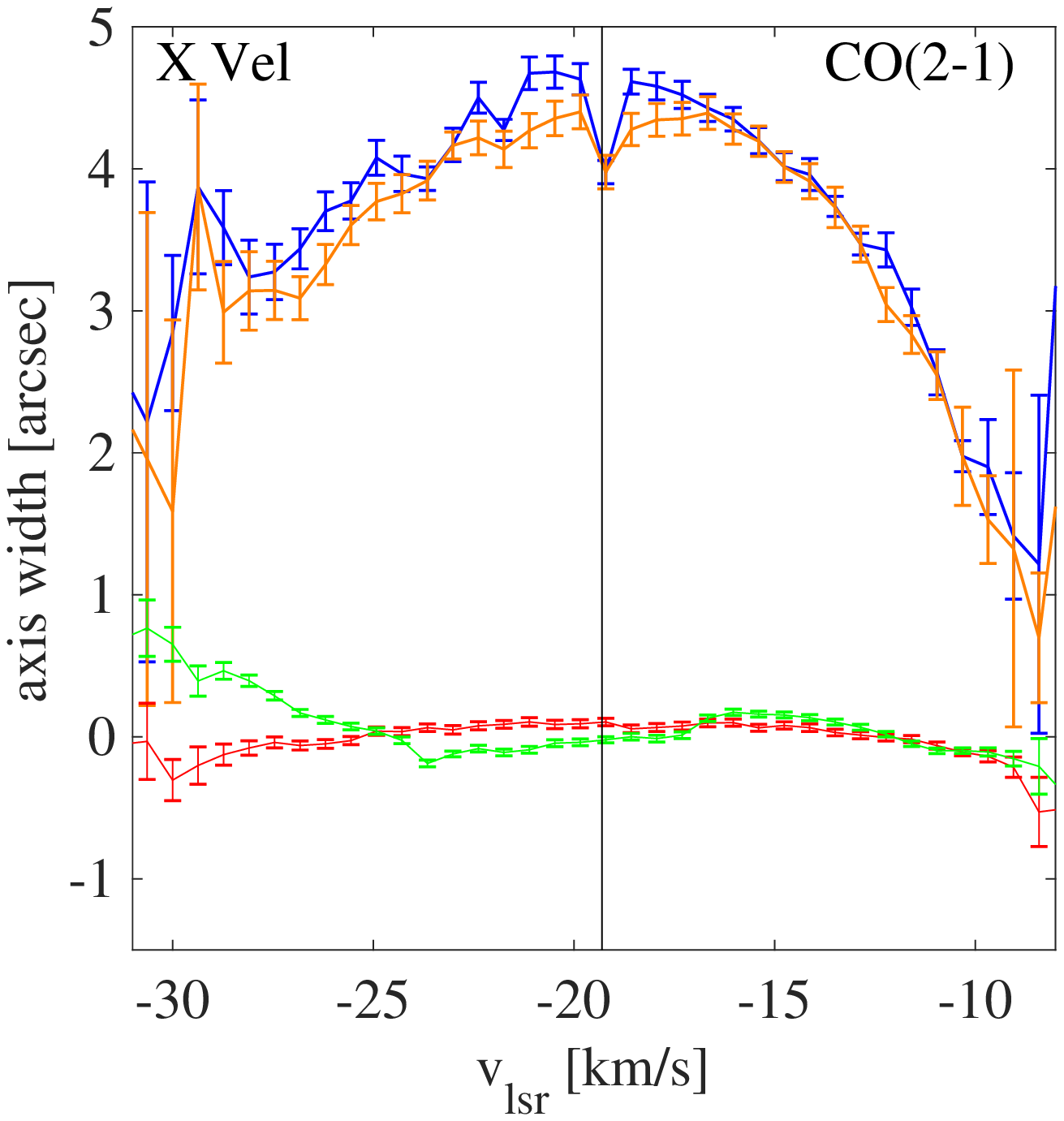}
\hspace{0.05cm}
\includegraphics[height=4.5cm]{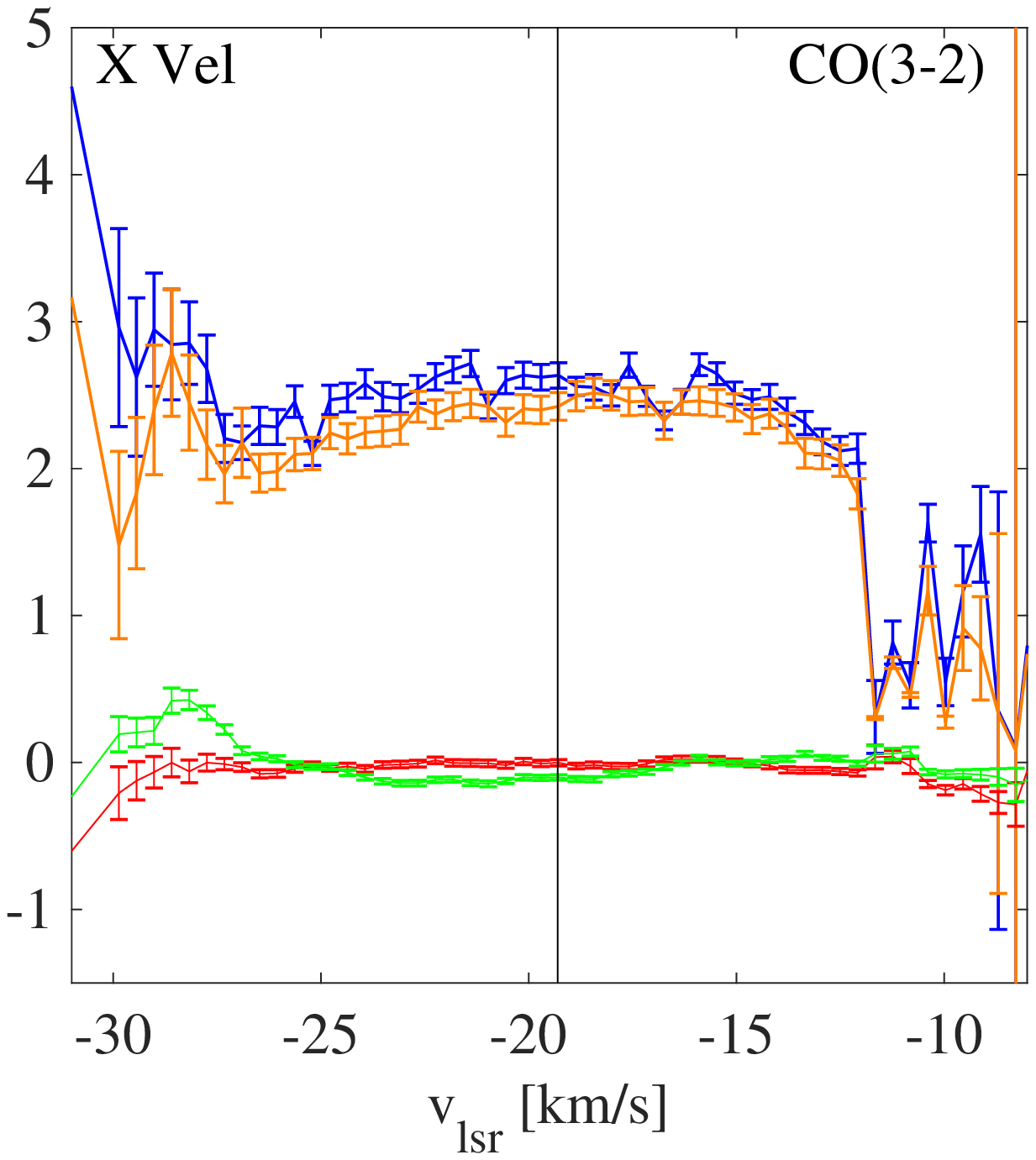}
\hspace{0.2cm}
\includegraphics[height=4.5cm]{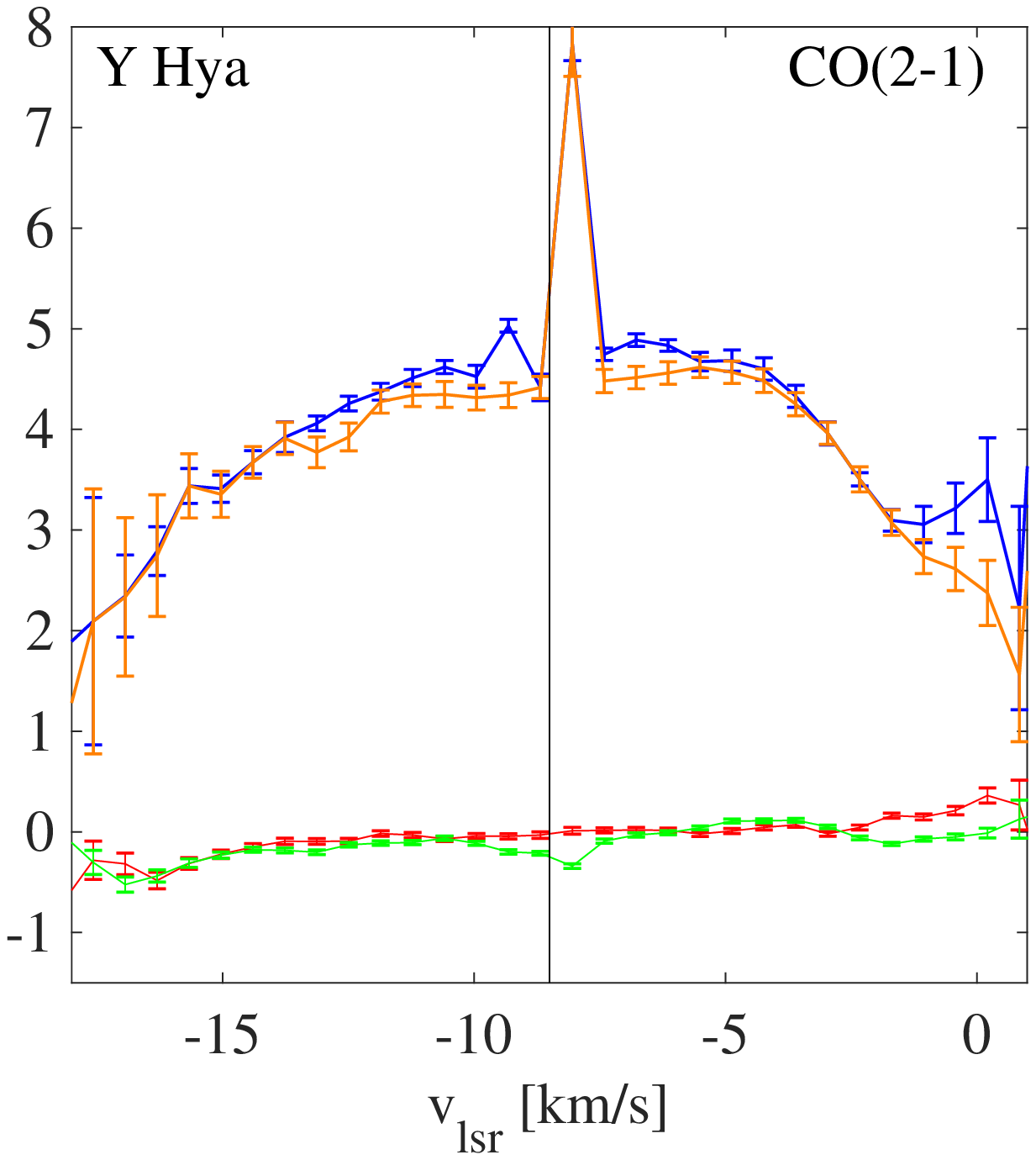}
\hspace{0.05cm}
\includegraphics[height=4.5cm]{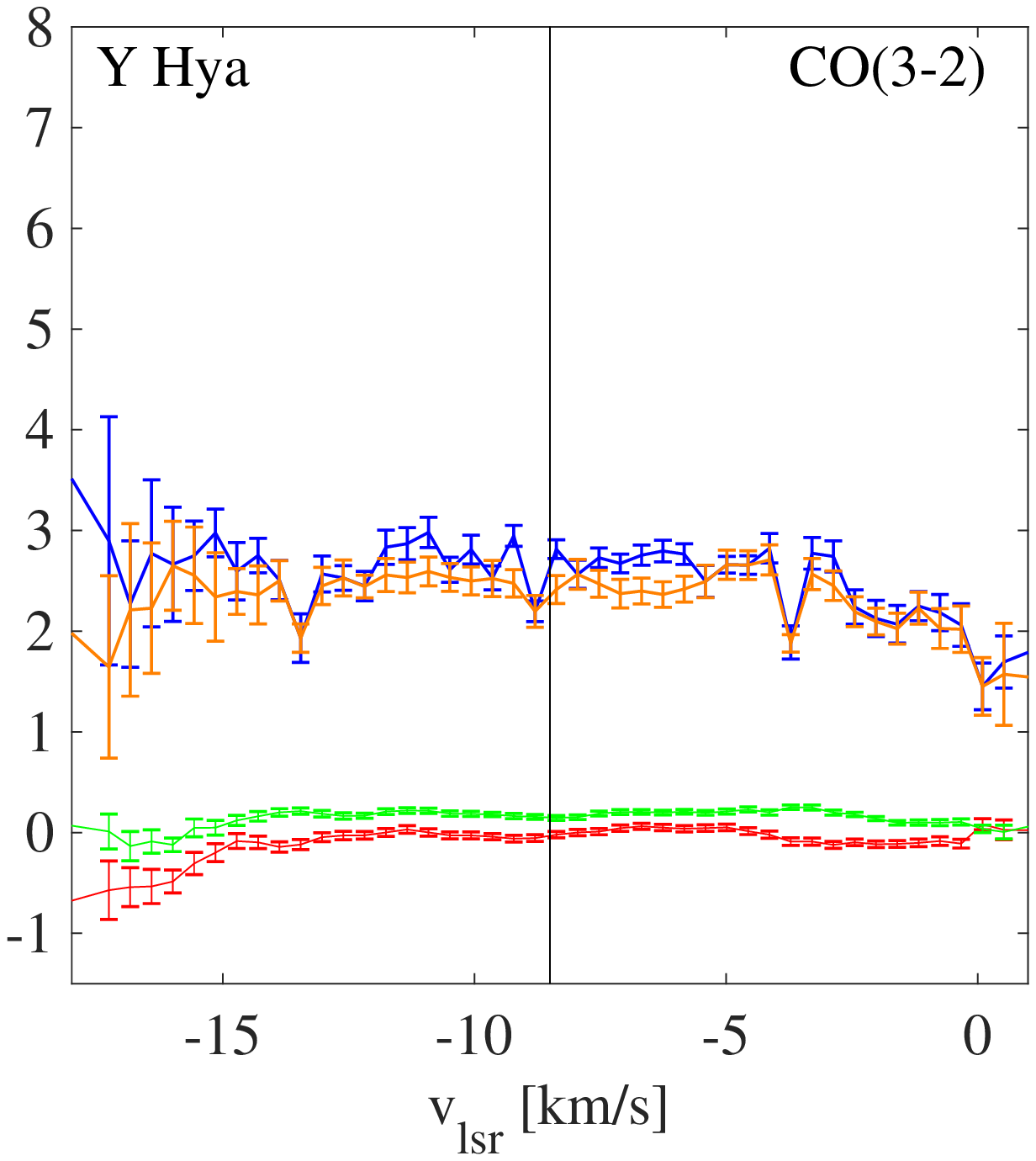}

\includegraphics[height=4.5cm]{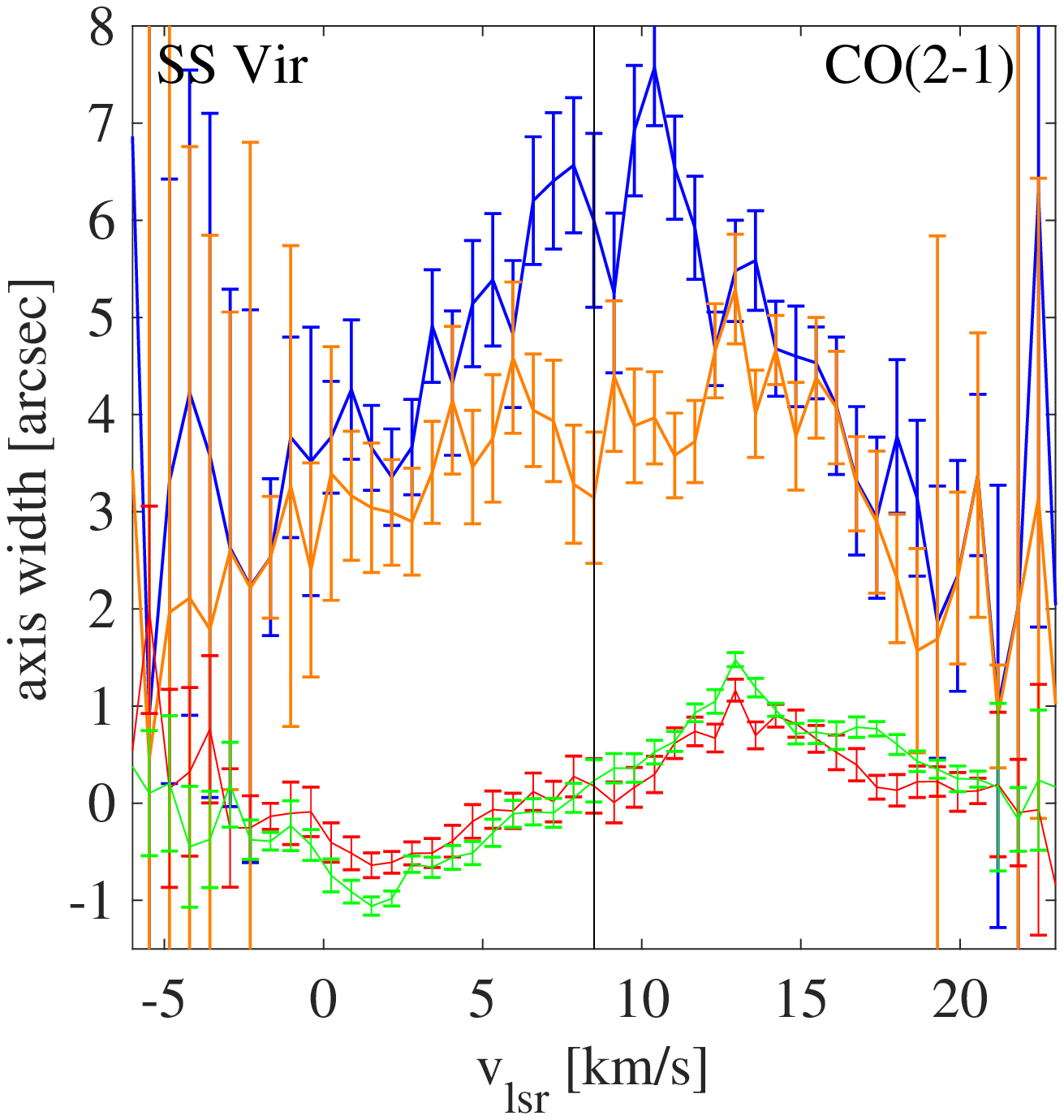}
\hspace{0.05cm}
\includegraphics[height=4.5cm]{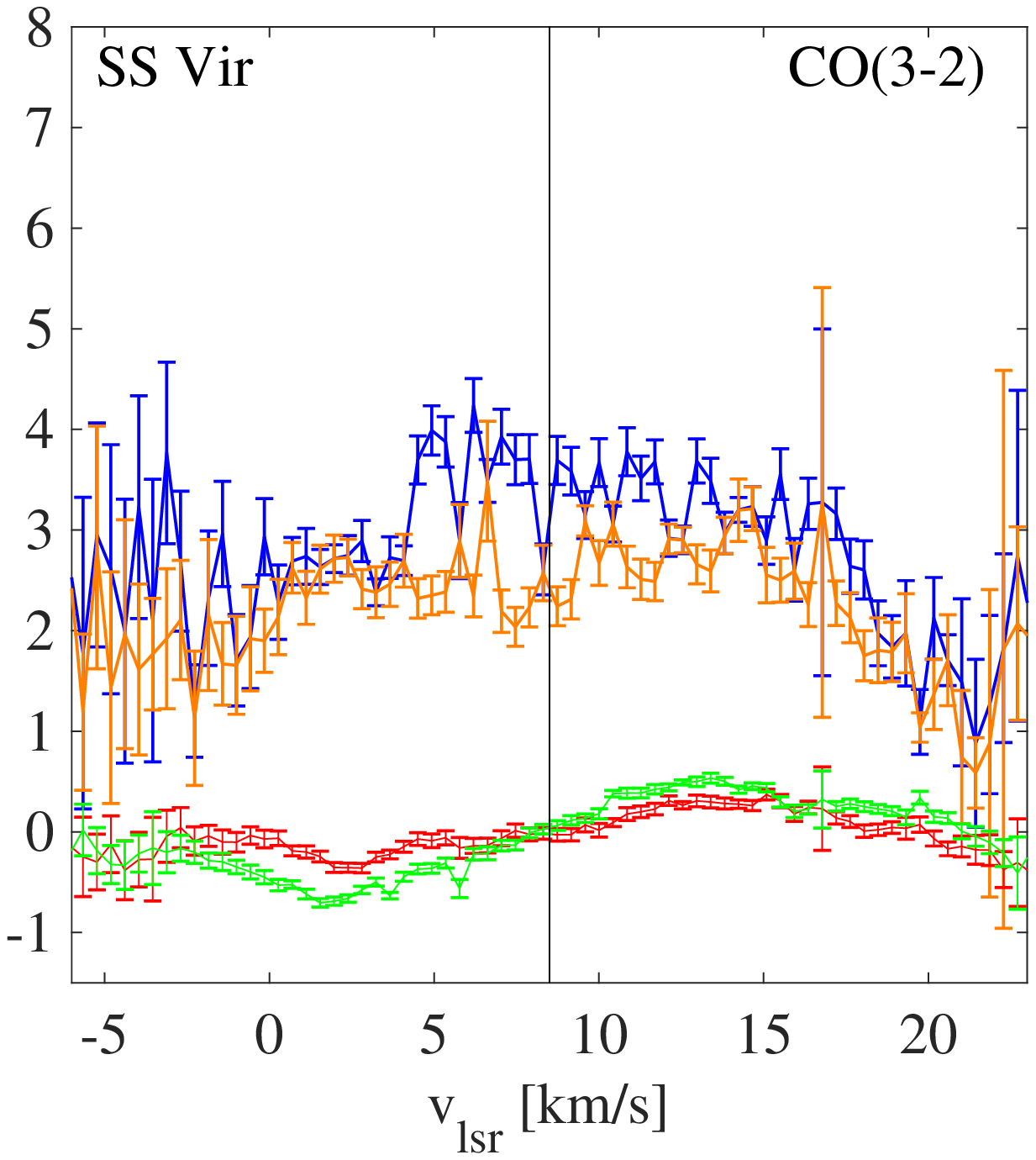}
\hspace{0.2cm}
\includegraphics[height=4.5cm]{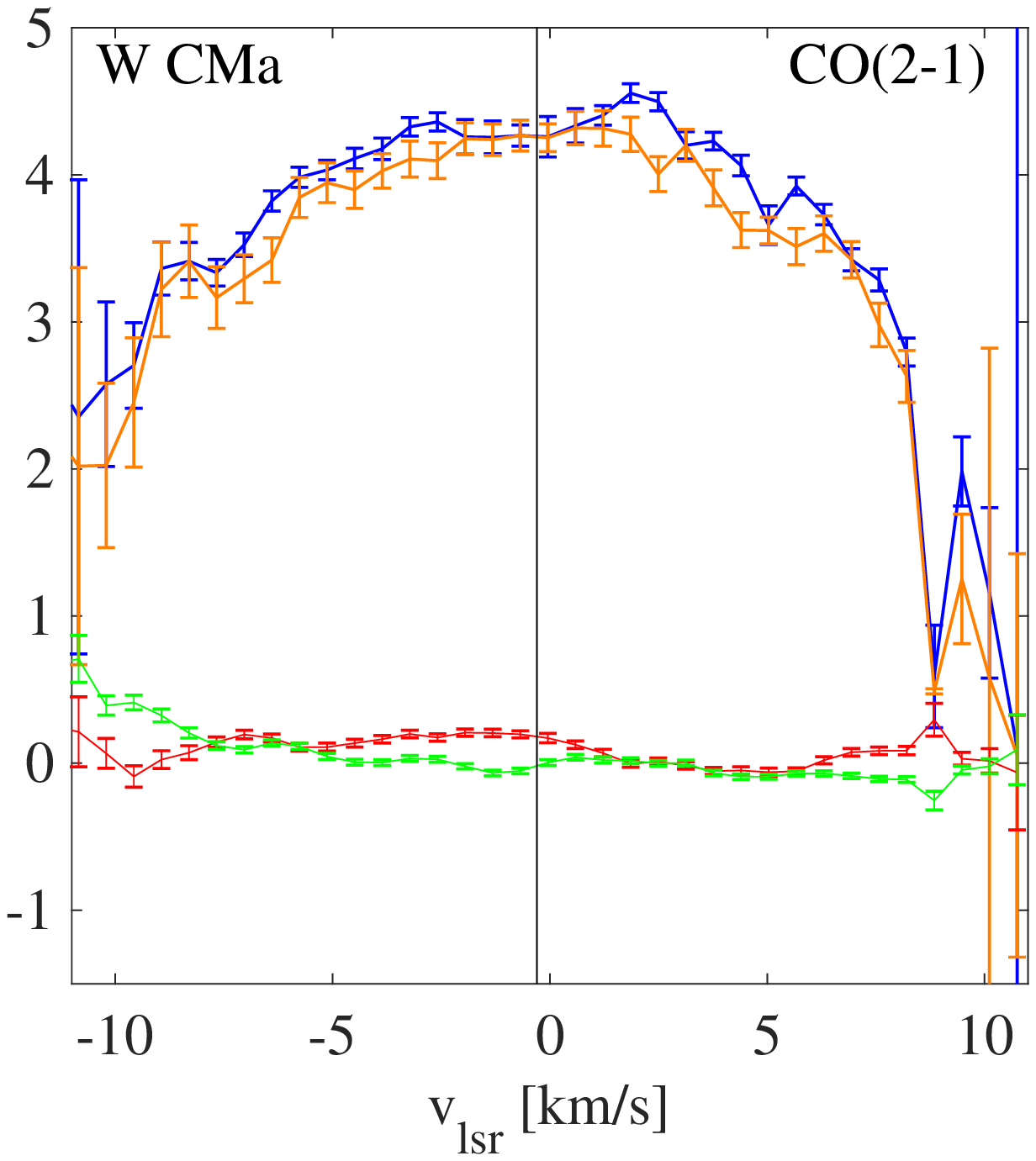}
\hspace{0.05cm}
\includegraphics[height=4.5cm]{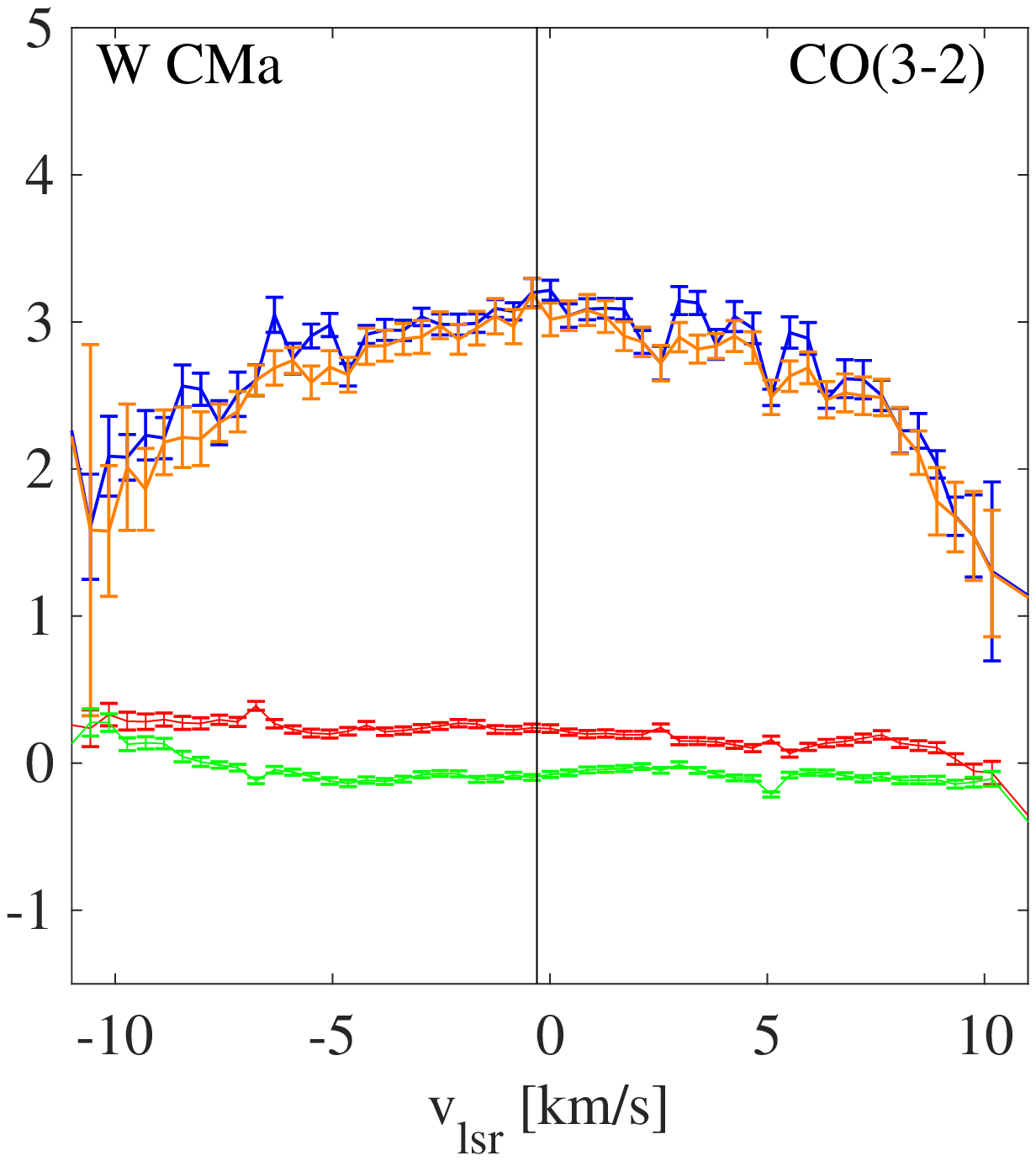}

\includegraphics[height=4.5cm]{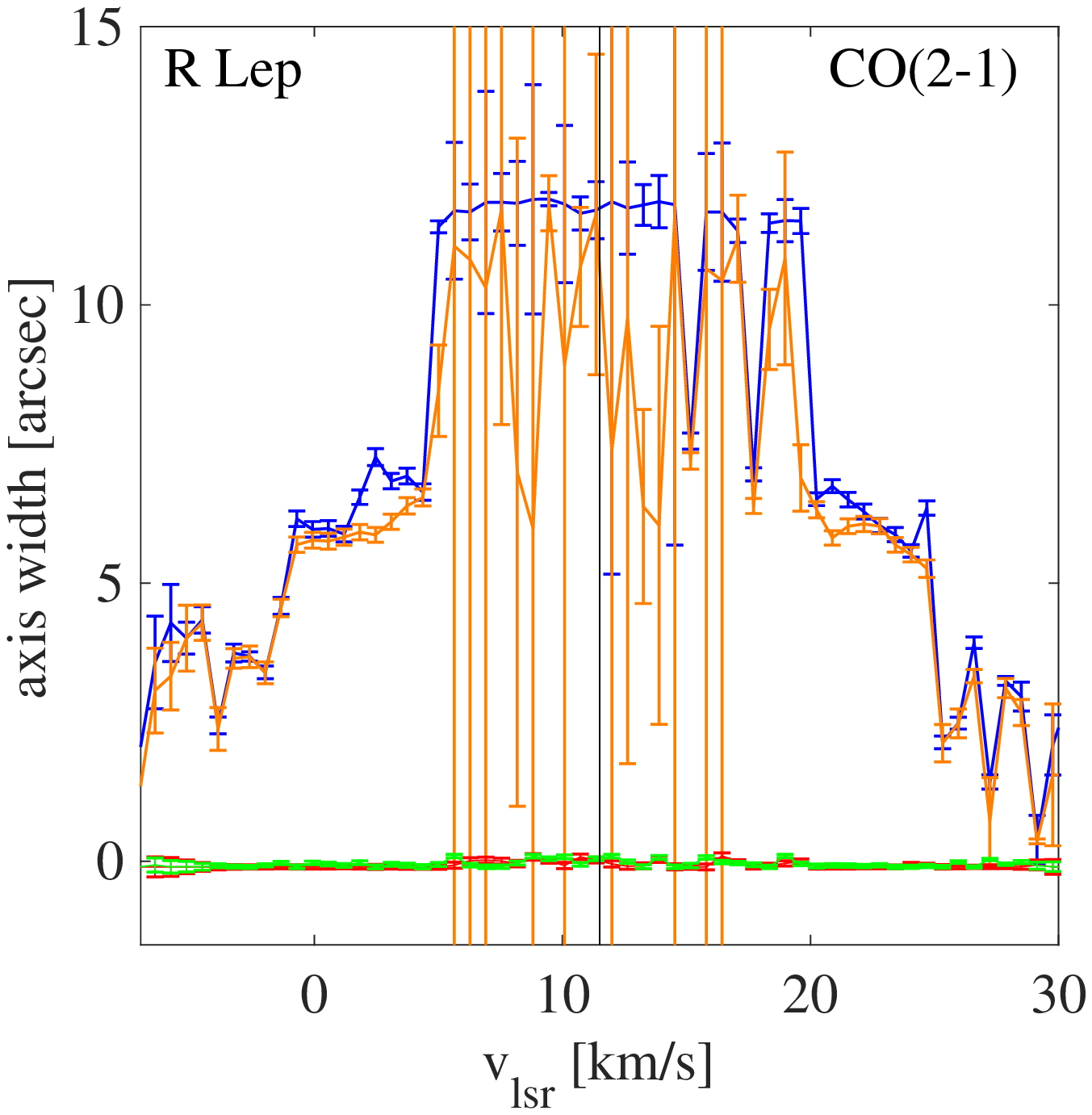}
\includegraphics[height=4.5cm]{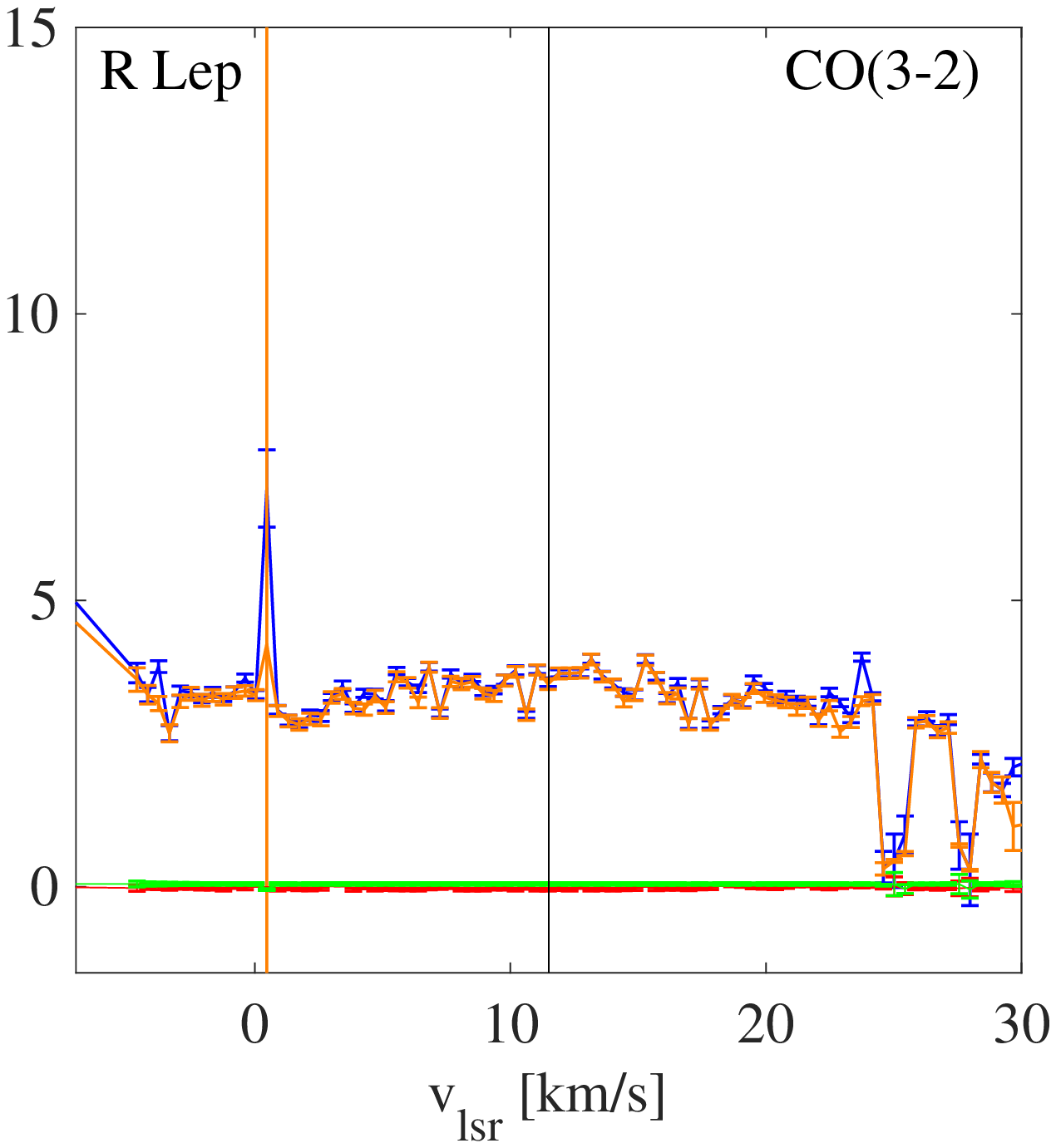}
\hspace{0.1cm}
\includegraphics[height=4.5cm]{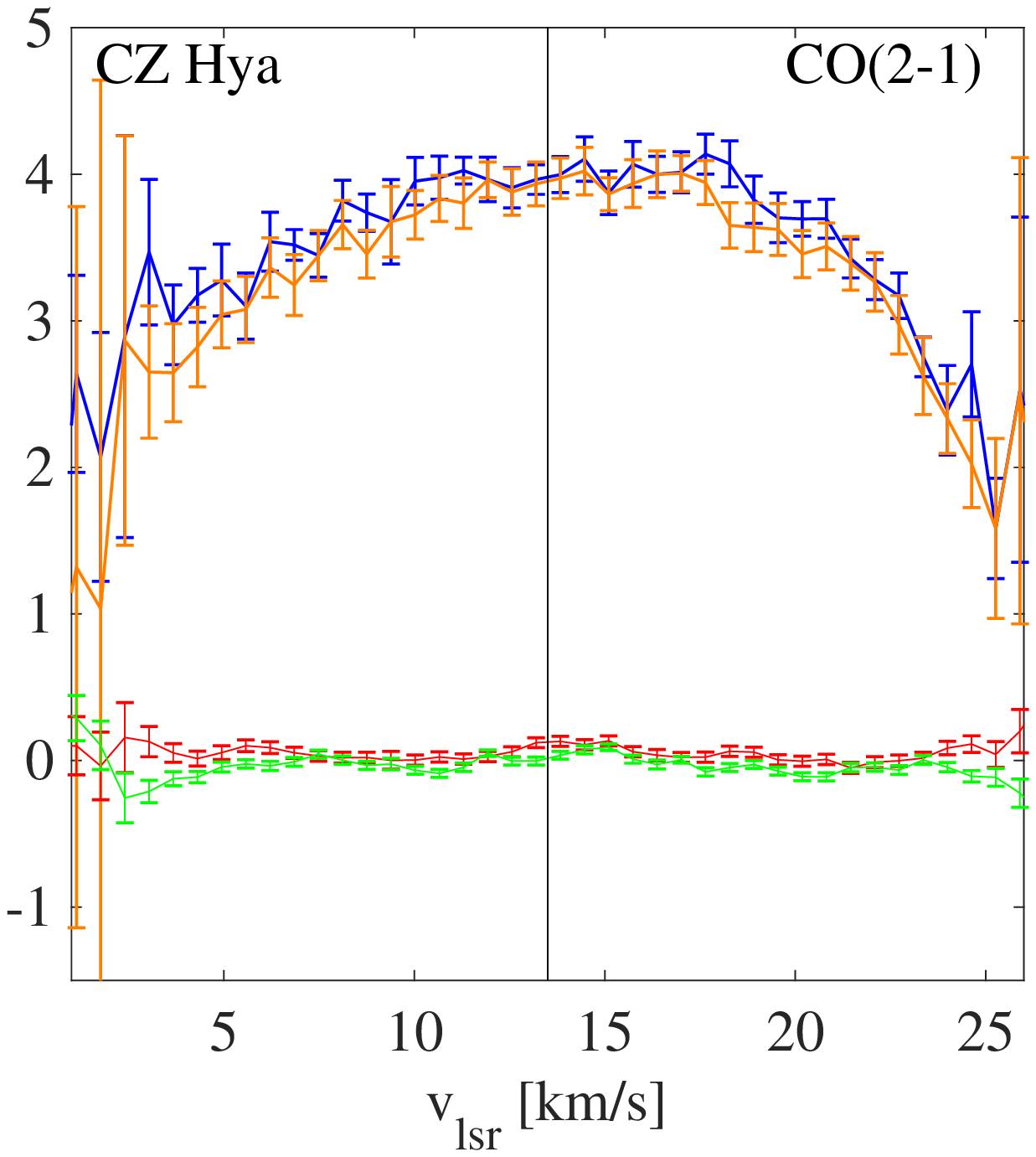}
\hspace{0.02cm}
\includegraphics[height=4.5cm]{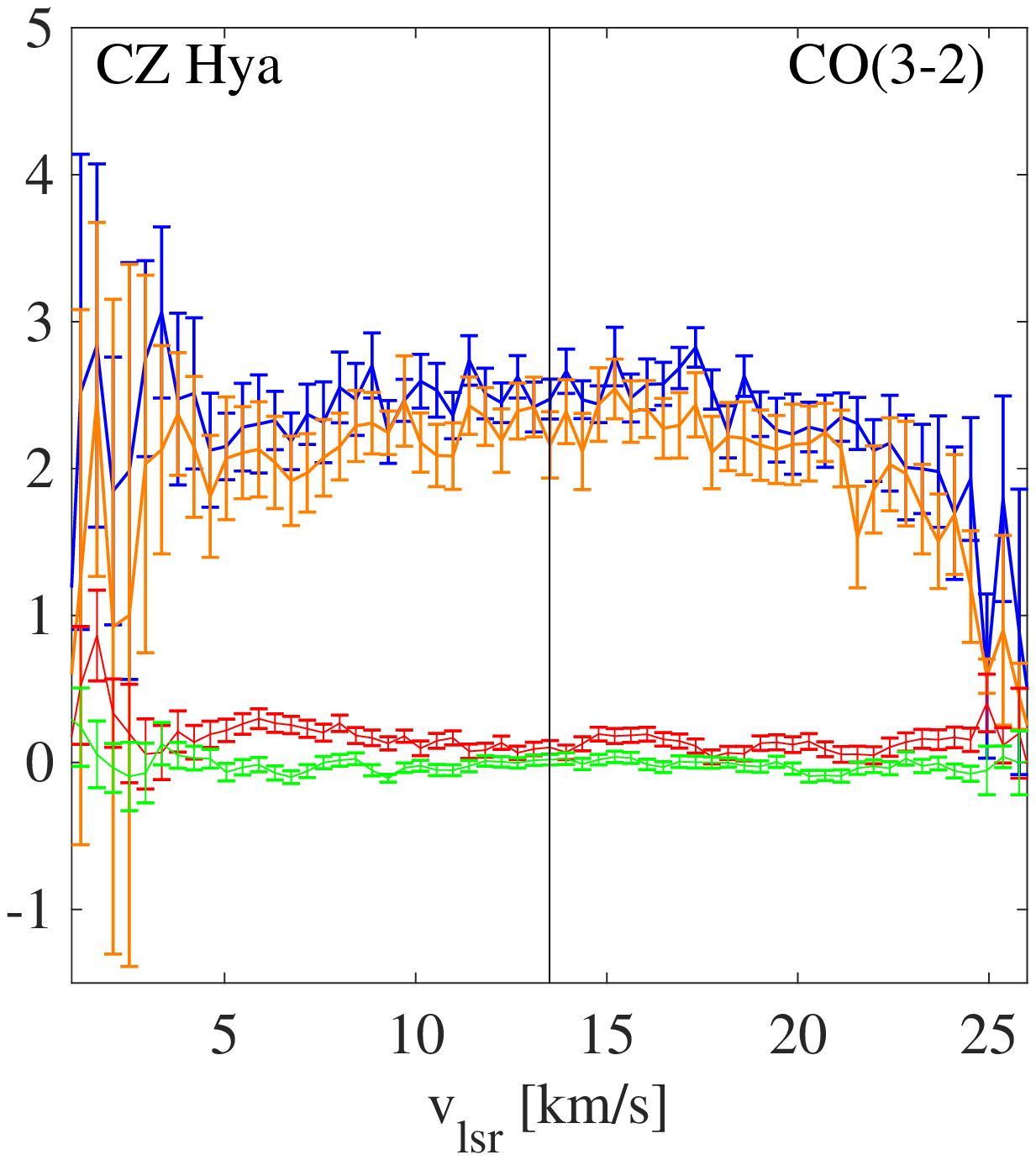}

\includegraphics[height=4.5cm]{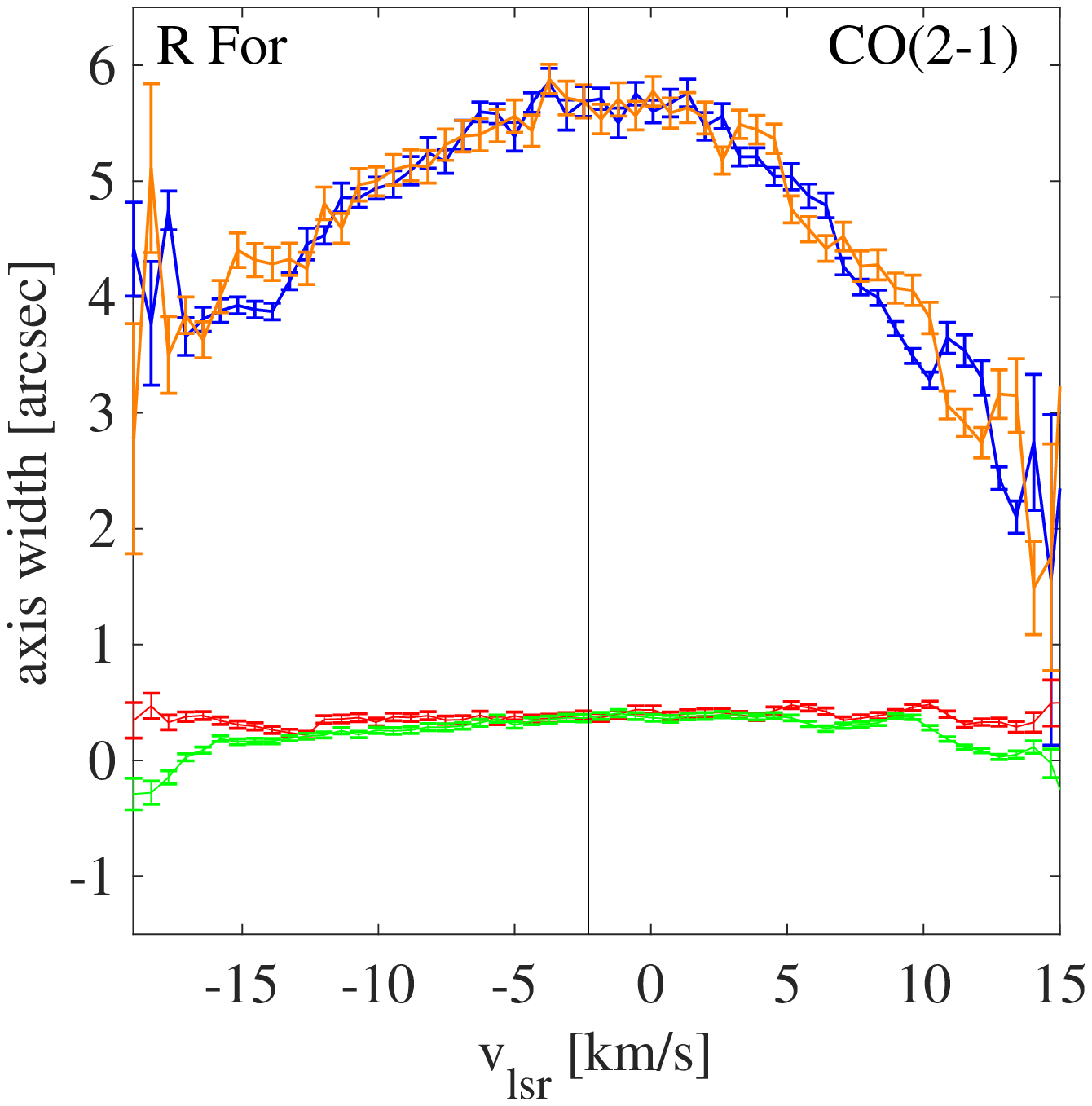}
\includegraphics[height=4.5cm]{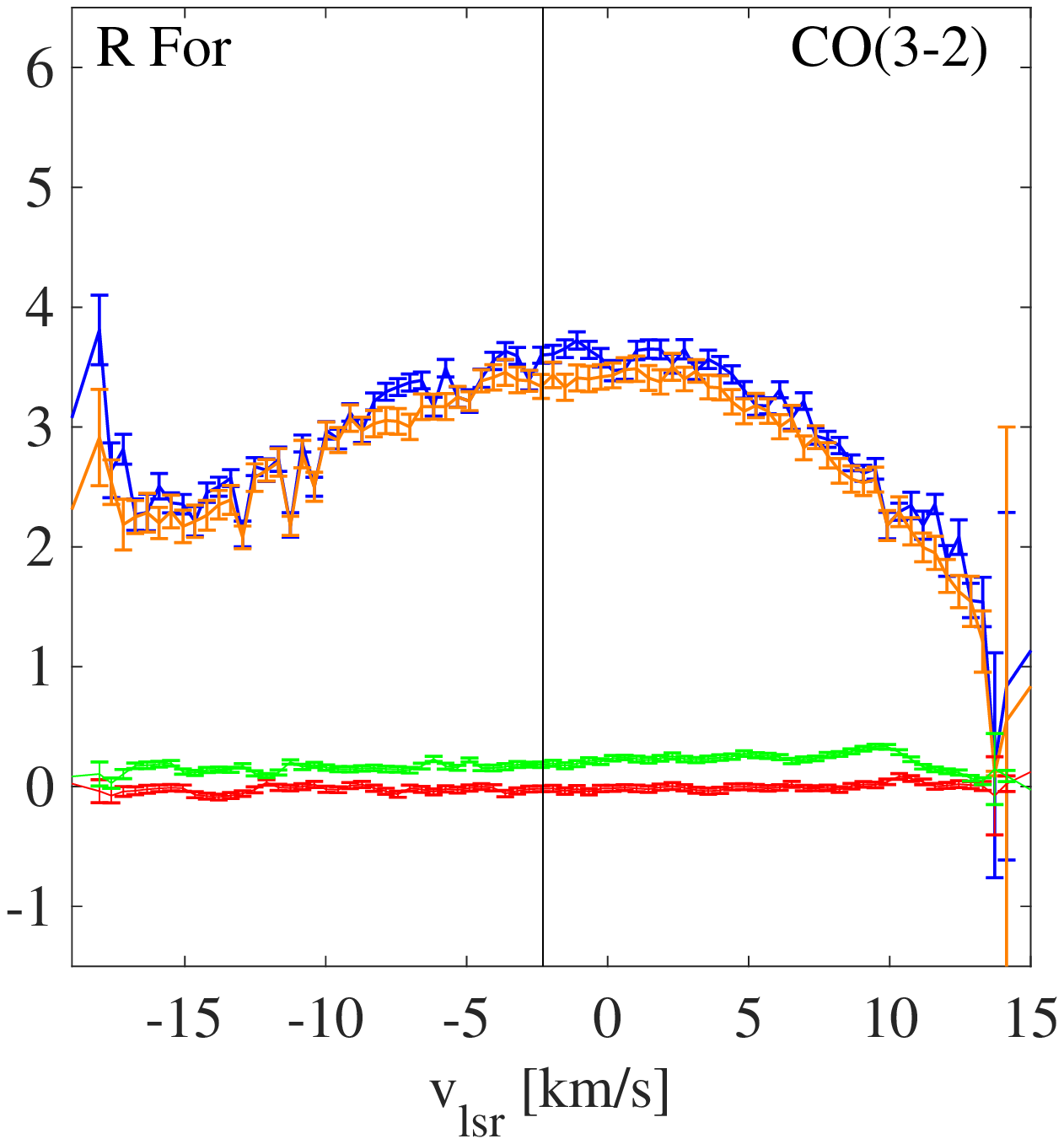}
\hspace{0.05cm}
\includegraphics[height=4.5cm]{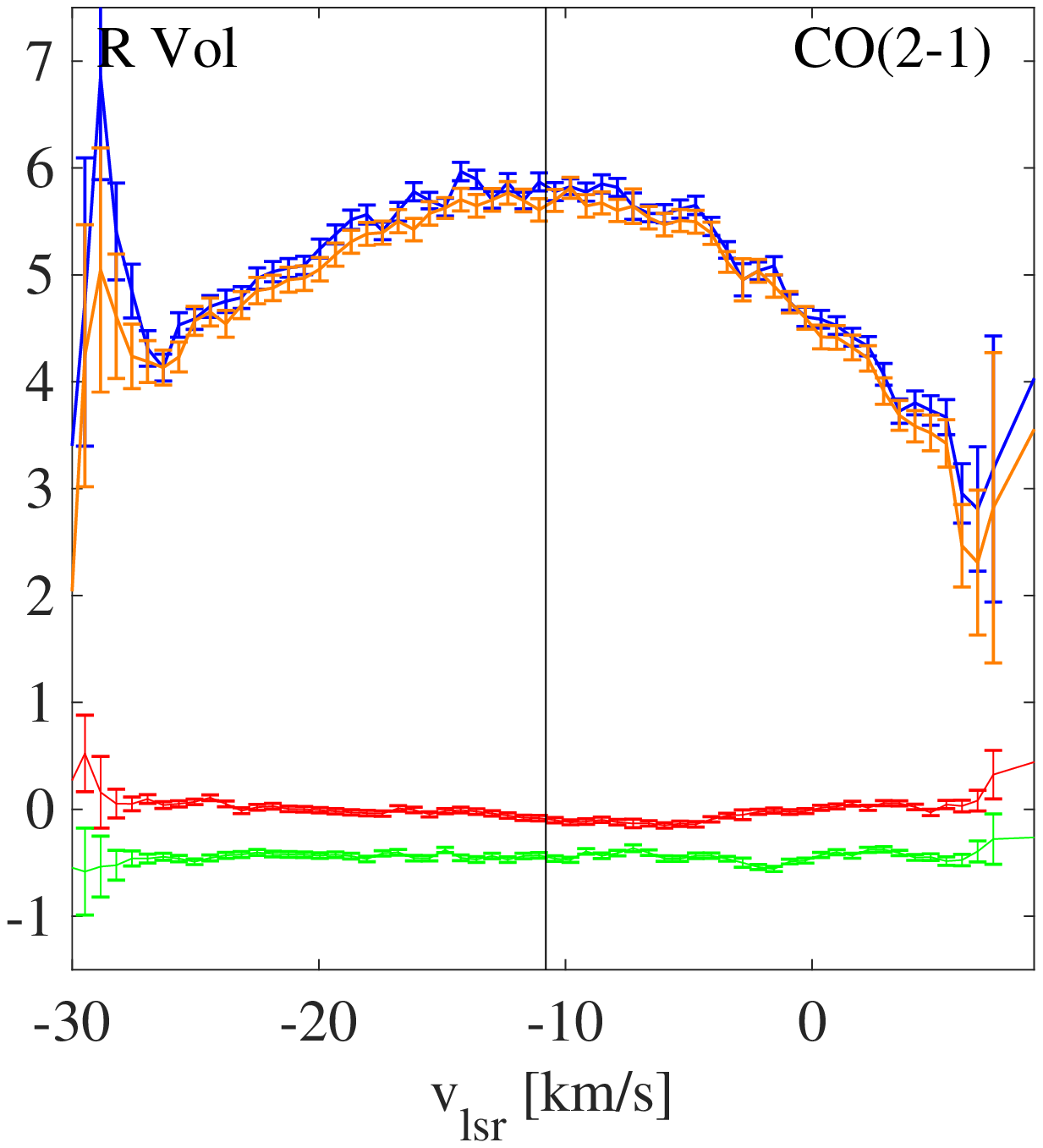}
\includegraphics[height=4.5cm]{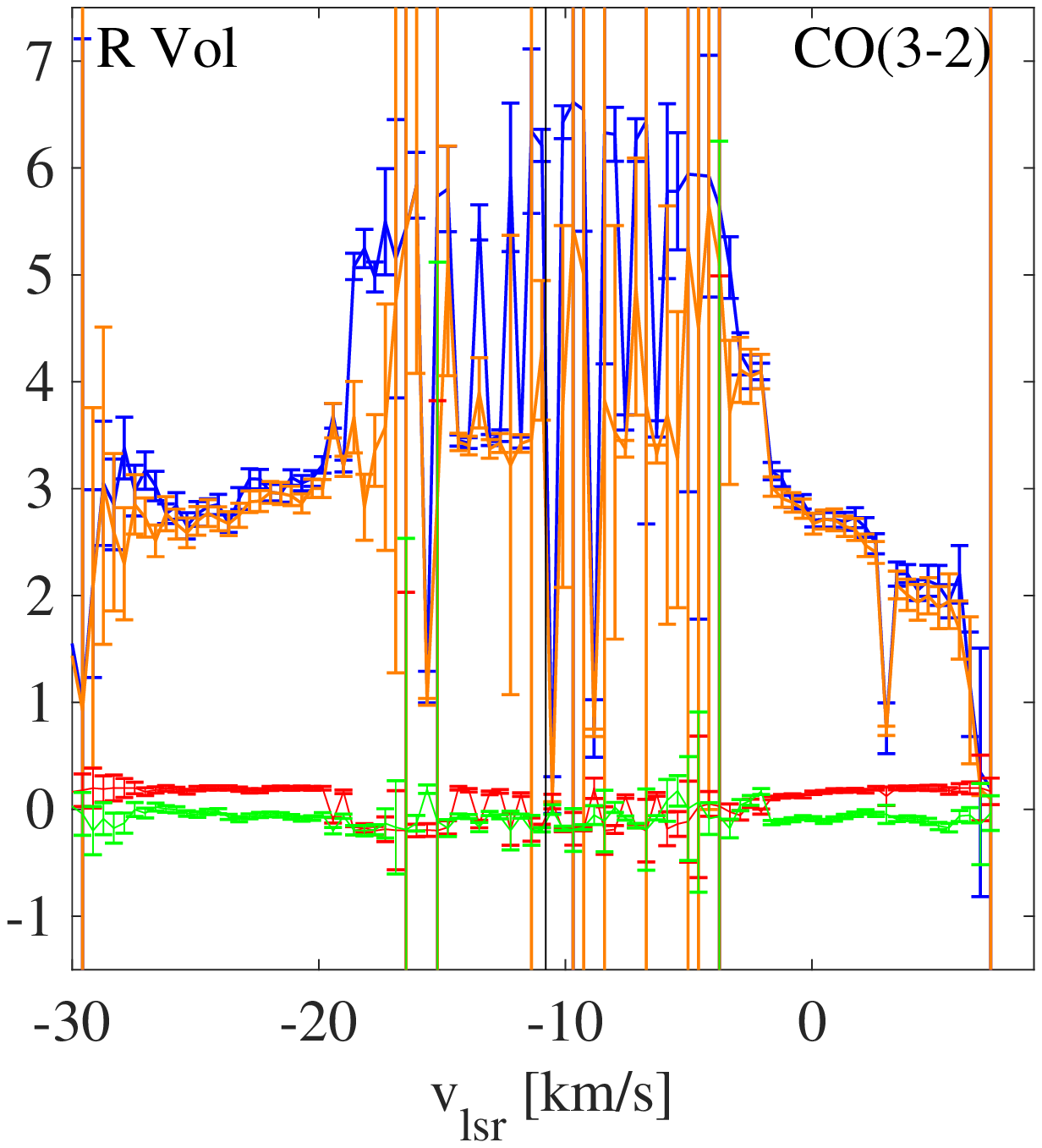}

\includegraphics[height=4.5cm]{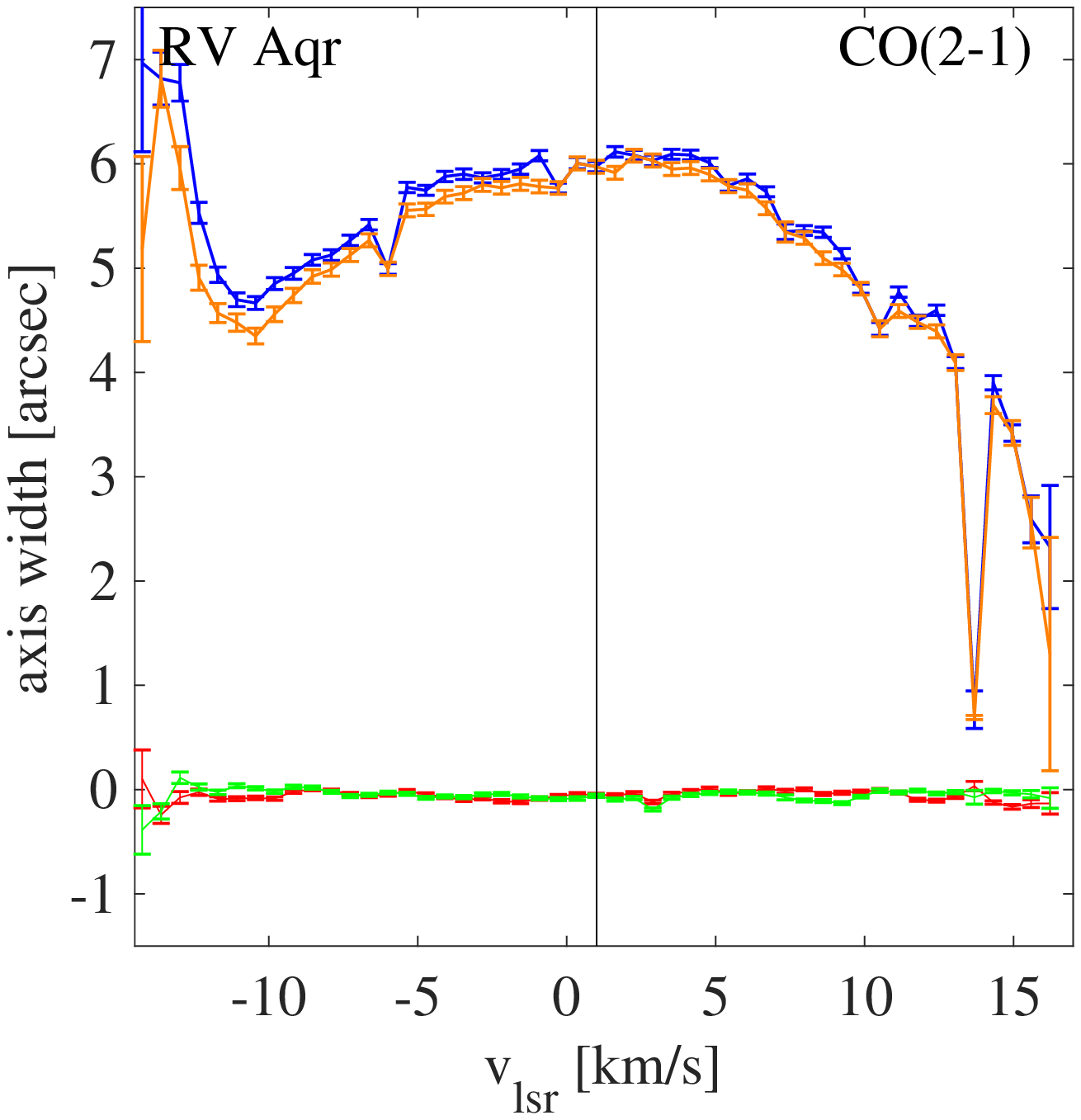}
\hspace{0.02cm}
\includegraphics[height=4.5cm]{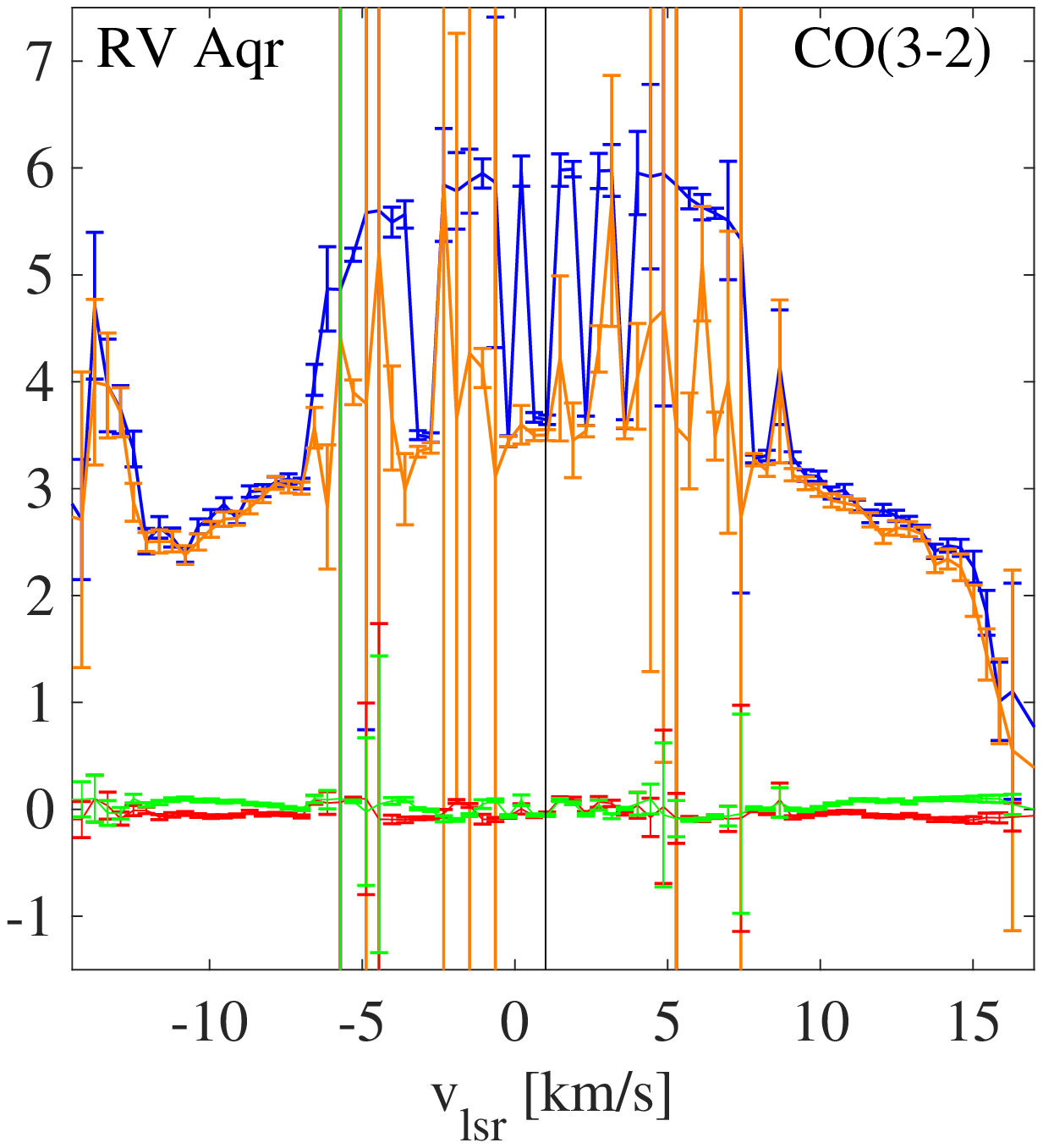}
\hspace{0.15cm}
\includegraphics[height=4.5cm]{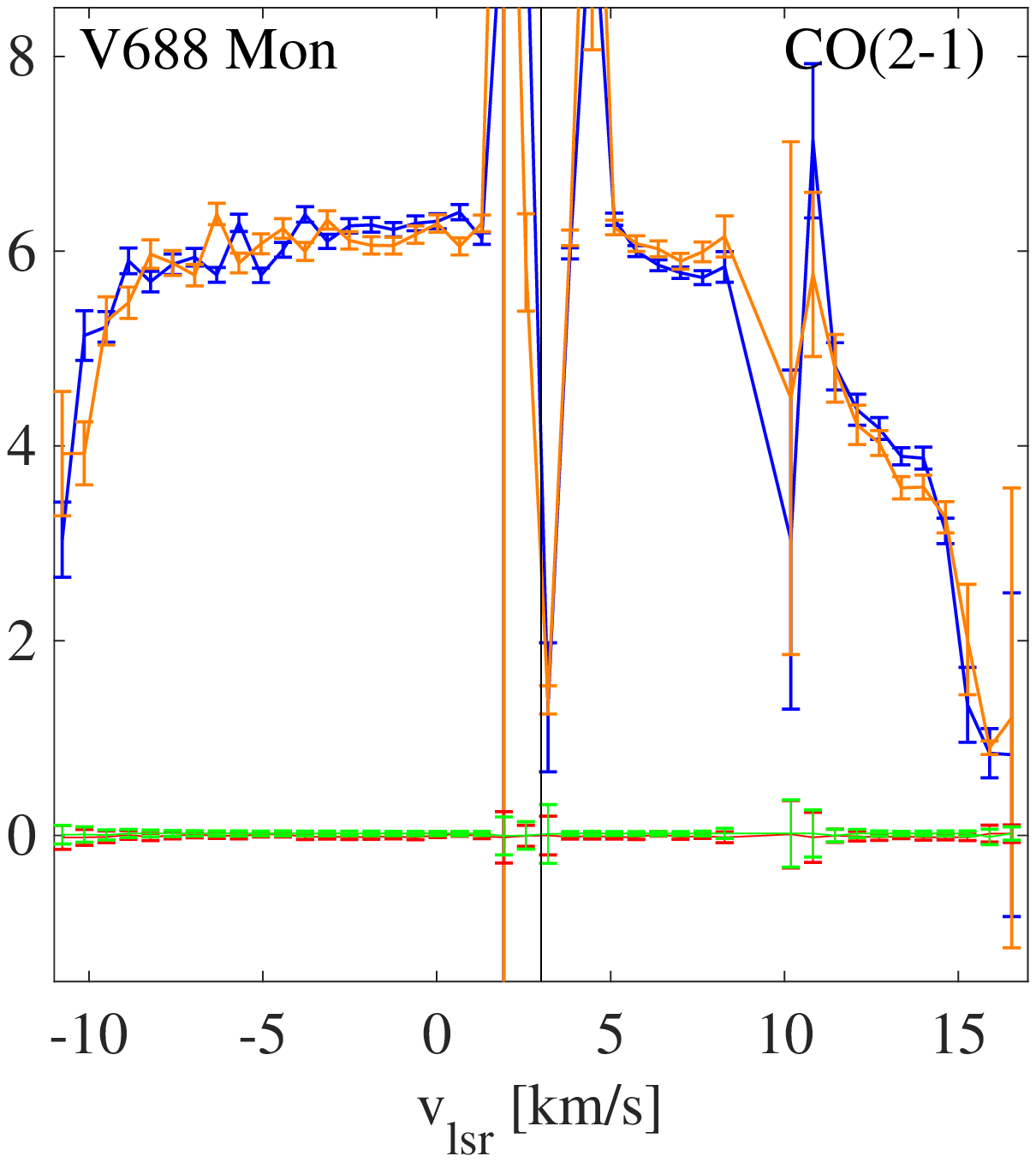}
\hspace{0.02cm}
\includegraphics[height=4.5cm]{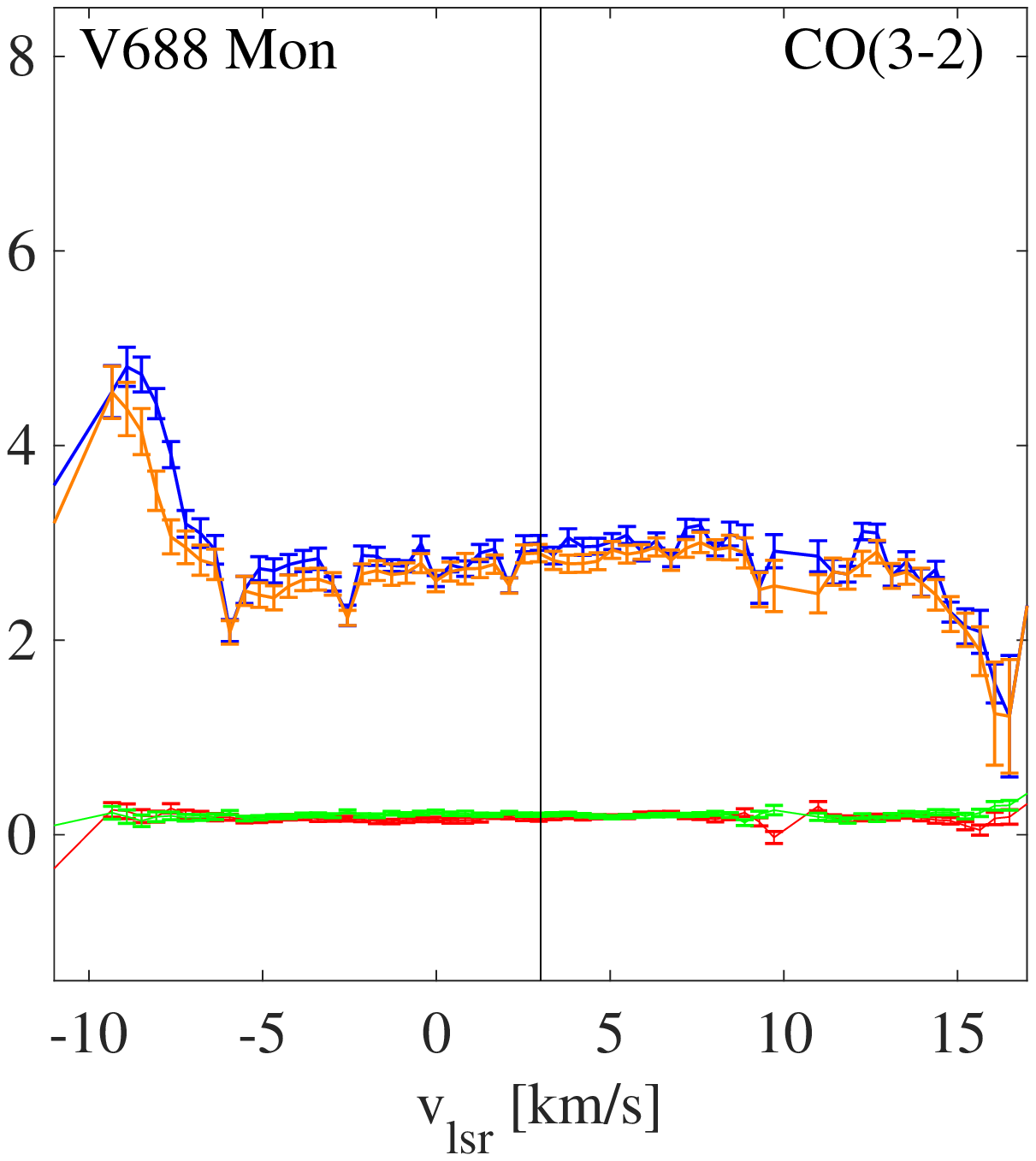}
\caption{Results from the visibility fitting to the data measured toward the C-type AGB stars of the sample discussed in this paper. The source name is given in the upper left corner and the transition is in the upper right corner of each plot. The upper blue and orange lines show the major and minor axis of the best-fit Gaussian in each channel, respectively. The lower red and green lines show the RA and Dec offset relative to the center position, respectively.}
\label{uvC_SRM}
\end{figure*}


\begin{figure*}[t]
\includegraphics[height=4.5cm]{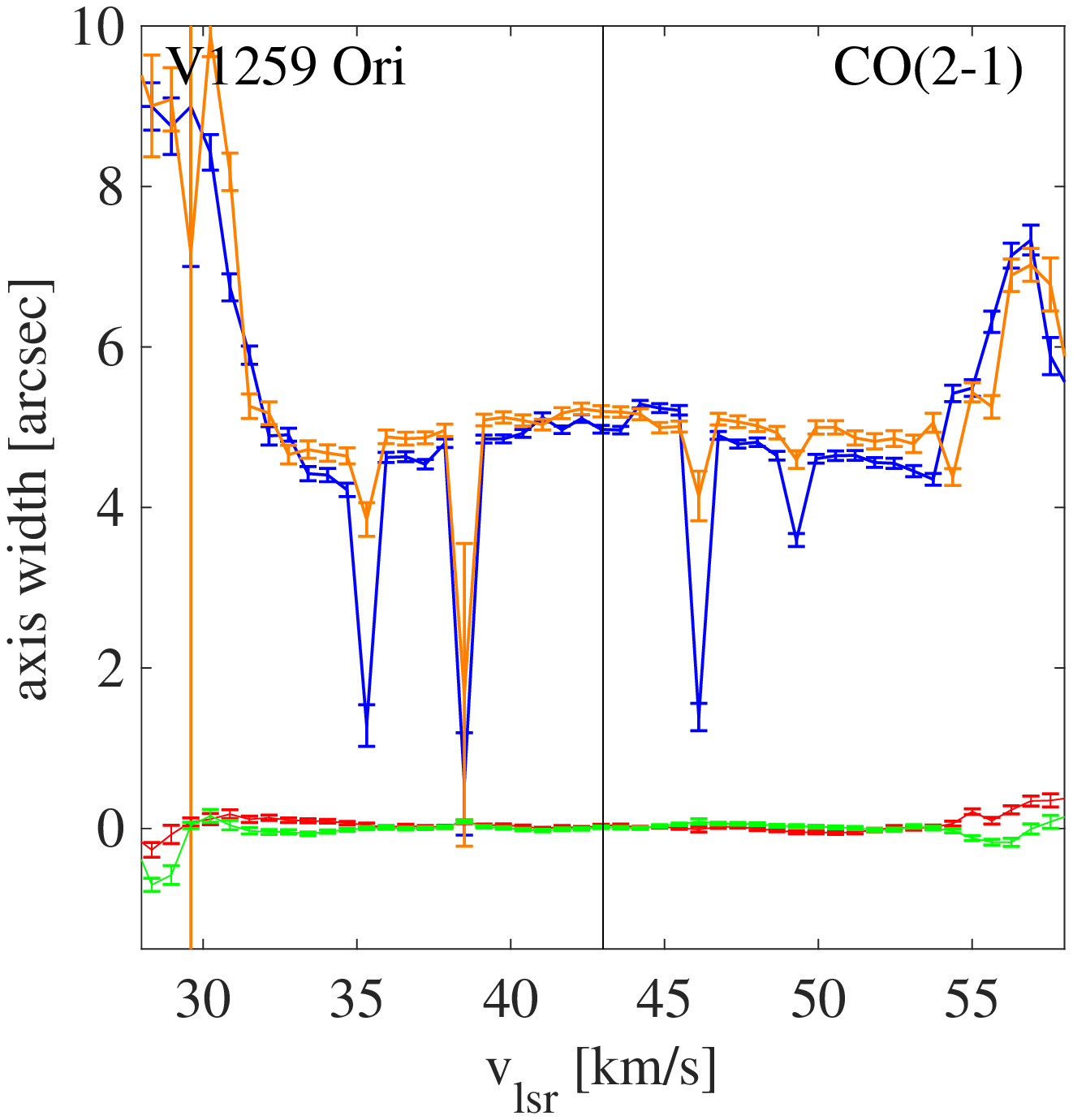}
\hspace{0.02cm}
\includegraphics[height=4.5cm]{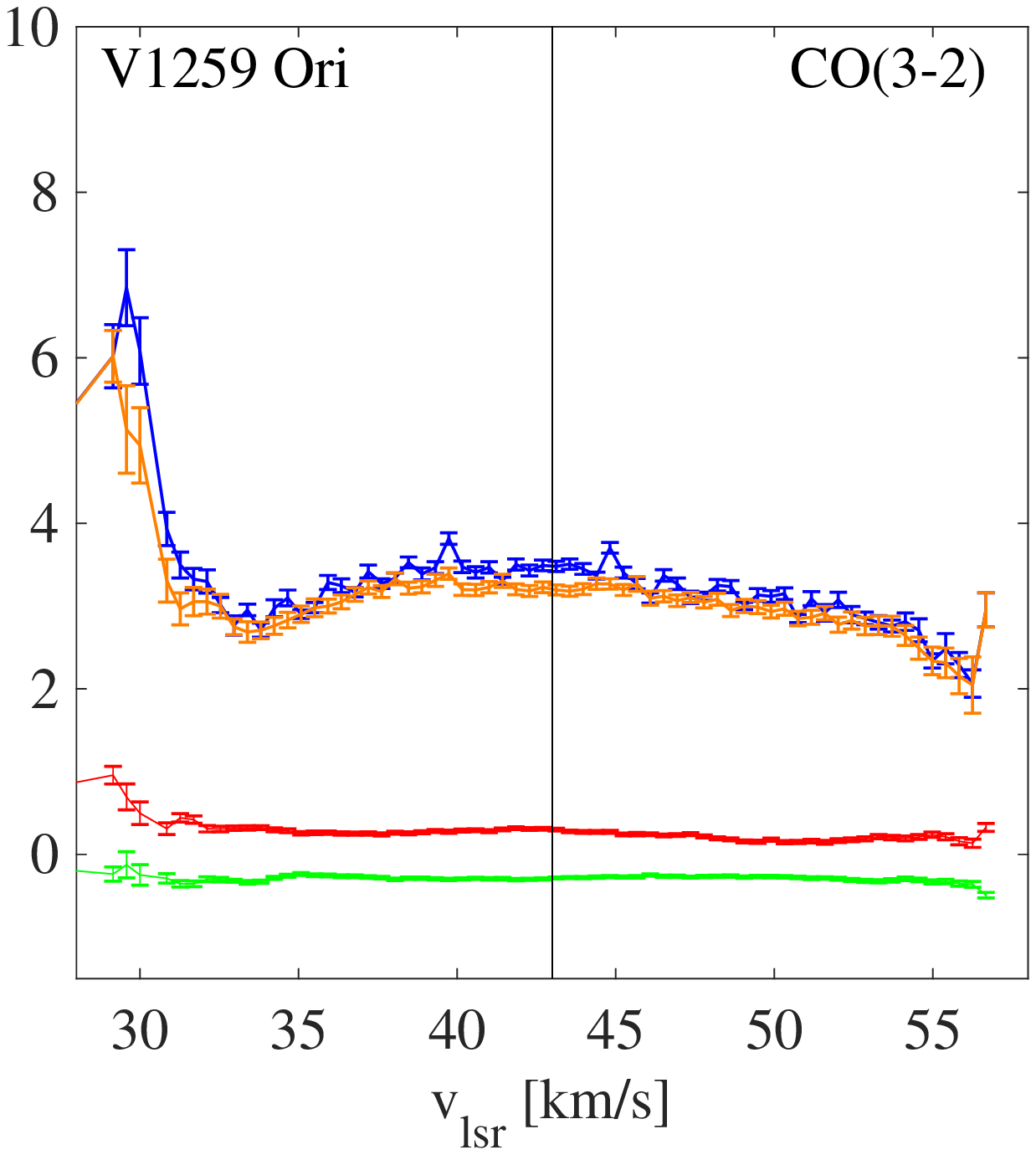}
\caption{Results from the visibility fitting to the data measured toward the C-type AGB stars of the sample discussed in this paper. The source name is given in the upper left corner and the transition is in the upper right corner of each plot. The upper blue and orange lines show the major and minor axis of the best-fit Gaussian in each channel, respectively. The lower red and green lines show the RA and Dec offset relative to the center position, respectively.}
\label{uvC_M}
\end{figure*}


\end{appendix}

\bibliographystyle{aa}
\bibliography{ds}

\end{document}